\documentclass[11pt]{article}
\usepackage[margin=1in]{geometry}
\usepackage[T1]{fontenc}
\usepackage{lmodern}
\usepackage{amsmath,amssymb,amsthm,mathtools}
\usepackage{microtype,booktabs,array}
\usepackage[numbers,sort&compress]{natbib}
\usepackage[colorlinks=true,allcolors=blue,hypertexnames=false]{hyperref}
\hypersetup{pdftitle={Sample complexity of learning fermionic and bosonic Gaussian states with single-copy and entangled measurements},pdfauthor={Ron Rubin}}
\newcommand{\doi}[1]{doi:\,\href{https://doi.org/#1}{\nolinkurl{#1}}}

\newtheorem{theorem}{Theorem}[section]
\numberwithin{equation}{section}
\newtheorem{lemma}[theorem]{Lemma}
\newtheorem{corollary}[theorem]{Corollary}
\newtheorem{proposition}[theorem]{Proposition}
\theoremstyle{remark}
\newtheorem{remark}[theorem]{Remark}
\theoremstyle{definition}
\newtheorem{definition}[theorem]{Definition}
\newtheorem{question}[theorem]{Question}

\newcommand{\Tr}{\operatorname{Tr}}
\newcommand{\tr}{\operatorname{tr}}
\newcommand{\ad}{\operatorname{ad}}
\newcommand{\E}{\mathbb E}
\newcommand{\Prob}{\mathbb P}
\newcommand{\R}{\mathbb R}
\newcommand{\C}{\mathbb C}
\usepackage{graphicx}
\usepackage{multicol}
\newcommand{\op}{\mathrm{op}}
\newcommand{\eps}{\epsilon}
\newcommand{\Fock}{\mathcal F}

\title{Sample complexity of learning fermionic and bosonic Gaussian states with single-copy and entangled measurements}
\author{Ron Rubin\\Rubin Anders Scientific, Inc.}
\date{November 2026}

\begin{document}
\maketitle

\begin{abstract}
With single-copy measurements, every mixed fermionic Gaussian state on $m$ modes, pure modes included, is learned to trace distance $\eps$ from $\Theta(m^3/\eps^2)$ copies, with no logarithmic factor, against the known $\Theta(m^2/\eps^2)$ for collective measurements. We prove the upper bound, resolving Conjecture~10.1 of arXiv:2609.23189 (lower bound for $\eps\le1/3600$), and fill most open entries of the sample-complexity table of fermionic and bosonic Gaussian states, for single-copy, $k$-copy and collective measurements. With blocks of $k$ copies, the covariant highest-weight measurement of arXiv:2611.07271, read out through the Weyl-shift estimator, is exactly unbiased for every state; it gives $\Theta(\max\{m^3/k,m^2\}/\eps^2)$ in Frobenius norm on all states and in trace distance on every class at a fixed distance from the pure boundary. The gain $1/k$ ($k\le m$) away from the pure boundary, against $1/\sqrt k$ on cold number-conserving states, is therefore attained. Learning a quadratic fermionic Hamiltonian in Frobenius norm from its Gibbs state at bounded $\beta\|h\|$ has the same rate (Corollary~\ref{cor:D-ham}). For bosonic Gaussian states with bounded covariance, any displacement and squeezing, and occupations $n_1\le\dots\le n_m$, $O((m^2+r^3+\sum_{i,j>r}1/(n_i+n_j))/\eps^2)$ single-copy measurements suffice, $r$ the largest index with $r\,n_r\le2$; this is the minimax rate on dyadic-interval classes, with no spectral gap or cold-mode count assumed, and holds with no covariance bound for displaced passive states. With no bound on energy or squeezing, non-Gaussian measurements learn centred one-mode states, mixed included, from $O(\eps^{-2})$ and centred pure $m$-mode states from $O(m^2/\eps^2)$ single copies, free of the $\log\log E$ term of Gaussian measurements, and $m$-mode states from $O(m^3/\eps^2)$ copies, entangled only in a first round on blocks of $4m+6$.
\end{abstract}

\section{Introduction}\label{sec:intro}

How many copies of an unknown Gaussian state are needed to learn it? Gaussian states are the ground and thermal states of free particles, the coherent and squeezed states of optics, and the states of mean-field approximations. The error is measured in trace distance $\eps$, the largest amount by which the true state and the learned one can disagree on the probability of any outcome of any measurement; no experiment can tell two states at trace distance $\eps$ apart with bias better than $\eps$.

The number of copies depends on how they may be measured. The central rate of this paper, $m^3/\eps^2$ copies for a single-copy experiment on a mixed fermionic state of $m$ modes ($m$ harmonic oscillators, or $m$ fermionic orbitals), has three factors: the $m^2$ unknowns, the second moments that fix a Gaussian state; a further factor $m$, because the quadratic observables do not commute and cannot all be measured on one copy; and $1/\eps^2$, because measurement outcomes are random and a finite number of them fixes an average only to about $1/\sqrt N$. The accuracy $\eps$ is set by what the learned state is for, and the number of copies is its price: $N$ copies buy an accuracy of order $\sqrt{m^3/N}$ when they are measured one at a time and of order $\sqrt{m^2/N}$ when they are measured together, so an accuracy $\eps$ costs of order $m^3/\eps^2$ or $m^2/\eps^2$ copies. Between these two lies a table whose rows are the states (fermionic or bosonic, cold or warm, pure or mixed) and whose columns are the measurement resources (one copy at a time, blocks of $k$ copies, or all copies at once); Table~\ref{tab:main} is that table as it now stands, every entry proved here or cited, the open cells marked. Against the table after \cite{CFGLMWW26,CMMP26,Rub26,Rub26b} (for background on the complexity of learning quantum states see \cite{AA24}), this paper adds the single-copy upper bound for all mixed fermionic Gaussian states (row F1), the single-scale bound (F5), the $k$-copy rate of the fixed-gap fermionic classes (F2, F3), the bosonic single-copy bounds for arbitrary temperature profiles and for displaced passive states (B3, B5), and energy-independent protocols with non-Gaussian measurements for mixed states, for pure states and for bounded thermal occupation (B6, B7).

An $m$-mode Gaussian state has about $2m^2$ real parameters, and a measurement on all copies at once learns them at the parametric rate, $\Theta(m^2/\eps^2)$ copies for trace distance $\eps$, for fermions and bosons alike \cite{CFGLMWW26} (for fermions the first protocol optimal in both $m$ and $\eps$ is \cite{WW25}; earlier trace-distance guarantees for Gaussian tomography are in \cite{BMEL25,MMBEGLLO25,BMMTL25}). This is the floor of every column. For states of bounded energy the ceiling of the single-copy column is $m^3/\eps^2$ (for squeezed bosonic states the general single-copy bound of \cite{BMEM25} carries an additive term doubly logarithmic in the energy). The extra factor $m$ is the ratio of a quantum Fisher information of order $m^2$ to the Fisher information of order $m$ that one copy can yield (cf.\ the Gill--Massar bound \cite{GM00}), because the quadratic observables cannot be measured jointly \cite[Lemma~3.1]{Rub26}. It is the Gaussian counterpart of the separation between $d^3/\eps^2$ single-copy \cite{HHJWY17,KRT17,GKKT20,CHLLS22,LN25} and $d^2/\eps^2$ collective \cite{HHJWY17,OW16} copies for unstructured $d$-dimensional states.

The trace distance between two nearby Gaussian states weights each entry of the covariance by the occupations of the two modes it couples, and a single-copy measurement learns cheaply when its noise is distributed the same way, \emph{matched to the metric}. Heterodyne detection is matched, up to a constant, on every pair of bosonic modes holding at least of order one photon between them, so once every mode is that warm (a classical field), bosons with bounded covariance are learned on single copies at the collective rate (Theorem~\ref{thm:Bos}). A fermionic mode holds at most one particle and has no such regime: a single-copy measurement can still be matched to the metric, which removes the factor $1/c$ that matchgate shadows pay at distance $c$ from the pure boundary, but the factor $m$ is then paid at every temperature. Blocks of $k$ copies measured jointly win back a factor $\min\{k,m\}$ of it at any fixed distance from the pure boundary (Theorem~\ref{thm:K}), but only $\min\{\sqrt k,m\}$ on the cold classes, which embed the tomography of a qudit \cite{CMMP26,Rub26b}. Outside this picture lies unbounded squeezing, the one regime in which no protocol with Gaussian measurements learns from a number of copies independent of the energy \cite{CGYZ26}; there a covariant non-Gaussian measurement finds the squeezing, on one copy for a centred mode, on single copies for a centred pure state of any number of modes, and on blocks of $4m+6$ copies for $m$ modes, and heterodyne detection on single copies does the rest (Theorems~\ref{thm:NG1}, \ref{thm:NG2} and~\ref{thm:NG-pure}).

For fermions the single-copy minimax rate of the class of all mixed Gaussian states is $\Theta(m^3/\eps^2)$ for $\eps\le1/3600$; the upper bound holds for every state, pure modes included, with no logarithmic factor (Theorem~\ref{thm:B}), while a promise that the state is globally pure gives a different rate (row F6). This was Conjecture~10.1 of \cite{Rub26}, where the lower bound was proved. The upper bound was known only as $O((m^4/\eps^2)\log(m^2/\delta))$, at failure probability $\delta$, in general \cite[Thm.~5]{BMEL25}, or as $O(m^3/(c\,\eps^2))$, by matchgate shadows \cite{ZRM21,WHLB23}, on states whose covariance stays a distance $c$ from the pure boundary \cite[Thm.~7.6]{Rub26} (for pure states, single-copy learners go back to \cite{AG23}; see also \cite{BMEL25,MH25}). The proof uses the \emph{pair-rotation measurement}: a random pairing of the modes, a two-mode rotation by $\pm\pi/4$ on every pair, and a readout of occupations. Its noise on a pair with occupations $p_k,p_l$ is proportional to $p_k+p_l$, and so is the sensitivity of the trace distance to a rotation of that pair (a fermionic Mandelstam--Tamm bound, Lemma~\ref{lem:MT}); the two match scale by scale. The absence of a warm regime was proved in \cite{Rub26b} on quadratic Gibbs balls; Theorem~\ref{thm:B} makes the upper bound $O(m^3/\eps^2)$ uniform over all states, so that every class containing the hard family of \cite{Rub26}, at any fixed distance from the pure boundary, has single-copy rate $\Theta(m^3/\eps^2)$ on the accuracies at which it contains that family.

\begin{figure}[t]
\centering
\includegraphics[width=\textwidth]{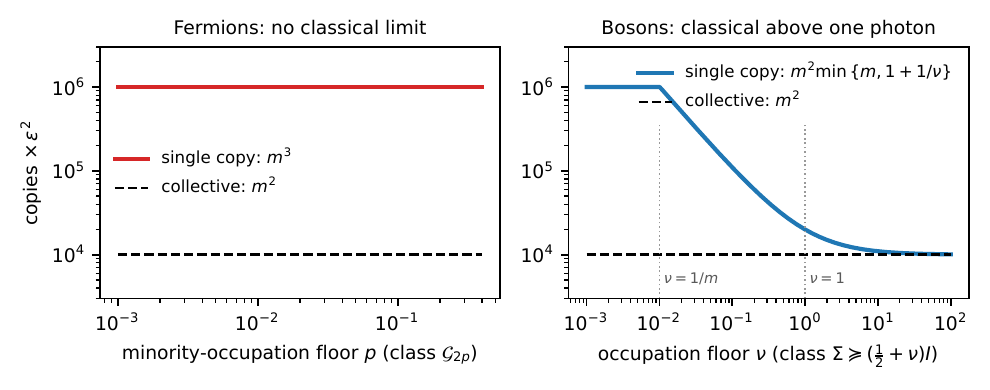}
\caption{Single-copy sample complexity for $m=100$ modes, in units of $1/\eps^2$ with all constants set to one (rows F2 and B2 of Table~\ref{tab:main}); each curve is the minimax rate of a class with a floor. Left, fermions, floor $p$ on the minority occupations, the class $\mathcal G_{2p}$: the rate is $m^3$ for every $p<\tfrac12$, on the accuracies $\eps\le\eps_0(2p)$, $\eps_0(c):=\min\{1/3600,\mathrm{artanh}(1-c)/(120\sqrt3)\}$, which shrink to zero as $p\to\tfrac12$, by the hard instances of \cite{Rub26} near the maximally mixed state and Theorem~\ref{thm:B}; a promise of purity (row F6) gives $\tilde\Theta(m^2)$ and is not on the curve. Right, bosons, the class $\Sigma\succeq(\tfrac12+\nu)I$: the rate $m^2\min\{m,1+1/\nu\}$ \cite{CMMP26} falls from $m^3$ for $\nu\lesssim1/m$ to the collective rate for $\nu\gtrsim1$. Dashed: the collective rate $m^2$ \cite{CFGLMWW26}. The curves show scaling, not copy counts.}
\label{fig:regimes}
\end{figure}

For bosons, heterodyne detection learns the warm class $\Sigma\succeq(\tfrac12+\nu)I$, a promise on the quadrature covariance and not only on the Williamson occupations (squeezing makes a quadrature variance arbitrarily small at fixed occupations), from $\Theta(m^2\min\{m,1+1/\nu\}/\eps^2)$ single copies \cite{CMMP26}, the collective order when $\nu$ is bounded below by a positive constant (heterodyne-based learning of Gaussian states is also analysed in \cite{FRS26}). Theorem~\ref{thm:Bos} treats a state whose modes are at many temperatures: for quadrature covariance of operator norm at most $K$, arbitrary displacement and squeezing, occupations $n_1\le\dots\le n_m$ and $r$ the largest index with $r\,n_r\le2$ ($r=0$ if there is none),
\[
O_K\Big(\frac{m^2+r^3+\sum_{i,j>r}1/(n_i+n_j)}{\eps^2}\Big)
\]
copies suffice for every such state, and on the classes whose occupations are promised to lie in dyadic intervals, under the room and accuracy conditions of Theorem~\ref{thm:Bos}(ii), this is the minimax rate, with constants that do not depend on the number of bands. Each warm band pays as if it were alone, the cold group pays $r^3$ as a near-vacuum class does, and neither a gap between cold and warm modes nor the value of $r$ is assumed. For displaced passive states the bound on the covariance can be dropped (Theorem~\ref{thm:Bos-energy}); for centred states it can be replaced by a bound $\bar n$ on the Williamson occupations, with no bound on energy or squeezing, at the cost of one round of a non-Gaussian single-copy measurement (Theorem~\ref{thm:NG-thermal}; pairs of copies for displaced states). Figure~\ref{fig:regimes} draws the two single-copy curves.

\begin{figure}[t]
\centering
\includegraphics[width=0.62\textwidth]{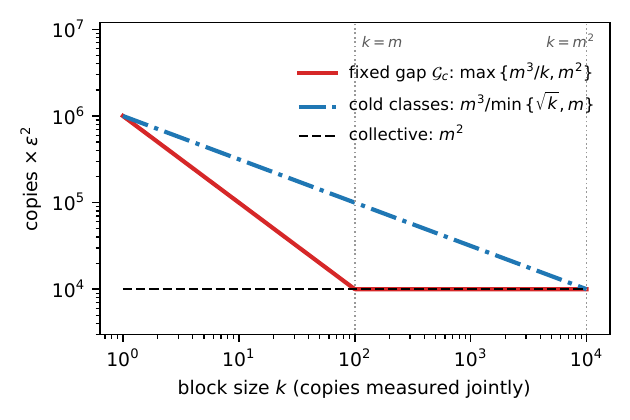}
\caption{Sample complexity against the block size $k$ of the entangled measurement, with the units and $m$ of Figure~\ref{fig:regimes}. Solid: the fixed-gap fermionic class $\mathcal G_c$ (row F2), $\Theta_c(\max\{m^3/k,m^2\})$ (Theorem~\ref{thm:K}; lower bound \cite{Rub26b}). Dash-dotted: the cold classes (rows F4, B1 and B4), $\Theta(m^3/\min\{\sqrt k,m\})$ \cite{Rub26b,CMMP26}. Dashed: the collective floor $m^2$ \cite{CFGLMWW26}. For $1<k<m^2$ the same block size buys a factor $\min\{\sqrt k,m/\sqrt k\}$ more on the fixed-gap class, largest, $\sqrt m$, at $k=m$.}
\label{fig:kcopy}
\end{figure}

With entangled measurements on blocks of $k$ copies the picture splits along the same line. On the cold number-conserving fermionic and all cold bosonic Gaussian states, which embed the tomography of an $m$-dimensional qudit, the rate is $m^3/(\eps^2\min\{\sqrt k,m\})$ \cite{CMMP26,Rub26b}, the gain $\sqrt k$ of unstructured states \cite{CLL24,PSW26,PSTW26,NZ26,KLMR26} (with pairing terms near the vacuum it is open). At the fermionic maximally mixed state, by contrast, one copy captures at most a fraction $1/(2m-1)$ of the quantum Fisher information, because the $m(2m-1)$ quadratic observables cannot be measured jointly (cf.\ the joint-measurability analyses of \cite{MMO23,MCO25}), and a block of $k\le m$ copies captures Fisher information of order $mk^2$, against $mk$ for $k$ separate copies \cite{Rub26b}. The quantum Fisher information of the block is additive, so this is not Heisenberg scaling \cite{GLM06,GLM11} (which does occur in learning fermionic linear-optical unitaries \cite{CZ26}) but the recovery, linear in $k$, of what single copies lose to incompatibility. Theorem~\ref{thm:K} attains it on every class a fixed distance $c$ from the pure boundary. The covariant highest-weight measurement of \cite{Rub26b}, read out as the rotated normal form of the highest weight shifted by twice the Weyl vector (the \emph{Weyl-shift estimator}), is an exactly unbiased estimator of the covariance matrix of every state, Gaussian or not; so $\Theta(\max\{m^3/k,m^2\}/\eps^2)$ copies learn the covariance in Frobenius norm on all states, and the same count, divided by $c$, learns the state in trace distance on the class with gap $c$. What happens when the gap closes with $m$ is open (Section~\ref{sec:open}). Figure~\ref{fig:kcopy} draws the two rates against the block size.

Several single-copy estimators of this paper are classical shadows in the sense of \cite{HKP20}: a random unitary from a group, a readout in a fixed basis, and an unbiased linear inversion. For fermions the group is that of Gaussian unitaries, which gives matchgate shadows \cite{ZRM21,WHLB23,OGo22,HCT25} and their number-conserving forms \cite{Low22,KWTY26}; for bosons, Gaussian unitaries followed by homodyne or heterodyne detection give continuous-variable shadows \cite{BDLR24,GAGTG24}. Two features differ. Shadows control the variance of each observable separately, while near the pure boundary the trace distance weights each covariance entry by its two occupations; the pair-rotation measurement shapes its noise to that weight, which removes the factor $1/c$ carried by the trace-distance guarantee of matchgate shadows \cite[Thm.~7.6]{Rub26}. And the estimator of Theorem~\ref{thm:K} is a covariant shadow on blocks of copies, the $\mathrm{Spin}(2m)$ counterpart of the staircase estimator of Pelecanos, Spilecki and Wright for Keyl's measurement \cite[Sec.~3.3]{PSW26}, whose stabiliser average is their debiased Keyl estimator (Remark~\ref{rem:K-minvar}).

\begin{table}[tp]
\centering\scriptsize
\newcommand{\new}{\leavevmode\llap{$\scriptscriptstyle\bigstar$\,}}
\newcommand{\nt}[1]{\textsuperscript{\itshape#1}}
\newcommand{\note}[2]{\par\noindent\hangindent=1.1em\hangafter=1\makebox[1.1em][l]{\nt{#1}}#2}
\setlength{\abovecaptionskip}{0pt}\setlength{\belowcaptionskip}{4pt}
\newcommand{\twoline}[2]{\begin{tabular}[t]{@{}l@{}}#1\\#2\end{tabular}}
\renewcommand{\arraystretch}{0.95}\setlength{\tabcolsep}{4pt}
\caption{Copies needed to learn an $m$-mode Gaussian state to trace distance $\eps$ with probability $2/3$, minimax over the class of each row, with upper bounds stated for every state holding state by state.}
\label{tab:main}
\begin{tabular*}{\textwidth}{@{\extracolsep{\fill}}l
  >{\raggedright\arraybackslash}p{0.162\textwidth}
  >{\raggedright\arraybackslash}p{0.240\textwidth}
  >{\raggedright\arraybackslash}p{0.300\textwidth}
  >{\raggedright\arraybackslash}p{0.165\textwidth}@{}}
\toprule
 & state class & one copy at a time & blocks of $k$ copies & all copies at once\\
\midrule
\multicolumn{5}{@{}l}{\textit{Fermions}}\\[-1pt]
\cmidrule(r){1-2}\addlinespace[-1.5pt]
F1 & all mixed states
   & \twoline{\new{}$\Theta(m^3/\eps^2)$}{Thm.~\ref{thm:B}; \cite{Rub26}}
   & \twoline{$O(m^3/\eps^2)$, $\Omega\big(m^3/(\eps^2\min\{\sqrt k,m\})\big)$\,\nt{a}}{Thm.~\ref{thm:B}; \cite{Rub26b}}
   & \twoline{$\Theta(m^2/\eps^2)$}{\cite{WW25,CFGLMWW26}}\\\addlinespace[0pt]
F2 & $\|\Gamma\|_\op\le1-c$, $c$ fixed\nt{b}
   & \twoline{$\Theta(m^3/\eps^2)$}{Thm.~\ref{thm:B}; \cite{Rub26}}
   & \twoline{\new{}$\Theta_c(\max\{m^3/k,m^2\}/\eps^2)$}{Thm.~\ref{thm:K}; \cite{Rub26b}}
   & \twoline{$\Theta(m^2/\eps^2)$}{\cite{CFGLMWW26}; \cite[Thm.~3.11]{Rub26b}}\\\addlinespace[0pt]
F3 & Gibbs ball, radius $\beta_0>0$\nt{c}
   & \twoline{$\Theta(m^3/\eps^2)$}{\cite{Rub26b}}
   & \twoline{\new{}$\Theta_{\beta_0}(\max\{m^3/k,m^2\}/\eps^2)$}{Thm.~\ref{thm:K}; \cite{Rub26b}}
   & \twoline{$\Theta(m^2/\eps^2)$}{\cite{CFGLMWW26}; \cite[Thm.~3.11]{Rub26b}}\\\addlinespace[0pt]
F4 & number-conserving, fugacity trace $\le C$
   & \twoline{$\Theta(m^3/\eps^2)$}{\cite{Rub26b}}
   & \twoline{$\Theta\big(m^3/(\eps^2\min\{\sqrt k,m\})\big)$}{\cite{Rub26b}}
   & \twoline{$\Theta(m^2/\eps^2)$}{\cite{CFGLMWW26}; \cite[Thm.~5.9]{Rub26b}}\\\addlinespace[0pt]
F5 & single scale $\mathcal U_m(x_c,\kappa)$\nt{d}
   & \twoline{\new{}$O(m^3/\eps^2)$\,\nt{e,f}}{Thm.~\ref{thm:A}; \cite{Rub26}, \cite{Rub26b}}
   & \twoline{\new{}$O\big(\kappa\max\{m^3/k,m^2\}/(x_c\eps^2)\big)$\,\nt{e,f}}{Thm.~\ref{thm:K}; \cite{Rub26b}}
   & \twoline{$O(m^2/\eps^2)$\,\nt{e}}{\cite{CFGLMWW26}; \cite[Thm.~3.11]{Rub26b}}\\\addlinespace[0pt]
F6 & pure states\nt{g}
   & \twoline{$\tilde\Theta(m^2/\eps^2)$}{\cite{ZLK24,CFGLMWW26}}
   & \twoline{$\tilde\Theta(m^2/\eps^2)$}{\cite{ZLK24,CFGLMWW26}}
   & \twoline{$\Theta(m^2/\eps^2)$}{\cite{CFGLMWW26}}\\\addlinespace[0pt]
\addlinespace[3pt]
\multicolumn{5}{@{}l}{\textit{Bosons}}\\[-1pt]
\cmidrule(r){1-2}\addlinespace[-1.5pt]
B1 & cold, $\Sigma\preceq(\frac12+\frac{c'}m)I$\nt{h,q}
   & \twoline{$\Theta(m^3/\eps^2)$}{\cite{CMMP26,Rub26b}}
   & \twoline{$\Theta\big(m^3/(\eps^2\min\{\sqrt k,m\})\big)$}{\cite{CMMP26,Rub26b}}
   & \twoline{$\Theta(m^2/\eps^2)$}{\cite{CFGLMWW26}}\\\addlinespace[0pt]
B2 & warm, $\Sigma\succeq(\frac12+\nu)I$
   & \twoline{$\Theta(m^2\min\{m,1+\nu^{-1}\}/\eps^2)$}{\cite{CMMP26}}
   & \twoline{$\Theta(m^2/\eps^2)$ for $\nu\gtrsim1$\,\nt{i}}{\cite{CMMP26}}
   & \twoline{$\Theta(m^2/\eps^2)$}{\cite{CFGLMWW26}}\\\addlinespace[0pt]
B3 & any profile, $\|\Sigma\|_\op\le K$\nt{j,r}
   & \twoline{\new{}$O_K(\mathcal R(\boldsymbol n)/\eps^2)$, $\Theta_K$ on bands\,\nt{k}}{Thm.~\ref{thm:Bos}, Cor.~\ref{cor:Bos-twoscale}}
   & \twoline{\new{}two bands: $\Theta(F_k/\eps^2)$\,\nt{s}}{Prop.~\ref{prop:NG-coldhot-k}}
   & \twoline{$\Theta(m^2/\eps^2)$}{\cite{CFGLMWW26}}\\\addlinespace[0pt]
B4 & passive, total mean photon number $\le1$
   & \twoline{$\Theta(m^3/\eps^2)$}{\cite{Rub26,CMMP26}}
   & \twoline{$\Theta\big(m^3/(\eps^2\min\{\sqrt k,m\})\big)$}{\cite{Rub26b}}
   & \twoline{$\Theta(m^2/\eps^2)$}{\cite{CFGLMWW26}; \cite[Cor.~9.5]{Rub26}}\\\addlinespace[0pt]
B5 & displaced passive, no energy bound\nt{m}
   & \twoline{\new{}$O(\mathcal R(\boldsymbol n)/\eps^2)$}{Thm.~\ref{thm:Bos-energy}; Rem.~\ref{rem:Bos-energy-scope}}
   & \twoline{---\,\nt{l}}{\phantom{[0]}}
   & \twoline{$\Theta(m^2/\eps^2)$}{\cite{CFGLMWW26}}\\\addlinespace[0pt]
B6 & all states, no energy bound
   & \begin{tabular}[t]{@{}l@{}}\new{}$m=1$, centred: $\Theta(\eps^{-2})$\,\nt{n}\\\new{}$n_j\le\bar n$, centred: $\Theta_{\bar n}(m^3/\eps^2)$\,\nt{t}\\Thm.~\ref{thm:NG1}; Thm.~\ref{thm:NG-thermal}\end{tabular}
   & \twoline{\new{}$O(m^3/\eps^2)$ with $k=4m+6$\,\nt{o}}{Thm.~\ref{thm:NG2}}
   & \twoline{$\Theta(m^2/\eps^2)$}{\cite{CFGLMWW26}}\\\addlinespace[0pt]
B7 & pure, no energy bound\nt{p}
   & \twoline{\new{}centred: $\Theta(m^2/\eps^2)$}{Thm.~\ref{thm:NG-pure}; \cite[Thm.~5.2]{CFGLMWW26}}
   & \twoline{\new{}$\Theta(m^2/\eps^2)$, $k\ge2$}{Thm.~\ref{thm:NG-pure}; \cite[Thm.~5.2]{CFGLMWW26}}
   & \twoline{$\Theta(m^2/\eps^2)$}{\cite{CGYZ26,CFGLMWW26}}\\\addlinespace[0pt]
\bottomrule
\end{tabular*}

\vspace{3pt}
\begin{minipage}{\textwidth}\scriptsize\setlength{\parskip}{0pt}
\setlength{\columnsep}{12pt}\begin{multicols}{2}\raggedright\sloppy
\noindent Classes, measurement models and accuracy ranges are stated after the table. $\scriptscriptstyle\bigstar$\,first proved in this paper. In each cell, line 1 is the rate and line 2 its source: upper bound; lower bound (one source: both). ---\,: open. A letter on a class applies to its row. $\Theta,O,\Omega$ hide constants depending only on the row's parameters ($c,c',C,K,\kappa,\beta_0,\bar n$); bounds hold for $\eps$ below (where the source requires it, $m$ above) such a threshold; single-copy lower bounds of \cite{Rub26}: $\eps\le1/3600$.
\note{a}{$k$-dependence $1/k$ on F2, $1/\sqrt k$ on F4 (Thm.~\ref{thm:K}(v)). Frobenius-norm covariance, all states: $\scriptstyle\bigstar$\,$\Theta(\max\{m^3/k,m^2\}/\eps^2)$ (Thm.~\ref{thm:K}(iii); lower bound for $\eps\le1/3600$).}
\note{b}{Lower bounds (all columns) for $\eps\le\eps_0(c):=\min\{\frac1{3600},\frac{\operatorname{artanh}(1-c)}{120\sqrt3}\}$; upper constants uniform in $c$ (one copy), $34/c$ (blocks).}
\note{c}{The ball lies in $\mathcal G_c$, $c=1-\tanh\beta_0$; lower bounds for $\eps\le\min\{1/3600,\beta_0/(120\sqrt3)\}$.}
\note{d}{Thm.~\ref{thm:A}: $\kappa\ge1$, $x_c\in[1/m,1]$, $m\ge4$, $\eps\le\eps_A(\kappa)$, constant independent of $x_c$.}
\note{e}{Lower bound at $x_c=1$ only, $\kappa>1$ ($\kappa=x_c=1$ is one state): $\mathcal U_m(1,\kappa)=\mathcal G_{2/(\kappa+1)}$, so F2 applies for $\eps\le\min\{\frac1{3600},\frac{\ln\kappa}{240\sqrt3}\}$.}
\note{f}{At $x_c=1/m$, $\kappa\ge4$: $\Omega(m^3/(\eps^2\min\{\sqrt k,m\}))$ for every $k$ \cite[Thm.~5.5]{Rub26b} (its $m$, $\eps$ and $L\eps\le\frac12$); matching rate open there, for fixed $x_c<1$ and as $x_c\to0$.}
\note{g}{$\tilde\Theta$ hides factors polylogarithmic in $m$ and $1/\eps$. Earlier single-copy learners: \cite{AG23,BMEL25}.}
\note{h}{Lower bounds for $c'\ge\frac23$.\qquad\nt{i}\,Open for $1/m\ll\nu\ll1$ \cite{CMMP26}.}
\note{j}{$\mathcal R(\boldsymbol n):=m^2+r^3+\sum_{i,j>r}1/(n_i+n_j)$ for occupations $n_1\le\dots\le n_m$, with $r:=\max(\{0\}\cup\{j:jn_j\le2\})$.}
\note{k}{Bands: occupations in dyadic intervals; lower bound for $m$ large, $\eps$ small depending on $K$, room $\nu_{\min}\ge1/r$, $2\nu_{\max}+1\le K$, constants free of the number of bands; two scales: $\Theta((m_c^3+m_w^2(1+\nu^{-1}))/\eps^2)$ for $m_c,m_w\ge m_\star$, $\eps\le\eps_\star$, $\nu\ge1/m_w$.}
\note{l}{As B1 when all modes are cold and as B2 when all occupations are $\gtrsim1$; open for $1/m\ll\nu\ll1$ \cite{CMMP26} and for mixed profiles.}
\note{m}{Absolute constant. Lower bounds as B3 on the subclasses $\|\Sigma\|_\op\le K$, with $K$-dependent constants.}
\note{n}{Adaptive; one non-Gaussian measurement. Gaussian measurements cannot be energy-independent on this row, even adaptive and entangled \cite{CGYZ26} (heterodyne detection is energy-independent on the subclass $\Sigma\succeq(\frac12+\nu)I$ \cite[Thm.~8]{CMMP26}); adaptive single-copy ones need $O(m^3/\eps^2)$ plus an $\eps$-independent term doubly logarithmic in the energy \cite{BMEM25}. Pure: B7. Occupations $n_j\le\bar n$: note~t. Else open.}
\note{o}{Entangled blocks, then single-copy heterodyne detection; $m=1$: $\Theta(\eps^{-2})$ with $k=2$ (Thm.~\ref{thm:NG1}); open for fixed $k$ as $m$ grows.}
\note{p}{One non-Gaussian round. Displaced states: blocks of two (one copy: open). For $\|\Sigma\|_\op\le\bar E$, adaptive single-copy Gaussian measurements need $O(m^2/\eps^2)$ plus a term doubly logarithmic in $\bar E$ \cite[Thm.~5]{CMFLMHCP26}, unavoidable for them \cite{CGYZ26}.}
\note{q}{$\scriptstyle\bigstar$\,The one-copy and block entries hold also for Williamson occupations $n_j\le c'/m$ with any squeezing and no energy bound: centred states, every $k\ge1$; displaced states, $k\ge2$ (Cor.~\ref{cor:NG-cold-k}).}
\note{r}{With notes s and t: in the paragraph ``Accuracy ranges and hypotheses'' after the table.}
\end{multicols}
\end{minipage}
\end{table}

Section~\ref{sec:background} fixes the conventions. Section~\ref{sec:theoremA} introduces the pair-rotation measurement and proves the single-scale case of the fermionic theorem, Section~\ref{sec:lemmaF} builds the measurement frame of the main theorem, and Section~\ref{sec:thmB} proves Theorem~\ref{thm:B}; Sections~\ref{sec:kcopy}, \ref{sec:bosonic} and~\ref{sec:unsqueezing} prove the $k$-copy, bosonic and energy-independent theorems, and Sections~\ref{sec:discussion} and~\ref{sec:open} discuss the table and what remains open. The proofs are in Appendices~\ref{app:proofs-A}--\ref{app:proofs-NG}, in the same order, and the derivations behind Section~\ref{sec:background} in Appendix~\ref{app:background}; the bosonic appendices use nothing from the fermionic ones. Most lower bounds are quoted from \cite{Rub26,Rub26b,CMMP26,CFGLMWW26,CGYZ26}; those proved here are the bosonic minimax bounds of Theorems~\ref{thm:Bos}(ii) and~\ref{thm:Bos-sep}(ii) and Corollary~\ref{cor:Bos-twoscale}(ii), assembled from the hard families of \cite{CMMP26} by a direct-sum argument (\S\ref{subsec:Bos-directsum}), the $k$-copy bound of Proposition~\ref{prop:NG-coldhot-k}(ii), assembled from the same families, and the one-mode bound of Remark~\ref{rem:NG-lower}.

\paragraph*{Notation and classes of Table~\ref{tab:main}.} A fermionic Gaussian state on $m$ modes is written $\rho_\Gamma$, with $\Gamma$ its real antisymmetric covariance matrix, $\|\Gamma\|_\op\le1$, normal-form values $|\lambda_k|\in[0,1]$, minority occupations $p_k=(1-|\lambda_k|)/2\in[0,\tfrac12]$ (the occupations in the empty-type normal form of \S\ref{subsec:bg-fermions}, in which $\lambda_k=2p_k-1\le0$) and fugacities $x_k=p_k/(1-p_k)\in[0,1]$; a bosonic Gaussian state is described by its quadrature covariance $\Sigma$ (vacuum $\Sigma=\tfrac12I$), its displacement, and its Williamson occupations $n_1\le\dots\le n_m$. ``Single copy'' means measurements on one copy at a time with classical memory between copies, ``blocks of $k$ copies'' means joint measurements on at most $k$ fresh copies with classical memory between blocks, and ``collective'' means one joint measurement on all copies; in every model the measurements are arbitrary POVMs, ancilla-assisted ones included, and may be chosen adaptively, except where a cell restricts the measurement class (Gaussian measurements in row B6). $\Theta$, $O$ and $\Omega$ hide constants that are absolute or depend only on the fixed class parameters of the row ($c,c',C,K,\kappa,\beta_0$); a subscript marks the dependence where it matters for the comparison; the tilde in row F6 hides factors polylogarithmic in $m$ and $1/\eps$. ``Open'' means that no bound beyond those displayed is known to us.

\paragraph*{Accuracy ranges and hypotheses of the lower bounds in Table~\ref{tab:main}.} Single-copy lower bounds taken from \cite{Rub26} hold for $\eps\le1/3600$, and the $\Theta$ entries that use them are claimed in that range. On $\mathcal G_c$ (F2) the lower bounds of all three columns hold for $\eps\le\eps_0(c):=\min\{1/3600,\mathrm{artanh}(1-c)/(120\sqrt3)\}$, and on the Gibbs balls of F3 for $\eps\le\min\{1/3600,\beta_0/(120\sqrt3)\}$; these windows shrink to zero as $c\to1$ or $\beta_0\to0$, where the class shrinks to the maximally mixed state, and the lower constants are uniform in $c$ and $\beta_0$ only on them. In F5, $\mathcal U_m(1,\kappa)=\mathcal G_{2/(\kappa+1)}$, a single state for $\kappa=1$, so the lower bounds at $x_c=1$ hold for $\kappa>1$ and $\eps\le\min\{1/3600,\ln\kappa/(240\sqrt3)\}$; those at $x_c=1/m$ are the bounds of \cite[Thm.~5.5]{Rub26b} on its local family, whose fugacities lie in $[1/(4m),3/(4m)]$ when $L\eps\le\tfrac12$ ($L$ the constant of that theorem), for $m$ and $1/\eps$ above the constants of that theorem. The Frobenius lower bound in F1 holds for $\eps\le1/3600$; the lower bounds in B1 hold for $c'\ge\tfrac23$; the $\Theta_K$ entries of B3 hold under the room conditions \eqref{eq:Bos-room} of Theorem~\ref{thm:Bos}(ii), for $m$ large and $\eps$ small depending on $K$, and the two-scale rate for $m_c,m_w\ge m_\star$, $\eps\le\eps_\star$ and $\nu\ge1/m_w$ (Cor.~\ref{cor:Bos-twoscale}). Notes r--t of Table~\ref{tab:main}: (r)~$\scriptstyle\bigstar$\,$\|\Sigma\|_\op\le K$ may be replaced by a known bound $n_j\le\bar n$, with no bound on energy or squeezing: $O_{\bar n}(\mathcal R(\boldsymbol n)/\eps^2)$, centred states on single copies with one non-Gaussian round, displaced states with pairs of copies (Thm.~\ref{thm:NG-thermal}); $\Theta_{\bar n}$ on the centred band classes whose occupations are all $\le\bar n$ (\S\ref{subsec:NG-thermal}). (s)~$F_k:=m_c^3/\min\{\sqrt k,m_c\}+m_w^2$ for $m_c$ occupations $\le c/m_c$, $c\ge\frac23$, and $m_w$ occupations $\ge\nu\ge1$, $K\ge2\nu+1$, any displacement and squeezing, every $k$, for $m_c,m_w\ge m_\star$; also with $n_j\le\bar n$, $\bar n\ge2\nu$, in place of $\|\Sigma\|_\op\le K$ (centred: every $k$; displaced: $k\ge2$). As B1 when all modes are cold and as B2 when all occupations are $\gtrsim1$; open for $1/m\ll\nu\ll1$ \cite{CMMP26} and for other mixed profiles. (t)~Adaptive, one non-Gaussian round; $O_{\bar n}(\mathcal R(\boldsymbol n)/\eps^2)$ state by state, as B3; lower bound for $\bar n>0$, $m\ge\max\{n_0,(1+\bar n^{-1})/2\}$, $\eps<\eps_0$ of \cite[Thm.~2]{CMMP26}; displaced states with pairs of copies.

\section{Background}\label{sec:background}\label{sec:conventions}

This section fixes the conventions and collects the facts about fermionic and bosonic Gaussian states, estimation and representation theory that the later sections use. Textbook facts are stated with a reference; facts that the proofs need in this paper's conventions are derived here or in Appendix~\ref{app:background}. Operator traces are written $\Tr$, traces of $2m\times2m$ coefficient matrices $\tr$; $\|\cdot\|_\op$, $\|\cdot\|_F$, $\|\cdot\|_1$ are the operator, Frobenius and trace norms, $\preceq$ is the Loewner order, and $T(\rho,\sigma):=\frac12\|\rho-\sigma\|_1$ is the trace distance.

\subsection{Fermionic Gaussian states}\label{subsec:bg-fermions}

\paragraph{Modes and Majorana operators.} An $m$-mode fermionic system has annihilation operators $a_1,\dots,a_m$, $\{a_k,a_l^\dagger\}=\delta_{kl}$, $\{a_k,a_l\}=0$, and occupations $n_k:=a_k^\dagger a_k$. The Majorana operators $c_{2k-1}:=a_k+a_k^\dagger$, $c_{2k}:=-i(a_k-a_k^\dagger)$ are Hermitian, $\{c_a,c_b\}=2\delta_{ab}$, and $Z_k:=ic_{2k-1}c_{2k}=2n_k-1$. We use the Jordan--Wigner representation on $(\C^2)^{\otimes m}$ \cite{JW28}, whose computational basis is the Fock basis ($|1\rangle_j$: mode $j$ occupied).

\paragraph{The covariance matrix.} The covariance matrix of a state $\rho$ is the real antisymmetric $\Gamma_{ab}:=\Tr(\rho\,ic_ac_b)$ ($a\ne b$), $\Gamma_{aa}:=0$. Positivity of $\Tr(\rho\,x^\dagger x)$, $x=\sum_az_ac_a$, gives $G:=[\Tr(\rho\,c_ac_b)]_{ab}=I-i\Gamma\succeq0$, so $\|\Gamma\|_\op\le1$ for every state, Gaussian or not. With $N_{kl}:=\langle a_k^\dagger a_l\rangle$ (Hermitian, diagonal $p_k=\langle n_k\rangle$) and $F_{kl}:=\langle a_ka_l\rangle$ (antisymmetric), substituting the Majoranas gives the $2\times2$ blocks \eqref{eq:blocks} of $\Gamma$, with $J=\begin{psmallmatrix}0&1\\-1&0\end{psmallmatrix}$: the \emph{hopping} part, in $\mathrm{span}\{I,J\}$, carries the coherences $a_k^\dagger a_l$, and the \emph{pairing} part, in $\mathrm{span}\{\sigma_x,\sigma_z\}$, carries $a_ka_l$ and vanishes in a number-conserving state. A second constraint is the single-copy budget behind the lower bounds of Table~\ref{tab:main} (\S\ref{subsec:bg-estimation}):

\begin{lemma}[{Single-copy Fisher budget; \cite[Lemma~3.1]{Rub26}; cf.\ \cite{GM00}}]\label{lem:bg-budget}
For every state on $m$ modes, $\sum_{a<b}\Gamma_{ab}^2\le m$, with equality on Fock states.
\end{lemma}

\paragraph{Normal form and the sign convention.} By the spectral theorem for the Hermitian $i\Gamma$, whose eigenvectors come in conjugate pairs,
\begin{equation}\label{eq:bg-normalform}
\Gamma=O\Big(\bigoplus_{k=1}^m\lambda_kJ\Big)O^T,\qquad O\in O(2m),\quad\lambda_k\in[-1,1] .
\end{equation}
Replacing $c_{2k}$ by $-c_{2k}$ (a particle--hole exchange of mode $k$) flips the sign of $\lambda_k$ and is absorbed into $O$. Throughout the paper the normal form is the \emph{empty-type} one, every $\lambda_k\le0$; the normal mode $k$ is then occupied with probability
\begin{equation}\label{eq:bg-sign}
p_k=\frac{1+\lambda_k}2=\frac{1-|\lambda_k|}2\in[0,\tfrac12],\qquad\lambda_k=2p_k-1\le0 ,
\end{equation}
the \emph{minority occupation}. The $|\lambda_k|$ are the normal-form values (the singular values of $\Gamma$, each twice), $\|\Gamma\|_\op=\max_k|\lambda_k|$, and the \emph{fugacity} is $x_k:=p_k/(1-p_k)\in[0,1]$.

\paragraph{Gaussian states.} For real antisymmetric $A$ put $K_A:=\frac i4\sum_{a,b}A_{ab}c_ac_b$. The Gibbs states $e^{-\beta K_h}/\Tr e^{-\beta K_h}$ of $K_h$, $h$ real antisymmetric, have $i\Gamma=-\tanh(\frac\beta2\,ih)$; the \emph{fermionic Gaussian states} are these and their limits. In the coordinates $\rho_\theta\propto e^{K_\theta}$, $K_\theta=\sum_{a<b}\theta_{ab}\,ic_ac_b$, of \cite{Rub26,Rub26b} (row F3), the Gibbs state of $K_h$ has $\theta=-\beta h/2$ and $i\Gamma=\tanh(i\theta)$. For $\|\Gamma\|_\op\le1$ with normal form \eqref{eq:bg-normalform}, rotated Majoranas $\tilde c:=O^Tc$ and $\tilde Z_k:=i\tilde c_{2k-1}\tilde c_{2k}$,
\begin{equation}\label{eq:bg-product}
\rho_\Gamma:=\prod_{k=1}^m\tfrac12\big(I+\lambda_k\tilde Z_k\big)
\end{equation}
is the Gaussian state with covariance $\Gamma$ (Appendix~\ref{app:bg-wick}); in its normal-mode frame $\rho_\Gamma=\bigotimes_k[(1-p_k)|0\rangle\langle0|+p_k|1\rangle\langle1|]$. It has full rank iff $\|\Gamma\|_\op<1$. The class $\mathcal G_c:=\{\Gamma:\|\Gamma\|_\op\le1-c\}$, $0<c<1$, is convex and closed, and on it every $p_k\ge c/2$.

\begin{proposition}[Wick's theorem; \cite{Bra05}]\label{prop:bg-wick}
Let $\rho_\Gamma$ be a fermionic Gaussian state, $G=I-i\Gamma$, and $x_j=\sum_au^{(j)}_ac_a$ ($j=1,\dots,2n$). Then $\langle x_ix_j\rangle=u^{(i)T}Gu^{(j)}$, odd products have zero expectation, and
\[
\langle x_1x_2\cdots x_{2n}\rangle=\sum_{j=2}^{2n}(-1)^j\langle x_1x_j\rangle\langle x_2\cdots\widehat{x_j}\cdots x_{2n}\rangle=\mathrm{Pf}\big[\langle x_ix_j\rangle\big]_{i<j}.
\]
In particular $\langle c_ac_bc_cc_d\rangle=G_{ab}G_{cd}-G_{ac}G_{bd}+G_{ad}G_{bc}$ for all index patterns, coincident indices included.
\end{proposition}

Appendix~\ref{app:bg-wick} derives it in the normalisation $G=I-i\Gamma$ used by the proofs.

\paragraph{Rotations of the Majoranas.} The unitary $e^{-iK_A}$ rotates the Majoranas by $O=e^A\in SO(2m)$ and maps $\rho_\Gamma$ to $\rho_{O\Gamma O^T}$ (Lemma~\ref{lem:A-rot}); $U_O:=e^{-iK_A}$ is fixed by $O$ up to a sign (the spin representation of $\mathrm{Spin}(2m)$, \S\ref{subsec:bg-reptheory}). An orthogonal $W$ with $\det W=-1$ is implemented by composing with conjugation by $c_1$.

\paragraph{The Nambu (Bogoliubov--de Gennes) picture.} Let $\alpha:=(a_1,\dots,a_m,a_1^\dagger,\dots,a_m^\dagger)^T$ and $S:=\begin{psmallmatrix}0&I\\I&0\end{psmallmatrix}$ (in Sections~\ref{sec:kcopy}--\ref{sec:unsqueezing}, $S$ is the spinor module or a symplectic matrix). A vector $c\in\C^{2m}$ stands for $\mathrm{op}(c):=\sum_ic_i\alpha_i$, and
\begin{equation}\label{eq:bg-nambu-car}
\mathrm{op}(S\bar c)=\mathrm{op}(c)^\dagger,\qquad\{\mathrm{op}(c),\mathrm{op}(d)^\dagger\}=d^\dagger c,\qquad\{\mathrm{op}(c),\mathrm{op}(d)\}=c^TSd .
\end{equation}
The \emph{Nambu matrix}, the generalised density matrix of \cite{BR86}, is $\mathcal N_{ij}:=\langle\alpha_i^\dagger\alpha_j\rangle$ \eqref{eq:nambu}; it satisfies $\langle\mathrm{op}(c)^\dagger\mathrm{op}(d)\rangle=c^\dagger\mathcal Nd$, $0\preceq\mathcal N\preceq I$ and $S\bar{\mathcal N}S=I-\mathcal N$ (Appendix~\ref{app:bg-tools}): each mode appears as a particle $c$ and as its hole $\vartheta c:=S\bar c$. A \emph{Bogoliubov transformation} $\alpha\mapsto\mathcal W\alpha$ preserving the anticommutation relations and $\alpha^\dagger=S\alpha$ is a unitary
\[
\mathcal W=\begin{pmatrix}u&v\\\bar v&\bar u\end{pmatrix},\qquad uu^\dagger+vv^\dagger=I,\qquad uv^T+vu^T=0
\]
($v=0$ for passive transformations), and $\mathcal N\mapsto\bar{\mathcal W}\mathcal N\mathcal W^T$. The two pictures are related by the fixed map $c=\Omega_{\mathrm M}\alpha$, $c_{2k-1}=\alpha_k+\alpha_{m+k}$, $c_{2k}=-i\alpha_k+i\alpha_{m+k}$ ($\Omega_{\mathrm M}/\sqrt2$ unitary): a rotation $R\in O(2m)$ of the Majoranas is the Bogoliubov unitary $\Omega_{\mathrm M}^{-1}R\,\Omega_{\mathrm M}$. In the empty-type normal-mode frame $\mathcal N=\mathrm{diag}(p_1,\dots,p_m,1-p_1,\dots,1-p_m)$, with particle eigenvectors $t_k$ (the normal modes) and holes $\vartheta t_k$.

\subsection{Bosonic Gaussian states}\label{subsec:bg-bosons}

\paragraph{Quadratures, covariance, energy.} We follow \cite[Ch.~3]{Ser17} (vacuum covariance $\frac12I$) and \cite[Sec.~5.6]{Rub26b}. Mode operators are $a_j=(q_j+ip_j)/\sqrt2$, $[a_j,a_k^\dagger]=\delta_{jk}$, and quadratures $R=(q_1,\dots,q_m,p_1,\dots,p_m)$, $[R_j,R_k]=i\Omega_{jk}$, $\Omega=\begin{psmallmatrix}0&I\\-I&0\end{psmallmatrix}$. A state has mean $v:=\Tr(\rho R)$, covariance $\Sigma_{jk}:=\frac12\Tr(\{R_j-v_j,R_k-v_k\}\rho)$, with $\Sigma+\frac i2\Omega\succeq0$ and vacuum $\Sigma=\frac12I$ (\cite{BMEM25,CGYZ26} use $V=2\Sigma$; formulas quoted from them are written in this normalisation), and occupation matrix $Q_{ij}:=\Tr(\rho\,a_j^\dagger a_i)$, the mean subtracted (\S\ref{subsec:Bos-conventions}). The mean photon number is $E=\frac12(\tr\Sigma-m)+\frac12\|v\|^2$. The realification $\Phi_{\mathrm{re}}(M)=\begin{psmallmatrix}\mathrm{Re}M&-\mathrm{Im}M\\\mathrm{Im}M&\mathrm{Re}M\end{psmallmatrix}$ is the real form of $a\mapsto Ma$.

\paragraph{Gaussian states and Wick's theorem.} A bosonic Gaussian state $\rho(v,\Sigma)$ is a displaced Gibbs state of a quadratic Hamiltonian or a limit of such states; its Wigner function is the density of $N(v,\Sigma)$. For a centred Gaussian state, with $\langle R_iR_j\rangle=\Sigma_{ij}+\frac i2\Omega_{ij}$ and all signs $+$ (Appendix~\ref{app:bg-wick}),
\begin{equation}\label{eq:bg-wick-bos}
\langle R_iR_jR_kR_l\rangle=\langle R_iR_j\rangle\langle R_kR_l\rangle+\langle R_iR_k\rangle\langle R_jR_l\rangle+\langle R_iR_l\rangle\langle R_jR_k\rangle ,
\end{equation}
so $\mathrm{Var}(\frac12R^ThR)=\frac12\tr((h\Sigma)^2)+\frac18\tr((h\Omega)^2)$ for real symmetric $h$.

\paragraph{Symplectic transformations.} Every $S\in Sp(2m,\R)=\{S:S\Omega S^T=\Omega\}$ is implemented by a Gaussian unitary $U_S$, unique up to a phase, with $U_S^\dagger RU_S=SR$ and $U_S\,\rho(v,\Sigma)\,U_S^\dagger=\rho(Sv,S\Sigma S^T)$; the displacement $D(v):=\exp(-iv^T\Omega R)$ gives $D(v)^\dagger RD(v)=R+v$. The orthogonal symplectic matrices are the $\Phi_{\mathrm{re}}(u)$, $u\in U(m)$, whose unitaries $\Gamma(u)$, $\Gamma(u)^\dagger a\Gamma(u)=ua$, are the \emph{passive} (number-conserving) transformations; a symmetric positive symplectic $P$ gives a squeezer $U_P$.

\begin{proposition}[Williamson; Bloch--Messiah {\cite[Ch.~3, Ch.~5 and App.~B]{Ser17}}]\label{prop:bg-williamson}
For every positive definite real $\Sigma$ there are $S\in Sp(2m,\R)$ and $d_j>0$ with $\Sigma=S\,\mathrm{diag}(d,d)\,S^T$; the \emph{symplectic eigenvalues} $d_j$ are the moduli of the eigenvalues of $i\Omega\Sigma$. If $\Sigma$ is the covariance of a state, every $d_j\ge\frac12$; we write $d_j=n_j+\frac12$ with the \emph{Williamson occupations} $n_1\le\dots\le n_m$. Moreover $S=PK$ with $P=(SS^T)^{1/2}$ symmetric positive symplectic and $K$ orthogonal symplectic (Appendix~\ref{app:bg-williamson}), and $S=O_1\,\mathrm{diag}(e^{r},e^{-r})\,O_2$ with $O_1,O_2$ orthogonal symplectic and $r\in\R^m$ (Bloch--Messiah, or Euler, decomposition).
\end{proposition}

\paragraph{Coherent states and measurements.} A coherent state $|\alpha\rangle=D(v)|0\rangle$, $\alpha_j=(v_j+iv_{m+j})/\sqrt2$, has $\Sigma=\frac12I$, $a_j|\alpha\rangle=\alpha_j|\alpha\rangle$ and $\pi^{-m}\int|\alpha\rangle\langle\alpha|\,d^{2m}\alpha=I$. A passive state is a Gaussian mixture of coherent states, $\rho_Q=\int\mathcal P_Q(\alpha)|\alpha\rangle\langle\alpha|\,d^{2m}\alpha$, $\mathcal P_Q$ the centred complex Gaussian density of covariance $Q$, a positive Glauber--Sudarshan $P$-function. \emph{Heterodyne detection} \cite{LR09} is the POVM $\{\pi^{-m}|\alpha\rangle\langle\alpha|\,d^{2m}\alpha\}$: each mode is mixed with a vacuum ancilla on a balanced beam splitter, and $q$ and $p$ are measured at the two output ports. On a Gaussian state, with $Y:=\sqrt2(\mathrm{Re}\,\alpha,\mathrm{Im}\,\alpha)$,
\begin{equation}\label{eq:bg-het}
Y\sim N\big(v,\Sigma+\tfrac12I\big),\qquad\E\big[(\alpha-\langle a\rangle)(\alpha-\langle a\rangle)^\dagger\big]=I+Q
\end{equation}
(derivation: \eqref{eq:Bos-het-moments}). The added $\frac12I$, the $1$ in $\E|\alpha_j|^2=n_j+1$, is one vacuum unit of noise, which enters through the second port of the beam splitter: the price of measuring the noncommuting $q$ and $p$ at once on one copy. A \emph{Gaussian measurement} on $k$ copies gives $N(v^{\oplus k},\Sigma^{\oplus k}+\Sigma^{(k)}_{\mathrm s})$ on $\rho(v,\Sigma)^{\otimes k}$ for a seed covariance $\Sigma^{(k)}_{\mathrm s}$ on $km$ modes fixed in advance (heterodyne detection: $k=1$, $\Sigma_{\mathrm s}=\frac12I$), and a \emph{Gaussian protocol} uses only such measurements, on blocks, randomised and adaptive \cite[Sec.~4.2, Def.~4.4]{CGYZ26}; these are the measurements of row B6.

\paragraph{Distances.} Section~\ref{sec:bosonic} compares passive states through the \emph{Bures--Wasserstein distance} \cite{BJL19} of their occupation matrices,
\begin{equation}\label{eq:bg-BW}
d_{\mathrm{BW}}(A,B)^2=\tr A+\tr B-2\tr\big(A^{1/2}BA^{1/2}\big)^{1/2} .
\end{equation}

\subsection{Quantum estimation: models, metrics, bounds}\label{subsec:bg-estimation}

\paragraph{The task and the models.} A learner receives $N$ copies of an unknown $\rho$ from a known class $\mathcal C$ and outputs $\hat\rho$; it succeeds if $T(\hat\rho,\rho)\le\eps$. The \emph{sample complexity} of $\mathcal C$ in a measurement model is the least $N$ for which some protocol succeeds with probability at least $2/3$ on every $\rho\in\mathcal C$ (by the median trick, $O(\log\delta^{-1})$ repetitions give $1-\delta$ at accuracy $3\eps$). The trace distance is contractive under channels and jointly convex \cite[\S9.2.1]{NC10}, and $1-\sqrt F\le T\le\sqrt{1-F}$ for the fidelity $F$ \cite{FvdG99}. The measurement models (single copy, blocks of $k$ copies, collective) are those of the paragraph after Table~\ref{tab:main}, as in \cite{BCL20,CCHL22,CHLLS22}; a block is a POVM on $t\le k$ fresh copies \cite[Sec.~2]{Rub26b}.

\paragraph{Fisher information and the Cram\'er--Rao bounds.} A POVM $\{M_y\}$ on a family $\rho_\theta$, $\theta\in\R^p$, has outcome law $q_\theta(y)=\Tr(M_y\rho_\theta)$ and Fisher information $I(\theta)_{jl}=\sum_y\partial_jq_\theta(y)\,\partial_lq_\theta(y)/q_\theta(y)$. For adaptive protocols with classical memory, $I_Y(\theta)=\sum_i\E[I_{M_i(y_{<i})}(\theta)]$ \cite[Lemma~4.1]{Rub26}: a bound $\tr I\le b$ per copy (or block) gives $\tr I_Y\le Nb$. An unbiased $\hat\theta$ has $\E\|\hat\theta-\theta\|^2\ge\tr I^{-1}\ge p^2/\tr I$ (Cram\'er--Rao); lower bounds on sample complexity use the Bayesian form:

\begin{proposition}[van Trees inequality {\cite[p.~72]{vT68}, \cite{GL95}}]\label{prop:bg-vantrees}
Let $\theta$ have a prior density $\varpi$ on a box, continuously differentiable and vanishing on its boundary, with prior information $J(\varpi):=\E_\varpi\|\nabla\log\varpi\|^2$. For a model whose outcome densities $q_\theta(y)$ are continuously differentiable in $\theta$ with $\E_\varpi\tr I_Y(\theta)<\infty$, every estimator $\hat\theta(Y)$ satisfies $\E\|\hat\theta-\theta\|^2\ge p^2/(\E_\varpi\tr I_Y(\theta)+J(\varpi))$. For the product prior with density $a^{-1}\cos^2(\pi u/(2a))$ on $[-a,a]$ in each coordinate, $a=r/\sqrt p$ (a cube in the ball of radius $r$), $J(\varpi)=\pi^2p^2/r^2$ \cite[Lemma~6.1]{Rub26}.
\end{proposition}

\paragraph{Quantum Fisher information.} The symmetric logarithmic derivative $L$ solves $\partial_\theta\rho=\frac12(\rho L+L\rho)$, and $F_Q=\Tr(\rho L^2)$. It bounds every measurement, $I_M\preceq F_Q$ \cite{BC94}, and is additive on tensor products. From $\|\rho L\|_1\le\|\rho^{1/2}\|_2\|\rho^{1/2}L\|_2$, along any differentiable path
\begin{equation}\label{eq:bg-path}
\|\partial_t\rho_t\|_1\le\sqrt{F_Q(t)},\qquad T(\rho_0,\rho_1)\le\tfrac12\int_0^1\sqrt{F_Q(t)}\,dt .
\end{equation}

\begin{proposition}[Gaussian quantum Fisher information]\label{prop:bg-qfi}
\emph{(Fermions \cite{CSV18}.)} For a full-rank fermionic Gaussian family with tangent $X=\dot\Gamma$,
\begin{equation}\label{eq:bg-qfi-ferm}
F_Q=\tfrac12\,\tr\big(X^T(\mathrm{id}-\mathcal C_\Gamma)^{-1}(X)\big),\qquad\mathcal C_\Gamma(X):=\Gamma X\Gamma^T .
\end{equation}
\emph{(Bosons \cite[Eqs.~(8)--(10), (14)]{Saf19}.)} For a bosonic Gaussian family whose symplectic eigenvalues all exceed $\frac12$, with tangent $(\dot v,\dot\Sigma)$,
\begin{equation}\label{eq:bg-qfi-bos}
F_Q=\tfrac12\,\mathrm{vec}(\dot\Sigma)^T\big(\Sigma\otimes\Sigma-\tfrac14\Omega\otimes\Omega\big)^{-1}\mathrm{vec}(\dot\Sigma)+\dot v^T\Sigma^{-1}\dot v .
\end{equation}
\end{proposition}

\paragraph{From a Fisher budget to a lower bound.} Near $I/2^m$ take $\rho_\theta\propto\exp(\sum_{a<b}\theta_{ab}\,ic_ac_b)$, so $\partial_{ab}\rho_\theta|_0=ic_ac_b/2^m$, with $p:=m(2m-1)$ parameters. \emph{Collective}: the SLD at $\theta=0$ is $ic_ac_b$, so $\tr F_Q=p$ per copy. \emph{Single copy}: a rank-one POVM element $w|\psi\rangle\langle\psi|$ contributes $2^{-m}w\sum_{a<b}\langle\psi|ic_ac_b|\psi\rangle^2\le2^{-m}wm$ to $\tr I$ (Lemma~\ref{lem:bg-budget}), and $\sum w=2^m$, so $\tr I\le m$ (at $\theta=0$; \cite[Prop.~3.4]{Rub26} gives $2m$ on a ball). \emph{Blocks}: a POVM on $t$ copies has $\tr I\le2mt^2$ on the ball \cite[Lemma~3.2]{Rub26b}. Since $N$ copies with per-copy budget $b$ estimate $p$ parameters with total squared error at least $p^2/(Nb)$, and near $I/2^m$ trace distance is comparable to the Euclidean distance of the parameters \cite[Prop.~5.4]{Rub26}, error $\eps$ needs $N\gtrsim p^2/(b\eps^2)$: $m^3/\eps^2$ for single copies, $m^3/(t\eps^2)$ with blocks of $t\le m$, $m^2/\eps^2$ collectively. Rigorously, by Proposition~\ref{prop:bg-vantrees} on a cube of radius $r\asymp\eps$, boosting and projection: $N>m^3/(8\,553\,600\,\eps^2)$ for $\eps\le1/3600$ \cite[Thm.~1.1]{Rub26}, with $k$-copy and collective analogues \cite[Thms.~3.4, 3.11]{Rub26b}.

\paragraph{Concentration.} For i.i.d.\ real $X_1,\dots,X_N$ with mean $\mu$ and average $\bar X$: \emph{Hoeffding}, $\Prob(|\bar X-\mu|\ge t)\le2e^{-2Nt^2}$ for $X_i\in[0,1]$ \cite[Thm.~2.2.6]{Ver18}; \emph{Bernstein} \cite[Thm.~2.8.4]{Ver18}, for $|X_i-\mu|\le b$ and $\mathrm{Var}X_i\le\sigma^2$,
\begin{equation}\label{eq:bg-bernstein}
\Prob(|\bar X-\mu|\ge\eta)\le2\exp\Big(-\frac{N\eta^2}{2(\sigma^2+b\eta/3)}\Big),
\end{equation}
so for $b\eta\le3\sigma^2$ the exponent is at least $N\eta^2/(4\sigma^2)$. In adaptive protocols these apply conditionally on the past. \emph{Nets}: the unit sphere of $\R^n$ has a $\delta$-net of at most $(1+2/\delta)^n$ points \cite[Cor.~4.2.13]{Ver18}, and for Hermitian $M$ on $\C^n\cong\R^{2n}$, $\|M\|_\op\le(1-2\delta)^{-1}\max_{w}|w^\dagger Mw|$ over the net \cite[Ex.~4.4.3]{Ver18}; at $\delta=\frac14$, at most $9^{2n}$ points and a factor $2$ (Lemma~\ref{lem:A2prime}).

\subsection{Representation theory for blocks of copies}\label{subsec:bg-reptheory}

Section~\ref{sec:kcopy} sets its own conventions (\S\ref{subsec:K-setting}); this subsection records the facts it quotes, and the second quantisation used in Appendices~\ref{app:proofs-K} and~\ref{app:proofs-Bos}. The general facts are in \cite{FH91} (highest weights: Lect.~14; spin representations: Lects.~18--20; Casimir: \S25.1).

\paragraph{The spinor representation.} On the Fock space $S:=(\C^2)^{\otimes m}$, $\pi(X):=\frac14\sum_{a,b}X_{ab}c_ac_b=-iK_X$ represents $\mathfrak{so}(2m)$ and integrates to $\mathrm{Spin}(2m)$, not $SO(2m)$: the \emph{spinor representation}. $S=S_+\oplus S_-$ splits into the \emph{half-spin} representations, spanned by the Fock states with an even ($S_+$) or odd ($S_-$) number of empty modes, whose highest-weight vectors are the filled state and $|1\cdots10\rangle$. Weights are taken in \emph{doubled} coordinates, twice those of \cite{FH91}: the Fock state $|n_1\cdots n_m\rangle$ has weight $(2n_j-1)_j$. In them, with the notation of \S\ref{subsec:K-setting}, the quadratic Casimir $C_2:=\sum_{a<b}G(E_{ab})^2$ acts on $V_\lambda$ as the scalar $\langle\lambda,\lambda+2\varrho\rangle=\|\lambda\|^2+4\lambda\cdot\delta$ \cite[\S25.1]{FH91}; in this normalisation it is the case $\varphi=d\pi$ of Lemma~\ref{lem:K3}.

\paragraph{Second quantisation.} For $X\succeq0$ on $\C^m$, $\Gamma(X)=\bigoplus_{\ell\ge0}\mathrm{Sym}^\ell(X)$ (bosons, $\|X\|_\op<1$) and $\Gamma_{\mathrm a}(X)=\bigoplus_{\ell=0}^m\Lambda^\ell(X)$ (fermions), with $\Tr\Gamma(X)=\det(I-X)^{-1}$ and $\Tr\Gamma_{\mathrm a}(X)=\det(I+X)$. On the one-particle sector both are $X$.

\section{The metric-matched measurement and the single-scale theorem}\label{sec:theoremA}

\subsection{Fermionic conventions}\label{subsec:conventions}

We use the conventions of \S\ref{subsec:bg-fermions}, where the formulas below are derived, with $p_k=\langle a_k^\dagger a_k\rangle$, $N_{kl}=\langle a_k^\dagger a_l\rangle$, $F_{kl}=\langle a_ka_l\rangle$ and $J=\begin{psmallmatrix}0&1\\-1&0\end{psmallmatrix}$. The $2\times2$ blocks of $\Gamma$ are
\begin{equation}\label{eq:blocks}
\begin{aligned}
\Gamma_{kk}&=(2p_k-1)J,\\
\Gamma_{kl}&=2\,\mathrm{Re}N_{kl}\,J-2\,\mathrm{Im}N_{kl}\,I+2\,\mathrm{Re}F_{kl}\,\sigma_x-2\,\mathrm{Im}F_{kl}\,\sigma_z\qquad(k\ne l).
\end{aligned}
\end{equation}
The first two terms form the \emph{hopping part} (superscript $+$, in $\mathrm{span}\{I,J\}$), the last two the \emph{pairing part} (superscript $-$, in $\mathrm{span}\{\sigma_x,\sigma_z\}$, symmetric blocks); right multiplication by $J$ preserves both types. For real antisymmetric $X$, $X=X^d+X^++X^-$ denotes its diagonal, off-diagonal hopping and off-diagonal pairing blocks; they are Frobenius-orthogonal, with $\|X^+_{kl}\|_F^2=8|N_{kl}|^2$ and $\|X^-_{kl}\|_F^2=8|F_{kl}|^2$.

A \emph{frame} is an orthogonal $W\in O(2m)$ acting on the Majoranas, implemented on each copy by a Gaussian unitary, composed when $\det W=-1$ with the conjugation by the Majorana operator $c_1$ (\S\ref{subsec:bg-fermions}). Rotating every copy by $W^T$ turns $\Gamma$ into $\Gamma'=W^T\Gamma W$, an estimate $\hat\Gamma'$ is returned as $W\hat\Gamma'W^T$, and the trace distance is invariant, so all statements below are made in the frame.

For $0<c<1$, $\mathcal G_c:=\{\Gamma:\|\Gamma\|_\op\le1-c\}$ is the convex class of \S\ref{subsec:bg-fermions}. For $\kappa\ge1$ and $0<x_c\le1$, $\mathcal U_m(x_c,\kappa)$ is the set of Gaussian states on $m$ modes all of whose normal-form fugacities satisfy $x_k\in[x_c/\kappa,x_c]$, in an arbitrary frame. Then $1-|\lambda_k|=2p_k\ge2x_c/(\kappa(1+x_c/\kappa))\ge x_c/\kappa$, so $\mathcal U_m(x_c,\kappa)\subset\mathcal G_{x_c/\kappa}$.
A state of $\mathcal U_m(x_c,\kappa)$ has expected number of excitations $\sum_kp_k$ between $mx_c/(2\kappa)$ and $mx_c$, so the range $x_c\in[1/m,1]$ of Theorem~\ref{thm:A} runs from the maximally mixed state down to about one excitation in the whole system.

\subsection{Two tools}\label{subsec:twotools}

For a real antisymmetric $A\in\mathbb R^{2m\times2m}$ let
\[
K_A:=\frac i4\sum_{a,b}A_{ab}\,c_ac_b,
\]
a Hermitian quadratic generator.

With $\lambda_k=2p_k-1$ the weights of the block form of Lemma~\ref{lem:wick} below are
\begin{equation}\label{eq:A-occweights}
1-\lambda_k\lambda_l=2\big[p_k(1-p_l)+p_l(1-p_k)\big],\qquad 1+\lambda_k\lambda_l=2\big[p_kp_l+(1-p_k)(1-p_l)\big].
\end{equation}
The first is twice the probability that one mode is filled and the other empty, the only configuration in which a particle can hop between them; the second is twice the probability that both are empty or both filled, in which a pair can be created or annihilated. Near the vacuum a hopping rotation by $\theta$ therefore moves the state by at most $\theta\sqrt{p_k+p_l}$ (Lemma~\ref{lem:MT}), while a pairing rotation moves it by order $\theta$.

\begin{lemma}[Wick variance of a quadratic generator]\label{lem:wick}
For a Gaussian state $\rho_\Gamma$ on $m$ modes and a real antisymmetric $A$, $\langle K_A\rangle=\tfrac14\tr(A^T\Gamma)$ and
\[
\mathrm{Var}_{\rho_\Gamma}(K_A)=\tfrac18\big(\|A\|_F^2-\tr(A\Gamma A\Gamma)\big)\le\tfrac14\|A\|_F^2 .
\]
In the normal form $\Gamma=\bigoplus_k\lambda_kJ$, writing $A$ in $2\times2$ blocks $A_{kl}=A^+_{kl}+A^-_{kl}$ with $A^+_{kl}\in\mathrm{span}\{I,J\}$ (hopping) and $A^-_{kl}\in\mathrm{span}\{\sigma_x,\sigma_z\}$ (pairing),
\[
\mathrm{Var}_{\rho_\Gamma}(K_A)=\tfrac18\sum_{k,l}\Big[\|A^+_{kl}\|_F^2\,(1-\lambda_k\lambda_l)+\|A^-_{kl}\|_F^2\,(1+\lambda_k\lambda_l)\Big]
\]
(sum over ordered pairs; the diagonal blocks $k=l$ are of hopping type and carry the weight $1-\lambda_k^2$). Moreover $e^{-iK_A}\rho_\Gamma e^{iK_A}=\rho_{e^A\Gamma e^{-A}}$.
\end{lemma}
The proof is in Appendix~\ref{app:bg-tools}.

\begin{lemma}[Fermionic Mandelstam--Tamm bound for rotations]\label{lem:MT}
Let $U=e^{-iK_A}$ with $A$ real antisymmetric and $R:=e^{A}$. Then $U\rho_\Gamma U^\dagger$ is Gaussian with covariance $R\Gamma R^T$, and
\[
T\big(\rho_\Gamma,\rho_{R\Gamma R^T}\big)\le\sqrt{\mathrm{Var}_{\rho_\Gamma}(K_A)}=\tfrac1{\sqrt8}\big(\|A\|_F^2-\tr(A\Gamma A\Gamma)\big)^{1/2}\le\tfrac12\|A\|_F .
\]
The bound is dimension-free and uniform in the spectrum of $\Gamma$.
\end{lemma}

\begin{proof}
The first statement is the last part of Lemma~\ref{lem:wick}. For any state $\rho$ and any Hermitian $H$ one has $T(\rho,e^{-iH}\rho e^{iH})\le\sqrt{\mathrm{Var}_\rho(H)}$ \cite[Lemma~5.12]{Rub26b} (the proof is reproduced in Lemma~\ref{lem:A-MTgen}); apply this with $H=K_A$ and insert Lemma~\ref{lem:wick}.
\end{proof}

\subsection{Further tools}\label{subsec:furthertools}

\begin{enumerate}
\item[(T1)] ($\chi^2$ continuity \cite[Prop.~7.1]{Rub26}; Lemma~\ref{lem:A-chi2} and the sketch after it.) For $0<c<1$ and $\Gamma,\Gamma'\in\mathcal G_c$: $\|\rho_\Gamma-\rho_{\Gamma'}\|_1\le K_c\|\Gamma-\Gamma'\|_F$ with $K_c\le(2c)^{-1/2}$. (The proof uses the segment from $\Gamma$ to $\Gamma'$, which stays in the convex set $\mathcal G_c$.)
\item[(T2)] (Davis--Kahan $\sin\Theta$ theorem \cite{DK70}, operator-norm form.) If $G,\hat G$ are Hermitian, $P$ is the spectral projector of $G$ for the eigenvalues in $[\alpha,\beta]$, $\hat P$ that of $\hat G$ for the eigenvalues in $[\alpha-\Delta,\beta+\Delta]$, and the rest of the spectrum of $\hat G$ is at distance $\ge\delta$ from $[\alpha,\beta]$, then $\|(I-\hat P)P\|_\op\le\|\hat G-G\|_\op/\delta$.
\item[(T3)] The metric projection onto a closed convex subset of a Hilbert space is non-expansive.
\item[(T4)] (Wick.) For two fermionic modes $A\ne B$ (anticommuting mode operators $a_A,a_B$) of a Gaussian state, with $n_A=a_A^\dagger a_A$, $n_B=a_B^\dagger a_B$:
$\mathrm{Cov}(n_A,n_B)=|\langle a_Aa_B\rangle|^2-|\langle a_A^\dagger a_B\rangle|^2$.
\end{enumerate}

Proofs of (T1) and (T2)--(T4), with the supporting Lemmas~\ref{lem:A-rot}, \ref{lem:A-MTgen}, \ref{lem:A-chi2} and~\ref{lem:A-DK} and the identity \eqref{eq:A-chi2id}, are in Appendix~\ref{app:bg-tools}.

\subsection{The Nambu matrix}\label{subsec:nambu}

The Nambu matrix $\mathcal N_{ij}=\langle\alpha_i^\dagger\alpha_j\rangle$ of \S\ref{subsec:bg-fermions}, $\alpha=(a_1,\dots,a_m,a_1^\dagger,\dots,a_m^\dagger)$, and its anomalous companion $\mathcal A_{ij}:=\langle\alpha_i\alpha_j\rangle$ are
\begin{equation}\label{eq:nambu}
\mathcal N=\begin{pmatrix}N&-\bar F\\ F&I-\bar N\end{pmatrix},\qquad
\mathcal A=\begin{pmatrix}F&I-N^T\\ N&-\bar F\end{pmatrix}.
\end{equation}
$\mathcal N$ has spectrum $\{p_k\}\cup\{1-p_k\}$ in every frame. For rotated modes $a'_A=c_A\cdot\alpha$ and $b'=c_B\cdot\alpha$, $\langle a'^\dagger_Ab'\rangle=c_A^\dagger\mathcal Nc_B$ and $\langle a'_Ab'\rangle=c_A^T\mathcal Ac_B$, and $\|\mathcal A\|_\op\le1$. We write $\mathcal N_{\mathrm{off}}$ for $\mathcal N$ with $\mathrm{diag}(p,1-p)$ removed. (Proofs, and the construction of a Bogoliubov frame from $m$ orthonormal rows used in Lemma~\ref{lem:A2}: Appendix~\ref{app:bg-tools}.)

\subsection{The pair-rotation measurement}\label{subsec:pairrotation}

One \emph{shot} of the pair-rotation measurement: draw a uniformly random perfect matching $\pi$ of the $m$ modes (for odd $m$ one uniformly random mode is left out) and, independently for each matched pair $P=(k,l)$, a label $\sigma\in\{\mathrm{hr},\mathrm{hi},\mathrm{pr},\mathrm{pi}\}$ and a sign $s\in\{+1,-1\}$, both uniform. Apply the two-mode matchgate $U_P^{\sigma,s}$ whose action on the mode operators is
\begin{equation}\label{eq:pairrot}
a_k'=\frac{a_k+s\,\omega_\sigma\,b_l}{\sqrt2},\qquad
b_l=\begin{cases}a_l&(\mathrm h)\\ a_l^\dagger&(\mathrm p)\end{cases},\qquad
\omega_\sigma=\begin{cases}1&(\mathrm r)\\ i&(\mathrm i)\end{cases},
\end{equation}
with partner mode $a_l'=(a_l-s\,\overline{\omega_\sigma}\,a_k)/\sqrt2$ (hopping labels) or $a_l'=(a_l-s\,\omega_\sigma\,a_k^\dagger)/\sqrt2$ (pairing labels); the rotation angle is $\theta=s\pi/4$. Both are Bogoliubov transformations ($\{a_k',a_k'^\dagger\}=1$, $\{a_k',a_l'\}=0$, etc.) with $\det=+1$ on the four Majoranas, so a two-mode Gaussian unitary implementing them exists. Then measure all occupations $n_k'=a_k'^\dagger a_k'$; both $n_k'$ and $n_l'$ are recorded, and the \emph{sample} of the pair is
\[
X_P:=2s\,n_k' .
\]

\begin{lemma}[Metric matching: unbiased, with noise matched to the occupations]\label{lem:A1}
In any frame, with $p_k,N_{kl},F_{kl}$ the moments of the true state in that frame, and with expectations taken over the measurement outcome \emph{and} the sign $s$:
\[
\begin{aligned}
\E[X_P\mid\mathrm{hr}]&=2\,\mathrm{Re}N_{kl},&
\E[X_P\mid\mathrm{hi}]&=-2\,\mathrm{Im}N_{kl},\\
\E[X_P\mid\mathrm{pr}]&=-2\,\mathrm{Re}F_{kl},&
\E[X_P\mid\mathrm{pi}]&=-2\,\mathrm{Im}F_{kl},
\end{aligned}
\]
and
\[
\begin{aligned}
\E[X_P^2\mid\mathrm{hr\ or\ hi}]&=2(p_k+p_l),\\
\E[X_P^2\mid\mathrm{pr\ or\ pi}]&=2(p_k+1-p_l)\le4,
\end{aligned}
\qquad\text{and}\qquad |X_P|\le2 .
\]
Conditionally on $s$, $\langle n_k'\rangle_s=\frac12(p_k+p_l)+s\,\mu$ for a hopping label (and $\frac12(p_k+1-p_l)+s\,\mu$ for a pairing label), where $\mu$ is the signed target entry; so the centred sample $X_P-2\mu$ has mean zero only after averaging over $s$. Accordingly, all conditioning below is on the matchings and labels $(\pi,\sigma)$ only, and the sign $s$ is part of the sample.
\end{lemma}

\begin{proof}
With $\theta=s\pi/4$: for $\mathrm{hr}$, $n_k'=\cos^2\theta\,n_k+\sin^2\theta\,n_l+\sin\theta\cos\theta\,(a_k^\dagger a_l+a_l^\dagger a_k)$, so $\langle n_k'\rangle=\frac12(p_k+p_l)+s\,\mathrm{Re}N_{kl}$; averaging $2s\langle n_k'\rangle$ over $s=\pm1$ gives $2\,\mathrm{Re}N_{kl}$. For $\mathrm{hi}$ the cross term is $i\sin\theta\cos\theta\,(a_k^\dagger a_l-a_l^\dagger a_k)$, with expectation $-s\,\mathrm{Im}N_{kl}$. For $\mathrm{pr}$, $n_k'=\cos^2\theta\,n_k+\sin^2\theta\,a_la_l^\dagger+\sin\theta\cos\theta\,(a_k^\dagger a_l^\dagger+a_la_k)$, with expectation $\frac12(p_k+1-p_l)+s(-\mathrm{Re}F_{kl})$, since $\langle a_k^\dagger a_l^\dagger\rangle+\langle a_la_k\rangle=-\overline{F_{kl}}-F_{kl}$. For $\mathrm{pi}$ the cross term is $i\sin\theta\cos\theta\,(a_k^\dagger a_l^\dagger-a_la_k)$, with expectation $-s\,\mathrm{Im}F_{kl}$. Since $n_k'$ is a projector, $X_P^2=4n_k'$ and $\E X_P^2=4\cdot\frac12\big(\langle n_k'\rangle_{s=+}+\langle n_k'\rangle_{s=-}\big)=2(p_k+p_l)$ (hopping) or $2(p_k+1-p_l)$ (pairing). The conditional means given $s$ were displayed along the way.
\end{proof}

We call a single-copy measurement \emph{metric-matched} on a class of states when its parameter variances compensate the corresponding local metric weights, so that their products remain bounded independently of the occupation scale. For a hopping label, $U_P^{\sigma,s}$ is a balanced beam splitter between modes $k$ and $l$, and as $n'_k$ is a projector the sample, an estimate of a covariance entry, has second moment $2(p_k+p_l)$; in these covariance coordinates the $\chi^2$ weight \eqref{eq:A-chi2id} of the entry is, for comparable occupations, of order $1/(p_k+p_l)$, and the product is of order one. In the generator coordinates of Lemma~\ref{lem:MT} the same scale enters as a sensitivity: a rotation of the pair by $\theta$ moves the state by at most $\theta$ times the square root of the weight \eqref{eq:A-occweights}, of order $\theta\sqrt{p_k+p_l}$. For a pairing label the partner is replaced by its hole, and noise and sensitivity are both of order one.

In one shot a given pair $P$ is matched with probability $1/(m-1)$ and then receives a given label with probability $\frac14$, so a given $(P,\sigma)$ occurs with probability $\frac1{4(m-1)}$; over $N$ shots the sample count of the entry is $S_{P,\sigma}\sim\mathrm{Bin}\big(N,\frac1{4(m-1)}\big)$ (for odd $m$: $\frac1{4m}$, and the Horvitz--Thompson weight $4(m-1)$ of \eqref{eq:HT} becomes $4m$). By a Chernoff bound, $S_{P,\sigma}\ge N/(8m)$ simultaneously for all $2m(m-1)$ entries with probability $\ge1-2m^2e^{-N/(32m)}\ge0.99$ as soon as $N\ge32m\ln(200m^2)$; this holds for $N\ge m^3$ when $m\ge19$ (for $4\le m<19$ a larger constant is needed). Conditionally on the sequence $(\pi_t,\sigma_t)_t$ of matchings and labels: the samples of a given entry are i.i.d.\ over its shots, bounded by $2$, with mean the entry (after centring by $s$) and variance $\le\E X_P^2$; samples in different shots are independent (different copies); and \emph{two entries sharing a mode are never sampled in the same shot}, because $\pi$ is a perfect matching.
(A uniformly random perfect matching of $m$ points, $m$ even, is one of $(m-1)!!$ equally likely matchings, and those containing a given pair are the matchings of the other $m-2$ points: $\Prob(P\in\pi)=(m-3)!!/(m-1)!!=1/(m-1)$, and $\Prob(P,P'\in\pi)=(m-5)!!/(m-1)!!=1/((m-1)(m-3))$ for disjoint $P,P'$. The Chernoff bound is $\Prob(S<\mu/2)\le e^{-\mu/8}$ for a binomial $S$ of mean $\mu$ \cite[Sec.~2.3]{Ver18}.)

In a given frame (the computational frame in Stage~0 of the protocol of Theorem~\ref{thm:A}, \S\ref{subsec:stage0}) the \emph{Horvitz--Thompson estimator} is built from $N_0$ pair-rotation shots. For shot $t$ let
\begin{equation}\label{eq:HT}
Z_t:=\sum_{P\in\pi_t}4(m-1)\,c_{\sigma_P}\,s_P\,n'_P\;E_{P,\sigma_P},
\end{equation}
where $c_{\mathrm{hr}}=+1$, $c_{\mathrm{hi}}=c_{\mathrm{pr}}=c_{\mathrm{pi}}=-1$, and $E_{P,\sigma}$ is the Hermitian pattern of the real parameter ($\mathrm{Re}N_{kl}$, $\mathrm{Im}N_{kl}$, $\mathrm{Re}F_{kl}$ or $\mathrm{Im}F_{kl}$) inside $\mathcal N$: four entries of modulus $1$ (for $\mathrm{Re}N_{kl}$: $+1$ at $(k,l),(l,k)$ and $-1$ at $(k+m,l+m),(l+m,k+m)$; similarly for the other three), with $\|E_{P,\sigma}\|_\op=1$, and patterns of disjoint pairs have disjoint supports. By Lemma~\ref{lem:A1}, $\E[c_\sigma s\,n'_P\mid\pi,\sigma]$ is the corresponding Nambu entry, so $\E Z_t=\mathcal N_{\mathrm{off}}$. Set $\hat{\mathcal N}_{\mathrm{off}}:=N_0^{-1}\sum_tZ_t$. The diagonal $p_k$ is estimated from $N_0'$ plain occupation shots by the empirical frequencies $\hat p_k$, and $\hat{\mathcal N}:=\hat{\mathcal N}_{\mathrm{off}}+\mathrm{diag}(\hat p,1-\hat p)$.

\subsection{The single-scale theorem}

\begin{theorem}[Single scale, unknown frame]\label{thm:A}
There are constants $C_A(\kappa),\eps_A(\kappa)>0$ such that for all $m\ge4$, all $x_c\in[1/m,1]$ and all $\eps\le\eps_A$, a single-copy adaptive protocol learns every $\rho\in\mathcal U_m(x_c,\kappa)$ to trace distance $\eps$ with probability $\ge2/3$ from $C_A\,m^3/\eps^2$ copies.
\end{theorem}

The protocol has two stages. Stage 0 estimates the Nambu matrix in the computational frame and diagonalises it to obtain a frame $W_0$ in which the state is cold and nearly diagonal (\S\ref{subsec:stage0}); it also learns a proxy $\hat x_c$ for the scale. Stage 1 runs the pair-rotation measurement in the frame $W_0$ on fresh copies, assembles a covariance estimate from the empirical means, projects it onto a class $\mathcal G_{c'}$ with $c'$ computed from $\hat x_c$, and returns the Gaussian state with the projected covariance rotated back to the original frame (\S\ref{subsec:stage1}).

Matchgate shadows give $12\kappa m^3/(x_c\eps^2)$ copies on this class \cite[Thm.~7.6]{Rub26}, a factor $m$ more at $x_c=1/m$. Theorem~\ref{thm:A} is implied, with an absolute constant, by Theorem~\ref{thm:B}; it is stated separately because at a single scale the cancellation is visible with one weight for the whole state, before the bands of Section~\ref{sec:thmB}.

Near a state whose modes are occupied with probability of order $x_c$, a change $\delta$ of a hopping or occupation entry costs about $\delta^2/x_c$ in the $\chi^2$ divergence, while the pair-rotation sample has second moment of order $x_c$ (Lemma~\ref{lem:A1}). Noise times weight is scale-free, as for a Poisson counter, and $2m^2$ entries, each read in about $N/(4m)$ shots, give $N\asymp m^3/\eps^2$. The proof (Appendix~\ref{app:proofs-A}) has to find a frame in which the state is cold, which Stage~0 does with an operator-norm error free of logarithms because a perfect matching never reads two entries that share a mode in the same shot (Lemma~\ref{lem:A2prime}), and to handle the pairing entries, whose error is removed by a rotation that Lemma~\ref{lem:MT} charges with no factor $1/x_c$. A second-order remainder of size $\eps^2/\sqrt m$ meets the weight $\sqrt{1/x_c}$, which is why the range of $x_c$ stops at $1/m$.

\section{The frame: relative-error bootstrapping}\label{sec:lemmaF}

Theorem~\ref{thm:B} measures in a frame that must itself be learned: a counting readout is quiet only in a frame where the state is cold, as $(|00\rangle+|11\rangle)/\sqrt2$ gives two fair coins in the computational frame and two certain zeros in its Bogoliubov frame. When the occupations span many unknown scales, one estimate of the Nambu matrix does not supply such a frame (at a single scale it does, Stage~0 of Theorem~\ref{thm:A}). Protocol~F$'$ (\S\ref{subsec:F-protocol}) runs $R+1$ rounds of fresh measurements, $R=\lceil\log_2\log_2m\rceil$, each estimating the Nambu matrix with relative error $\eta$ at a decreasing scale $\mu_r$ and passing to the eigenframe of the estimate, at a total cost of $O(m^3)$ copies (Theorem~\ref{thm:Fprime}). Theorem~\ref{thm:B} uses the output only through the frame condition $G(1/m,3)$.

\subsection{The Nambu picture}\label{subsec:F-nambu}

Let $\mathcal N$ be the Nambu matrix \eqref{eq:nambu}, with the particle--hole symmetry $S\bar{\mathcal N}S=I-\mathcal N$ ($S$ the swap of the two halves) and the notation $\mathrm{op}(c)$, $\vartheta c=S\bar c$ of \S\ref{subsec:bg-fermions}; every matrix in this section, and in the Nambu parts of Section~\ref{sec:thmB}, acts on $\C^{2m}$.

The true particle vectors $t_k$ below are the normal modes $b_k=\mathrm{op}(t_k)$, in which the state is a product of independent modes with occupations $p_k$; the hole $\bar t_k=\vartheta t_k$ represents $b_k^\dagger$, with $\langle b_kb_k^\dagger\rangle=1-p_k$.

\begin{itemize}
\item True particle vectors $t_k$ satisfy $\mathcal Nt_k=p_kt_k$, with $p_k\in[0,\frac12]$ the occupations of the empty-type normal form $\Gamma=ODO^T$ \eqref{eq:bg-sign}. The holes $\bar t_k$ have eigenvalue $1-p_k$. When $p_k=\frac12$, the normal form fixes which vector is the particle and which the hole.
\item $\Pi_p$ and $\Pi_h$ are the true particle and hole projectors; both commute with $\mathcal N$.
\item $\mathcal K:=\sum_kp_k\big(|t_k\rangle\langle t_k|+|\bar t_k\rangle\langle\bar t_k|\big)$, the \emph{minority part} of $\mathcal N$: it gives the particle and the hole of the true mode $k$ the same weight $p_k$, the probability that the mode is found in its minority (occupied) state. Then $\mathcal K\preceq\mathcal N$, $\mathcal N\Pi_p=\mathcal K\Pi_p$, $(I-\mathcal N)\Pi_h=\mathcal K\Pi_h$, and $\mathcal N\Pi_h\succeq\frac12\Pi_h$.
\item $\Omega_\mu:=\mathcal N+\mu I$ for $\mu>0$: the occupation with a floor $\mu$, the weight against which errors and frame occupations are measured at the scale $\mu$.
\end{itemize}

\subsection{Frames, the frame condition, relative error}\label{subsec:F-frames}

A frame $W$ (\S\ref{subsec:conventions}) acts on Nambu vectors by a Bogoliubov unitary, with particle vectors $f_a$ and holes $\bar f_a:=S\overline{f_a}$, $a=1,\dots,m$; its particle and hole projectors are $\hat\Pi_p$ and $\hat\Pi_h$. Define:
\begin{itemize}
\item $p'_a:=\langle f_a|\mathcal N|f_a\rangle$, the true occupation of the frame mode $a$, and $h_a:=\|\Pi_hf_a\|^2$;
\item $\mathcal Q_W:=\sum_ap'_a|f_a\rangle\langle f_a|$;
\item the pinching $\mathcal C_W(X):=\sum_a\big(\langle f_a|X|f_a\rangle\,|f_a\rangle\langle f_a|+\langle\bar f_a|X|\bar f_a\rangle\,|\bar f_a\rangle\langle\bar f_a|\big)$, the part of $X$ diagonal in the frame basis $\{f_a,\bar f_a\}$; $\mathcal C_W(\mathcal N)$ lists the frame occupations $p'_a$ and $1-p'_a$.
\end{itemize}

\begin{definition}[Frame condition and relative error]\label{def:G}
A frame $W$ satisfies the \emph{frame condition} $G(\mu,C)$ if
\[
\mathcal Q_W\preceq C\,\Omega_\mu\qquad\text{and}\qquad\hat\Pi_h\preceq C\,\Omega_\mu .
\]
A Nambu-symmetric estimate $\hat{\mathcal N}$ (i.e.\ $S\overline{\hat{\mathcal N}}S=I-\hat{\mathcal N}$) satisfies $\mathrm{REL}(\mu,\eta)$ if $-\eta\,\Omega_\mu\preceq\hat{\mathcal N}-\mathcal N\preceq\eta\,\Omega_\mu$.
\end{definition}

The computational frame satisfies $G(1,1)$, since $\mathcal Q_W\preceq I\preceq\Omega_1$ and $\hat\Pi_h\preceq I\preceq\Omega_1$.

For a unit vector $w$, $\langle w|\mathcal Q_W|w\rangle=\sum_ap'_a|\langle f_a|w\rangle|^2$ is the occupation that the frame modes place along $w$, and $\langle w|\Omega_\mu|w\rangle$ is the true occupation of $\mathrm{op}(w)$ plus the floor $\mu$. So $G(\mu,C)$ says that, above the floor, no direction carries more than $C$ times its true occupation in frame occupation or in frame-hole weight, and $\mathrm{REL}(\mu,\eta)$ bounds the error along $w$ by $\eta$ times the true occupation plus $\eta\mu$: the accuracy that a Pauli-blocked counting measurement delivers in a frame satisfying $G$ (Remark~\ref{F-rem-blocking}, Lemma~\ref{lem:R0}). Being Loewner inequalities, both are blind to rotations inside a cluster of nearly equal occupations.

\paragraph{The eigenframe $\hat W$ of a Nambu-symmetric $\hat{\mathcal N}$.}
Its particles are eigenvectors of $\hat{\mathcal N}$ with eigenvalues $\hat\lambda_a\le\frac12$, one from each particle--hole pair; its holes then have eigenvalue $1-\hat\lambda_a\ge\frac12$. If $\hat\lambda_a=\frac12$ exactly, the $\frac12$-eigenspace of $\hat{\mathcal N}$ is invariant under particle--hole conjugation; a particle--hole paired orthonormal basis of it is chosen by a fixed deterministic rule, so that the choice is a measurable function of $\hat{\mathcal N}$, and one vector of each pair is declared a particle. The $\hat\lambda_a$ of a raw estimate may be negative; nothing below uses $\hat\lambda_a\ge0$. The eigenframe may have $\det=-1$; it is then implemented with the conjugation by $c_1$ of \S\ref{subsec:conventions}.

\begin{remark}[Frames are particle--hole paired bases]\label{F-rem-bog}
(i) By \eqref{eq:bg-nambu-car} (\S\ref{subsec:bg-fermions}), $\tilde b_a:=\mathrm{op}(f_a)$, $a=1,\dots,m$, obey the canonical anticommutation relations if and only if $\{f_a\}\cup\{\bar f_a\}$, $\bar f_a=\vartheta f_a$, is an orthonormal basis of $\C^{2m}$; this is a frame in Nambu form, $\bar f_a$ represents $\tilde b_a^\dagger$, $p'_a=\langle\tilde b_a^\dagger\tilde b_a\rangle$, and $\langle\bar f_a|\mathcal N|\bar f_a\rangle=1-p'_a$.
(ii) Nambu symmetry of $\hat{\mathcal N}$ reads $\hat{\mathcal N}\vartheta=\vartheta(I-\hat{\mathcal N})$, so $\hat{\mathcal N}v=\lambda v$ implies $\hat{\mathcal N}\vartheta v=(1-\lambda)\vartheta v$: eigenvectors with $\lambda\ne\frac12$ come in orthogonal particle--hole pairs. The $\frac12$-eigenspace is $\vartheta$-invariant and of even dimension $2d$, since the other eigenvalues pair off. As $\vartheta$ is an antiunitary involution, the $\frac12$-eigenspace has an orthonormal basis $e_1,\dots,e_{2d}$ of $\vartheta$-fixed vectors (an orthonormal basis of the real subspace of fixed vectors, on which inner products are real), and $z_j:=(e_{2j-1}+ie_{2j})/\sqrt2$, $\vartheta z_j=(e_{2j-1}-ie_{2j})/\sqrt2$, $j\le d$, is a paired orthonormal basis of it. By (i) the eigenframe is a frame.
(iii) The tie-breaking rule only has to be deterministic: $W_r$ is then a function of the data of rounds $1,\dots,r$, which is what allows Lemma~\ref{lem:FR} to be applied conditionally on the past in the proof of Theorem~\ref{thm:Fprime}.
\end{remark}

\subsection{Relative error gives a good frame}\label{subsec:F-FS}

\begin{lemma}\label{lem:FS}
Let $\hat{\mathcal N}$ be Nambu-symmetric with $\mathrm{REL}(\mu,\eta)$, $\mu\in(0,1]$, $\eta\le\frac12$, and let $\hat W$ be its eigenframe. Put $C_\eta:=(1+\eta)/(1-\eta)$ ($\le3$; $=9/7$ at $\eta=1/8$). Then:
\begin{enumerate}
\item[(FS0)] $(\hat\lambda_a-\eta\mu)/(1+\eta)\le p'_a\le(\hat\lambda_a+\eta\mu)/(1-\eta)$ for every $a$;
\item[(FS1)] $\sum_a(p'_a+\mu)|f_a\rangle\langle f_a|+\frac12\hat\Pi_h\preceq(1-\eta)^{-1}(\hat{\mathcal N}+\mu)\preceq C_\eta\Omega_\mu$; hence $\hat W$ satisfies $G(\mu,2C_\eta)$, which at $\eta=1/8$ is $G(\mu,18/7)\subset G(\mu,3)$;
\end{enumerate}
\end{lemma}

\begin{proof}
(FS0): $p'_a=\hat\lambda_a-\langle f_a|E|f_a\rangle$ with $E:=\hat{\mathcal N}-\mathcal N$, and $|\langle f_a|E|f_a\rangle|\le\eta(p'_a+\mu)$; solve for $p'_a$.

(FS1): both operators in the first inequality are diagonal in the basis $\{f_a,\bar f_a\}$, so it suffices to compare coefficients. On $f_a$: $p'_a+\mu\le(\hat\lambda_a+\eta\mu)/(1-\eta)+\mu=(\hat\lambda_a+\mu)/(1-\eta)$, by (FS0), with no sign condition on $\hat\lambda_a$. On $\bar f_a$: $\frac12\le1-\hat\lambda_a\le(1-\hat\lambda_a+\mu)/(1-\eta)$, since $\hat{\mathcal N}\bar f_a=(1-\hat\lambda_a)\bar f_a$ by Nambu symmetry. For the second inequality, $\mathrm{REL}$ gives $\hat{\mathcal N}+\mu\preceq(1+\eta)\mathcal N+\eta\mu+\mu=(1+\eta)\Omega_\mu$. Finally $\mathcal Q_{\hat W}\preceq\sum_a(p'_a+\mu)|f_a\rangle\langle f_a|$ and $\hat\Pi_h\preceq2\cdot(\text{left side})$.
\end{proof}

Lemma~\ref{lem:FS} uses no eigenvalue gap and never $\hat{\mathcal N}\succeq0$: by (FS0) the eigenframe calls a mode cold only if it is cold, whether or not $f_a$ is close to a true normal mode; inside a cluster of nearly equal eigenvalues the Davis--Kahan bound (T2) would give nothing.

\subsection{Shots in a frame}\label{subsec:F-shots}

The notation is that of the Horvitz--Thompson shot \eqref{eq:HT} (\S\ref{subsec:pairrotation}), in the frame $W$, for even $m\ge4$ (odd $m$: \S\ref{subsec:thmB-assembly}); the Horvitz--Thompson patterns $E_{P,\sigma}$ of \eqref{eq:HT} are written $\mathcal E_{P,\sigma}$ here, since $E$ denotes an error.
\begin{itemize}
\item \emph{Horvitz--Thompson shot:} $Z:=\sum_{P\in\pi}4(m-1)\,c_{\sigma_P}s_Pn'_P\,\mathcal E_{P,\sigma_P}$. By Lemma~\ref{lem:A1} in the frame $W$, $\E Z=\mathcal N_{\mathrm{off}}:=\mathcal N-\mathcal C_W(\mathcal N)$.
\item \emph{Occupation shot:} $Z_{\mathrm{occ}}:=\sum_an_a\big(|f_a\rangle\langle f_a|-|\bar f_a\rangle\langle\bar f_a|\big)+\sum_a|\bar f_a\rangle\langle\bar f_a|$, with $n_a$ the measured occupation of the frame mode $a$; $\E Z_{\mathrm{occ}}=\mathcal C_W(\mathcal N)$, since $\E n_a=p'_a$ and $\langle\bar f_a|\mathcal N|\bar f_a\rangle=1-p'_a$ (Remark~\ref{F-rem-bog}).
\item The estimate $N^{-1}\sum_tZ_t+N'^{-1}\sum_tZ_{\mathrm{occ},t}$ is Nambu-symmetric and unbiased: each pattern satisfies $S\overline{\mathcal E_{P,\sigma}}S=-\mathcal E_{P,\sigma}$, being the derivative of the Nambu-symmetric $\mathcal N$ along one real parameter, and $\vartheta$ exchanges $|f_a\rangle\langle f_a|$ and $|\bar f_a\rangle\langle\bar f_a|$.
\end{itemize}

\subsection{Relative-error concentration}\label{subsec:F-FR}

The concentration rests on a per-shot variance bound in the frame, Lemma~\ref{lem:R0} of Appendix~\ref{app:proofs-F}, whose large terms are all weighted by frame occupations.

\begin{lemma}[Relative-error concentration]\label{lem:FR}
Let $W$ satisfy $G(\mu_0,C)$ with $C\le3$ and $\mu_0\le1$. Let $\mu\in[1/m,\mu_0]$ and $\eta\in(0,1]$, and put
\[
K_1:=1664C+3(1+2C)^2+12\ (=5151\text{ at }C=3),\qquad K_2:=10(C+C^2)\ (=120\text{ at }C=3).
\]
Suppose $n\ge640K_1m^2\mu_0/(\eta^2\mu^2)$ Horvitz--Thompson shots and $n'\ge640K_2m\mu_0/(\eta^2\mu^2)$ occupation shots are taken in $W$ with fresh copies, and put $\hat{\mathcal N}:=n^{-1}\sum_tZ_t+n'^{-1}\sum_tZ_{\mathrm{occ},t}$. Then $\Prob(\mathrm{REL}(\mu,\eta)\text{ fails}\mid\text{past})\le4e^{-m}$.
\end{lemma}

\subsection{Protocol F\texorpdfstring{$'$}{'} and the frame theorem}\label{subsec:F-protocol}

\paragraph{Protocol F$'$.}
\emph{Parameters.} $\eta:=1/8$ and $R:=\lceil\log_2\log_2m\rceil$; $\mu_0:=1$, $\mu_r:=\min\{1,4^{R-r}m^{2^{-r}}/m\}$ for $1\le r\le R$, and $\mu_{R+1}:=1/m$; $K_1:=K_1(C=3)=5151$ and $K_2:=K_2(C=3)=120$ with the formulas of Lemma~\ref{lem:FR}. These dominate the values for the computational frame ($C=1$) and for every later frame ($C=18/7$), by monotonicity in $C$.

\emph{Rounds.} $W_0:=$ the computational frame. Round $r=1,\dots,R+1$ uses fresh copies: in $W_{r-1}$ take
\[
N_r:=\Big\lceil\frac{640K_1m^2\mu_{r-1}}{\eta^2\mu_r^2}\Big\rceil\ \text{Horvitz--Thompson shots and}\quad N'_r:=\Big\lceil\frac{640K_2m\mu_{r-1}}{\eta^2\mu_r^2}\Big\rceil\ \text{occupation shots}
\]
(the shots of \S\ref{subsec:F-shots}), form $\hat{\mathcal N}_r:=N_r^{-1}\sum_tZ_t+N_r'^{-1}\sum_tZ_{\mathrm{occ},t}$, and let $W_r$ be the eigenframe of $\hat{\mathcal N}_r$ (\S\ref{subsec:F-frames}).

\emph{Output:} $W:=W_{R+1}$.

\begin{theorem}[The frame]\label{thm:Fprime}
Let $m\ge16$. With probability $\ge1-4(R+1)e^{-m}$ ($\ge0.99$), Protocol~F$'$ gives $\mathrm{REL}(\mu_r,1/8)$ in every round $r$, and every $W_r$ satisfies $G(\mu_r,18/7)$. In particular the output frame $W$ satisfies $G(1/m,18/7)\subset G(1/m,3)$. The total cost is
\[
\sum_r(N_r+N'_r)\le640\cdot64\cdot\tfrac{22}3\,(K_1m^3+K_2m^2)+2(R+1)\le1.6\cdot10^9m^3+3.7\cdot10^7m^2+2(R+1)=O(m^3),
\]
with no logarithm ($640\cdot64\cdot\frac{22}3\cdot5151=1.55\cdot10^9$).
\end{theorem}

In the computational frame, which satisfies only $G(1,1)$, $m^3$ copies give relative accuracy only at the scale $m^{-1/2}$ (Remark~\ref{rem:Fprime}(ii)). In a frame satisfying $G(\mu_0,C)$ the shots probing a direction of true occupation $p$ have variance of order $m(p+\mu_0)$, so relative accuracy at a finer scale $\mu$ costs $m^2\mu_0/\mu^2$ copies (Lemma~\ref{lem:FR}), and the new eigenframe satisfies $G(\mu,3)$ (Lemma~\ref{lem:FS}); the proof (Appendix~\ref{app:proofs-F}) is an induction over the rounds on these two lemmas. Keeping each round at $O(m^3)$ copies requires $\mu_{r-1}/\mu_r^2\lesssim m$: each round may at most halve the number of decades left to the floor $1/m$, and the factors $4^{R-r}$ of the schedule make the round budgets grow geometrically towards the last round, which dominates. What Theorem~\ref{thm:B} needs is extracted as Corollary~\ref{cor:Fprime} in Appendix~\ref{app:proofs-B}.

\begin{remark}\label{rem:Fprime}
(i) The $\log\log m$ adaptive rounds are free in copies, because the budgets decay geometrically backwards from the last round; the early rounds need relative accuracy only at the levels $\mu_r\gg1/m$. (ii) Lemma~\ref{lem:FR} with $\mu_0=1$ gives, from a single round of $Cm^3$ copies in the computational frame, $\mathrm{REL}$ at $\mu\asymp m^{-1/2}$; its count for the floor $\mu=1/m$ in one round from the computational frame is of order $m^4$. (iii) Any schedule with $\mu_{r-1}/\mu_r^2\le Cm$ over $R=O(\log\log m)$ rounds gives $O(m^3\log\log m)$ copies.
\end{remark}

\section{All mixed states: the main theorem}\label{sec:thmB}

\begin{theorem}[All mixed fermionic Gaussian states, single copy]\label{thm:B}
There are absolute constants $C_B$ and $m_B$ such that for all $m\ge m_B$ and $\eps\in(0,1]$ a single-copy adaptive protocol learns every fermionic Gaussian state on $m$ modes (any mixedness, pure modes allowed) to trace distance $\eps$ with probability $\ge2/3$ from
\[
N=C_B\big(m^3/\eps^2+m^3\big)\le2C_B\,m^3/\eps^2
\]
copies; in fact the output satisfies $\|\hat\rho-\rho\|_1\le\eps$, i.e.\ trace distance $\eps/2$. One may take $m_B=17$. By \cite[Thm.~1.1]{Rub26}, every single-copy adaptive protocol (no quantum memory, arbitrary output) that reaches trace distance $\eps$ with probability $2/3$ on this class needs $N>m^3/(8\,553\,600\,\eps^2)$ for $0<\eps\le1/3600$, for the same loss, success probability and measurement model. Hence the single-copy sample complexity of mixed fermionic Gaussian tomography is $\Theta(m^3/\eps^2)$ for all $0<\eps\le1/3600$, uniformly in the covariance spectrum, with no logarithmic factor and no condition relating $\eps$ to $m$; this resolves Conjecture~10.1 of \cite{Rub26}. (No lower bound is claimed for $\eps>1/3600$.)
\end{theorem}
For $m<17$, adjoin $17-m$ vacuum modes to every copy (a known ancilla, so the measurement is still single-copy), run the protocol on the Gaussian state $\rho\otimes|0\rangle\langle0|^{\otimes(17-m)}$, which Theorem~\ref{thm:B} admits since pure modes are allowed, and output the reduced state of the estimate on the first $m$ modes: it is Gaussian, and its trace distance to $\rho$ is at most that of the padded estimate, because the trace distance does not increase under the partial trace. Hence $N\le2\cdot17^3C_B/\eps^2$ copies suffice for $m<17$, and, with the theorem, $O(m^3/\eps^2)$ copies suffice for every $m\ge1$.

Protocol~F$'$ builds a frame from $O(m^3)$ copies (Theorem~\ref{thm:Fprime}), and Stage~1 (\S\ref{subsec:thmB-protocol}) runs the pair-rotation measurement in that frame on $O(m^3/\eps^2)$ fresh copies and clips the eigenvalues of the assembled covariance. In that frame the readout of a pair $(k,l)$ has noise proportional to $\max(p_k,1/m)+\max(p_l,1/m)$ (Lemma~\ref{lem:B1}), so the matching of Theorem~\ref{thm:A} holds pair by pair, at every scale at once. The analysis (Appendix~\ref{app:proofs-B}) sorts the modes into dyadic occupation bands and a cold group with $p\le1/m$. Between bands far apart an error mainly rotates the eigenbasis and is absorbed into a Gaussian unitary, whose generator is the error divided by the occupation gap and which Lemma~\ref{lem:MT} charges with the weight $1-\lambda_k\lambda_l\approx2(p_k+p_l)$, again scale-free; this rotation is a Schrieffer--Wolff transformation carried to all orders, found as a fixed point in a norm weighted band by band (Lemma~\ref{lem:L1}). Between neighbouring bands the error is compared directly in $\chi^2$ (Lemma~\ref{lem:W}), and in the cold group by $\|\rho_\Gamma-\rho_{\Gamma'}\|_1\le\frac12\|\Gamma-\Gamma'\|_1$ \cite{BMEL25}, whose loss of a factor $\sqrt m$ is paid for by the noise of order $1/m$ per entry. No logarithm survives: the frame's $\log\log m$ rounds are free in copies, a $\frac14$-net whose exponent $15m$ beats its entropy $13.2m$ replaces matrix Bernstein, and the triangular truncation in scale, whose norm grows like $\log\log m$ in a whole-matrix norm \cite{KP70} (Proposition~\ref{prop:obstruction}), costs a constant in the band-weighted norm (\S\ref{subsec:thmB-nolog}).

\begin{remark}[The explicit count]\label{rem:B-count}
The proof (\S\ref{subsec:thmB-assembly}) gives $N=N_{F'}+N_2+N_3$ with
\[
N_{F'}\le1.6\cdot10^9m^3+3.7\cdot10^7m^2+2(R+1),\qquad
N_2=C_N\frac{m^3}{\eps^2}+K_\triangle\frac{m^3}{\tau_0^2},\qquad
N_3=C_N\frac m{\eps^2}+K'_\triangle\frac{m^2}{\tau_0^2},
\]
where $K_\triangle=960\cdot2.62\cdot10^5\le2.6\cdot10^8$ and $K'_\triangle=960\cdot1.44\cdot10^4\le1.4\cdot10^7$ (Lemma~\ref{lem:Htri}); $\tau_0:=e_0/\sqrt5$, with $e_0:=\min\{1/5760,\,1/(34\,304\,C_D^2)\}=1/2\,195\,456\approx4.55\cdot10^{-7}$ the absolute tolerance of the far decomposition at the dictionary constant $C_D=8$ \eqref{eq:NF4}; and $C_N:=\max\{C_{\mathrm I},C_{\mathrm{II}},C_{\mathrm{III}},C_\star'\}$ (each taken as the larger of its even-$m$ and odd-$m$ values; for odd $m$ they grow by at most a factor $2$, \S\ref{subsec:thmB-protocol}) is the absolute constant of the second-moment legs, $O(C_1')$ with $C_1'\le10^6$ (Lemma~\ref{lem:B1}), not computed explicitly here. The $\eps$-free term in front of $m^3$ is $K_\triangle/\tau_0^2=5K_\triangle/e_0^2\le6.3\cdot10^{21}$ ($6.06\cdot10^{21}$ with the unrounded $K_\triangle=2.5152\cdot10^8$), too large for a practical sample guarantee, but independent of dimension, accuracy and spectrum; the one in front of $m^2$ is $\le3.4\cdot10^{20}$. Putting the $m/\eps^2$ term of $N_3$ under $m^3/\eps^2$, and the Stage-F$'$ terms and the $m^2$ terms under $m^3$, one may take $C_B:=\max\{2C_N,\,6.7\cdot10^{21}\}$. The constants are not optimised. The threshold $m_B=17$ serves every condition used: $J\ge5$; $m\ge16$ (Theorem~\ref{thm:Fprime} and the union bound of Lemma~\ref{lem:Htri}); and $4(R+1)e^{-m}\le0.01$.
\end{remark}

The $\eps$-free term of Remark~\ref{rem:B-count} is a localisation cost: the perturbative steps are proved only once the estimate is within a relative error $e_0\approx4.55\cdot10^{-7}$ of the state, band by band, and reaching a fixed relative accuracy costs a multiple of $m^3$ copies, which changes the constant, not the rate. The constant is a product of loose factors that enter squared: $e_0^{-1}=536\cdot64\,C_D^2$ collects the commutator constant $\kappa_{\mathrm M}\le67$, the radius $8$ of the fixed-point ball, a margin $64$ and the dictionary constant $C_D^2=64$, and $K_\triangle=960K_5$ the per-shot variance constant $K_5\approx2.6\cdot10^5$ of \eqref{eq:R0} and the factor $960=16\cdot4\cdot15$ from the Bernstein deviation $\tau/4$ and the net exponent $15m$.

\subsection{Protocol}\label{subsec:thmB-protocol}

\emph{Stage~F$'$.} Protocol~F$'$ (\S\ref{subsec:F-protocol}) produces the frame $W$.

\emph{Stage~1} (in the frame $W$, fresh copies): $N_2$ Horvitz--Thompson pair-rotation shots (Lemma~\ref{lem:A1}): a random perfect matching $\pi$ of the $m$ frame modes, per matched pair $P=(a,b)$ a label $\sigma\in\{\mathrm{hr},\mathrm{hi},\mathrm{pr},\mathrm{pi}\}$ and a sign $s$, the two-mode $\pm\pi/4$ rotation, occupation readout $n'_P\in\{0,1\}$; the per-shot matrix $Y_t$ carries the block
\[
4(m-1)\,c_\sigma X_P\,B_{P,\sigma}=8(m-1)\,c_\sigma s_Pn'_P\,B_{P,\sigma}
\]
at $(a,b)$ and its negative transpose at $(b,a)$, where $X_P:=2s_Pn'_P$ is the sample of Lemma~\ref{lem:A1}, whose mean is the covariance entry ($2\,\mathrm{Re}N'_{ab}$, etc.), $4(m-1)$ is the Horvitz--Thompson inverse of the probability $1/(4(m-1))$ of a given (pair, label), $c_\sigma$ is the sign in the Horvitz--Thompson estimator \eqref{eq:HT} (\S\ref{subsec:pairrotation}), and $B_{P,\sigma}\in\{J,-I,\sigma_x,-\sigma_z\}$; thus $\E Y_t=\Gamma'_{\mathrm{off}}$. (For odd $m$ one mode is left out of the matching, with probability $1/(4m)$ and weight $4m$ for a given (pair, label), as in \S\ref{subsec:pairrotation}; all constants below are for even $m$ and change by at most a factor $2$ otherwise.) Further, $N_3$ occupation shots read all frame occupations; $\tilde\Gamma':=N_2^{-1}\sum_tY_t+\mathrm{diag}(2\hat p'_a-1)J$. Output $\hat\Gamma':=\operatorname{Clip}(\tilde\Gamma')$ (the eigenvalues of $i\tilde\Gamma'$ clipped to $[-1,1]$); return $\rho_{W\hat\Gamma'W^T}$. Bands are an analysis device; nothing in Stage~1 uses them.

\section{Blocks of \texorpdfstring{$k$}{k} copies: Frobenius estimation on all states, trace distance on every fixed-gap class}\label{sec:kcopy}

The conventions are those of \S\ref{subsec:bg-fermions}, on $S:=(\C^2)^{\otimes m}$, with $B_{ab}:=ic_ac_b$ ($a<b$), $\Gamma_{ab}=\Tr(\rho B_{ab})$, $\|X\|_F^2=\sum_{a,b}X_{ab}^2=2\sum_{a<b}X_{ab}^2$ and $p:=m(2m-1)$; a $k$-copy protocol measures blocks of at most $k$ fresh copies with arbitrary joint POVMs and keeps classical memory only \cite[Sec.~2]{Rub26b}. Let $\delta:=(m-1,m-2,\dots,1,0)\in\R^m$ and, for $\xi\in\R^m$, let $X_\xi:=\bigoplus_{j=1}^m\xi_jJ$ with $J=\begin{psmallmatrix}0&1\\-1&0\end{psmallmatrix}$, the real antisymmetric matrix with $(X_\xi)_{2j-1,2j}=\xi_j$. The remaining notation is set in \S\ref{subsec:K-setting}.

\begin{theorem}[the $k$-copy rate of fermionic Gaussian states]\label{thm:K}
Fix $m\ge1$ and a block size $t\ge1$. There is a POVM on $S^{\otimes t}$ (the covariant highest-weight POVM of \cite[Thm.~3.6]{Rub26b}, \S\ref{subsec:K-povm}) with outcomes $(\lambda,i,O)$, $\lambda\in\mathbb Z^m$ a highest weight of $S^{\otimes t}$ and $O\in SO(2m)$, such that the estimator (the \emph{Weyl-shift estimator})
\begin{equation}\label{eq:K-est}
\widehat\Gamma(\lambda,i,O):=\frac1t\,O\,X_{\lambda+4\delta}\,O^{T},\qquad\text{i.e.}\quad\widehat\Gamma_{2j-1,2j}=\frac{\lambda_j+4(m-j)}{t}\ \text{ in the frame }O,
\end{equation}
has the following properties for every state $\rho$ on $m$ modes (Gaussian or not, mixed or pure).
\begin{enumerate}
\item[(i)] (Exact unbiasedness.) $\E_{\rho^{\otimes t}}[\widehat\Gamma]=\Gamma$, where $\Gamma_{ab}=\Tr(\rho B_{ab})$.
\item[(ii)] (Deterministic second moment.) $\sum_{a<b}\widehat\Gamma_{ab}^2=\|\lambda+4\delta\|^2/t^2\le V(m,t)/t^2$ for every outcome, where
\begin{equation}\label{eq:K-V}
V(m,t):=\sum_{i=0}^{m-1}(t+4i)^2\ \le\ mt^2+4m^2t+\tfrac{16}{3}m^3,\qquad V(m,t)\le11m^3\ \text{ for }t\le m .
\end{equation}
\end{enumerate}
Consequently, with $N=tB$ copies in $B$ blocks and $\bar\Gamma$ the average of the block estimates:
\begin{enumerate}
\item[(iii)] (Frobenius rate on all states.) $\E\|\bar\Gamma-\Gamma\|_F^2\le 2V(m,t)/(tN)$, and $N\ge6V(m,t)/(t\eps^2)$ copies give $\|\bar\Gamma-\Gamma\|_F\le\eps$ with probability $\ge2/3$. With $t=\min\{k,m\}$, $0<\eps\le1$ and $N=t\lceil6V(m,t)/(t^2\eps^2)\rceil$: $N\le66\max\{m^3/k,m^2\}/\eps^2+m\le67\max\{m^3/k,\,m^2\}/\eps^2$. This is optimal: for $0<\eps\le1/3600$ every $k$-copy protocol that outputs $\widehat\Gamma$ with $\|\widehat\Gamma-\Gamma\|_F\le\eps$ with probability $\ge2/3$ on all mixed fermionic Gaussian states uses $N=\Omega(\max\{m^3/k,m^2\}/\eps^2)$ copies. Hence the $k$-copy sample complexity of estimating the covariance matrix of a fermionic Gaussian state in Frobenius norm is $\Theta(\max\{m^3/k,m^2\}/\eps^2)$, uniformly over the whole class (pure states included), for all $k\ge1$ and $\eps\le1/3600$.
\item[(iv)] (Trace-distance rate on the fixed-gap classes.) Let $0<c<1$ and $\Gamma\in\mathcal G_c$. Let $\tilde\Gamma$ be the Frobenius projection of $\bar\Gamma$ onto $\mathcal G_c$ and output $\rho_{\tilde\Gamma}$. Then $N\ge3V(m,t)/(tc\,\eps^2)$ copies give $\|\rho_{\tilde\Gamma}-\rho_\Gamma\|_1\le\eps$ (trace distance $\le\eps/2$) with probability $\ge2/3$; with $t=\min\{k,m\}$, $0<\eps\le1$ and $N=t\lceil3V(m,t)/(t^2c\,\eps^2)\rceil$, $N\le34\max\{m^3/k,m^2\}/(c\,\eps^2)$. Matching lower bound \cite[Thms.~3.4 and~3.11]{Rub26b}: for $0<\eps\le\eps_0(c):=\min\{1/3600,\ \operatorname{artanh}(1-c)/(120\sqrt3)\}$ every $k$-copy protocol learning $\mathcal G_c$ to trace distance $\eps$ with probability $\ge2/3$ uses $N>\max\{m^3/(8{,}553{,}600\,k),\,m^2/4{,}276{,}800\}/\eps^2$. Hence on every fixed-gap class $\mathcal G_c$,
\[
N_{k\text{-copy}}(\mathcal G_c,\eps)=\Theta_c\Big(\frac{\max\{m^3/k,\,m^2\}}{\eps^2}\Big)\qquad(k\ge1,\ \eps\le\eps_0(c)),
\]
with the upper constant $34/c$ and the lower constant absolute.
\item[(v)] (The unrestricted mixed class.) Combining (iv) with \cite[Thm.~5.5]{Rub26b} (the near-vacuum number-conserving states $\rho^{\rm F}_w$, $w\preceq2I/m$, are mixed fermionic Gaussian states and force $N=\Omega(m^3/(\eps^2\min\{\sqrt k,m\}))$) and Theorem~\ref{thm:B} (single copies, $N=O(m^3/\eps^2)$): the $k$-copy trace-distance rate of the unrestricted mixed class lies in
\[
\big[\Omega\big(m^3/(\eps^2\min\{\sqrt k,m\})\big),\ O\big(m^3/\eps^2\big)\big],
\]
and the $k$-dependence is provably not uniform over the class: it is $1/k$ on every fixed-gap class by (iv) and $1/\sqrt k$ on the cold number-conserving classes by \cite[Thm.~5.9]{Rub26b}. By (iii) the covariance itself is learned in Frobenius norm at the rate $\max\{m^3/k,m^2\}/\eps^2$ on all states; the difference between the subclasses therefore lies in the conversion from Frobenius error to trace distance, which is uniform on $\mathcal G_c$ and has no dimension-free form near pure states.
\end{enumerate}
\end{theorem}

On $t$ copies the quadratic Majorana operators add up to collective generators, as spins add up to a total angular momentum. The measurement finds the irreducible block, labelled by a highest weight $\lambda$ (the analogue of the total spin), and makes a coherent-state measurement inside it; at $t=1$ it is the measurement of matchgate shadows \cite{ZRM21,WHLB23}. The naive readout $OX_\lambda O^T/t$ is biased because coherent states overlap, and the exact correction is the shift $j^2\to j(j+1)$ of the Casimir, $\lambda\mapsto\lambda+2\varrho$ (Proposition~\ref{prop:K4}). In the bound (ii) the offset $\|2\varrho\|^2\approx\frac{16}3m^3$, the noise of the coherent-state measurement itself, does not grow with $t$, so shared among the copies of a block it gives the factor $1/t$, until at $t\approx m$ it reaches the collective floor. This attains the block Fisher budget of \cite{Rub26b} up to a constant (\S\ref{subsec:K-why}), and answers \cite[Question~7.2]{Rub26b} affirmatively on every fixed-gap class and for every $k$, which proves the conjecture at the end of \cite[Rem.~3.9]{Rub26b}; part (iv) answers the fixed-gap case of the first part of \cite[Question~7.1]{Rub26b} and fills the ``Blocks of $\le k$ copies'' cell of the $\mathcal G_c$ row of \cite[Table~1]{Rub26b}. \cite[Question~7.3]{Rub26b}, on the crossover in $c$, is answered for single copies by Theorem~\ref{thm:B}; its $k$-copy form is open (Section~\ref{sec:open}).

\subsection{Setting and conventions}\label{subsec:K-setting}

On $S$, $\{c_a,c_b\}=2\delta_{ab}$. For $a<b$, $B_{ab}=ic_ac_b$ is Hermitian, traceless, $B_{ab}^2=1$; set $B_{ba}:=-B_{ab}$. $\mathfrak{so}(2m)$ denotes the real antisymmetric $2m\times2m$ matrices with the inner product $\langle X,Y\rangle:=\sum_{a<b}X_{ab}Y_{ab}=\tfrac12\Tr(X^TY)$; $E_{ab}:=e_ae_b^T-e_be_a^T$ ($a<b$) is an orthonormal basis. For $X\in\mathfrak{so}(2m)$ put $B(X):=\sum_{a<b}X_{ab}B_{ab}$, so $B(E_{ab})=B_{ab}$ and $\Tr(\rho B(X))=\langle\Gamma,X\rangle$. The Cartan subalgebra is $\mathfrak t:=\mathrm{span}_{\R}\{T_j:=E_{2j-1,2j}\}_{j=1}^m\cong\R^m$, with $X_\xi=\sum_j\xi_jT_j$ for $\xi\in\R^m$ and $\langle X_\xi,X_\eta\rangle=\xi\cdot\eta$.

\paragraph{Gaussian unitaries.} For $O\in SO(2m)$ there is a unitary $U_O$ on $S$, unique up to a phase (up to sign if taken in the image of $\mathrm{Spin}(2m)$, e.g.\ $U_{e^{X}}=e^{B(X)/(2i)}$), with $U_Oc_aU_O^\dagger=\sum_bO_{ba}c_b$; then $U_OB(X)U_O^\dagger=B(OXO^T)=:B(\mathrm{Ad}_OX)$, $\mathrm{Ad}_O$ is orthogonal for $\langle\cdot,\cdot\rangle$, and the covariance of $U_O\rho U_O^\dagger$ is $O\Gamma O^T$. The map $O\mapsto U_O|\psi\rangle\langle\psi|U_O^\dagger$ is well defined and continuous, so integrals $\int(\cdot)\,dO$ over the Haar probability measure of $SO(2m)$ of such expressions make sense.
(A derivation of these facts is in Appendix~\ref{app:proofs-K}.)

\paragraph{\texorpdfstring{$t$}{t} copies.} On $S^{\otimes t}$ let $B^{(c)}_{ab}:=I^{\otimes(c-1)}\otimes B_{ab}\otimes I^{\otimes(t-c)}$ \cite[Lemma~3.1]{Rub26b}, $G_{ab}:=\sum_{c=1}^tB^{(c)}_{ab}$, $G(X):=\sum_{a<b}X_{ab}G_{ab}=\sum_cB^{(c)}(X)$, and $H_j:=G(T_j)=G_{2j-1,2j}$. Then $U_O^{\otimes t}G(X)U_O^{\dagger\otimes t}=G(\mathrm{Ad}_OX)$, and $\Tr(\rho^{\otimes t}G(X))=t\langle\Gamma,X\rangle$ for every state $\rho$ (each $B^{(c)}(X)$ acts on one factor). We write $U_O$ also for $U_O^{\otimes t}$.

\paragraph{Lie algebra structure.} $\pi(X):=G(X)/(2i)$ is a Lie algebra homomorphism $\mathfrak{so}(2m)\to\mathfrak u(S^{\otimes t})$, $[\pi(X),\pi(Y)]=\pi([X,Y])$, with $U_{e^{sX}}=e^{s\pi(X)}$ up to sign; equivalently $[G(Y),G(X)]=2i\,G([Y,X])$. (At $t=1$, $\pi(X)=\tfrac14\sum_{a,b}X_{ab}c_ac_b$, and $[\pi(X),\pi(Y)]=\pi([X,Y])$ is a direct computation from $\{c_a,c_b\}=2\delta_{ab}$: the quadratic Majorana operators represent $\mathfrak{so}(2m)$, which is the statement that the Gaussian unitaries act on the Majoranas by $SO(2m)$ rotations, as recalled in \cite[Sec.~2]{Rub26}; for general $t$ it follows by summing over the commuting factors.) $S^{\otimes t}=\bigoplus_\lambda V_\lambda\otimes\C^{m_\lambda}$ decomposes into irreducible $\mathfrak{so}(2m)$-modules; fixing an orthonormal basis of each multiplicity space gives irreducible subspaces $V_{\lambda,i}$, $1\le i\le m_\lambda$, with orthogonal projectors $\Pi_{\lambda,i}$, $\sum_{\lambda,i}\Pi_{\lambda,i}=I$, each $V_{\lambda,i}$ invariant under all $G(X)$ and all $U_O$. The $H_j$ commute and are diagonalisable with integer eigenvalues; a weight is a joint eigenvalue vector $\mu\in\mathbb Z^m$, $H_jw=\mu_jw$. The weights of $S$ are $\{\pm1\}^m$, so every weight of $S^{\otimes t}$ has $|\mu_j|\le t$. The roots, in these coordinates, are the weight shifts of the adjoint action: $\Phi=\{\pm2e_i\pm2e_j:i<j\}$ (the vector representation $\Lambda^1\subset S\otimes S$ has weights $\pm2e_j$). For $\beta\in\Phi$ there is a root vector $E_\beta\in\mathfrak{so}(2m)_{\C}$, unique up to scale, with $[H_j,G(E_\beta)]=\beta_jG(E_\beta)$, equivalently $[T_j,E_\beta]=\tfrac{\beta_j}{2i}E_\beta$; $\mathfrak{so}(2m)_{\C}=\mathfrak t_{\C}\oplus\bigoplus_{\beta\in\Phi}\C E_\beta$, and the complex-bilinear extension of $\langle\cdot,\cdot\rangle$ is $\mathrm{Ad}$-invariant and satisfies $\langle E_\beta,E_\gamma\rangle=0$ unless $\gamma=-\beta$. Normalise $\langle E_\beta,E_{-\beta}\rangle=1$. The positive roots are $\Phi^+:=\{2e_i\pm2e_j:i<j\}$, and we write $\varrho$ for the Weyl vector (to distinguish it from a state $\rho$):
\[
2\varrho:=\sum_{\beta\in\Phi^+}\beta=\sum_{i<j}4e_i=4\delta,\qquad\delta=(m-1,\dots,0).
\]
Each $V_{\lambda,i}$ has a highest weight vector $v=v_{\lambda,i}$ (unit): $H_jv=\lambda_jv$, $G(E_\beta)v=0$ for $\beta\in\Phi^+$; $\lambda$ is the highest weight, $\lambda_1\ge\dots\ge\lambda_{m-1}\ge|\lambda_m|$, $|\lambda_j|\le t$, and every weight $\mu$ of $V_\lambda$ satisfies $\lambda-\mu\in\mathbb N\Phi^+$; in particular $\lambda+\beta$ is not a weight of $V_\lambda$ for $\beta\in\Phi^+$. Write $n_\lambda=\dim V_\lambda$. We use these standard facts and nothing else from representation theory.

\paragraph{Casimir.} The Casimir $C_2=\sum_{a<b}G_{ab}^2$ acts on $V_\lambda$ as $C_2(\lambda):=\langle\lambda,\lambda+2\varrho\rangle=\|\lambda\|^2+4\lambda\cdot\delta$ (\S\ref{subsec:bg-reptheory}); the proof of Proposition~\ref{prop:K4} uses its generalisation, Lemma~\ref{lem:K3}, rather than this value.

\subsection{The measurement and the estimator}\label{subsec:K-povm}

\paragraph{The POVM.} This is the attainment part of \cite[Thm.~3.6]{Rub26b}. By Schur's lemma
\[
n_\lambda\int U_O|v_{\lambda,i}\rangle\langle v_{\lambda,i}|U_O^\dagger\,dO=\Pi_{\lambda,i},
\]
so
\begin{equation}\label{eq:K-povm}
M_{\lambda,i,O}\,dO:=n_\lambda\,U_O|v_{\lambda,i}\rangle\langle v_{\lambda,i}|U_O^\dagger\,dO,\qquad\sum_{\lambda,i}\int M_{\lambda,i,O}\,dO=I,
\end{equation}
is a POVM on $S^{\otimes t}$ with outcome space $\{(\lambda,i,O)\}$. Its outcome law on a state $\sigma$ of $S^{\otimes t}$ has density $n_\lambda\langle v_{\lambda,i}|U_O^\dagger\sigma U_O|v_{\lambda,i}\rangle$ with respect to counting measure in $(\lambda,i)$ times Haar measure in $O$.

At $t=1$ every Fock state of a given parity is a Gaussian rotation of the highest weight vector of its sector $S_\pm$, so \eqref{eq:K-povm} has the outcome law of a Haar-random Gaussian unitary followed by a readout of all occupations: the measurement of matchgate classical shadows \cite{ZRM21,WHLB23}, whose random Gaussian Clifford has the same first three moments \cite{WHLB23,HCT25}. For $t\ge2$ it is the collective version, a measurement of $\lambda$ followed by a coherent-state measurement in that component, and a random rotation followed by a fixed basis does not in general realise it. A Haar-random rotation $U_g^{\otimes t}$ followed by a finite POVM does, if every element of the latter is a positive multiple $w_a\,U_{h_a}|v_{\lambda,i}\rangle\langle v_{\lambda,i}|U_{h_a}^\dagger$ of a coherent state in one $V_{\lambda,i}$ and the outcome is recorded as $(\lambda,i,gh_a)$: completeness on $V_{\lambda,i}$ gives $\sum_aw_a=n_\lambda$ there, and the substitution $O=gh_a$ in the Haar integral gives exactly the outcome law of \eqref{eq:K-povm}, so every readout below keeps its law. At $m=t=2$ the parities of the two copies split $S^{\otimes2}$ into $S_+\otimes S_+$ and $S_-\otimes S_-$, each a triplet and a singlet (whose one vector is trivially coherent), and $S_+\otimes S_-$ and $S_-\otimes S_+$, each a copy of the four-dimensional vector representation, whose weight vectors are coherent states; on a triplet, a spin-one representation of one factor of $\mathrm{Spin}(4)\cong SU(2)\times SU(2)$, the four spin-coherent states along the vertices of a regular tetrahedron, with weights $\tfrac34$, form such a POVM, the tetrahedron being a spherical $2$-design. Without the random rotation this finite POVM is not covariant, and the readout $2\mathbf n_a$ of the tetrahedral outcome $\mathbf n_a$ is biased on the triplet: on the spin-coherent state along $(1,1,0)/\sqrt2$ its $z$-component has mean $1/(2\sqrt3)$ when the vertices are $(1,1,1)/\sqrt3$ and its images under sign changes of two coordinates, while $\langle J_z\rangle=0$.

\paragraph{The estimator.} $\widehat\Gamma(\lambda,i,O):=\tfrac1t\,OX_{\lambda+2\varrho}O^T=\tfrac1t\mathrm{Ad}_OX_{\lambda+4\delta}$; entrywise $\widehat\Gamma_{ab}=\tfrac1t\langle\mathrm{Ad}_OX_{\lambda+2\varrho},E_{ab}\rangle$. It does not depend on $i$ or on the choice of bases of the multiplicity spaces. The block protocol measures $B$ independent blocks of $t$ copies and averages. The shift by the Weyl vector makes it exactly unbiased on every state; Remark~\ref{rem:K-unshifted} compares it with the unshifted moment map.
For an operator $A$ on $V=V_{\lambda,i}$, the Q-symbol is $O\mapsto\langle v|U_O^\dagger AU_O|v\rangle$, and a P-symbol is a function $f$ with $A=n_\lambda\int f(O)\,U_O|v\rangle\langle v|U_O^\dagger\,dO$ \cite{Ber75,Per86}; for a state $\sigma$ on $V$, $\Tr(\sigma A)=\E_\sigma[f(O)]$ under the coherent-state POVM, so the P-symbol is an unbiased readout. The unshifted $\lambda_j/t$ is the Q-symbol of $G/t$, shrunk towards zero by $\|\lambda\|^2/C_2(\lambda)$ along $d\pi$ (Lemmas~\ref{lem:K2}(b) and~\ref{lem:K3}), as the outcome $\mathbf n$ of one qubit measured along a Haar-random direction has $\E[\mathbf n]=\mathbf r/3$ and unbiased shadow $3\mathbf n$ \cite{HKP20}, $3=(j+1)/j$ at $j=\tfrac12$. As the shrinking depends on $\lambda$, no global rescaling is exact on every state (Remark~\ref{rem:K-unshifted}).

\subsection{The P-symbol identity}\label{subsec:K-lemmas}

Fix one irreducible subspace $V=V_{\lambda,i}$, $n=n_\lambda$, $v=v_{\lambda,i}$, $\Pi=\Pi_{\lambda,i}$, and write $d\pi(X):=\Pi G(X)\Pi|_V\in\mathrm{Herm}(V)$ (the Hermitian generators restricted to $V$). Let $\mathcal E$ be the real vector space of linear maps $\varphi:\mathfrak{so}(2m)\to\mathrm{Herm}(V)$ that are equivariant: $U_O\varphi(X)U_O^\dagger=\varphi(\mathrm{Ad}_OX)$ for all $O\in SO(2m)$. Then $d\pi\in\mathcal E$. Differentiating along $O=e^{sY}$ gives the infinitesimal form $[G(Y),\varphi(X)]=2i\,\varphi([Y,X])$ for all $X,Y\in\mathfrak{so}(2m)$ (as for $\varphi=d\pi$). Extend $\varphi$ complex-linearly to $\mathfrak{so}(2m)_{\C}$; then $[H_j,\varphi(E_\beta)]=2i\varphi([T_j,E_\beta])=\beta_j\varphi(E_\beta)$: $\varphi(E_\beta)$ shifts weights by $\beta$. On $\mathcal E$ put $\langle\varphi,\psi\rangle_{\mathcal E}:=\sum_{a<b}\Tr(\varphi(E_{ab})\psi(E_{ab}))$, a positive definite inner product ($\varphi(E_{ab})$ Hermitian). Define the \emph{symbol} of $\varphi\in\mathcal E$ as $\xi_\varphi\in\R^m$, $(\xi_\varphi)_j:=\langle v|\varphi(T_j)|v\rangle$; so $\xi_{d\pi}=\lambda$.

For $\xi\in\R^m$ let $Y_\xi(X):=n\int\langle\mathrm{Ad}_OX_\xi,X\rangle\,U_O|v\rangle\langle v|U_O^\dagger\,dO$ ($X\in\mathfrak{so}(2m)$), the coherent-state twirl of the readout $X_\xi$; it lies in $\mathcal E$ (Lemma~\ref{lem:K2}).

For $SU(2)$ the Wigner--Eckart theorem makes every vector operator on a spin-$j$ block proportional to $\mathbf J$; for $\mathrm{Spin}(2m)$ the adjoint representation can occur several times in the operators on $V_\lambda$, so the proof (Appendix~\ref{app:proofs-K}) works with all of $\mathcal E$ at once. The symbol determines every expectation of $\varphi$ in $v$, since root operators move $v$ to orthogonal weights (Lemma~\ref{lem:K1}); the twirl $Y_\xi$ pairs with every $\varphi$ as $n\,\xi\cdot\xi_\varphi$ (Lemma~\ref{lem:K2}); and the generators pair with every $\varphi$ as $n\,(\lambda+2\varrho)\cdot\xi_\varphi$, the Casimir computation with one factor replaced by $\varphi$, in which moving lowering past raising operators produces $\sum_{\beta\in\Phi^+}\beta=2\varrho$, as $\mathbf J^2=J_z^2+J_z+J_-J_+$ gives $j^2+j$ on $|j,j\rangle$ (Lemma~\ref{lem:K3}). Hence $Y_{\lambda+2\varrho}-d\pi$ is orthogonal to all of $\mathcal E$, itself included, and vanishes:
\begin{proposition}[P-symbol identity]\label{prop:K4}
$Y_{\lambda+2\varrho}=d\pi$ on $V$. Explicitly, for every $X\in\mathfrak{so}(2m)$,
\begin{equation}\label{eq:K-Psymbol}
n_\lambda\int\langle\mathrm{Ad}_OX_{\lambda+4\delta},X\rangle\,U_O|v_{\lambda,i}\rangle\langle v_{\lambda,i}|U_O^\dagger\,dO=\Pi_{\lambda,i}\,G(X)\,\Pi_{\lambda,i}.
\end{equation}
More generally $Y_\xi=d\pi$ for every $\xi$ with $(\xi-\lambda-2\varrho)\cdot\xi_\varphi=0$ for all $\varphi\in\mathcal E$.
\end{proposition}
Summing \eqref{eq:K-Psymbol} over the blocks gives part (i) for every input on $S^{\otimes t}$, Gaussian or not; part (ii) is Parseval for the rotated diagonal matrix $X_{\lambda+2\varrho}$ (both in Appendix~\ref{app:proofs-K}). Parts (iii) and (iv) are averaging, Markov's inequality, the non-expansive projection onto $\mathcal G_c$ and the conversion $\|\rho_\Gamma-\rho_{\Gamma'}\|_1\le(2c)^{-1/2}\|\Gamma-\Gamma'\|_F$ of \cite[Prop.~7.1]{Rub26}. The lower bounds are reductions to \cite[Thms.~3.4 and~3.11]{Rub26b}: a Frobenius learner, projected onto $\mathcal G_{1/2}$, is a trace-distance learner on their hard family. Part (v) combines (iv) with the near-vacuum family of \cite[Thm.~5.5]{Rub26b} and with Theorem~\ref{thm:B}.

\begin{remark}[the $SU(2)$ case]\label{rem:K-SU2}
For $SU(2)$ and spin-$j$ coherent states the Q-symbol of $J_z$ is $j\cos\theta$ and its P-symbol is $(j+1)\cos\theta$ \cite{Lie73,Per86}; $\lambda\mapsto\lambda+2\varrho$ is the compact-group form of the shift $j\mapsto j+1$ (in our doubled coordinates $\lambda=2j$, $2\varrho=2$). Proposition~\ref{prop:K4} is this statement for the spinor tensor powers of $\mathrm{Spin}(2m)$, and Lemma~\ref{lem:K3} is the standard highest-weight computation of the Casimir eigenvalue with one of the two factors replaced by an arbitrary equivariant map. For $U(d)$ and Keyl's measurement the same shift gives the staircase estimator of \cite{PSW26}, which they prove unbiased (Section~\ref{sec:discussion}).
The short derivation is in Appendix~\ref{app:proofs-K}.
\end{remark}

\paragraph{Signal and zero-point fluctuation.} By Lemma~\ref{lem:K1} with $\varphi=d\pi$, $\langle v|G_{ab}|v\rangle=(X_\lambda)_{ab}$, and $\sum_{a<b}G_{ab}^2=C_2(\lambda)$ on $V_\lambda$ (Lemma~\ref{lem:K3}); so at every coherent state
\begin{equation}\label{eq:K-fluct}
\sum_{a<b}\langle G_{ab}\rangle^2=\|\lambda\|^2,\qquad\sum_{a<b}\mathrm{Var}(G_{ab})=C_2(\lambda)-\|\lambda\|^2=4\lambda\cdot\delta ,
\end{equation}
the squared classical moment and its zero-point fluctuation, as $j^2$ and $j(j+1)-j^2$ for a spin $j$ at $|j,j\rangle$. At $t=1$ the zero-point part is $2m(m-1)$ against $\|\lambda\|^2=m$, so one copy reveals the fraction $1/(2m-1)$ of its quadratic data; on $t\le m$ copies near the top weight, $\|\lambda\|^2\approx mt^2$ is quadratic in $t$ while $4\lambda\cdot\delta\le2tm(m-1)$ is linear: the copies add coherently in the signal and incoherently in the noise.

\subsection{The block Fisher budget}\label{subsec:K-why}

For the Gibbs family $\rho_\theta\propto e^{\sum_{a<b}\theta_{ab}B_{ab}}$ \cite[Eq.~(1)]{Rub26b}, $\partial_{\theta_{ab}}\rho_\theta^{\otimes t}|_0=G_{ab}/2^{mt}$, and the Fisher information of a POVM $\{M_x\}$ on $t$ copies, with $q(x)=\Tr M_x/2^{mt}$ and $\sigma_x=M_x/\Tr M_x$, is
\begin{equation}\label{eq:K-fisher}
\tr I^{(t)}_M(0)=\sum_xq(x)\sum_{a<b}\Tr(\sigma_xG_{ab})^2 ,
\end{equation}
the average squared length of the moment vector of the normalised POVM elements. A state in $V_\lambda$ has squared moment length at most $\|\lambda\|^2$, with equality at coherent states, so the POVM \eqref{eq:K-povm} is optimal, with $F_{\max}(m,t)=\sum_\lambda P(\lambda)\|\lambda\|^2\le mt^2$, $P(\lambda)=m_\lambda n_\lambda/2^{mt}$, close to $mt^2$ for $t\ll m$ \cite[Thms.~3.6 and~3.8]{Rub26b}, against the quantum Fisher information $tp$ \cite[Lemma~3.10]{Rub26b}. Summed over blocks and inserted in the van Trees inequality, each power of $t$ in the budget beyond the first becomes a power of $k$ in the rate \cite[Thms.~3.4 and~3.11]{Rub26b} (Appendix~\ref{app:proofs-K}). Near the vacuum the hard states of (v) encode an $m$-level qudit near $I/m$, whose block budget grows only like $t^{3/2}$, whence $\sqrt k$ \cite{Rub26b,CMMP26}.

The budget is also a variance floor, the one statement of this subsection not in \cite{Rub26b}. Let $\widehat\Gamma$, computed from one block of $t$ copies, be unbiased for $\Gamma$ near $\theta=0$. As $\partial_{\theta_{cd}}\Tr(\rho_\theta B_{ab})|_0=\delta_{(ab),(cd)}$, the Cram\'er--Rao inequality and $\tr A^{-1}\ge p^2/\tr A$ ($A\succ0$ of size $p$) give $\E_{I/2^m}\sum_{a<b}\widehat\Gamma_{ab}^2\ge\tr I^{(t)}_M(0)^{-1}\ge p^2/F_{\max}(m,t)\ge p^2/(mt^2)$; averaging $N/t$ blocks, the error per copy $N\,\E\sum_{a<b}(\bar\Gamma_{ab}-\Gamma_{ab})^2$ is at least $p^2/(mt)$. At $t=1$ this is $p(2m-1)$, attained by matchgate shadows \cite{ZRM21,WHLB23}; for $t\le m$ the estimator \eqref{eq:K-est} has error per copy at most $V(m,t)/t\le11m^3/t$ by (ii), within a constant factor of $p^2/(mt)\ge\tfrac94m^3/t$ ($m\ge2$), and on every state.

\subsection{Remarks}\label{subsec:K-remarks}

\begin{remark}[minimal-variance variant]\label{rem:K-minvar}
By Proposition~\ref{prop:K4}, every $\xi\in(\lambda+2\varrho)+\Sigma_\lambda^\perp$, $\Sigma_\lambda:=\{\xi_\varphi:\varphi\in\mathcal E\}\subset\R^m$, gives an unbiased estimator $\tfrac1tOX_\xi O^T$ with second moment $\|\xi\|^2/t^2$; the minimum is attained at $\xi^*=\Pi_{\Sigma_\lambda}(\lambda+2\varrho)$, and $\|\xi^*\|^2\ge C_2(\lambda)^2/\|\lambda\|^2$ for $\lambda\ne0$, with equality if and only if $\lambda+2\varrho\in\mathrm{span}(\lambda)+\Sigma_\lambda^\perp$ (Cauchy--Schwarz, since $\lambda=\xi_{d\pi}\in\Sigma_\lambda$ and $\lambda\cdot\xi^*=\lambda\cdot(\lambda+2\varrho)=C_2$). $\Sigma_\lambda$ is contained in the fixed space of the stabiliser of $v$, $\{\xi:\xi\cdot\beta=0\text{ for all }\beta\in\Phi\text{ with }\lambda\cdot\beta=0\}$. At $t=1$: $\xi^*=(2m-1)(1,\dots,1,\pm1)$ for $\lambda=(1,\dots,1,\pm1)$ and $\|\xi^*\|^2=p(2m-1)$, exactly the shadow norm of matchgate classical shadows \cite{ZRM21,WHLB23,OGo22}, while the unprojected $\xi=\lambda+2\varrho$ gives $\sum_{i<m}(4i+1)^2\approx\tfrac{16}3m^3$.
Projecting helps because the outcome law is unchanged if $O$ is replaced by $Oh$ with $h$ in the stabiliser $H_\lambda$ of the line $\C v_{\lambda,i}$, so averaging the estimate over $H_\lambda$ is a Rao--Blackwellisation: it keeps the mean, does not increase the second moment, and projects $\xi$ onto the $H_\lambda$-fixed vectors of $\mathfrak t$, which lie in the space displayed above. For the case $t=1$ above, $\lambda=(1,\dots,1,\pm1)$, the roots orthogonal to $\lambda$ are, up to sign, $2e_i-2e_j$ ($i<j<m$) and $2e_i-2\lambda_me_m$ ($i<m$), with common orthogonal complement $\mathrm{span}(\lambda)$; so $\Sigma_\lambda=\mathrm{span}(\lambda)$, equality holds, and $\xi^*=(C_2/\|\lambda\|^2)\lambda=(2m-1)\lambda$, $\|\xi^*\|^2=p(2m-1)$: $2m-1$ times the covariance of the detected Fock state, the matchgate-shadow estimator, with $2m-1$ the ratio of total to classical moment \eqref{eq:K-fluct}. For $U(d)$ and Keyl's measurement, the same stabiliser average turns the staircase estimator $\lambda_i+d+1-2i$ of \cite{PSW26} into their debiased Keyl estimator, which averages it over each block of equal rows of $\lambda$.
\end{remark}

\begin{remark}[the unshifted moment map]\label{rem:K-unshifted}
The same POVM can be read out through the rescaled moment map, $c\,OX_\lambda O^T\mathbf 1[\|\lambda\|^2\ge mt^2/4]$ with a single global constant $c=p/\E_0[\|\lambda\|^2\mathbf 1[\|\lambda\|^2\ge mt^2/4]]$, $\E_0$ the outcome law at $I/2^m$; that estimator is unbiased only to first order at $I/2^m$. The estimator \eqref{eq:K-est} replaces $c\,\lambda$ by $(\lambda+2\varrho)/t$ and drops the indicator; the shift by $2\varrho$, twice the Weyl vector, makes it exactly unbiased on every state, and its sum of squared entries $\sum_{a<b}\widehat\Gamma_{ab}^2=\|\lambda+4\delta\|^2/t^2$, which depends on the observed $\lambda$, is bounded outcome by outcome by $V(m,t)/t^2\le11m^3/t^2$ for $t\le m$ (part (ii)).
\end{remark}

A worked example for two modes and two copies is Remark~\ref{rem:K-m2} in Appendix~\ref{app:proofs-K}.
\begin{remark}[non-Gaussian inputs]\label{rem:K-nonGaussian}
Part (i) holds for any state on $m$ modes: the estimator learns the two-point function $\Tr(\rho B_{ab})$ at the $k$-copy rate $\max\{m^3/k,m^2\}/\eps^2$ in Frobenius norm regardless of Gaussianity; only the conversion to trace distance in (iv) and (v) uses that $\rho$ is Gaussian.
\end{remark}

\begin{remark}[implementation]\label{rem:K-implementation}
The POVM is that of \cite[Thm.~3.6]{Rub26b}; implementing it requires the isotypic projection (a measurement of $\lambda$), a basis of the multiplicity space, and the coherent-state POVM within $V_\lambda$. The post-processing is the explicit $2m\times2m$ matrix $OX_{\lambda+4\delta}O^T/t$. No knowledge of the state or of $c$ (except for the final projection in (iv)) is needed.
\end{remark}

\section{Bosons: arbitrary temperature profiles}\label{sec:bosonic}

\subsection{Bosonic conventions}\label{subsec:Bos-conventions}

We use the conventions of \S\ref{subsec:bg-bosons} (those of \cite[Sec.~5.6]{Rub26b}): quadratures $R$, covariance $\Sigma$ with vacuum $\tfrac12I$, $\rho(v,\Sigma)$, Williamson occupations $n_j$ ($d_j=n_j+\tfrac12$), $\Phi_{\mathrm{re}}$, $\Gamma(u)$, $U_P$ and $D(v)$; $T$ is the trace distance and $\langle A,B\rangle=\Tr(A^\dagger B)$.
Since $\Omega=\Phi_{\mathrm{re}}(-iI)$, a real $2m\times2m$ matrix commutes with $\Omega$ if and only if it lies in the image of $\Phi_{\mathrm{re}}$.

\paragraph{Occupation matrices and passive states.}
Let $a=(a_1,\dots,a_m)^T$ be the annihilation operators, $[a_i,a_j^\dagger]=\delta_{ij}$, and $a^\#:=(a_1^\dagger,\dots,a_m^\dagger)^T$. The (centred, normal) occupation matrix of a state $\rho$ of mean zero is the Hermitian matrix $Q\succeq0$ with $Q_{ij}=\Tr(\rho\,a_j^\dagger a_i)$; for a state of nonzero mean the same expression is taken after subtracting the mean. For $Q\succ0$ the zero-mean, gauge-invariant Gaussian state of occupation $Q$ is
\[
\rho_Q=\det(I+Q)^{-1}\exp\big[-a^\dagger\log(I+Q^{-1})\,a\big].
\]
In an eigenbasis of $Q$, with eigenvalues $q_i>0$, it is a product of thermal states with number probabilities
\[
p_{\boldsymbol s}=\prod_i\frac{q_i^{s_i}}{(1+q_i)^{s_i+1}} .
\]
For singular $Q\succeq0$, $\rho_Q$ is the limit (the modes of zero occupation are in the vacuum). Its covariance is $\Sigma=\Phi_{\mathrm{re}}(Q+\tfrac12I)$, its fugacity matrix is $X=Q(I+Q)^{-1}$, so that $\rho_Q$ is the state $\rho_X$ of \cite[Sec.~5]{Rub26b}, and $\Gamma(u)\rho_Q\Gamma(u)^\dagger=\rho_{uQu^\dagger}$. A Gaussian state is of this form if and only if its mean is zero and $[\Sigma,\Omega]=0$. Throughout this section a \emph{passive state} means such a centred, gauge-invariant Gaussian state $\rho_Q$, and a \emph{displaced passive state} means $D(v)\rho_QD(v)^\dagger$; this is not passivity with respect to a Hamiltonian.
(Derivations of these facts: Appendix~\ref{app:Bos-conv}.)

\paragraph{The local metric.}
For $Q\succeq0$ define the positive superoperators
\[
\mathcal D_Q(X)=QX+XQ+2QXQ,\qquad \mathcal S_Q(X)=QX+XQ,
\]
so that in an eigenbasis of $Q$, $\mathcal D_Q$ acts entrywise by $d_{ij}=q_i+q_j+2q_iq_j$, and for $Q\succ0$ and Hermitian $H$ put
\begin{equation}\label{eq:Bos-WQ}
W_Q(H)^2:=\langle H,\mathcal D_Q^{-1}H\rangle=\sum_{i,j}\frac{|H_{ij}|^2}{d_{ij}} .
\end{equation}
The sum includes both ordered off-diagonal entries, and all constants below use that convention.

\paragraph{Heterodyne data.}
A heterodyne measurement on $\rho(v,\Sigma)$ has real outcome $Y\sim N(v,\Sigma+\tfrac12I)$. We also use its complex form $z\in\C^m$, $z_j=(Y_j+iY_{m+j})/\sqrt2$, the Husimi variable of $a$: for a Gaussian state of mean zero, $z$ is a centred complex Gaussian vector with covariance $\E zz^\dagger=I+Q$, $Q$ the normal occupation matrix, and pseudocovariance $\mathsf P:=\E zz^T=(\Tr(\rho\,a_ia_j))_{ij}$; for a passive state $\mathsf P=0$ and $z$ is a proper complex Gaussian. If the mean is unknown we use disjoint pair differences $(z_{2l}-z_{2l-1})/\sqrt2$, which are independent centred Gaussian samples with the same covariance and pseudocovariance, at a factor-two copy cost.
(The heterodyne outcome law \eqref{eq:Bos-het-moments} is derived in Appendix~\ref{app:Bos-conv}.)

\paragraph{The measurement model.}
Single-copy adaptive protocols are as in \S\ref{subsec:bg-estimation}. Constants written $C_K$, $c_K$ depend only on the parameter $K$ and may change from line to line; $C_0$ denotes an absolute constant that may also change from line to line. The proofs use \cite[Lemmas~5.8, 5.11(i) and~5.12, Thms.~5.9 and~5.13]{Rub26b}, \cite[Thms.~2, 5 and~9]{CMMP26}, \cite[Thm.~4]{KL17}, \cite[Thm.~4.7.1]{Ver18}, \cite{BJL19}, \cite{PW94} and \cite{Saf19}; Appendix~\ref{app:Bos-conv} states four of them, with the reason each holds, and says where the others are stated.

\subsection{The bosonic theorems}\label{subsec:Bos-statements}

\paragraph{The profile quantities.}
For an $m$-mode Gaussian state with Williamson occupations $0\le n_1\le\dots\le n_m$ define the following quantities; we call $\mathcal R(\boldsymbol n)$ the \emph{occupation-profile rate}:
\begin{equation}\label{eq:Bos-R}
\begin{gathered}
r:=\max\big(\{j:jn_j\le2\}\cup\{0\}\big),\qquad r_0:=\max\{r,1\},\qquad\mu:=2/r_0,\\
S_w:=\sum_{i,j>r}\frac1{n_i+n_j},\qquad\mathcal R(\boldsymbol n):=m^2+r^3+S_w .
\end{gathered}
\end{equation}
An empty sum is zero. If $r\ge1$, then $n_i\le\mu$ for $i\le r$. If $r<m$, then $n_{r+1}>2/(r+1)\ge\mu/2$; if $r=0$, every $n_i>2=\mu$. Thus every denominator in $S_w$ is strictly positive.

\begin{theorem}[Arbitrary profiles, single copy]\label{thm:Bos}
Fix $K\ge1$.
\begin{enumerate}
\item[(i)] (Upper bound for every state.) There are constants $C(K)$ and $\eps_\star(K)>0$ and a family of single-copy adaptive protocols indexed only by the number of modes $m\ge1$, $K$, the accuracy $\eps$ and the total copy budget $N$, with the following property. For every $m$-mode Gaussian state $\rho$ of arbitrary displacement, with quadrature covariance $\|\Sigma\|_\op\le K$ and Williamson occupations $n_1\le\dots\le n_m$, the protocol outputs a valid Gaussian state $\hat\rho$ and
\begin{equation}\label{eq:Bos-thm3}
\Pr[T(\rho,\hat\rho)\le\eps]\ge\tfrac23\quad\text{whenever}\quad N\ge C(K)\frac{\mathcal R(\boldsymbol n)}{\eps^2},\qquad0<\eps\le\eps_\star(K).
\end{equation}
There is no cold--warm spectral-gap promise and no supplied cold-group size.
\item[(ii)] (Matching minimax rate on interval classes.) Let $r\ge1$ be an integer, $\nu_b=\nu_02^b$ dyadic scales (empty bands omitted) with multiplicities $m_b\ge1$, and $m=r+\sum_bm_b$, with
\begin{equation}\label{eq:Bos-room}
\nu_{\min}\ge1/r,\qquad2\nu_{\max}+1\le K .
\end{equation}
An empty warm group is allowed. Let $\mathcal P_K(r;(m_b,\nu_b)_b)$ be the class of all Gaussian states with arbitrary displacement, $\|\Sigma\|_\op\le K$, the smallest $r$ Williamson occupations in $[0,2/r]$, and the remaining occupations assigned to the intervals $[\nu_b,2\nu_b]$ with the prescribed multiplicities $m_b$, under a fixed endpoint convention. For $m$ sufficiently large and $\eps$ sufficiently small, depending only on $K$, the single-copy adaptive minimax sample complexity of $\mathcal P_K(r;(m_b,\nu_b)_b)$ is
\begin{equation}\label{eq:Bos-thm4}
\Theta_K\Big(\frac{m^2+r^3+\sum_bm_b^2(1+\nu_b^{-1})}{\eps^2}\Big).
\end{equation}
If $r=0$, the class is defined instead with all occupations in dyadic warm intervals with $\nu_{\min}\ge1$ and $2\nu_{\max}+1\le K$, and its rate is $\Theta_K(m^2/\eps^2)$. The constants do not depend on $r$, on the number of bands, on their multiplicities or on their lowest scale.
\end{enumerate}
\end{theorem}

The quantum Fisher metric of thermal states weights the $(i,j)$ entry of an occupation error by $1/(q_i+q_j+2q_iq_j)$ (Proposition~\ref{prop:Bos-QFI}), and heterodyne detection estimates that entry with variance $(1+q_i)(1+q_j)$ per copy \eqref{eq:Bos-risk}; the ratio is at most $1+1/(q_i+q_j)$ \eqref{eq:Bos-scalar}, a constant as soon as one partner is warm. This is Einstein's split of thermal photon-number fluctuations, $\bar n+\bar n^2$, into a particle and a wave term, read pairwise: heterodyne detection is matched in the wave regime and swamped by its own vacuum noise in the particle regime, and for fermions Pauli blocking reverses the sign of the wave term. Summed over pairs the ratio gives $m^2+S_w$, plus cold--warm terms absorbed by $r^3+S_w$ \eqref{eq:Bos-44}. The cold group is instead read by counting photons, its one-particle block an $r$-level system learned from $r^3/\eps^2$ copies \cite[Thm.~5.9]{Rub26b}; at the threshold $jn_j=2$ the two descriptions agree up to a constant ($j^3$ as a qudit, $j^3/4$ as classical modes), so neither a gap nor a known $r$ is needed. The protocol of (i) (Appendix~\ref{app:proofs-Bos}) undoes the estimated displacement and squeezing (Lemmas~\ref{lem:Bos0} and~\ref{lem:Bos-undo}), chooses the split from a pilot estimate (Lemmas~\ref{lem:Bos-effrank} and~\ref{lem:Bos-split}). It then replaces the state by the product of its cold and warm marginals at a cost $2\sqrt\ell$, $\ell=\Tr(BC^{-1}B^\dagger)$ with $B$, $C$ the cold--warm and warm blocks of $Q$ (Lemma~\ref{lem:Bos-split-schur}), and reads the cold group by photon counting and the warm block by heterodyne detection (Lemma~\ref{lem:Bos-PSD}). The lower bound of (ii) assembles the hard families of \cite{CMMP26}, one of them made Bayesian (Lemma~\ref{lem:Bos-Bayes}); the costs of independent bands add, because the trace distance of a tensor product is at least a constant times the quadrature sum of the blockwise distances (Lemmas~\ref{lem:Bos-tensor} and~\ref{lem:Bos-directsum}). In (ii) the cold interval $[0,2/r]$ and the first possible warm interval $[1/r,2/r]$ overlap, so no positive spectral gap is imposed. Part~(i) is proved in \S\ref{subsec:Bos-proof-i} and part~(ii) in \S\ref{subsec:Bos-proof-ii}.

\begin{remark}[Budget-indexed guarantee]\label{rem:Bos-scope}
The protocol of (i) uses the budget $N$ supplied to it and chooses its split from the data. The theorem does not say that it can certify that $N$ was sufficient, or stop at an unknown optimal $\mathcal R(\boldsymbol n)$ without additional work: it is an instance-dependent upper guarantee for one budget-indexed procedure. No lower bound is claimed for a fixed, exactly known spectrum: an exactly pure promised family, for example, can have a cheaper learning rate. At the exactly vacuum spectrum, $\mathcal R=m^2+m^3$ is a valid upper bound but is not asserted to be optimal if purity is promised in advance.
\end{remark}

\begin{remark}[Endpoints and coverage]\label{rem:Bos-endpoints}
If every $n_i\le2/m$, then $r=m$ and $\mathcal R=m^2+m^3$. If $n_i=\nu>0$ is uniform, then $r=\min\{m,\lfloor2/\nu\rfloor\}$ (with $r=0$ if $\nu>2$); for bounded $\nu$ and $m\nu$ larger than a constant, $\mathcal R$ is $O_K(m^2(1+\nu^{-1}))$, and in the cold regime it is $O(m^3)$. These recover the known single-scale orders. Every bounded occupation profile has the canonical $r$ of \eqref{eq:Bos-R}. For $r\ge1$, $n_{r+1}>2/(r+1)\ge1/r$, so the warm occupations can be put in dyadic bands based at $1/r$; for $r=0$ they all exceed $2$ and can be put in bands based at $1$. The upper theorem always uses the actual covariance bound $K$. If enlarging the bands violates the room condition \eqref{eq:Bos-room} for that exact $K$, a larger \emph{constant} covariance bound, for example $2K+1$, contains the entire corresponding interval class and its lower-bound families; we do not transfer that lower bound back to the narrower class.
\end{remark}

\paragraph{Two temperatures.}
Fix integers $m_c,m_w\ge1$, $m:=m_c+m_w$, constants $c\ge\tfrac23$ and $K\ge1$, and a temperature $\nu$ with
\[
\nu\ge\frac{4c}{m_c},\qquad\frac{c}{m_c}\le\frac14,\qquad K\ge2\nu+1 .
\]
Let $\mathcal C:=\mathcal C(m_c,c;m_w,\nu;K)$ be the set of Gaussian states $\rho(v,\Sigma)$ on $m$ modes such that the Williamson occupations of $\Sigma$ satisfy $n_1\le\dots\le n_{m_c}\le c/m_c$ (the \emph{cold modes}) and $n_{m_c+1},\dots,n_m\ge\nu$ (the \emph{warm modes}), $\|\Sigma\|_\op\le K$, and $v\in\R^{2m}$ is arbitrary. The first condition makes the gap between the cold and the warm occupations at least $3\nu/4$; the third makes $\mathcal C$ non-empty with room for the lower-bound families, whose occupations reach $2\nu$.

\begin{corollary}[A cold group and a warm group, single copy]\label{cor:Bos-twoscale}
There are $m_\star$, $\eps_\star>0$ and $C_\star<\infty$ depending only on $(c,K)$ such that for $m_c,m_w\ge m_\star$ and $\eps\le\eps_\star$:
\begin{enumerate}
\item[(i)] (upper bound) a single-copy protocol learns every state of $\mathcal C$ to trace distance $\eps$ with probability $\ge2/3$ from
\[
N=C_\star\,\frac{m_c^3+m_w^2(1+1/\nu)}{\eps^2}
\]
copies;
\item[(ii)] (lower bound) every single-copy protocol that does so needs $\Omega\big(\max\{m_c^3,\,m_w^2(1+1/\nu)\}/\eps^2\big)$ copies when $\nu\ge1/m_w$, and $\Omega(m_c^3/\eps^2)$ always.
\end{enumerate}
Hence the single-copy sample complexity of $\mathcal C$ is $\Theta\big((m_c^3+m_w^2(1+1/\nu))/\eps^2\big)$ for $\nu\ge1/m_w$, and $\Theta(m_c^3/\eps^2)$ for $4c/m_c\le\nu<1/m_w$, where $m_w<m_c/(4c)$ forces $m_w^2(1+1/\nu)\le2m_w^2m_c/(4c)\le m_c^3/(32c^3)$, so that the warm term is dominated. The collective floor $m^2/\eps^2$ is dominated by both terms, $m^2\le2(m_c^2+m_w^2)\le2(m_c^3+m_w^2(1+1/\nu))$, and is not displayed.
\end{corollary}

The upper bound is Theorem~\ref{thm:Bos}(i), since $\mathcal R(\boldsymbol n)\le C_c(m_c^3+m_w^2(1+1/\nu))$ on the class. The lower bound embeds the cold hard family of \cite[Thm.~2]{CMMP26} with the warm modes fixed, and the warm families of \cite[Thms.~5 and~9]{CMMP26} with the cold modes fixed at the vacuum; a single-copy measurement on a product with a known factor is one on the other factor, and the trace distance contracts under the partial trace (\S\ref{subsec:Bos-proof-cor}).

\paragraph{Separated bands with a known cold group.}
Fix $c\ge\tfrac23$ and $K\ge1$. Let $m_c$ be an integer, and let the warm bands have nonnegative integer sizes $m_b$ and dyadic scales $\nu_b=\nu_02^b$; omit empty bands. Assume $\nu_{\min}\ge4c/m_c$ and $K\ge2\nu_{\max}+1$, and let $m=m_c+\sum_bm_b$ and $m_w=\sum_bm_b$. Let $\mathcal C_{\rm sep}$ consist of all $m$-mode Gaussian states with arbitrary displacement, quadrature covariance $\|\Sigma\|_\op\le K$, $m_c$ Williamson occupations in $[0,c/m_c]$, and exactly $m_b$ occupations in each interval $[\nu_b,2\nu_b]$; endpoint coincidences are assigned to either neighbouring band once and for all. Write
\begin{equation}\label{eq:Bos-Rsep}
\mathcal R_{\rm sep}=m_c^3+m^2+\sum_bm_b^2(1+\nu_b^{-1}).
\end{equation}

\begin{theorem}[Separated multiband class]\label{thm:Bos-sep}
There are $m_\star$, $C$, $c_0>0$ and $\eps_\star>0$ depending only on $(c,K)$ such that for $m_c\ge m_\star$ and $0<\eps\le\eps_\star$:
\begin{enumerate}
\item[(i)] a single-copy adaptive protocol, which uses $m_c$ and $\mathcal R_{\rm sep}$ but never the true occupation eigenbasis or the individual occupations, learns every state of $\mathcal C_{\rm sep}$ to trace distance $\eps$ with probability at least $2/3$ from at most $C\mathcal R_{\rm sep}/\eps^2$ copies, and outputs a valid Gaussian state;
\item[(ii)] every single-copy adaptive protocol that learns every state of $\mathcal C_{\rm sep}$ to trace distance $\eps$ with probability $2/3$ uses at least $c_0\mathcal R_{\rm sep}/\eps^2$ copies.
\end{enumerate}
Thus the rate of $\mathcal C_{\rm sep}$ is $\Theta_{c,K}(\mathcal R_{\rm sep}/\eps^2)$, with constants independent of the number of bands, of $m$ and of the profile. Here $m_\star(c,K)$ is at least the threshold $n_0$ of \cite[Thm.~2]{CMMP26} and large enough that $\mu=21c/(20m_c)<\tfrac14$ and that the cold occupation bound $\mu+\ell<\tfrac13$ of \S\ref{subsec:Bos-sep-upper} holds; $\eps_\star(c,K)$ is at most $\eps_0$ of \cite[Thm.~2]{CMMP26}, $1/1800$ (the range of \cite[Thm.~9]{CMMP26}), $e_\star/100$ with $e_\star$ the absolute accuracy threshold of Lemma~\ref{lem:Bos-Bayes}, and below $1/18$, together with the smallness conditions of Lemma~\ref{lem:Bos0} and of the cold readout in \S\ref{subsec:Bos-sep-upper}. The absolute dimension threshold $n_\star$ of Lemma~\ref{lem:Bos-Bayes} is not a hypothesis: bands below it are absorbed in the proof.
\end{theorem}

With a known cold group separated by a factor $4$, the $m_c$ lowest eigenvectors of a heterodyne estimate span an estimated cold space, whose misalignment with the true one costs, in trace distance, an amount governed by the warm occupation $\ell$ it carries (Lemmas~\ref{lem:Bos-coldframe} and~\ref{lem:Bos-cross}); the cold modes are then read by photon counting and the warm modes by heterodyne detection (\S\ref{subsec:Bos-sep-upper}). Part~(ii) combines the band priors of Lemma~\ref{lem:Bos-Bayes} and the families of \cite[Thms.~2 and~9]{CMMP26} by Lemma~\ref{lem:Bos-directsum} (\S\ref{subsec:Bos-bands}). An empty warm group is covered by omitting the warm stages. Theorem~\ref{thm:Bos} removes the gap and the knowledge of $m_c$ by deleting the cross block instead of estimating a rotation.

\paragraph{Unbounded thermal energy.}

\begin{theorem}[Displaced passive states without an energy bound]\label{thm:Bos-energy}
Define $r$, $r_0$, $\mu$, $S_w$ and $\mathcal R(\boldsymbol n)$ from the eigenvalues $n_i$ of the occupation matrix $Q$ exactly as in \eqref{eq:Bos-R}. There are absolute constants $C,\eps_\star>0$ and a single-copy adaptive protocol, with inputs $m,\eps,N$ only, that learns every displaced passive $m$-mode Gaussian state $D(v)\rho_QD(v)^\dagger$, that is, every Gaussian state whose covariance commutes with $\Omega$, to trace distance $\eps$ with probability at least $2/3$ whenever
\begin{equation}\label{eq:Bos-thm5}
N\ge C\,\mathcal R(\boldsymbol n)/\eps^2,\qquad0<\eps\le\eps_\star .
\end{equation}
There is no upper bound on $Q$ or on the displacement $v$, and no squeezing transformation is estimated. The hypothesis that $\Sigma$ commutes with $\Omega$ is used exactly; tolerance of an unknown residual squeezing with absolute constants is not claimed.
\end{theorem}

The energy entered the proof of Theorem~\ref{thm:Bos} in three places, each replaced here by a statement uniform in the temperature (\S\ref{subsec:Bos-energy}). A displacement error is measured against the thermal noise it hides in, $T\le\frac12\sqrt{d^T\Sigma^{-1}d}$, a metric in which the heterodyne mean has mean-square error $4m/L$ at every temperature \eqref{eq:Bos-57}; the warm reconstruction keeps the wave term of the metric, so that a one-sided comparison (Proposition~\ref{prop:Bos-onesided}) needs no upper bound on the occupations; and the data-chosen split may select no cold mode, so a hot mode is never forced into photon counting. Squeezing would bring the energy back, since an unknown squeezer cannot be estimated with Gaussian measurements from a number of copies independent of the energy \cite{CGYZ26}.

\begin{remark}[Scope of Theorem~\ref{thm:Bos-energy}]\label{rem:Bos-energy-scope}
All constants in the proof of Theorem~\ref{thm:Bos-energy} (\S\ref{subsec:Bos-energy}) are absolute. The theorem removes the \emph{thermal-energy} restriction on the profile upper bound; it does not estimate an unknown active squeezer, and arbitrary unknown passive rotations are included. The coarser worst-case upper bound $O(m^3/\eps^2)$ on unsqueezed states was already accessible by the heterodyne theorem of \cite{BMEM25}; the contribution here is the profile refinement and its proof. Energy-independent single-copy heterodyne tomography is also \cite[Thm.~8]{CMMP26}: every Gaussian state with $\Sigma\succeq(\tfrac12+\nu)I$, of any mean and with no upper bound on $\Sigma$, is learned from $O((1+\nu^{-1})\,m(m+\log\delta^{-1})/\eps^2)$ copies with probability $1-\delta$. That class and the displaced passive states are not comparable: the former allows squeezing as long as every quadrature variance stays at least $\tfrac12+\nu$ and excludes Williamson occupations below $\nu$, in particular pure modes; the latter allows cold and pure modes and excludes squeezing. Matching lower bounds on every unbounded-temperature interval profile would additionally need an energy-uniform treatment of the $m^2$ term and of arbitrarily many hot bands; Theorem~\ref{thm:Bos}(ii) does not establish that by letting $K\to\infty$. Every hard family in the proof of Theorem~\ref{thm:Bos}(ii) is passive and centred, so its lower bound holds, with the same $K$-dependent constants, on the displaced passive part of each class $\mathcal P_K(r;(m_b,\nu_b)_b)$.
\end{remark}

\subsection{The passive metric and finite trace distance}\label{subsec:Bos-metric}

By Proposition~\ref{prop:Bos-QFI}, $2W_Q(H)^2$ is the quantum Fisher information of the occupation family in the direction $H$, and \eqref{eq:Bos-path} integrates it into trace distance (tight two-sided bounds between the trace distance of bosonic Gaussian states and the distance of their moments are proved in \cite{BMMTL25}). The off-diagonal weight $d_{ij}=q_i(1+q_j)+q_j(1+q_i)$ is the variance of the beam-splitter generator between modes $i$ and $j$ (proof of Lemma~\ref{lem:Bos3}, Appendix~\ref{subsec:Bos-CS}); it is $\approx q_i+q_j$, a $\chi^2$-type weight, near the vacuum, and $\approx2q_iq_j$ for warm modes, where $F_Q(Q;H)\approx\Tr(Q^{-1}HQ^{-1}H)$ is the Fisher information of a classical complex Gaussian field of covariance $Q$.

\begin{proposition}[Symmetric logarithmic derivative]\label{prop:Bos-QFI}
Let $Q\succ0$ and let $H$ be a Hermitian occupation tangent. The symmetric logarithmic derivative of $\rho_{Q+tH}$ at $t=0$ is
\[
L=a^\dagger Ba-\Tr(BQ),\qquad B=2\mathcal D_Q^{-1}H,
\]
and the quantum Fisher information is
\begin{equation}\label{eq:Bos-QFI}
F_Q(Q;H)=\Tr(\rho_QL^2)=2W_Q(H)^2 .
\end{equation}
\end{proposition}
The proof is in Appendix~\ref{app:Bos-QFI}.

The Gaussian symmetric logarithmic derivative and quantum Fisher information are established results \cite[Eqs.~(8)--(10) and~(14)]{Saf19}; with the symplectic eigenvalues $\lambda_i=2q_i+1$ of that reference, the passive part agrees with \eqref{eq:Bos-QFI}. The calculation in Appendix~\ref{app:Bos-QFI} fixes the occupation and ordered-pair conventions used here.

The path bound \eqref{eq:bg-path} and \eqref{eq:Bos-QFI} give, for an occupation path $Q_t\succ0$,
\begin{equation}\label{eq:Bos-path}
T(\rho_{Q_0},\rho_{Q_1})\le\frac1{\sqrt2}\int_0^1\sqrt{\langle\dot Q_t,\mathcal D_{Q_t}^{-1}\dot Q_t\rangle}\,dt .
\end{equation}

\subsection{Heterodyne risk}\label{subsec:Bos-risk}

Let the state be passive and centred. For $L$ independent complex heterodyne samples put
\[
\widetilde Q=\frac1L\sum_{a=1}^Lz_az_a^\dagger-I,\qquad H=\widetilde Q-Q .
\]
In the eigenbasis of $Q$ the complex Gaussian fourth moment gives exactly
\begin{equation}\label{eq:Bos-risk}
\E|H_{ij}|^2=\frac{(1+q_i)(1+q_j)}L,\qquad\E W_Q(H)^2=\frac1L\sum_{ij}\frac{(1+q_i)(1+q_j)}{q_i+q_j+2q_iq_j}.
\end{equation}
This is a statement about the raw Hermitian estimate, which need not be positive.
 $H_{ij}=L^{-1}\sum_a(z_{a,i}\bar z_{a,j}-\E z_i\bar z_j)$ is an average of independent centred terms, so $\E|H_{ij}|^2=L^{-1}(\E|z_i|^2|z_j|^2-|\E z_i\bar z_j|^2)$. In the eigenbasis of $Q$ the components of the proper complex Gaussian $z$ are independent with $\E|z_i|^2=1+q_i$. For $i\ne j$ this gives $(1+q_i)(1+q_j)$; for $i=j$, $\E|z_i|^4=2(1+q_i)^2$ gives $(1+q_i)^2$, the same formula. Each factor $1+q_i$ is the signal $q_i$ of one mode plus the vacuum unit added by heterodyne detection; an off-diagonal entry is a product of two independent amplitudes, so the two noises multiply.

For $x,y>0$ put $s=x+y$ and $p=xy$. Then
\begin{equation}\label{eq:Bos-scalar}
\frac14\Big(1+\frac1s\Big)\le\frac{1+s+p}{s+2p}\le1+\frac1s .
\end{equation}
For the upper bound multiply by $s(s+2p)$; the difference of the right and the left numerators is $p(s+2)\ge0$. For the lower bound, the ratio is at least $\tfrac12$. If $s\le1$, then $p\le s^2/4$ and the ratio is also at least $1/(s+s^2/2)\ge2/(3s)$, which is at least $(1+1/s)/3$. If $s\ge1$, the bound $\tfrac12$ is at least $(1+1/s)/4$.
Group the positive occupations into disjoint dyadic bands $[\nu_b,2\nu_b)$, $\nu_b=\nu_02^b$, of sizes $m_b$ (empty bands allowed). Then
\begin{equation}\label{eq:Bos-dyadic}
\sum_{ij}\frac1{q_i+q_j}\asymp\sum_{b,c}\frac{m_bm_c}{\max(\nu_b,\nu_c)}\asymp\sum_b\frac{m_b^2}{\nu_b}.
\end{equation}
The first comparison has absolute constants because $\max(\nu_b,\nu_c)\le q_i+q_j<4\max(\nu_b,\nu_c)$; the same holds for closed intervals $[\nu_b,2\nu_b]$ with $\le4\max$. For the second, set $u_b=m_b/\sqrt{\nu_b}$; the middle quadratic form is $\sum_{b,c}u_bu_c\,2^{-|b-c|/2}$. Its diagonal is $\sum_bu_b^2$, and by $2u_bu_c\le u_b^2+u_c^2$ and summation of the geometric kernel it is at most $(1+2/(\sqrt2-1))\sum_bu_b^2$. Consequently, with $m_w$ the number of positive occupations,
\begin{equation}\label{eq:Bos-riskdyadic}
\E W_Q(H)^2\asymp\frac{m_w^2+\sum_bm_b^2/\nu_b}L .
\end{equation}
No factor equal to the number of occupied bands appears in \eqref{eq:Bos-dyadic}.
Within a dyadic band all pairs cost within a factor four of $1/\nu_b$, and the cross-band costs decay geometrically in $|b-c|$; the same intervals $[\nu,2\nu]$ carry the hard families of \cite{CMMP26}, which makes the dyadic classes the natural ones for the matching lower bounds.

\section{Energy-independent learning with few-copy non-Gaussian measurements}\label{sec:unsqueezing}

The conventions are those of \S\ref{subsec:Bos-conventions}, and a state is \emph{centred} if $v=0$. The references \cite{BMEM25,CGYZ26} normalise the vacuum covariance to $I$: their covariance matrix is $V=2\Sigma$. The sample complexity counts copies of the unknown state only; gate precision, detector dynamic range, ancillary energy and implementation complexity are not counted, and ``energy-independent'' refers to the number of copies.

Let $E$ be the energy (mean photon number) of the unknown state. With Gaussian measurements, adaptive and entangled ones included, learning pure Gaussian states to trace distance $\eps$ needs $\Omega(\log\log E+1/\eps^2)$ copies, and $\Omega(\ell^{-1}\eps^{-2^\ell/(2^\ell-1)}E^{1/(2^\ell-1)})$ with $\ell$ rounds of adaptivity \cite[Thms.~1.1 and~1.2]{CGYZ26}; with an energy bound, Gaussian measurements need $\Omega(m^3/\eps^2)$ copies on $m$ modes and arbitrary ones $\Omega(m^2/\eps^2)$, and non-Gaussian measurements are provably required for the optimal learning of passive states \cite{CMFLMHCP26}; on pure states with $\|\Sigma\|_\op\le\bar E$, adaptive single-copy Gaussian measurements, with an estimate projected to a pure Gaussian state, use $O(m^2/\eps^2+(m+\log\log\log\bar E)\log\log\bar E)$ copies \cite[Thm.~5]{CMFLMHCP26}. The adaptive single-copy Gaussian protocol of \cite[Thm.~3 (supplement, Thm.~14)]{BMEM25} (auxiliary squeezed states, passive optics and homodyne detection) uses $O((m+\log\log\log E)\log\log E+m^3/\eps^2)$ copies, and access to the transposed state removes the energy dependence \cite[Thm.~4]{BMEM25}. Non-Gaussian measurements remove it without the transpose: for pure $m$-mode states with a collective measurement on $O((m^2+\log\delta^{-1})/\eps^2)$ copies \cite[Thm.~1.4]{CGYZ26}; for pure one-mode states with nonadaptive two-copy measurements, single-copy ones for centred states, built on the canonical phase POVM \cite[Thm.~1.5]{CGYZ26}; and for all mixed $m$-mode Gaussian states, of arbitrary mean, at the optimal rate $O((m^2+\log\delta^{-1})/\eps^2)$ (in purified distance, which bounds the trace distance) with a fully collective measurement \cite[Thm.~5.1]{CFGLMWW26}, whose matching lower bound is proved on pure states \cite[Thm.~5.2]{CFGLMWW26}. Left open there are an energy-independent rate for general $m$-mode Gaussian states with few-copy measurements \cite[Sec.~1]{CGYZ26}, and whether entangled measurements are fundamentally required for optimal tomography \cite[Sec.~6]{CFGLMWW26}.

The protocols below contain a non-Gaussian measurement on one copy or on few copies, so these lower bounds, which concern protocols all of whose measurements are Gaussian, do not apply. Their hard family is the one-mode squeezed vacua of energy $E$ at a uniformly random angle, against which a Gaussian measurement, whose noise ellipse is fixed in advance \cite[Lemma~6.1]{CGYZ26}, can at best square-root the residual misalignment per round, as the adaptive unsqueezing of \cite{BMEM25} does (Appendix~\ref{app:proofs-NG}). The measurements below are instead \emph{covariant}: a symmetry applied to the state transports the outcome law by the same symmetry (squeezings and shears in \S\ref{subsec:NG-wavelet}, quadratic phases in \S\ref{subsec:NG-radialPOVM}, all centred Gaussian unitaries in \S\ref{subsec:NG-moments}), so the error of the inferred squeezer, in the units of the state, has a law independent of the squeezing. On the hard family one wavelet outcome does what costs Gaussian protocols $\log\log E$ rounds: if a pure $P$ satisfies $P\preceq C\Sigma_\theta$ (Lemma~\ref{lem:NG-dom}), then $\det P=\det\Sigma_\theta$ gives $P\succeq C^{-1}\Sigma_\theta$, and with $f_\mp$ unit vectors along the thin and thick axes of $\Sigma_\theta$ and $e$ a top eigenvector of $P$, $\lambda/(2C)\le\lambda_{\max}(P)$ and $\lambda_{\max}(P)\sin^2\angle(e,f_+)\le f_-^TPf_-\le C/(2\lambda)$, so the angle is known to $O(C/E)$.

\subsection{The energy-independence theorems}\label{subsec:NG-statements}

\begin{theorem}[One mode]\label{thm:NG1}
Let $0<\eps<1$ and
\begin{equation}\label{eq:NG-N1}
N_1>\Big(\frac{4.3}{\eps}\,(2+2\cdot10^6)\big(\sqrt2+\sqrt{2\log40}\big)\Big)^2\approx1.26\cdot10^{15}\,\eps^{-2}.
\end{equation}
\begin{enumerate}
\item[(i)] (Centred states, adaptive single copies.) There is a protocol that measures one copy with the non-Gaussian POVM $\mathsf M$ of \S\ref{subsec:NG-wavelet} and then measures $N_1$ further copies one at a time, each by a Gaussian unitary chosen from the first outcome followed by heterodyne detection, such that for every centred one-mode Gaussian state $\rho(0,\Sigma)$, with no bound on its energy, squeezing or temperature, the output $\hat\rho$ satisfies $T(\hat\rho,\rho(0,\Sigma))\le\eps$ with probability greater than $0.91$.
\item[(ii)] (Arbitrary displacement, blocks of two copies.) The same holds for every one-mode Gaussian state $\rho(v,\Sigma)$, with no bound on $v$, using $2+N_1$ copies: the first two copies are measured jointly (a balanced beam splitter followed by $\mathsf M$ on one output port), and every other copy is measured alone.
\end{enumerate}
The constant is absolute; it is very large and is not optimised (Remark~\ref{rem:NG-constant}). The order $\eps^{-2}$ cannot be improved, even with collective measurements (Remark~\ref{rem:NG-lower}).
\end{theorem}

\begin{theorem}[$m$ modes, blocks linear in $m$]\label{thm:NG2}
Let $m\ge1$, $0<\eps<1$, $L_0:=\lceil64000\,m^2\rceil$ and
\begin{equation}\label{eq:NG-N2}
N_2>\Big(\frac{34.4\,m}{\eps}\big(\sqrt{2m}+\sqrt{2\log40}\big)\Big)^2 .
\end{equation}
There is a protocol with two rounds, a nonadaptive round of $L_0$ joint measurements on disjoint blocks of $4m+6$ copies followed by $N_2$ copies measured one at a time, each by a Gaussian unitary chosen from the first round followed by heterodyne detection, which learns every $m$-mode Gaussian state $\rho(v,\Sigma)$, of arbitrary displacement, thermal energy and correlated multimode squeezing, to trace distance $\eps$ with probability at least $0.94$. It uses $L_0(4m+6)+N_2=O(m^3/\eps^2)$ copies, with an absolute constant. For centred states, blocks of $4m+5$ copies suffice. There is no promise on the energy or on the squeezing and no access to the transposed state. The block size does not depend on $\eps$ or on the energy; it grows linearly with $m$.
\end{theorem}

Heterodyne tomography pays for the vacuum unit it adds to every quadrature through a factor $\Tr V^{-1}$, unbounded for strongly squeezed states \cite[Thm.~10]{BMEM25}. Both protocols therefore first find a symplectic $S$ with $P=\frac12SS^T\preceq C\Sigma$ \eqref{eq:NG-target} and apply $U_S^\dagger$ to the later copies, after which heterodyne detection costs $O(C^2m^3/\eps^2)$ copies (Lemma~\ref{lem:NG-target}). For Theorem~\ref{thm:NG1}, the scale of a wavelet outcome has an energy-free law (Lemma~\ref{lem:NG-scale}), the shear is centred on the state's own shear (Lemma~\ref{lem:NG-shear}), and these give $\E[\Tr(\Sigma^{-1}P)\mid\tau=u]\le2u+\frac94+\frac9{4u}$, whence $P\preceq10^6\Sigma$ on a fixed quantile event of $\tau$ (Lemma~\ref{lem:NG-dom}). For Theorem~\ref{thm:NG2}, an orthogonal mixing of the $4m+6$ copies of a block moves the displacement into one copy; the quasi-purification channel of \cite{CFGLMWW26} and the covariant measurement of \S\ref{subsec:NG-moments} then reduce every moment to the vacuum, where the Siegel-disk density has a finite relative second moment exactly for $k\ge2d+5$, $d=2m$ (Lemmas~\ref{lem:NG-radial} and~\ref{lem:NG-pure}), and averaging $\lceil64000m^2\rceil$ blocks gives $\Sigma\preceq\hat\Sigma\preceq3\Sigma$ (Proposition~\ref{prop:NG-mixed}). The proofs are in Appendix~\ref{app:proofs-NG}.

Theorem~\ref{thm:NG1}(i) extends the energy-independent one-mode result of \cite[Thm.~1.5]{CGYZ26} from pure to mixed states, with adaptive (two-round) instead of nonadaptive single-copy measurements and an affine-wavelet POVM, its only non-Gaussian step, in place of the canonical phase POVM. For Theorem~\ref{thm:NG2} the comparison is with \cite[Thm.~5.1]{CFGLMWW26}, which learns every mixed $m$-mode Gaussian state with no energy dependence from $O((m^2+\log\delta^{-1})/\eps^2)$ copies with one fully collective measurement, and whose multiplicative covariance lemma \cite[App.~D.6, Lemma~D.14]{CFGLMWW26} already yields, from a collective measurement on $O(m^2)$ copies, the intermediate target \eqref{eq:NG-target}. The new content of Theorem~\ref{thm:NG2} is the measurement resource, entangled blocks of $4m+6$ copies independent of $\eps$ and of the energy followed by single-copy heterodyne detection, at the price of the rate $m^3/\eps^2$; it answers the question of \cite[Sec.~1]{CGYZ26} partially, with blocks of $O(m)$ rather than $O(1)$ copies, and bears on, without settling, that of \cite[Sec.~6]{CFGLMWW26}.

\begin{theorem}[Pure states without an energy bound]\label{thm:NG-pure}
Let $m\ge1$, $0<\eps<1$, $n:=m+4$, and
\[
N_q:=\lceil1.6\cdot10^5(m^2+m)\rceil,\qquad N_r:=\lceil10^{14}n^2\rceil,\qquad N_h:=\lceil3.6\cdot10^5m^2/\eps^2\rceil,\qquad N_v:=\lceil10^4m/\eps^2\rceil .
\]
\begin{enumerate}
\item[(i)] (Centred states, adaptive single copies.) There is a protocol with three rounds, each measuring copies one at a time: homodyne detection of the $m$ position quadratures on $N_q$ copies; then, on each of $N_r$ copies, a Gaussian unitary chosen from the first round, four ancillary modes in the vacuum and the radial POVM $\mathsf R_n$ of \S\ref{subsec:NG-radialPOVM} on the $n$ modes; then, on each of $N_h$ copies, a Gaussian unitary chosen from the earlier outcomes followed by heterodyne detection. For every centred pure $m$-mode Gaussian state $\rho(0,\Sigma)$, with arbitrary correlated squeezing and no bound on its energy or squeezing, it outputs a pure Gaussian state $\hat\rho$ with $T(\hat\rho,\rho(0,\Sigma))\le\eps$ with probability at least $0.9$.
\item[(ii)] (Arbitrary displacement, blocks of two copies.) The same holds for every pure $m$-mode Gaussian state $\rho(v,\Sigma)$, with no bound on $v$, using $2N_q+2N_r+2N_h+N_v$ copies: in the second round $\mathsf R_n$ is applied to the centred output port of a balanced beam splitter acting on two copies, and every other copy is measured alone.
\end{enumerate}
Both protocols use $O(m^2/\eps^2)$ copies, with an absolute constant. Conversely, there is an absolute $\eps_0>0$ such that for $0<\eps\le\eps_0$ every protocol, collective ones included, that learns all centred pure $m$-mode Gaussian states to trace distance $\eps$ with probability $2/3$ uses $\Omega(m^2/\eps^2)$ copies \cite[Thm.~5.2]{CFGLMWW26}. The rate is therefore $\Theta(m^2/\eps^2)$, for centred pure states with single copies and for all pure states with blocks of two.
\end{theorem}

For pure states with $\|\Sigma\|_\op\le\bar E$, adaptive single-copy Gaussian measurements already reach $O(m^2/\eps^2+(m+\log\log\log\bar E)\log\log\bar E)$ copies \cite[Thm.~5]{CMFLMHCP26}; no Gaussian measurement removes the dependence on the energy \cite[Thm.~1.1]{CGYZ26}, and \cite[Thm.~1.5]{CGYZ26} removed it for one mode with the canonical phase POVM. The radial POVM is its multimode analogue: it reads the direction $u$ of the position vector and the Fourier conjugate of its squared length, so that a quadratic phase $e^{ix^TBx/2}$ only translates the outcome, by $u^TBu$, and least squares recovers $B$ with an error whose law does not involve $B$. One such round, between a homodyne and a heterodyne round, removes the term for every $m$ and answers the question of \cite[Sec.~1]{CGYZ26} for pure states, at the optimal rate. Theorem~\ref{thm:NG-thermal} extends the round to mixed states of bounded thermal occupation.

\subsection{Bounded thermal occupation}\label{subsec:NG-thermal}

\begin{theorem}[Bounded thermal occupation, single copies]\label{thm:NG-thermal}
Fix $\bar n\ge0$ and put
\[
K_{\bar n}:=4\big(\bar n+\tfrac12\big)^2\Big(\frac{11}{10}+4\big(\bar n+\tfrac12\big)^2\Big)^2,\qquad N_r(\bar n):=\big\lceil10^{14}(1+2\bar n+2\bar n^2)^3(m+4)^2\big\rceil .
\]
There are constants $C(\bar n)$ and $\eps_\star(\bar n)>0$ and a family of adaptive protocols, indexed by $m$, $\bar n$, $\eps$ and the budget $N$, with the following property. Let $\rho(v,\Sigma)$ be an $m$-mode Gaussian state whose Williamson occupations satisfy $n_1\le\dots\le n_m\le\bar n$, with no bound on $\|\Sigma\|_\op$, on its energy or on its squeezing, and let $\mathcal R(\boldsymbol n)$ be its occupation-profile rate \eqref{eq:Bos-R}.
\begin{enumerate}
\item[(i)] (Centred states, single copies.) If $v=0$, the protocol measures one copy at a time: homodyne detection of the $m$ position quadratures on $N_q$ copies, as in Theorem~\ref{thm:NG-pure}; then, on each of $N_r(\bar n)$ copies, a Gaussian unitary chosen from the first round, four ancillary vacuum modes and the radial POVM $\mathsf R_{m+4}$ of \S\ref{subsec:NG-radialPOVM}; then $54$ independent runs of the protocol of Theorem~\ref{thm:Bos}(i) with $K=K_{\bar n}$, at accuracy $\eps/3$, on copies transformed by the resulting known Gaussian unitary, and a selection among their outputs. It outputs a Gaussian state $\hat\rho$ with $\Pr[T(\rho,\hat\rho)\le\eps]\ge\tfrac23$ whenever $N\ge C(\bar n)\,\mathcal R(\boldsymbol n)/\eps^2$ and $0<\eps\le\eps_\star(\bar n)$.
\item[(ii)] (Arbitrary displacement, blocks of two.) The same holds for every $v$ when the radial POVM is applied to the centred output port of a balanced beam splitter acting on two copies \cite[Lemma~8.1]{CGYZ26}; every other copy is measured alone.
\end{enumerate}
In particular $O_{\bar n}(m^3/\eps^2)$ copies suffice for every such state, and $O_{\bar n,g}(m^2/\eps^2)$ when every $n_j\ge g>0$. For every $\bar n>0$, $m\ge\max\{n_0,(1+\bar n^{-1})/2\}$ and $0<\eps<\eps_0$ with $\eps\le\eps_\star(\bar n)$, $n_0$ and $\eps_0$ being the constants of \cite[Thm.~2]{CMMP26}, every single-copy adaptive protocol that learns every centred state of the class to trace distance $\eps$ with probability $\tfrac23$ uses $\Omega(m^3/\eps^2)$ copies, since the class contains the passive family of \cite[Thm.~2]{CMMP26}; the single-copy rate of the centred states of the class is therefore $\Theta_{\bar n}(m^3/\eps^2)$.
\end{theorem}

The radial round is that of Theorem~\ref{thm:NG-pure}. On a mixed state its outcome law is a negative-binomial mixture of the Student laws of Lemma~\ref{lem:NG-student}, and the antisymmetric part of $\Sigma_{pq}\Sigma_{qq}^{-1}$, which the radial data cannot see, is bounded by the occupation cap through the uncertainty relation (Lemma~\ref{lem:NG-mixedlaw}); after the shear the covariance satisfies $\|\Sigma'\|_\op\le K_{\bar n}$ (Lemma~\ref{lem:NG-window}), the hypothesis of Theorem~\ref{thm:Bos}(i), and symplectic maps preserve the occupations and with them $\mathcal R(\boldsymbol n)$. The first round of Corollary~\ref{cor:NG-gap}, on blocks of $12m+7$ copies, already gives $\Sigma\preceq\bar W\preceq2\Sigma$ (Remark~\ref{rem:NG-12m}), and a Williamson form $\bar W=SDS^T$ gives $\frac14I\preceq S^{-1}\Sigma S^{-T}\preceq(2\bar n+1)I$, symplectic eigenvalues being monotone in the Loewner order, so that Theorem~\ref{thm:Bos}(i) yields the same rate; what is new is the measurement resource, single copies for centred states and pairs for displaced ones. For states of bounded occupation the theorem removes the covariance bound of Theorem~\ref{thm:Bos}. On the interval classes $\mathcal P_K(r;(m_b,\nu_b)_b)$ of Theorem~\ref{thm:Bos}(ii) with $K=2\bar n+1$ and all prescribed occupations at most $\bar n$ ($2/r\le\bar n$ if $r\ge1$, and $2\nu_{\max}\le\bar n$) the hard families are passive and centred (Remark~\ref{rem:Bos-energy-scope}), so the centred states with the same occupation intervals and no covariance bound have, for $m$ large and $\eps$ small depending on $\bar n$, the single-copy rate \eqref{eq:Bos-thm4}, with constants depending on $\bar n$; the upper bound follows from the theorem as in the proof of Theorem~\ref{thm:Bos}(ii). The class contains the squeezed vacua, on which Gaussian measurements cannot learn independently of the energy \cite[Thm.~1.1]{CGYZ26}, so a non-Gaussian round is necessary; the adaptive single-copy Gaussian protocol of \cite[Thm.~3]{BMEM25} pays an additive term doubly logarithmic in the energy and gives no profile rate. The theorem answers the question of \cite[Sec.~1]{CGYZ26} for mixed states of bounded thermal occupation, with single copies, and does not compete with the collective $O(m^2/\eps^2)$ of \cite[Thm.~5.1]{CFGLMWW26}. At $\bar n=0$ it gives $m^3$ through $\mathcal R$, where Theorem~\ref{thm:NG-pure} gives $\Theta(m^2/\eps^2)$. The guarantee is budget-indexed in the sense of Remark~\ref{rem:Bos-scope}.

\begin{corollary}[Cold states with any squeezing, blocks of $k$]\label{cor:NG-cold-k}
Fix $c>0$. There are $m_\star(c)$ and $\eps_\star(c)>0$ such that for $m\ge m_\star$, $0<\eps\le\eps_\star$ and every $k\ge1$, every centred $m$-mode Gaussian state with Williamson occupations $n_j\le c/m$, with no bound on its squeezing or energy, is learned to trace distance $\eps$ with probability $\tfrac23$ by a protocol whose joint measurements act on at most $k$ copies, with classical memory between blocks, from $O_c\big(m^3/(\eps^2\min\{\sqrt k,m\})\big)$ copies; the same holds for arbitrary displacement when $k\ge2$. For $c\ge\tfrac23$, $m\ge n_0$ and $\eps<\eps_0$ this is the $k$-copy rate of the class \cite[Thm.~2]{CMMP26}.
\end{corollary}

The class contains the cold class $\Sigma\preceq(\frac12+c/m)I$ of \cite[Thm.~5.13]{Rub26b}, whose Loewner bound limits the squeezing to $O(1/m)$ per mode; here the squeezing is arbitrary. For centred states with no energy bound, exact purity gives the single-copy rate $\Theta(m^2/\eps^2)$ (Theorem~\ref{thm:NG-pure}), and occupations at most $c/m$ per mode, $c\ge\frac23$, give $\Theta(m^3/\eps^2)$ ($k=1$ here).

\begin{proposition}[A cold group and modes of occupation at least one, blocks of $k$]\label{prop:NG-coldhot-k}
Fix $c\ge\tfrac23$, $\nu\ge1$ and $K\ge2\nu+1$, and let $\mathcal C=\mathcal C(m_c,c;m_w,\nu;K)$ be the class of Corollary~\ref{cor:Bos-twoscale}: arbitrary displacement and squeezing, $\|\Sigma\|_\op\le K$, $m_c$ Williamson occupations at most $c/m_c$ and $m_w$ at least $\nu$. Put $F_k:=m_c^3/\min\{\sqrt k,m_c\}+m_w^2$. There are $m_\star$, $\eps_\star>0$ and $0<c_\star\le C_\star$ depending only on $(c,K)$ such that for $m_c,m_w\ge m_\star$, $0<\eps\le\eps_\star$ and every $k\ge1$:
\begin{enumerate}
\item[(i)] a protocol whose joint measurements act on at most $k$ copies, with classical memory between blocks, learns every state of $\mathcal C$ to trace distance $\eps$ with probability at least $\tfrac23$ from $C_\star F_k/\eps^2$ copies;
\item[(ii)] every such protocol uses at least $c_\star F_k/\eps^2$ copies, and every protocol, collective ones included, at least $c_\star m_w^2/\eps^2$.
\end{enumerate}
If the bound $\|\Sigma\|_\op\le K$ is replaced by $n_j\le\bar n$ for all $j$, with $\bar n\ge2\nu$ and no bound on energy or squeezing, (i) and (ii) hold with constants depending only on $(c,\bar n)$, for centred states with every $k\ge1$ and for arbitrary displacement with every $k\ge2$.
\end{proposition}

The proposition settles the $k$-copy rate of the simplest mixed profile: the cold group gains $\min\{\sqrt k,m_c\}$, as on row B1, and the modes holding at least one photon cost $m_w^2$, as with collective measurements. Its proof is that of Theorem~\ref{thm:Bos-sep}(i) with the cold frame of Lemma~\ref{lem:Bos-coldframe} taken at $\mu=\frac14$, which costs $O(m^2)$ copies rather than $m_c^3$ because the warm occupations are bounded below by a constant, and with the $k$-copy photon-number readout of \cite[Thm.~5.13]{Rub26b}. Occupations between $1/m$ and $1$ remain open (Question~\ref{q:NG-bosk}).
\subsection{The intermediate target}\label{subsec:NG-target}

The proofs of Theorems~\ref{thm:NG1} and~\ref{thm:NG2} produce, from a first round of measurements, a symplectic matrix $S$ whose \emph{pure} covariance $P:=\tfrac12SS^T$ is dominated by the unknown covariance,
\begin{equation}\label{eq:NG-target}
P\preceq C\,\Sigma ,
\end{equation}
with an absolute constant $C$. This is weaker than an accurate estimate of a squeezer: it gives no upper bound on the transformed covariance, and none is needed.

Heterodyne outcomes have covariance $\Gamma:=\Sigma+\frac12I$, and the empirical covariance $\hat\Gamma$ of $N$ samples has a relative error $W_N:=\Gamma^{-1/2}(\hat\Gamma-\Gamma)\Gamma^{-1/2}$ whose law does not depend on $\Gamma$, since $\hat\Gamma=\Gamma^{1/2}\hat G\Gamma^{1/2}$ with $\hat G$ the empirical covariance of standard Gaussian vectors. The relative error of $\hat\Sigma:=\hat\Gamma-\frac12I$ is $\Sigma^{-1/2}\Gamma^{1/2}W_N\Gamma^{1/2}\Sigma^{-1/2}$, amplified with respect to $W_N$ by at most $\|\Sigma^{-1/2}\Gamma\Sigma^{-1/2}\|_\op=1+1/(2\lambda_{\min}(\Sigma))$: a quadrature much narrower than the vacuum is drowned in vacuum noise, while large variances are harmless. After $U_S^\dagger$, \eqref{eq:NG-target} gives $\lambda_{\min}(\Sigma')\ge1/(2C)$ and an amplification at most $1+C$, which is why domination suffices.

\begin{lemma}[domination suffices]\label{lem:NG-target}
Let $\rho=\rho(v,\Sigma)$ be an $m$-mode Gaussian state and $S$ symplectic with \eqref{eq:NG-target} for some $C\ge1$. Then $\rho':=U_S^\dagger\rho U_S=\rho(S^{-1}v,\Sigma')$ with
\[
\Sigma':=S^{-1}\Sigma S^{-T}\succeq\frac1{2C}\,I .
\]
Let $0<\delta<1$, $\chi:=\sqrt{2m}+\sqrt{2\log(2/\delta)}$ and $N\ge100\chi^2$. Let $\hat\rho'$ be the Gaussian state estimated by \cite[Thm.~10]{BMEM25} from heterodyne measurements on $N$ copies of $\rho'$, replaced by a fixed Gaussian state if its estimated covariance violates the uncertainty relation, and $\hat\rho:=U_S\hat\rho'U_S^\dagger$. Then $\hat\rho$ is a Gaussian state and, with probability at least $1-\delta$,
\begin{equation}\label{eq:NG-het}
T(\rho,\hat\rho)=T(\rho',\hat\rho')\le4.3\,(2m+2mC)\,\frac{\sqrt{2m}+\sqrt{2\log(2/\delta)}}{\sqrt N}.
\end{equation}
In particular, for fixed $\delta$, $N=O(C^2m^3/\eps^2)$ copies give $T(\rho,\hat\rho)\le\eps$.
\end{lemma}

\subsection{The measurements}

The measurement of Theorem~\ref{thm:NG1} is a one-mode POVM $\mathsf M$ with outcomes $(s,a,b)\in\{+,-\}\times(0,\infty)\times\R$ and effects
\[
\mathsf M(s,da\,db)=|\chi_{s,a,b}\rangle\langle\chi_{s,a,b}|\,\frac{5\,da\,db}{8\pi a^2},\qquad\chi_{s,a,b}(x)=\sqrt{\frac2{\Gamma(7/2)}}\;\mathbf 1_{\{sx>0\}}\,a^{7/4}|x|^3e^{-(a-ib)x^2/2}.
\]
It is not Gaussian (the half-line support and the factor $|x|^3$ are essential), and it is covariant under dilations and quadratic shears: the outcome $(a,b)$ is read as the wave packet $e^{-(a-ib)x^2/2}$, and for a centred state of position variance $\sigma$ the scale $\tau:=2\sigma a$ has the energy-free law $\mathrm{BetaPrime}(5/2,1/2)$ (Lemma~\ref{lem:NG-scale}). The measurement of Theorem~\ref{thm:NG2} and Corollary~\ref{cor:NG-gap} acts on $k$ copies of a $d$-mode state: with $|\psi_K\rangle=\det(I-KK^\dagger)^{1/4}\exp(\frac12a^{\dagger T}Ka^\dagger)|0\rangle$ the centred pure Gaussian state of a complex symmetric $d\times d$ matrix $K$ with $KK^\dagger\prec I$, and $k>2d$,
\[
M_k(dK)=\frac1Z\,|\psi_K\rangle\langle\psi_K|^{\otimes k}\,\frac{dK}{\det(I-KK^\dagger)^{d+1}},
\]
completed by the projector onto the orthogonal complement of the span of the $|\psi_K\rangle^{\otimes k}$; Theorem~\ref{thm:NG2} ($k=4m+5$) and Corollary~\ref{cor:NG-gap} ($k=12m+6$) apply it with $d=2m$ to the output of the quasi-purification channel of \cite{CFGLMWW26}. The construction of both measurements, their exact outcome laws and covariance moments, and the whitened continuity bound used for the corollary are in \S\S\ref{subsec:NG-wavelet}--\ref{subsec:NG-continuity} of Appendix~\ref{app:proofs-NG}.

The measurement of Theorem~\ref{thm:NG-pure} acts on one copy together with four ancillary vacuum modes, $n=m+4$ modes in all, in the position representation. With $x=\sqrt t\,u$, $t>0$, $u\in S^{n-1}$ and $dS$ the surface measure of the sphere,
\[
(\mathcal V\psi)(u,b)=\frac1{\sqrt{8\pi}}\int_0^\infty e^{-ibt/2}\,t^{(n-2)/4}\,\psi(\sqrt t\,u)\,dt,\qquad\mathsf R_n(E)=\mathcal V^\dagger\mathbf 1_E\mathcal V,
\]
for Borel sets $E\subseteq S^{n-1}\times\R$: $\mathsf R_n$ reads the direction $u$ of the position vector and the Fourier conjugate $b$ of $t=|x|^2$. $\mathcal V$ is an isometry that is not onto, so $\mathsf R_n$ is a POVM that is not projective, as is the canonical phase POVM of \cite[Thm.~1.5]{CGYZ26}; on a centred pure state with wave function proportional to $e^{-x^TAx/2+ix^TBx/2}$ the outcome $b$, given $u$, has a Student law with $n/2$ degrees of freedom centred at $u^TBu$, and its scale $u^TAu/\sqrt{n/2}$ does not involve $B$ (Lemmas~\ref{lem:NG-Rn} and~\ref{lem:NG-student}).

\subsection{A fixed symplectic gap}\label{subsec:NG-gap}

The relative error of an $N$-sample covariance estimate behaves like a random symmetric $2m\times2m$ matrix with entries of order $N^{-1/2}$, of Frobenius norm of order $m/\sqrt N$ (exactly, $\E\|W_N\|_F^2=((2m)^2+2m)/N$, from $\E(g_ig_j-\delta_{ij})^2=1+\delta_{ij}$ for a standard Gaussian vector $g$) and trace norm of order $m^{3/2}/\sqrt N$. With a gap, Lemma~\ref{lem:NG-cont} controls the trace distance by the Frobenius norm, and $m^2/\eps^2$ copies suffice; without one the bound of \cite{BMEM25} charges the trace norm, and near pure modes errors do add linearly: the vacuum and the product of $m$ thermal modes of occupation $n$ are at trace distance $1-(1+n)^{-m}\approx mn$, while their whitened Frobenius distance is $\|2nI_{2m}\|_F=2n\sqrt{2m}$.

\begin{corollary}[fixed symplectic gap, blocks linear in $m$]\label{cor:NG-gap}
Fix $g>0$. There are a constant $C_g$, depending only on $g$, and a protocol using joint measurements on $\lceil9600m\rceil$ blocks of $12m+7$ copies followed by single-copy heterodyne measurements, which learns every $m$-mode Gaussian state $\rho(v,\Sigma)$ with $\Sigma\in\mathcal K_g$ (all symplectic eigenvalues at least $(1+g)/2$), of arbitrary displacement, energy and squeezing, to trace distance $\eps\in(0,1)$ with probability at least $2/3$ from $C_gm^2/\eps^2$ copies.
\end{corollary}

The first round is the variant of Remark~\ref{rem:NG-12m}, blocks of $12m+6$ centred copies with $\Sigma\preceq\bar W\preceq2\Sigma$ after $\lceil9600m\rceil$ blocks; after it $\Sigma'\succeq\frac14I$, the heterodyne relative error is at most three times the standard Gaussian one, and Lemma~\ref{lem:NG-cont} with $c=g/2$ converts it into trace distance (Appendix~\ref{app:proofs-NG}), with a constant growing like $c^{-1/2}$ as the gap closes, as in the Frobenius-to-trace conversion on the fermionic side.

\begin{remark}[measurement resource of Corollary~\ref{cor:NG-gap}]\label{rem:NG-gap}
The rate $O_g(m^2/\eps^2)$ is the order of the collective rate of \cite[Thm.~5.1]{CFGLMWW26}, which holds for all Gaussian states with no gap promise; the corollary is new only in its measurement resource, blocks of $12m+7$ copies followed by single-copy heterodyne detection (Theorem~\ref{thm:NG2} alone needs only blocks of $4m+6$, at the price of a first round of $O(m^3)$ copies). The cited lower bound is the collective $\Omega((m^2+\log\delta^{-1})/\eps^2)$ of \cite[Thm.~5.2]{CFGLMWW26}, which is proved on pure states; these lie outside $\mathcal K_g$, and no lower bound on $\mathcal K_g$ itself is claimed here. The corollary is not a single-copy theorem: the blocks are entangled. On the subclass $\Sigma\succeq(\tfrac12+\nu)I$ no blocks are needed: single-copy heterodyne detection alone gives $O((1+\nu^{-1})m^2/\eps^2)$ with no bound on the energy \cite[Thm.~8]{CMMP26}. That subclass lies in $\mathcal K_{2\nu}$, which is larger: a squeezed state can have all symplectic eigenvalues at least $\tfrac12+\nu$ while one of its quadrature variances is below the vacuum level, and the blocks of the corollary deal with such states.
\end{remark}

\subsection{Scope}\label{subsec:NG-scope}

\begin{remark}[what Theorems~\ref{thm:NG1}--\ref{thm:NG-thermal} settle]\label{rem:NG-scope}
Theorem~\ref{thm:NG1} removes the obstruction of arbitrary unknown squeezing for centred mixed one-mode Gaussian states with adaptive single-copy measurements, and for arbitrary one-mode Gaussian states with two-copy preprocessing; it does not assert single-copy learning of displaced states. Theorem~\ref{thm:NG2} handles arbitrary unknown correlated multimode squeezing and arbitrary means and temperatures with entangled blocks whose size is linear in the number of modes. Its total $m^3/\eps^2$ has the order of the general single-copy rate (row B1 of Table~\ref{tab:main}), but its measurement resource is stronger than single copies, and it is worse than the collective $m^2/\eps^2$ of \cite{CFGLMWW26}. It is not claimed optimal for blocks of $k$ copies: the lower bounds on the cold class, $m^3/(\eps^2\min\{\sqrt k,m\})$ \cite{CMMP26,Rub26b}, are smaller when $k$ grows with $m$. Theorem~\ref{thm:NG-pure} gives the rate $\Theta(m^2/\eps^2)$ of centred pure states with single copies, and of all pure states with blocks of two copies. Theorem~\ref{thm:NG-thermal} gives single-copy learning of centred states whose Williamson occupations are at most a known $\bar n$, at the profile rate of Theorem~\ref{thm:Bos} and $\Theta_{\bar n}(m^3/\eps^2)$ in the worst case, and Corollary~\ref{cor:NG-cold-k} and Proposition~\ref{prop:NG-coldhot-k} give the $k$-copy rates of cold states and of a cold group with modes of occupation at least one, with any squeezing (for the latter, under a known bound $\bar n$ on the occupations).

The following are not settled here:
\begin{itemize}
\item energy-independent learning of general correlated mixed $m$-mode Gaussian states, $m\ge2$, with no bound on their Williamson occupations, with single-copy measurements (on the subclass $\Sigma\succeq(\tfrac12+\nu)I$ heterodyne detection does this \cite[Thm.~8]{CMMP26});
\item energy-independent learning of Gaussian states of arbitrary displacement with single-copy measurements only, already for one mode and for pure states;
\item energy-independent learning of mixed states with no bound on their Williamson occupations with blocks of every fixed small $k$ as $m$ grows, and the sharp $k$-dependent rate on the unrestricted mixed class;
\item the profile rate of Theorem~\ref{thm:Bos} for states with unknown unbounded squeezing and no bound on their Williamson occupations: a coarse frame with $\Sigma'\succeq\frac14I$ does not by itself identify the passive profile frame that its proof uses, and the window $\Sigma'\preceq K_{\bar n}I$ of Theorem~\ref{thm:NG-thermal} needs the bound $n_j\le\bar n$;
\item an efficient finite-resource implementation of the non-Gaussian POVMs used here. These resources are not counted: the inverse squeezer $U_S^\dagger$ must apply squeezing comparable to that of the state, and no implementation of $\mathcal Q^{\rm Bose}_{k,m}$, which rests on the Kashiwara--Vergne decomposition, is known \cite[Sec.~6]{CFGLMWW26}.
\end{itemize}
\end{remark}

\section{Discussion: what the table says about measurement resources}\label{sec:discussion}

The columns of Table~\ref{tab:main} bound the number of input copies processed coherently in one measurement: one (single copy, the setting without quantum memory of \cite{BCL20,CCHL22,CHLLS22}), at most $k$ (blocks of $k$ copies), or all of them (collective). Within a block the copies may be stored and processed jointly; no quantum state passes between blocks. The columns are not a theorem about general streaming models of quantum memory, and one bosonic copy is already infinite-dimensional.

Measuring one copy at a time costs a factor $m$, and for mixed fermionic states the factor is paid at every temperature. The lower bound of \cite{Rub26}, whose hard instances sit at the maximally mixed state, holds for every single-copy protocol, adaptive or not, with or without ancillas; Theorem~\ref{thm:B} shows that nothing more is lost towards the pure boundary, and the quadratic Gibbs balls of every radius $\beta_0>0$ keep the single-copy rate $\Theta(m^3/\eps^2)$ for $\eps\le\min\{1/3600,\beta_0/(120\sqrt3)\}$ \cite{Rub26b}, so warmth does not help either (a promise of purity does: row F6). For bosons the factor is $\min\{m,1+\nu^{-1}\}$ on the class $\Sigma\succeq(\tfrac12+\nu)I$ \cite{CMMP26}, and in a mixed profile each warm band pays according to its own occupation (Theorem~\ref{thm:Bos}).

Coherent processing of a bounded number of copies is worth more at high temperature than at low. That joint measurements on several copies can extract more than separate ones goes back to \cite{MP95} (for mixed qubit states see \cite{HM08,BBGMM06}, and for shadow tomography \cite{KGKB25}). On every fixed-gap class $\mathcal G_c$ blocks of $m$ copies already reach the collective order $m^2/\eps^2$, and the covariance itself is learned in Frobenius norm at that order on every state (Theorem~\ref{thm:K}(iii)): the block does not need to contain all copies. For small blocks the gain is a constant factor: at the maximally mixed state the best measurement on two copies has Fisher-information trace $4m(1-\binom{2m}{m}4^{-m})$ against $2m$ for two single copies \cite[Cor.~3.7]{Rub26b}, a Cram\'er--Rao ratio of per-copy variances of $4/5$ at $m=2$ that tends to $1/2$ as $m$ grows. On the cold classes the gain is only $\min\{\sqrt k,m\}$ \cite{CMMP26,Rub26b}, so for $1<k<m^2$ the same block size buys more for a warm class than for a cold one, although with single copies the warm class is no cheaper.

Matching the noise to the metric scale by scale is what removes the factor $1/c$ from Theorem~\ref{thm:B}; the route through a Frobenius-norm covariance estimate loses a factor that grows as the state approaches purity, since the conversion from Frobenius error to trace distance has constant of order $c^{-1/2}$ and no dimension-free form near pure states \cite{Rub26}. We expect the principle to be the right starting point for the tomography of other structured families near their pure boundary.

The exact unbiasedness of the $k$-copy estimator has a different source, the shift by twice the Weyl vector, the compact-group form of the standard fact that the $P$-symbol of $J_z$ in spin coherent states is $(j+1)\cos\theta$ while its $Q$-symbol is $j\cos\theta$ \cite{Lie73,Ber75,Per86}; upper and lower symbols for the coherent states of general compact groups were studied in \cite{Sim80}. For $U(d)$ the shift is already in the literature: Keyl's measurement \cite{Keyl06} returns a Young diagram $\lambda$ and a unitary $U$, and reading it out as $U\,\mathrm{diag}(\lambda+2\rho)\,U^\dagger/n$, with $\rho$ the Weyl vector of $U(d)$, i.e.\ $(\lambda+2\rho)_i=\lambda_i+d+1-2i$, is the staircase estimator of Pelecanos, Spilecki and Wright \cite[Sec.~3.3]{PSW26}, which they prove unbiased. Their debiased Keyl estimator \cite[Defs.~1.1 and~1.2]{PSW26} averages $\lambda+2\rho$ over each block of equal rows of $\lambda$ and is the minimum-variance member of a family of unbiased readouts that contains the staircase estimator; the stabiliser average of Remark~\ref{rem:K-minvar} is its $\mathrm{Spin}(2m)$ counterpart. The proof of Proposition~\ref{prop:K4} uses only highest-weight theory and equivariance, and we expect the shift to give exactly unbiased covariant shadows for the highest-weight coherent-state measurements of every compact group.

For bosonic devices, such as the multimode squeezed-light sources of Gaussian boson sampling \cite{Zho20,Mad22}, the energy is often unknown and the squeezing may be large. Energy-independent tomography then needs non-Gaussian measurements or access to the transposed state \cite{BMEM25,CGYZ26}, and non-Gaussian measurements matter even with an energy bound \cite{CMFLMHCP26}. Where the collective protocol of \cite{CFGLMWW26} uses a joint measurement on all copies, Section~\ref{sec:unsqueezing} needs entangled measurements only once, on blocks of $4m+6$ copies, and for one centred mode or a centred pure state none; displaced passive states need single copies alone (Theorem~\ref{thm:Bos-energy}).

The constant of Theorem~\ref{thm:B} is absolute but very large (Remark~\ref{rem:B-count}), and we have made no attempt to optimise it; for a device with tens of modes the practical single-copy estimator remains matchgate shadows \cite{ZRM21,WHLB23}, with $\lceil12m^3/(c\eps^2)\rceil$ copies on $\mathcal G_c$ \cite[Thm.~7.6]{Rub26}. The $k$-copy estimator of Theorem~\ref{thm:K}, by contrast, has small explicit constants.

Fermionic Gaussian states are prepared on superconducting processors as Hartree--Fock states \cite{Aru20} (readouts built from free evolution and occupation measurements are also analysed for analogue simulators \cite{DMDGE24}), classical shadows of fermionic states have been measured on such processors \cite{HOR22}, and site-resolved occupation readout of fermionic atoms is available in quantum-gas microscopes \cite{Che15}. Multimode squeezed light on up to a few hundred modes is the resource of Gaussian boson sampling, and continuous-variable states are characterised by homodyne or heterodyne detection \cite{LR09}. For such devices the table separates three regimes: mixed fermionic states, where every single-copy readout pays a factor $m$ in the worst case over the class when nothing beyond Gaussianity or a temperature bound is promised (certifying a known pure target is a different task \cite{GKEA18}); warm bosonic modes, where heterodyne detection is optimal up to a constant factor, so that coherent processing of several copies can save at most a constant factor; and cold bosonic modes, which cost $m^3/\eps^2$ with single copies as fermions do \cite{CMMP26}. Collective measurements on two and three copies have been demonstrated \cite{Hou18,Con23,ZYY25}. For $m=50$ modes and $\eps=0.05$ the single-copy scaling $m^3/\eps^2$ is $5\cdot10^7$ copies and the collective scaling $10^6$, about $80$ minutes and under two minutes at $10^4$ shots per second; these figures set every constant to one, and $\eps=0.05$ lies outside the range of the lower bounds, so they illustrate the factor $m$ rather than predict a run time. For mixed fermionic states at a fixed distance from purity, blocks of ten copies, $500$ qubits measured jointly, lower the rate by a factor $10$ (Theorem~\ref{thm:K}), and with the explicit constants the guaranteed count improves on matchgate shadows by $12k/34\approx3.5$. The $k$-copy model stores copies only within a block, and no quantum state passes between blocks; the copies are a product state, the entanglement is in the measurement, and the lower bounds, which count copies consumed, allow arbitrary preprocessing that respects the single-copy or block restriction, while the upper bounds assume noiseless copies.

\paragraph{Learning a quadratic Hamiltonian from its thermal state.}
The Gibbs states of row F3 are parametrised by their Hamiltonians: $\rho_\theta\propto\exp\big(\sum_{a<b}\theta_{ab}\,ic_ac_b\big)$ is the Gibbs state $e^{-\beta K_h}/\Tr e^{-\beta K_h}$ with $\theta=-\beta h/2$, and $i\Gamma=\tanh(i\theta)$ (\S\ref{subsec:bg-fermions}). On the ball $\mathcal C(\beta_0):=\{\rho_\theta:\|\theta\|_2\le\beta_0\}$, $\|\theta\|_2^2:=\sum_{a<b}\theta_{ab}^2$ \cite[Sec.~4]{Rub26b}, the map $\theta\mapsto\Gamma$ and its inverse are Lipschitz in Frobenius norm, so Theorem~\ref{thm:K}(iii) is also a statement about learning the coefficient matrix of the Hamiltonian, in units of the temperature.

\begin{corollary}[learning a quadratic fermionic Hamiltonian from its Gibbs state]\label{cor:D-ham}
Let $\beta_0>0$ and $k\ge1$. Call a $k$-copy protocol an \emph{$\eta$-learner on $\mathcal C(\beta_0)$} if from copies of any $\rho_\theta\in\mathcal C(\beta_0)$ it outputs a matrix $\hat\theta$ with $\|\hat\theta-\theta\|_F\le\eta$ with probability at least $2/3$.
\begin{enumerate}
\item[(i)] For $0<\eta\le\cosh^2\beta_0$ there is an $\eta$-learner on $\mathcal C(\beta_0)$ that uses
\[
N\le67\cosh^4(\beta_0)\,\frac{\max\{m^3/k,\,m^2\}}{\eta^2}
\]
copies: take the estimate $\bar\Gamma$ of Theorem~\ref{thm:K}(iii) with $t=\min\{k,m\}$ and accuracy $\eta/\cosh^2\beta_0$, its Frobenius projection $\tilde\Gamma$ onto $\mathcal G_{1-\tanh\beta_0}$, and $\hat\theta:=-i\operatorname{artanh}(i\tilde\Gamma)$.
\item[(ii)] For $0<\eta\le\min\{1/3600,\ \beta_0/(120\sqrt3)\}$ every $\eta$-learner on $\mathcal C(\beta_0)$ uses
\[
N>\frac1{\eta^2}\max\Big\{\frac{m^3}{8{,}553{,}600\,k},\ \frac{m^2}{4{,}276{,}800}\Big\}
\]
copies.
\end{enumerate}
Hence, for fixed $\beta_0$ and $\eta$ in the window of (ii), the $k$-copy sample complexity of learning $\theta$ in Frobenius norm is $\Theta_{\beta_0}(\max\{m^3/k,m^2\}/\eta^2)$: of order $m^3/\eta^2$ with single copies, and of the collective order $m^2/\eta^2$ exactly when the blocks hold of order $m$ copies or more.
\end{corollary}

\begin{proof}
We use two facts. First, if $A,B$ are Hermitian with spectra in an interval on which $f$ is $L$-Lipschitz, then $\|f(A)-f(B)\|_F\le L\|A-B\|_F$: with spectral decompositions $A=\sum_ia_i|u_i\rangle\langle u_i|$ and $B=\sum_jb_j|v_j\rangle\langle v_j|$,
\[
\|f(A)-f(B)\|_F^2=\sum_{i,j}|f(a_i)-f(b_j)|^2\,|\langle u_i|v_j\rangle|^2\le L^2\sum_{i,j}|a_i-b_j|^2\,|\langle u_i|v_j\rangle|^2=L^2\|A-B\|_F^2 .
\]
Second, the eigenvalues of $i\theta$ are $\pm\lambda_j$ with $\sum_j\lambda_j^2=\|\theta\|_2^2$, so on $\mathcal C(\beta_0)$ the spectrum of $i\theta$ lies in $[-\beta_0,\beta_0]$, that of $i\Gamma=\tanh(i\theta)$ in $[-\tanh\beta_0,\tanh\beta_0]$, and $\Gamma\in\mathcal G_{1-\tanh\beta_0}$ \cite[Thm.~4.1(ii)]{Rub26b}.

(i) Put $\eps:=\eta/\cosh^2\beta_0\le1$. By Theorem~\ref{thm:K}(iii), at most $67\max\{m^3/k,m^2\}/\eps^2$ copies give $\|\bar\Gamma-\Gamma\|_F\le\eps$ with probability at least $2/3$. The projection onto the closed convex set $\mathcal G_{1-\tanh\beta_0}$ fixes $\Gamma$ and is non-expansive ((T3) of \S\ref{subsec:furthertools}), so $\|\tilde\Gamma-\Gamma\|_F\le\eps$. Since $\tilde\Gamma$ is real antisymmetric with $\|\tilde\Gamma\|_\op\le\tanh\beta_0$ and $\operatorname{artanh}$ is odd, $\hat\theta$ is well defined, real and antisymmetric. On $[-\tanh\beta_0,\tanh\beta_0]$ the function $\operatorname{artanh}$ has derivative at most $1/(1-\tanh^2\beta_0)=\cosh^2\beta_0$, and $i\theta=\operatorname{artanh}(i\Gamma)$, so the first fact gives $\|\hat\theta-\theta\|_F\le\cosh^2(\beta_0)\,\|\tilde\Gamma-\Gamma\|_F\le\eta$.

(ii) Let $\mathcal P$ be an $\eta$-learner on $\mathcal C(\beta_0)$. Replacing $\hat\theta$ by the antisymmetric part of its real part does not increase the Frobenius error. Then $\widehat\Gamma:=-i\tanh(i\hat\theta)$ is real and antisymmetric, and since $\tanh$ is $1$-Lipschitz the first fact gives $\|\widehat\Gamma-\Gamma\|_F\le\|\hat\theta-\theta\|_F\le\eta$ with probability at least $2/3$ on $\mathcal C(\beta_0)$. The lower-bound argument in the proof of Theorem~\ref{thm:K}(iii) (\S\ref{subsec:K-iii}) uses its hypothesis only on the hard family $\{\rho_\theta:\|\theta\|_2\le120\sqrt3\,\eta\}$ of \cite[Thms.~3.4 and~3.11]{Rub26b}, which lies in $\mathcal C(\beta_0)$ when $120\sqrt3\,\eta\le\beta_0$; for $\eta\le1/3600$ it gives $N>m^3/(8{,}553{,}600\,k\,\eta^2)$ and $N>m^2/(4{,}276{,}800\,\eta^2)$.
\end{proof}

Three comparisons place the corollary. For local Hamiltonians at high temperature, single-copy measurements of the local terms attain the optimal number of copies, up to a constant for error in the maximum norm and up to a logarithm in $\ell_2$; the matching lower bound is proved on commuting Hamiltonians, whose Gibbs states are diagonal, so it holds for every measurement \cite{HKT24}. No gap of this kind arises there. For the quadratic Majorana observables themselves, estimating every expectation $\Tr(\rho\,ic_ac_b)$ of an arbitrary state to additive accuracy $\eps$ needs $\Omega(m/\eps^2)$ copies with single-copy measurements, adaptive ones included, and $O(\log m/\eps^4)$ with two-copy measurements \cite{KGKB24}: entry by entry, blocks of two copies remove the dependence on $m$ up to a logarithm, at the price of $\eps^{-4}$, whereas in Frobenius norm blocks of $k$ copies gain the factor $\min\{k,m\}$, up to constants. For bosonic Gaussian states, the Hamiltonian is learned from heterodyne data on interaction graphs of bounded degree, under bounded temperature, squeezing and displacement, with a number of copies logarithmic in $m$, in \cite{FRS24}, which lists as open the fermionic case, lower bounds, and the possible benefit of entangled measurements. Corollary~\ref{cor:D-ham} gives a fermionic counterpart of all three in a different setting, without locality, in Frobenius norm and for $\|\theta\|_2\le\beta_0$; the questions of \cite{FRS24} in their own, local, setting are not addressed here. Work by Escobar Fernandez-Marcote, Fanizza and Stilck Fran\c{c}a on the high-temperature regime of Hamiltonian learning of Gaussian states is announced in \cite[Note added]{CMMP26}.

\paragraph{Arrays of cold sensors.}
The cold bosonic rows describe an array of $m$ receivers of faint thermal radiation, such as the telescopes of an optical interferometer. The modes are the receivers and a copy is one coherence time of the field. The state is a passive centred Gaussian state with occupation matrix $Q$ (\S\ref{subsec:bg-bosons}); the trace $s$ of its fugacity matrix $X=Q(I+Q)^{-1}$ is, to first order in $s$, the mean number of photons per coherence time, and its one-photon block, normalised, is $X/s$, to first order the normalised matrix of mutual coherences $Q/\Tr Q$ from which an image is formed. We consider the weak-source regime $s\ll1$, where the total mean photon number across the array in one coherence time is small, and ideal independent temporal copies. The protocol of \cite[Thm.~5.9]{Rub26b} estimates $s$ from photon counts on single copies, then projects each copy onto a total-photon-number sector without resolving the occupied receiver, and applies the joint measurement of \cite{NZ26} to groups of one-photon copies within each block. For a mean photon number per coherence time at most a constant it learns such a state to trace distance $\eps$, and with probability at least $0.95$ it uses
\[
O\Big(\frac1{\eps^2}+\frac{m^3\,s}{\eps^2\max\{1,\min\{\sqrt{sk},m\}\}}\Big)
\]
copies when the blocks span $k$ coherence times. In the second-stage term of this bound, coherent blocks give no asymptotic improvement for $sk\lesssim1$, a factor of order $\sqrt{sk}$ for $1\lesssim sk\lesssim m^2$, and a factor of order $m$ beyond that range. The first-stage term $1/\eps^2$ can dominate, so these factors need not be gains in the total copy count. On the constant-brightness cold class, the worst-case gain of a $k$-copy protocol is at most of order $\min\{\sqrt k,m\}$ \cite[Thm.~2]{CMMP26}. For telescope arrays, a Fisher-information counterpart of the same phenomenon, a penalty that grows with the number of stations when photons are measured one at a time and that joint measurement of many photons removes up to a constant factor, was obtained with a receiver design by Guo, Miao, Cai and Yang \cite{GMCY26}; see also \cite{GJC12} for quantum networks in optical interferometry and \cite{MSLB25} for quantum computation in optical imaging. We do not determine the rate at fixed brightness $s\ll1$: no lower bound is proved there (that of \cite{CMMP26} is for $s$ of order one), and the term $1/\eps^2$ of the display, which dominates for $s<1/m^3$, is not optimised. For $s\lesssim\eps$ the vacuum is already an $\eps$-accurate output \cite[Rem.~5.10]{Rub26b}, and the task relevant to imaging there is to learn $X/s$.

The smallest instance is within reach of present devices: for two fermionic modes, two qubits per copy, the largest Fisher-information trace of a measurement on two copies at the maximally mixed state is $5$, against $4$ for two single copies \cite[Cor.~3.7]{Rub26b}. \cite{Rub26} reports a hardware calibration of a two-copy measurement of this quantity: a pilot gave a one-sided $95\%$ shot-noise lower bound of $3.99$, against the separable value $4$, and an independent replication gave $4.12$, both assuming that the inputs were prepared as intended; preparation error and drift are not bounded by the stated confidence. The estimator is a separate test. By the Cram\'er--Rao bound the per-copy variance of an unbiased two-copy estimator at $m=2$ cannot be smaller than $4/5$ of the single-copy optimum, which matchgate shadows attain; the estimator \eqref{eq:K-est} with $t=2$ reaches only $35/36$ of it, its minimal-variance variant (Remark~\ref{rem:K-minvar}) $5/6$, and the Fisher-symmetric two-copy measurement used in \cite{Rub26} $4/5$. The ratio of \eqref{eq:K-est} to matchgate shadows improves with $m$ (to $0.83$ at $m=4$ and towards $2/3$), so an estimator-level test at $m=2$ has to resolve a variance reduction of $17$--$20\%$ with the minimal-variance or Fisher-symmetric readout. Since $\Gamma=\theta+O(\theta^3)$ at high temperature, the same experiment is also the smallest instance of the task of Corollary~\ref{cor:D-ham}, though not a test of its scaling in $m$: programme a quadratic Hamiltonian on two modes, prepare its Gibbs state as the mixture \eqref{eq:bg-product}, drawing for each copy independently an occupation-number state of the normal modes and applying the corresponding Gaussian unitary from the normal-mode basis to the laboratory mode basis, with the sampled occupation labels withheld from the estimator, and compare the mean squared errors of $\hat\theta$ from two-copy and from single-copy measurements at equal numbers of copies; in the local large-sample limit at $\theta=0$, the leading mean-squared-error ratio is that of the covariance estimates above. For the cold bosonic rows, a receiver for four stations and blocks of two photons is designed in \cite{GMCY26}.

\section{Open questions}\label{sec:open}

\begin{question}[blocks of $k$ copies when the gap closes]
Theorem~\ref{thm:K} gives the $k$-copy rate $\Theta_c(\max\{m^3/k,m^2\}/\eps^2)$ on every class $\mathcal G_c$ of covariance matrices with $\|\Gamma\|_\op\le1-c$, and \cite{Rub26b} gives $\Theta(m^3/(\eps^2\min\{\sqrt k,m\}))$ on the cold number-conserving classes. Between the two lies the regime $c=c(m)\to0$: the upper constant $34/c$ of Theorem~\ref{thm:K} and the absolute lower constant separate, and for $c\le1/(2m+1)$ the class $\mathcal G_c$ contains the local near-vacuum family of \cite[Thm.~5.5]{Rub26b}, so the lower bound becomes $m^3/(\eps^2\min\{\sqrt k,m\})$ while the best upper bound, the smaller of $34\max\{m^3/k,m^2\}/(c\eps^2)$ and the single-copy count $O(m^3/\eps^2)$ of Theorem~\ref{thm:B}, does not match it. What is the $k$-copy rate on $\mathcal G_c$ as a function of $c$? The same question for fermionic Gaussian states near the vacuum with pairing terms, where there is no heterodyne stage to remove them, is the remaining case of the first part of Question~7.1 of \cite{Rub26b}.
\end{question}

\begin{question}[bosons with blocks of $k$ copies]\label{q:NG-bosk}
The $k$-copy rate at intermediate bosonic temperatures $1/m\ll\nu\ll1$ is the second part of Question~7.1 of \cite{Rub26b} and the first open problem of \cite{CMMP26}: nothing beyond the trivial bracket $[\Omega(m^2/\eps^2),O(m^2/(\nu\eps^2))]$ is known for $k>1$. The profile rate of Theorem~\ref{thm:Bos} is proved for single copies; its $k$-copy form is open for the same reason, except for a cold group together with modes of occupation at least one (Proposition~\ref{prop:NG-coldhot-k}).
\end{question}

\begin{question}[energy and squeezing]
Theorem~\ref{thm:Bos-energy} removes the bound on the covariance for displaced passive states. For states with unknown squeezing, no protocol using Gaussian measurements, adaptive and entangled ones included, learns Gaussian states with a number of copies independent of their energy \cite{CGYZ26}; the collective non-Gaussian protocol of \cite{CFGLMWW26} does, at the parametric rate; the adaptive single-copy Gaussian protocol of \cite{BMEM25} pays an additive term doubly logarithmic in the energy $E$, which no Gaussian protocol avoids \cite{CGYZ26}; with non-Gaussian measurements, \cite{CGYZ26} learn one-mode pure Gaussian states from $O(1/\eps^2)$ copies independently of $E$, with single copies for centred states and two-copy measurements otherwise. Theorem~\ref{thm:NG2} learns every Gaussian state with no energy or squeezing promise from $O(m^3/\eps^2)$ copies with non-Gaussian measurements on blocks of at most $4m+6$ copies, against the collective $\Omega(m^2/\eps^2)$; whether blocks of boundedly many copies, or single copies for $m\ge2$ modes as asked in \cite[Question~10.9]{Rub26}, can do the same for mixed states with no bound on their Williamson occupations is open. The class contains the cold passive states, on which single copies need $\Omega(m^3/\eps^2)$ and blocks of $k$ copies $\Omega(m^3/(\eps^2\min\{\sqrt k,m\}))$ \cite[Thms.~1 and~2]{CMMP26}; so for single copies and for blocks of bounded size the question is whether $O(m^3/\eps^2)$ can be attained, and a rate below $m^3$ is possible only with blocks that grow with $m$. Theorem~\ref{thm:NG1} answers the one-mode case of \cite[Question~10.9]{Rub26}: a centred one-mode Gaussian state is learned from $O(\eps^{-2})$ adaptive single-copy measurements uniformly in its squeezing and energy, and Theorem~\ref{thm:NG-pure} answers it for centred pure states of any number of modes, at the rate $\Theta(m^2/\eps^2)$, and Theorem~\ref{thm:NG-thermal} for centred states whose Williamson occupations are at most a known $\bar n$, at the profile rate of Theorem~\ref{thm:Bos}, $\Theta_{\bar n}(m^3/\eps^2)$ in the worst case.
\end{question}

\begin{question}[other resources]
The rates for Jordan--Wigner-local measurements (\cite[Question~10.4]{Rub26} asks whether $\Theta(m^4/\eps^2)$ is needed), for protocols with $s$ qubits of persistent quantum memory, and for interacting (quadratic plus quartic) Gibbs states remain as stated in \cite{Rub26}; learning fermionic Lindbladians under parity superselection \cite{MSR26} is a neighbouring problem.
\end{question}

\section*{Acknowledgements}
I thank Senrui Chen, Antonio Anna Mele, Francesco Anna Mele and John Preskill for correspondence about the relation between \cite{CMMP26} and \cite{Rub26}.

\section*{Statement on the use of artificial-intelligence tools}
The results in this paper were obtained by the author. LLM systems were used to check work and help with calculations and to prepare the manuscript text, as well as search the literature and check references. No language model is an author, and the author takes full responsibility for every statement in this paper.

\appendix
\section{Background: derivations}\label{app:background}

This appendix contains the derivations behind Section~\ref{sec:background} that are not available in the textbooks in the conventions of this paper: Wick's four-point formula with $G=I-i\Gamma$, the polar decomposition of a symplectic matrix used in Appendix~\ref{app:proofs-Bos}, and, in its last subsection, the proofs of the fermionic tools stated in Section~\ref{sec:theoremA}.

\subsection{Wick's theorem}\label{app:bg-wick}

\begin{proof}[Proof of Proposition~\ref{prop:bg-wick}]
\emph{Full-rank states.} Let $\rho=e^{-\beta K_h}/\Tr e^{-\beta K_h}$, $h$ real antisymmetric. The second identity of Lemma~\ref{lem:A-rot} holds for complex antisymmetric $A$ by the same proof, and $e^{-\beta K_h}=e^{-iK_{-i\beta h}}$; hence, with $c(u):=\sum_au_ac_a$ for $u\in\C^{2m}$, $\rho^{-1}c(u)\rho=c(M^{-1}u)$ with $M^{-1}=e^{\beta\,ih}\succ0$, and $\{c(u),c(w)\}=2u^Tw$. Let $x_1=c(u)$, $x_j=c(w_j)$ for $j\ge2$, $Y:=x_2\cdots x_{2n}$, and $Y_{\hat j}$ the product with $x_j$ omitted. Moving $x_1$ to the right through the $2n-1$ factors of $Y$,
\[
x_1Y=\sum_{j=2}^{2n}(-1)^j\{x_1,x_j\}\,Y_{\hat j}-Yx_1 .
\]
Take expectations and use cyclicity, $\Tr(\rho Yx_1)=\Tr\big(\rho\,(\rho^{-1}x_1\rho)\,Y\big)$. With $\varphi(u):=\langle c(u)Y\rangle$, linear in $u$,
\[
\varphi(u)+\varphi(M^{-1}u)=\sum_{j=2}^{2n}(-1)^j\,2u^Tw_j\,\langle Y_{\hat j}\rangle .
\]
Since $I+M^{-1}$ is invertible, $\varphi(u)=\sum_j(-1)^j\,2\big((I+M^{-1})^{-1}u\big)^Tw_j\,\langle Y_{\hat j}\rangle$. For $n=1$ this reads $\langle x_1x_2\rangle=2((I+M^{-1})^{-1}u)^Tw_2$, so the coefficient of $\langle Y_{\hat j}\rangle$ is $\langle x_1x_j\rangle$: the recursion of the proposition, which is also the expansion of a Pfaffian along its first row. Odd products have zero expectation because $\rho$ commutes with the parity $P=(-i)^mc_1\cdots c_{2m}=\prod_k(-Z_k)$, which anticommutes with them. The two-point function is $\langle c(u)c(w)\rangle=u^TGw$ by definition of $G$; the four-point formula is the case $n=2$ with $x_j$ basis Majoranas, and the derivation never used that the indices are distinct.

\emph{All Gaussian states.} By the relation $i\Gamma=-\tanh(\frac\beta2\,ih)$ of \S\ref{subsec:bg-fermions}, the Gibbs states realise every $\Gamma$ with $\|\Gamma\|_\op<1$. For them every moment $\Tr(\rho\,c_S)$ of a product $c_S$ of distinct Majoranas is a Pfaffian of a submatrix of $G$; the $4^m$ products $c_S$ are a trace-orthogonal basis of all operators, so $\rho=2^{-m}\sum_S\Tr(c_S^\dagger\rho)\,c_S$ is a polynomial in $\Gamma$ and extends continuously to $\|\Gamma\|_\op\le1$; limits of Gibbs states are exactly these extensions, and the identities, polynomial in $\Gamma$, hold for them by continuity. The product \eqref{eq:bg-product} is the Gibbs state $\propto\prod_k\exp(\operatorname{artanh}(\lambda_k)\tilde Z_k)$ for $|\lambda_k|<1$, hence equals $\rho_\Gamma$ there and, by continuity, for all $|\lambda_k|\le1$.
\end{proof}

\emph{Bosons.} The same argument with commutators, $\{c(u),c(w)\}$ replaced by $[R_i,R_j]=i\Omega_{ij}$, gives \eqref{eq:bg-wick-bos} with all signs $+$: moving $x_1$ through $Y$ by commutation produces no signs.

\subsection{Polar decomposition of a symplectic matrix}\label{app:bg-williamson}

The Williamson decomposition of Proposition~\ref{prop:bg-williamson} is textbook \cite[Ch.~3, Ch.~5 and App.~B]{Ser17}; Appendix~\ref{app:proofs-Bos} uses in addition that the polar factors of a symplectic matrix are symplectic, in the form below. Let $S$ be symplectic and $M:=SS^T$, symmetric, positive and symplectic. From $M\Omega M^T=\Omega$, $M^{-1}=\Omega M\Omega^{-1}$, and taking square roots, $M^{-1/2}=\Omega M^{1/2}\Omega^{-1}$; with $P:=M^{1/2}$ this says $\Omega P=P^{-1}\Omega$, i.e.\ $P\Omega P^T=\Omega$. So $P$ is symplectic, and $K:=P^{-1}S$ is symplectic and orthogonal ($KK^T=P^{-1}MP^{-1}=I$). An orthogonal symplectic $K$ commutes with $\Omega$, hence is a realification $\Phi_{\mathrm{re}}(u)$ with $u$ unitary: a passive transformation.

\subsection{The fermionic tools of Section~\ref{sec:theoremA}}\label{app:bg-tools}

This subsection proves the tools stated in \S\S\ref{subsec:twotools}--\ref{subsec:nambu}.

\begin{proof}[Proof of Lemma~\ref{lem:wick}]
The four-point formula of Proposition~\ref{prop:bg-wick}, $\langle c_ac_bc_cc_d\rangle=G_{ab}G_{cd}-G_{ac}G_{bd}+G_{ad}G_{bc}$ for all index patterns, gives
\begin{align*}
\langle K_A^2\rangle-\langle K_A\rangle^2
&=\tfrac1{16}\Big[\sum_{a,b,c,d}A_{ab}A_{cd}G_{ac}G_{bd}-\sum_{a,b,c,d}A_{ab}A_{cd}G_{ad}G_{bc}\Big]\\
&=\tfrac1{16}\big[\tr(AGA^TG^T)-\tr(A^TGA^TG^T)\big]
=-\tfrac18\tr(AGAG^T),
\end{align*}
using $A^T=-A$, and $\tr(AGAG^T)=\tr(A^2)+\tr(A\Gamma A\Gamma)$ because $\tr(A^2\Gamma)=0$ ($A^2$ symmetric, $\Gamma$ antisymmetric). This is the first formula; the bound $\le\frac14\|A\|_F^2$ follows from $|\tr(A\Gamma A\Gamma)|\le\|A\Gamma\|_F^2\le\|A\|_F^2$. The block form is the first formula at $\Gamma=\bigoplus_k\lambda_kJ$: hopping blocks commute with $J$ and pairing blocks anticommute with it. The last statement is Lemma~\ref{lem:A-rot}.
\end{proof}

\paragraph{Proofs of the imported tools.} The tools (T1)--(T4) are quoted from \cite{Rub26,Rub26b,DK70} or are standard; we include the short derivations, and a sketch for (T1).

\begin{lemma}[Gaussian rotations {\cite[Lemma~A.2]{Rub26}}]\label{lem:A-rot}
For real antisymmetric $A$, $[K_A,c_d]=i\sum_ac_aA_{ad}$, $e^{-iK_A}c_de^{iK_A}=\sum_ac_a(e^{A})_{ad}$, and $e^{-iK_A}\rho_\Gamma e^{iK_A}=\rho_{e^{A}\Gamma e^{-A}}$.
\end{lemma}

\begin{proof}
From $[c_ac_b,c_d]=2\delta_{bd}c_a-2\delta_{ad}c_b$ and $A^T=-A$, $[K_A,c_d]=i\sum_ac_aA_{ad}$. Hence $f(t):=e^{-itK_A}c_de^{itK_A}$ solves $f'(t)=\sum_af_a(t)A_{ad}$, so $f(1)=\sum_ac_a(e^{A})_{ad}$. With $U=e^{-iK_A}$, $U^\dagger c_aU=\sum_ec_e(e^{-A})_{ea}$, so the covariance of $U\rho_\Gamma U^\dagger$ is $e^{A}\Gamma e^{-A}$; and $U\rho_\Gamma U^\dagger$ is a product state in the rotated normal-mode frame, hence Gaussian, hence equal to $\rho_{e^A\Gamma e^{-A}}$.
\end{proof}

\begin{lemma}[Mandelstam--Tamm inequality; {\cite{MT45}, \cite[Lemma~5.12]{Rub26b}}]\label{lem:A-MTgen}
For every state $\rho$ and Hermitian $H$ on a finite-dimensional space, $T(\rho,e^{-iH}\rho e^{iH})\le\sqrt{\mathrm{Var}_\rho(H)}$.
\end{lemma}

\begin{proof}
Let $\rho=\sum_jq_j|\psi_j\rangle\langle\psi_j|$, $h:=\Tr(\rho H)$ and $\phi_j:=e^{-i(H-h)}\psi_j$, which differs from $e^{-iH}\psi_j$ by a phase. By joint convexity, $T(\rho,e^{-iH}\rho e^{iH})\le\sum_jq_jT(\psi_j,\phi_j)$. For unit vectors and $z:=\langle\psi,\phi\rangle$, $T(\psi,\phi)^2=1-|z|^2\le2(1-\mathrm{Re}\,z)=\|\phi-\psi\|^2$, since the difference of the two sides is $|1-z|^2\ge0$. Next, $\|\phi_j-\psi_j\|\le\|(H-h)\psi_j\|$ by the spectral theorem and $|e^{-iy}-1|\le|y|$. Finally $\sum_jq_j\|(H-h)\psi_j\|\le\big(\Tr\rho(H-h)^2\big)^{1/2}$ by Cauchy--Schwarz.
\end{proof}

\begin{lemma}[$\chi^2$ bound on the trace norm; {\cite[Lemma~A.1]{Rub26}}]\label{lem:A-chi2}
For Hermitian $Y$ and a full-rank state $\rho$, $\|Y\|_1\le\big(\Tr(\rho^{-1/2}Y\rho^{-1/2}Y)\big)^{1/2}$.
\end{lemma}

\begin{proof}
Let $W:=\mathrm{sgn}(Y)$ (with $\mathrm{sgn}\,0:=1$), a Hermitian unitary with $\Tr(WY)=\|Y\|_1$. By Cauchy--Schwarz for the Hilbert--Schmidt inner product, $\Tr(WY)=\Tr\big((\rho^{1/4}W\rho^{1/4})(\rho^{-1/4}Y\rho^{-1/4})\big)\le\|\rho^{1/4}W\rho^{1/4}\|_2\,\|\rho^{-1/4}Y\rho^{-1/4}\|_2$. The second factor squared is the stated trace; the first squared is $\Tr(\rho^{1/2}W\rho^{1/2}W)\le\|\rho^{1/2}\|_2\|W\rho^{1/2}W\|_2=1$.
\end{proof}

\emph{How (T1) follows (sketch of \cite[App.~A]{Rub26}).} Let $D_X\rho_\Gamma$ be the derivative of the polynomial map $\Gamma\mapsto\rho_\Gamma$ in the direction of a real antisymmetric $X$, and work in the normal-form frame with (by density) distinct, nonzero $\lambda_k$. The diagonal blocks $\xi_kJ$ of $X$ change the occupations; every off-diagonal part is a commutator $[A,\Gamma]$, since $[A,\Gamma]_{kl}=(\lambda_l-\lambda_k)A^+_{kl}J+(\lambda_l+\lambda_k)A^-_{kl}J$, and by Lemma~\ref{lem:A-rot} it acts on the state as the commutator with $K_A$. Since $\rho_\Gamma$ is diagonal in the occupation basis and each pair term flips the occupations of exactly two modes, all cross terms vanish, and a direct computation gives \cite[Eq.~(7.2)]{Rub26}
\begin{equation}\label{eq:A-chi2id}
\begin{split}\Tr\big(\rho_\Gamma^{-1/2}D_X\rho_\Gamma\,\rho_\Gamma^{-1/2}D_X\rho_\Gamma\big)&=\sum_k\frac{\xi_k^2}{1-\lambda_k^2}+\sum_{k<l}\frac{\|X_{kl}\|_F^2}{\sqrt{(1-\lambda_k^2)(1-\lambda_l^2)}}\\&=\tfrac12\big\|(I+\Gamma^2)^{-1/4}X(I+\Gamma^2)^{-1/4}\big\|_F^2 .\end{split}
\end{equation}
On $\mathcal G_c$ every $1-\lambda_k^2\ge c(2-c)$, so Lemma~\ref{lem:A-chi2} gives $\|D_X\rho_\Gamma\|_1\le\|X\|_F/\sqrt{2c(2-c)}$; integrating along the segment from $\Gamma$ to $\Gamma'$, which stays in the convex set $\mathcal G_c$, gives (T1) with $K_c=(2c(2-c))^{-1/2}\le(2c)^{-1/2}$.

\begin{lemma}[Davis--Kahan, (T2); {\cite{DK70}}]\label{lem:A-DK}
In the setting of (T2), $\|(I-\hat P)P\|_\op\le\|\hat G-G\|_\op/\delta$. If moreover $\mathrm{rank}\,P=\mathrm{rank}\,\hat P$, then $\|(I-P)\hat P\|_\op=\|(I-\hat P)P\|_\op$.
\end{lemma}

\begin{proof}
This is the $\sin\Theta$ theorem of \cite{DK70} with a widened window for $\hat P$; we include the short proof in this form. Put $Q:=I-\hat P$, $X:=QP$, $\gamma:=(\alpha+\beta)/2$ and $r:=(\beta-\alpha)/2$. Since $Q$ commutes with $\hat G$ and $P$ with $G$, $Q(\hat G-G)P=(\hat G-\gamma)X-X(G-\gamma)P$. On the range of $Q$ every eigenvalue of $\hat G$ is at distance at least $r+\delta$ from $\gamma$, so $\|(\hat G-\gamma)Xv\|\ge(r+\delta)\|Xv\|$; and $\|(G-\gamma)P\|_\op\le r$. Hence $(r+\delta)\|X\|_\op\le\|\hat G-G\|_\op+r\|X\|_\op$. For the second statement, $\|(I-\hat P)P\|_\op^2$ and $\|(I-P)\hat P\|_\op^2$ are $1$ minus the smallest eigenvalue of $B^\dagger B$ on the range of $P$, respectively of $BB^\dagger$ on the range of $\hat P$, with $B=\hat PP$; these have the same nonzero eigenvalues and, the ranges having equal dimension, the same multiplicity of $0$.
\end{proof}

\emph{Proof of (T3).} If $\Pi$ is the metric projection onto a closed convex set, then $\langle x-\Pi x,y-\Pi x\rangle\le0$ for every $y$ in the set; using this at $x$ with $y=\Pi x'$ and at $x'$ with $y=\Pi x$ and adding, $\|\Pi x-\Pi x'\|^2\le\langle x-x',\Pi x-\Pi x'\rangle\le\|x-x'\|\,\|\Pi x-\Pi x'\|$. We use it for the Frobenius inner product on real antisymmetric matrices, where $\mathcal G_c$ is closed and convex.

\emph{Proof of (T4).} The four-point formula of Proposition~\ref{prop:bg-wick} for mode operators gives $\langle n_An_B\rangle=\langle n_A\rangle\langle n_B\rangle-\langle a_A^\dagger a_B^\dagger\rangle\langle a_Aa_B\rangle+\langle a_A^\dagger a_B\rangle\langle a_Aa_B^\dagger\rangle$, and since the two modes are orthogonal ($a_A$ and $a_A^\dagger$ anticommute with $a_B$ and $a_B^\dagger$), $\langle a_A^\dagger a_B^\dagger\rangle=-\overline{\langle a_Aa_B\rangle}$ and $\langle a_Aa_B^\dagger\rangle=-\overline{\langle a_A^\dagger a_B\rangle}$.

\paragraph{The Nambu matrix.}
The components of $\alpha$ satisfy $\{\alpha_i,\alpha_j^\dagger\}=\delta_{ij}$. For $c\in\C^{2m}$ and $b:=c\cdot\alpha$, $c^\dagger\mathcal Nc=\langle b^\dagger b\rangle$ and $\{b,b^\dagger\}=\|c\|^2$, so $0\le b^\dagger b\le\|c\|^2$, which is $0\preceq\mathcal N\preceq I$, and $\|b\|_\op\le\|c\|$, so $|c_A^T\mathcal Ac_B|=|\langle b_Ab_B\rangle|\le\|c_A\|\|c_B\|$, which is $\|\mathcal A\|_\op\le1$; $\langle bb^\dagger\rangle=\|c\|^2-\langle b^\dagger b\rangle$, read with $b^\dagger=\mathrm{op}(S\bar c)$, is the Nambu symmetry $S\bar{\mathcal N}S=I-\mathcal N$. A change of frame $\alpha\mapsto\mathcal W\alpha$, $\mathcal W$ unitary (\S\ref{subsec:bg-fermions}), sends $\mathcal N$ to $\bar{\mathcal W}\mathcal N\mathcal W^T$; in the normal-form frame $F=0$ and $N=\mathrm{diag}(p)$. Conversely, rows $r_a=(u_a,v_a)$ define modes $a''_a=\sum_k(u_{ak}b_k+v_{ak}b_k^\dagger)$, with $\{a''_a,a''^\dagger_b\}=\langle r_b,r_a\rangle$ and $\{a''_a,a''_b\}=(uv^T+vu^T)_{ab}=\langle S\bar r_b,r_a\rangle$; so $m$ rows define a Bogoliubov frame exactly when they form an orthonormal basis of a subspace $\mathcal P$ with $\mathcal P\perp S\bar{\mathcal P}$, as used in Lemma~\ref{lem:A2}. If the $b_k$ are the normal modes ($\langle b_k^\dagger b_l\rangle=p_k\delta_{kl}$, $\langle b_kb_l\rangle=0$), expanding $\langle a''^\dagger_aa''_b\rangle$ and $\langle a''_aa''_b\rangle$ gives the formulas for $N'$ and $F'$ in Lemma~\ref{lem:A2}(ii)--(iii).

\section{Proofs for Section~\ref{sec:theoremA}}\label{app:proofs-A}

\emph{Warm case.} Let $x_\star=x_\star(\kappa)$ be the constant fixed in \S\ref{subsec:bookkeeping}. For $x_c\ge x_\star$ the class lies in $\mathcal G_{x_\star/\kappa}$, and \cite[Thm.~7.6]{Rub26} already gives $12\kappa m^3/(x_\star\eps^2)=O_\kappa(m^3/\eps^2)$ copies. From now on $x_c<x_\star$ (the cold regime). Below $x_\star$ the matchgate-shadow estimator of \cite[Thm.~7.6]{Rub26} loses the factor $m$ noted in Section~\ref{sec:theoremA} because its $\pm1$ readout never vanishes: its noise per covariance entry stays of order $m$ while the weight $K_c^2\approx1/(2c)$ of (T1) grows as the state cools. The pair-rotation measurement removes exactly this mismatch.

\subsection{Stage 0: a cold frame}\label{subsec:stage0}

\begin{lemma}[Operator-norm concentration, log-free]\label{lem:A2prime}
For $m\ge4$ and $N_0\ge m^2$, with probability $\ge0.95$,
\[
\big\|\hat{\mathcal N}_{\mathrm{off}}-\mathcal N_{\mathrm{off}}\big\|_\op\le C_3\,\frac{m}{\sqrt{N_0}},\qquad C_3\ \text{an absolute constant}.
\]
\end{lemma}

\begin{proof}
$M:=\hat{\mathcal N}_{\mathrm{off}}-\mathcal N_{\mathrm{off}}=N_0^{-1}\sum_t(Z_t-\mathcal N_{\mathrm{off}})$ is a sum of independent zero-mean Hermitian summands. Fix a unit vector $w\in\C^{2m}$ and put $q_t:=w^\dagger Z_tw$.

\emph{(1) Range and variance.} The shots $Z_t$ are the Horvitz--Thompson shots $Z$ of \S\ref{subsec:F-shots} in the computational frame, so the per-shot bound \eqref{eq:R0} of Lemma~\ref{lem:R0} applies (its proof uses nothing from this appendix). In the computational frame every frame occupation is at most $1$, so $Q_w\le\|w\|^2=1$, $B_w\le1$ and $N_w\le1$, and \eqref{eq:R0} gives, for even $m\ge4$,
\[
|q_t-w^\dagger\mathcal N_{\mathrm{off}}w|\le4m,\qquad\mathrm{Var}(q_t)\le640m+1024m+27+12\le c_Vm,\qquad c_V:=1674 .
\]
For odd $m\ge5$ the same holds with the range $4m+2$ (\S\ref{subsec:thmB-assembly}, \emph{Odd $m$}). In both cases the range is at most $b:=5m$.

\emph{(2) Bernstein for the scalar $w^\dagger Mw$.} By \eqref{eq:bg-bernstein}, $\Prob\big(|w^\dagger Mw|\ge\eta\big)\le2\exp\!\big(-N_0\eta^2/(2(c_Vm+b\eta/3))\big)$. Put $C_3^2:=160\,c_V$ and $\eta:=C_3m/(2\sqrt{N_0})$. The range condition $b\eta/3\le c_Vm$ reads $N_0\ge\frac{1000}{9c_V}m^2$, which is implied by the hypothesis $N_0\ge m^2$ (automatic in the application, since $N_0=C_0'm^2/x_c\ge m^2$ below); then the exponent is $\ge\frac{N_0\eta^2}{4c_Vm}=\frac{C_3^2m}{16c_V}=10m$.

\emph{(3) Net.} The $\frac14$-net $\mathcal W$ of the unit sphere of $\C^{2m}\cong\R^{4m}$ of \S\ref{subsec:bg-estimation} has $|\mathcal W|\le9^{4m}\le e^{9m}$, and $\|M\|_\op\le2\max_{w\in\mathcal W}|w^\dagger Mw|$ for Hermitian $M$. By a union bound, $\Prob(\|M\|_\op>2\eta)\le2e^{9m}e^{-10m}\le0.05$ for $m\ge4$.
\end{proof}

\begin{lemma}[The frame is cold]\label{lem:A2}
Let $N_0:=C_0'm^2/x_c\le C_0'm^3$, with $C_0'$ a fixed multiple of $4C_3^2C_1$ chosen so that $\eta^2:=\|\hat{\mathcal N}-\mathcal N\|_\op^2\le x_c/(4C_1)$ on the event of Lemma~\ref{lem:A2prime}, and let $N_0'$ be the number of occupation shots, chosen so that $\max_k|\hat p_k-p_k|\le\sqrt{x_c/(16C_1)}$ (Hoeffding and a union bound over the $m$ modes: $N_0'\ge8C_1\ln(120m)/x_c$, which is $\le8C_1m\ln(120m)$, with failure probability $\le1/60$; the cost is negligible). Let $\hat{\mathcal P}$ be the span of the $m$ eigenvectors of $\hat{\mathcal N}$ with the smallest eigenvalues. By the particle--hole symmetry of the Nambu form (after symmetrising $\hat{\mathcal N}$ to that form), $\hat{\mathcal P}$ defines a Bogoliubov frame $W_0$ whose modes are $a_a''=\sum_k(u_{ak}b_k+v_{ak}b_k^\dagger)$ in terms of the true empty-type normal modes $b_k$. Concretely, on the event of Lemma~\ref{lem:A2prime} the $m$ smallest eigenvalues of $\hat{\mathcal N}$ are $<\frac12$ and the $m$ largest are $>\frac12$ (Weyl, below), so $\hat{\mathcal P}$ is the spectral subspace of $\hat{\mathcal N}$ for $(-\infty,\frac12)$ (the raw estimate need not be positive semidefinite) and $\hat{\mathcal P}\perp S\overline{\hat{\mathcal P}}$, where $S$ is the particle--hole swap, because $\hat{\mathcal N}v=\lambda v$ implies $\hat{\mathcal N}S\bar v=(1-\lambda)S\bar v$; any orthonormal basis of $\hat{\mathcal P}$ fixed by a deterministic rule (the eigenvectors of $\hat{\mathcal N}$ in increasing order of eigenvalue, ties broken lexicographically) forms the rows $(u_a,v_a)$ of a Bogoliubov unitary, and the residual freedom is a passive rotation of the new modes, which affects none of (i)--(iv) below. Off that event the output of the protocol is irrelevant. Then, with probability $\ge0.9$:
\begin{enumerate}
\item[(i)] hole weights: $\|v_a\|^2\le4\eta^2\le x_c/C_1$ for every $a$, and $\|v\|_\op^2\le4\eta^2$;
\item[(ii)] occupations: $p_a'\le x_c+x_c/C_1=:\kappa'x_c$, and $\|N'\|_\op\le\kappa'x_c$, where $N'=\bar u\,\mathrm{diag}(p)\,u^T+\bar v\,(I-\mathrm{diag}(p))\,v^T$;
\item[(iii)] $\|F'\|_\op\le2\|v\|_\op\le4\eta\le2\sqrt{x_c/C_1}$, where $F'=u\,(I-\mathrm{diag}(p))\,v^T+v\,\mathrm{diag}(p)\,u^T$;
\item[(iv)] in the Stage-0 frame, writing $\Gamma'=D'+E'$ with $D'$ the diagonal blocks and $E'=E'^++E'^-$: $\|E'^+\|_\op=2\|N'-\mathrm{diag}N'\|_\op\le4\kappa'x_c$, $\|E'^-\|_\op=2\|F'\|_\op\le4\sqrt{x_c/C_1}$, $\lambda'_k=2p'_k-1\in[-1,-1+2\kappa'x_c]$, and $\Gamma'\in\mathcal G_{x_c/\kappa}$.
\end{enumerate}
\end{lemma}

\begin{proof}
By Weyl's inequality the $m$ smallest eigenvalues of $\hat{\mathcal N}$ are $\le x_c+\eta$ and the others are $\ge1-x_c-\eta$. Apply (T2) with the particle interval $[0,x_c]$ and $\delta\ge1-2x_c-\eta\ge\frac12$: $\|(I-\hat P)P\|_\op\le2\eta$, where $P$ is the spectral projector of $\mathcal N$ onto $[0,x_c]$ and $\hat P$ that of $\hat{\mathcal N}$ onto $\hat{\mathcal P}$; equal ranks give $\|(I-P)\hat P\|_\op=\|(I-\hat P)P\|_\op$. Each row $(u_a,v_a)$ of the Bogoliubov matrix is a unit vector of $\hat{\mathcal P}$, whose hole component $v_a$ satisfies $\|v_a\|^2\le\|(I-P)\hat P\|_\op^2\le4\eta^2$, and $\|v\|_\op=\|(I-P)\hat P\|_\op$; this is (i). (ii) and (iii) follow from the displayed formulas, using $\|u\|_\op\le1$ and $\mathrm{diag}(p)\preceq x_cI$. (iv) follows from \eqref{eq:blocks}; $\Gamma'$ has the same spectrum as $\Gamma$, so $\Gamma'\in\mathcal G_{x_c/\kappa}$.
\end{proof}

\emph{Cost of Stage 0 and the scale proxy.} The protocol also learns $\hat x_c:=\max_a\hat p'_a$ from $N_0''\ge8\kappa^2\ln(120m)/x_c^2$ occupation shots taken in the new frame $W_0$ (Hoeffding and a union bound over the $m$ modes: all $|\hat p'_a-p'_a|\le x_c/(4\kappa)$ with failure probability $\le1/60$); since $N_0''\le8\kappa^2m^2\ln(120m)$, $N_0+N_0'+N_0''=O(m^3)$ for $x_c\ge1/m$. Upper end: $p'_a\le\kappa'x_c$ by (ii). Lower end: every true occupation satisfies $p_k\ge(x_c/\kappa)/(1+x_c/\kappa)$ (class definition) and $\|u_a\|^2\ge1-4\eta^2$, so $\max_ap'_a\ge(x_c/\kappa)(1-x_\star/C_1)/(1+x_\star)\ge3x_c/(4\kappa)$; subtracting the precision,
\[
\hat x_c\in\big[x_c/(2\kappa),\,2\kappa'x_c\big].
\]
The protocol sets
\[
c':=\frac{\hat x_c}{8\kappa\kappa'}\le\frac{x_c}{4\kappa},
\]
so that $\mathcal G_{c'}\supset\mathcal G_{x_c/\kappa}$ deterministically on the good event, and
\[
K_{c'}\le(2c')^{-1/2}\le\sqrt{8\kappa\kappa'/\hat x_c}\le\sqrt{16\kappa^2\kappa'/x_c}=:c_\kappa\sqrt{\kappa/x_c},\qquad c_\kappa=4\sqrt{\kappa\kappa'} .
\]

\subsection{Stage 1 and the error}\label{subsec:stage1}

Stage 1 takes $N_2$ pair-rotation shots and $N_3$ occupation shots in the Stage-0 frame; $\tilde\Gamma$ is assembled from the empirical means by \eqref{eq:blocks}; write $\Delta:=\tilde\Gamma-\Gamma'=\Delta^d+\Delta^++\Delta^-$. The output is $\hat\Gamma:=\Pi_{\mathcal G_{c'}}(\tilde\Gamma)$, the Frobenius projection onto $\mathcal G_{c'}$, and the protocol returns $\rho_{W_0\hat\Gamma W_0^T}$.

\begin{lemma}[Stage-1 noise]\label{lem:A3}
On the $0.99$-event of \S\ref{subsec:pairrotation}, with $N_2\ge C_2m^3/\eps^2$ and $N_3\ge C_3'm/\eps^2$, where $C_2\ge30\cdot2^{17}c_\kappa^2\kappa\kappa'$ and $C_3'\ge30\cdot2^{11}c_\kappa^2\kappa\kappa'$, with probability $\ge0.9$:
(a) $\|\Delta^-\|_F\le\eps/8$; (b) $\|\Delta^+\|_F\le\dfrac{\eps\sqrt{x_c/\kappa}}{16c_\kappa}$; (c) $\|\Delta^d\|_F\le\dfrac{\eps\sqrt{x_c/\kappa}}{16c_\kappa}$.
\end{lemma}

\begin{proof}
With two blocks per unordered pair and the noise figures of Lemma~\ref{lem:A1} and Lemma~\ref{lem:A2}(ii): $\E\|\Delta^-\|_F^2\le512m^3/N_2$, $\E\|\Delta^+\|_F^2\le512\kappa'x_cm^3/N_2$ and $\E\|\Delta^d\|_F^2\le8m\kappa'x_c/N_3$. Markov's inequality at level $1/30$ for each of the three gives (a) once $C_2\ge30\cdot512\cdot64$, and (b), (c) once $C_2\ge30\cdot512\cdot256\,c_\kappa^2\kappa\kappa'$ and $C_3'\ge30\cdot8\cdot256\,c_\kappa^2\kappa\kappa'$; the factor $c_\kappa^2$ is the price of the $\chi^2$ weight $K_{c'}$ in Lemma~\ref{lem:A6}.
\end{proof}

\emph{The pairing rotation.} Let $A$ be the pairing-type real antisymmetric matrix with blocks
\[
A_{kl}:=-\frac{\Delta^-_{kl}J}{\lambda'_k+\lambda'_l}\quad(k\ne l),\qquad A^d=A^+=0 .
\]
It is antisymmetric as a whole because $\Delta^-_{kl}$ is symmetric and anticommutes with $J$, and $[A,D']=\Delta^-$ exactly: for a pairing-type block, $[A,D']_{kl}=(\lambda'_k+\lambda'_l)A_{kl}J$. Each entry of $A$ is $\pm$ the block scalar $-1/(\lambda'_k+\lambda'_l)\in[\frac12,\frac{16}{31}]$ (for $x_c\le1/(64\kappa\kappa'^2)$) times one entry of $\Delta^-$. Hence
\[
\|A\|_F\le\tfrac{16}{31}\|\Delta^-\|_F\le\eps/15 .
\]
With $R:=e^{A}$, Lemma~\ref{lem:MT} gives $T(\rho_{\Gamma'},\rho_{R\Gamma'R^T})\le\frac12\|A\|_F\le\eps/30$. Lemma~\ref{lem:A4} uses $R\Gamma'R^T=e^{A}\Gamma'e^{-A}=\sum_{n\ge0}\ad_A^n\Gamma'/n!$, where $\ad_AX:=[A,X]$.

Pairing errors are rotated, and hopping errors are not, because the $\chi^2$ bound would have charged the pairing error about $\|A\|_F/\sqrt{x_c}$: by \eqref{eq:A-chi2id} it weighs the squared coherence between $|00\rangle$ (probability close to one) and $|11\rangle$ (probability about $p_kp_l$) of two cold modes by $(p_kp_l)^{-1/2}$, although the trace norm of that coherence is set by its size alone. Hopping errors cannot be rotated away in this manner: their denominator $\lambda'_l-\lambda'_k$ vanishes for two modes of equal occupation, where a hopping change of the covariance splits the common occupation and is not a rotation at all; they stay in the $\chi^2$ bound. (Theorem~\ref{thm:B} rotates away the hopping errors between occupation bands far apart, where this denominator is large.)

The second-order remainder is small because $A$ is built from measurement noise, an incoherent random rotation: its $\approx4m^2$ entries are nearly independent with variance of order $s^2$, so $\|A\|_F\approx ms$ but $\|A\|_\op\approx\sqrt m\,s$, and $\|A^2\|_F\approx m^{3/2}s^2$ (Lemma~\ref{lem:A5}), a factor $\sqrt m$ below the value $\approx\|A\|_F^2$ for a rotation of the same size in a single plane; the only correlations come from pairs read in the same shot, which the matching makes rare. That factor $\sqrt m$ pays for the weight $\sqrt{\kappa/x_c}\le\sqrt{\kappa m}$ in \S\ref{subsec:assembly}.

\begin{lemma}[Remainder of the rotation]\label{lem:A4}
Let $\mathcal R:=R\Gamma'R^T-\Gamma'-\Delta^-=[A,E']+\sum_{n\ge2}\ad_A^n\Gamma'/n!$ and $a:=\|A\|_\op$. Then
\[
\|\mathcal R\|_F\le2\|A\|_F\|E'\|_\op+2e^{2a}\|A^2\|_F .
\]
\end{lemma}

\begin{proof}
(i) For real antisymmetric $A$ and any matrix $B$, $\|ABA\|_F\le\|B\|_\op\|A^2\|_F$. Indeed $A$ is normal with eigenvalues $ia_i$, $a_i\in\mathbb R$; writing $\tilde B$ for $B$ in a unitary eigenbasis of $A$, $(ABA)_{ij}=-a_ia_j\tilde B_{ij}$; by $a_i^2a_j^2\le\frac12(a_i^4+a_j^4)$, and since every row and every column of $\tilde B$ has squared norm $\le\|B\|_\op^2$, $\|ABA\|_F^2\le\|B\|_\op^2\sum_ia_i^4=\|B\|_\op^2\|A^2\|_F^2$.
(ii) For $n\ge2$, $\ad_A^n\Gamma'=\sum_{j=0}^n\binom nj(-1)^jA^{n-j}\Gamma'A^j$. If $n-j\ge2$ or $j\ge2$, then $\|A^{n-j}\Gamma'A^j\|_F\le\|A^2\|_F\,a^{n-2}\,\|\Gamma'\|_\op\le a^{n-2}\|A^2\|_F$; the only remaining term is $A\Gamma'A$ ($n=2$, $j=1$), which (i) bounds by $\|A^2\|_F$ since $\|\Gamma'\|_\op\le1$. Hence $\|\ad_A^n\Gamma'\|_F\le2^na^{n-2}\|A^2\|_F$ for all $n\ge2$.
(iii) $\|[A,E']\|_F\le2\|A\|_F\|E'\|_\op$, and $\sum_{n\ge2}2^na^{n-2}/n!=(e^{2a}-1-2a)/a^2\le2e^{2a}$, comparing the two series termwise. The series converges because $a\le\|A\|_F\le\eps/15<1$; this is where $\eps\le1$ and Lemma~\ref{lem:A3}(a) are used.
\end{proof}

\begin{lemma}[Fourth moment of $A^2$]\label{lem:A5}
Let $s^2:=32m/N_2=32\eps^2/(C_2m^2)$ and let $\mathcal C$ be the Chernoff event $\{S_{P,\sigma}\ge N_2/(8m)\ \ \forall P,\sigma\}$ of \S\ref{subsec:pairrotation}. Then
\[
\E\big[\|A^2\|_F^2\,\mathbf 1_{\mathcal C}\big]\le C_5\,m^3s^4
\]
with an absolute constant $C_5$. The expectation is unconditional, over outcomes, matchings and labels; this is the form in which Markov's inequality is applied. The co-measurement counts below are controlled only in expectation.
\end{lemma}

\begin{proof}
Block norms $\|\cdot\|$ below are Frobenius norms. Each $2\times2$ block $A_{kj}$ ($k\ne j$) is a fixed linear image, of norm $\le16/31$, of the two empirical-mean errors $e_{kj}=(e^{\mathrm r}_{kj},e^{\mathrm i}_{kj})$ of the pairing entries of the pair $(k,j)$; $A_{kk}=0$. Condition on $(\pi_t,\sigma_t)_t$ (not on the signs); the event $\mathcal C$ is measurable with respect to this conditioning. Then:
\begin{enumerate}
\item[(P1)] by the sampling facts of \S\ref{subsec:pairrotation} and Lemma~\ref{lem:A1}, the raw samples of a given entry are i.i.d.\ over its shot set $T_{kj}$, bounded by $2$ in absolute value, and independent across shots; their centred versions $\xi$ are zero-mean, bounded by $4$ in absolute value, with variance $\le4$;
\item[(P2)] $\E[e_{kj}^2\mid\pi,\sigma]\le4/S_{kj}\le s^2$ and $\E[e_{kj}^4\mid\pi,\sigma]=\big(S\,\E\xi^4+3S(S-1)(\E\xi^2)^2\big)/S^4\le3s^4+s^6\le4s^4$ on $\mathcal C$ (using $\E\xi^4\le16\,\E\xi^2$ and $1/S\le s^2/4$);
\item[(P3)] two pairs that share a mode have disjoint shot sets (\S\ref{subsec:pairrotation}), hence independent errors;
\item[(P4)] two disjoint pairs $P,P'$ (with their labels) are co-measured in the shots $T_P\cap T_{P'}$, with $\E_\pi|T_P\cap T_{P'}|=N_2\,q$, $q:=1/(16(m-1)(m-3))$ (odd $m$: $q=1/(16m(m-2))$), and
\[
\E[e_Pe_{P'}\mid\pi,\sigma]=\frac{1}{S_PS_{P'}}\sum_{t\in T_P\cap T_{P'}}\mathrm{Cov}(\xi_t,\xi'_t)\qquad\text{with }|\mathrm{Cov}(\xi_t,\xi'_t)|\le4
\]
(the cross-shot terms vanish: independent shots, zero mean after $s$-averaging); so on $\mathcal C$, $|\E[e_Pe_{P'}\mid\pi,\sigma]|\le4|T_P\cap T_{P'}|\cdot64m^2/N_2^2=256m^2|T_P\cap T_{P'}|/N_2^2$;
\item[(P5)] for two such co-measurement counts, splitting $\sum_{t,t'}$ into $t\ne t'$ (independent shots) and $t=t'$,
$\E\big[|T_P\cap T_{P'}|\,|T_Q\cap T_{Q'}|\big]=N_2(N_2-1)\,p_{PP'}p_{QQ'}+N_2\,p_{PP'QQ'}$, where $p$ denotes the per-shot probability of the indicated co-measurement event; in particular it is $\le\E|T_P\cap T_{P'}|\cdot\E|T_Q\cap T_{Q'}|$ whenever the two events are mutually exclusive in one shot (e.g.\ when they require one mode to have two different partners). Compatible disjoint edges of a uniform perfect matching are \emph{positively} correlated ($1/((m-1)(m-3))>1/(m-1)^2$); the inequality rests only on the independence of distinct shots.
\end{enumerate}

\emph{Expansion.} $Z:=A^2$ has $Z_{kl}=\sum_{j\notin\{k,l\}}A_{kj}A_{jl}$ and
\[
\E\|Z\|_F^2=\sum_{k,l}\sum_{j,j'}\E\,\tr\big((A_{kj}A_{jl})^T(A_{kj'}A_{j'l})\big).
\]
Each summand is a fixed real multilinear form in the components of $e_{kj},e_{jl},e_{kj'},e_{j'l}$, with an absolute number of monomials and coefficients bounded by $(16/31)^4$; we expand into these signed monomials \emph{before} taking any absolute value. Cases:
($\alpha$) $j=j'$, $k\ne l$; ($\beta$) $k=l$, $j\ne j'$; ($\delta$) $k=l$, $j=j'$: at most $2m^3+m^2$ terms in all, each bounded by $\E\big[\|A_{kj}\|\|A_{jl}\|\|A_{kj'}\|\|A_{j'l}\|\mathbf 1_{\mathcal C}\big]\le c\,s^4$ by H\"older and (P2). (In ($\beta$) the two factors are independent squares with nonzero means: $(A^2)_{kk}=-\frac12\|A_{k\cdot}\|_F^2\,I$, and these $m^3$ terms carry about half of the total.)
($\gamma$) $k\ne l$, $j\ne j'$: four distinct pairs. Every pair of $G_1:=\{(k,j),(j',l)\}$ shares a mode with every pair of $G_2:=\{(j,l),(k,j')\}$, so by (P3) the shot sets of $G_1$ and of $G_2$ are disjoint and, conditionally on $(\pi,\sigma)$, every monomial factorises:
\[
\E[e_ae_be_ce_d\mid\pi,\sigma]=\E[e_ae_d\mid\pi,\sigma]\;\E[e_be_c\mid\pi,\sigma],
\]
$a,d$ being components of the two entries of $G_1$ and $b,c$ of $G_2$. By (P4), on $\mathcal C$ each factor is at most $256m^2K/N_2^2$ in absolute value, $K$ the corresponding co-measurement count. The two co-measurement events are mutually exclusive in one shot ($k$ cannot be matched to both $j$ and $j'$), so by (P5) $\E[K_{ad}K_{bc}]=N_2(N_2-1)q^2$. Since $\mathbf 1_{\mathcal C}$ is $(\pi,\sigma)$-measurable,
\[
\big|\E[e_ae_be_ce_d\,\mathbf 1_{\mathcal C}]\big|\le\frac{256^2m^4}{N_2^4}\,N_2^2q^2=\frac{256\,m^4}{N_2^2(m-1)^2(m-3)^2}\le\frac{8\,s^4}{m^2}\qquad(m\ge4,\ \text{both parities}),
\]
and the $m^4$ terms of case ($\gamma$) contribute $\le8m^2s^4$. In total, $\E[\|A^2\|_F^2\mathbf 1_{\mathcal C}]\le C_5m^3s^4$.
\end{proof}

\emph{Consequence.} Markov's inequality at level $1/40$ applied to $\|A^2\|_F^2\mathbf 1_{\mathcal C}$ gives
\begin{equation}\label{eq:fourthmoment-consequences}
\|A^2\|_F\le\sqrt{40C_5}\,m^{3/2}s^2=\sqrt{40C_5}\cdot\frac{32\,\eps^2}{C_2\sqrt m}.
\end{equation}

\begin{lemma}[The $\chi^2$ step]\label{lem:A6}
$R\Gamma'R^T\in\mathcal G_{x_c/\kappa}\subset\mathcal G_{c'}$ (it has the same spectrum as $\Gamma'$) and $\hat\Gamma=\Pi_{\mathcal G_{c'}}(\tilde\Gamma)\in\mathcal G_{c'}$; by (T3),
$\|\hat\Gamma-R\Gamma'R^T\|_F\le\|\tilde\Gamma-R\Gamma'R^T\|_F=\|\Delta^++\Delta^d-\mathcal R\|_F$, and by (T1) on the segment in $\mathcal G_{c'}$,
\[
T\big(\rho_{R\Gamma'R^T},\rho_{\hat\Gamma}\big)\le\tfrac12K_{c'}\big(\|\Delta^+\|_F+\|\Delta^d\|_F+\|\mathcal R\|_F\big),\qquad K_{c'}\le c_\kappa\sqrt{\kappa/x_c}.
\]
\end{lemma}

\noindent($R$ and $A$ are analysis devices only; the protocol outputs $\Pi_{\mathcal G_{c'}}(\tilde\Gamma)$ with $c'$ computed from $\hat x_c$.)

\subsection{Assembly}\label{subsec:assembly}

\begin{proof}[Proof of Theorem~\ref{thm:A}]
By the triangle inequality, the bound after the pairing rotation and Lemma~\ref{lem:A6},
\[
T(\rho_{\Gamma'},\rho_{\hat\Gamma})\le\frac{\eps}{30}+\frac{c_\kappa}2\sqrt{\kappa/x_c}\Big(\|\Delta^+\|_F+\|\Delta^d\|_F+\|\mathcal R\|_F\Big),
\]
and the three contributions are bounded as follows.
\begin{itemize}
\item The Stage-1 noise: by Lemma~\ref{lem:A3}(b,c),
\[
\frac{c_\kappa}2\sqrt{\kappa/x_c}\,\big(\|\Delta^+\|_F+\|\Delta^d\|_F\big)\le\frac{\eps}{16};
\]
the $x_c$ cancels against the hopping and diagonal noise, which is $\propto\kappa'x_c$.
\item The first-order term of $\|\mathcal R\|_F$ (Lemma~\ref{lem:A4}):
\[
\sqrt{\kappa/x_c}\cdot2\|A\|_F\|E'\|_\op\le\sqrt{\kappa/x_c}\cdot\frac{2\eps}{15}\big(4\kappa'x_c+4\sqrt{x_c/C_1}\big)=\frac{8\eps}{15}\sqrt\kappa\,\big(\kappa'\sqrt{x_c}+C_1^{-1/2}\big);
\]
the $x_c$ cancels, and this is $\le\eps/(8c_\kappa)$ for $x_c\le x_\star:=1/(256c_\kappa^2\kappa\kappa'^2)$ and $C_1\ge256c_\kappa^2\kappa$; after the weight $c_\kappa/2$ the contribution is $\le\eps/16$.
\item The second-order term of $\|\mathcal R\|_F$, with $a=\|A\|_\op\le\eps/15\le1/15$, \eqref{eq:fourthmoment-consequences} and $\sqrt{\kappa/x_c}\le\sqrt{\kappa m}$:
\[
\frac{c_\kappa}2\sqrt{\kappa/x_c}\cdot2e^{2a}\|A^2\|_F\le c_\kappa\sqrt{\kappa m}\,e^{2/15}\sqrt{40C_5}\cdot\frac{32\eps^2}{C_2\sqrt m}=\frac{32e^{2/15}c_\kappa\sqrt{40C_5\kappa}}{C_2}\,\eps^2\le\frac{\eps}{16}
\]
for $\eps\le\eps_A(\kappa):=C_2/\big(512e^{2/15}c_\kappa\sqrt{40C_5\kappa}\big)$; the $\sqrt m$ cancels because $mx_c\ge1$, which is why the range of the theorem stops at $x_c=1/m$.
\end{itemize}
Hence $T\le\eps(1/30+3/16)<\eps/2$, i.e.\ $\|\rho-\hat\rho\|_1\le\eps$, and the number of copies is $N_0+N_0'+N_0''+N_2+N_3=O_\kappa(m^3/\eps^2)$. The success probability follows from a union bound over all failure events: Lemma~\ref{lem:A2prime} ($0.05$), the two Hoeffding events for $\hat p$ and $\hat x_c$ ($2/60$), the Stage-0 Chernoff event on the sample counts ($0.01$), Lemma~\ref{lem:A3} (three Markov events at $1/30$: $0.10$), the Stage-1 Chernoff event ($0.01$), and Lemma~\ref{lem:A5} (one Markov event at $1/40$: $0.025$); in total $\le0.23<1/3$. This proves Theorem~\ref{thm:A}.
\end{proof}

\subsection{Bookkeeping}\label{subsec:bookkeeping}

\begin{itemize}
\item \emph{Order of the constants.} $C_1:=8192\kappa^2$ is fixed first; then $\kappa':=1+1/C_1\le2$ and $c_\kappa:=4\sqrt{\kappa\kappa'}$, so that $C_1\ge4096\kappa^2\kappa'=256c_\kappa^2\kappa$, as the cold branch requires; then $x_\star:=1/(256c_\kappa^2\kappa\kappa'^2)$; then $c_V=1674$ and $C_3=\sqrt{160c_V}$ (absolute; Lemma~\ref{lem:A2prime}), $C_0'$ (from $\eta^2\le x_c/(4C_1)$), $C_5$ (absolute), $C_2(\kappa)$, $C_3'(\kappa)$ and $\eps_A(\kappa)$. Stage 1 uses fresh copies and conditions on the Stage-0 frame.
\item \emph{Small $m$.} The cold branch runs only when $1/m<x_\star$, i.e.\ $m>256c_\kappa^2\kappa\kappa'^2\ge256$; there $N\ge m^3\ge32m\ln(200m^2)$, so the Chernoff condition of \S\ref{subsec:pairrotation} holds. For smaller $m$ the class lies in $\mathcal G_{x_\star/\kappa}$ and \cite[Thm.~7.6]{Rub26} applies with no restriction on $m$ or $\eps$.
\item $\eps\le\min\{1,\eps_A(\kappa)\}$; odd $m$ changes the matching probabilities by a factor $(m-1)/m$ (constants only); full rank is automatic ($x_k\ge x_c/\kappa>0$).
\item Conditioning on the Chernoff events inflates the Markov expectations by a factor $\le1/0.99$, which is absorbed into the constants.
\item \emph{The protocol is single-copy:} one Gaussian unitary per copy ($W_0^T$ composed with the commuting two-mode rotations), a computational-basis occupation measurement, classical randomness for the matchings, labels and signs, and classical feedforward of $W_0$ and $\hat x_c$ from Stage 0; no ancillas and no quantum memory.
\end{itemize}

\section{Proofs for Section~\ref{sec:lemmaF}}\label{app:proofs-F}

\subsection{The hole bound}

The next lemma says that a cold frame mode cannot hide a creation operator.

\begin{lemma}\label{lem:F0}
$h_a\le2p'_a$ for every frame mode $a$ of every frame.
\end{lemma}

\begin{proof}
$\Pi_p$ commutes with $\mathcal N$, so $p'_a=\langle f_a|\Pi_p\mathcal N\Pi_p|f_a\rangle+\langle f_a|\Pi_h\mathcal N\Pi_h|f_a\rangle\ge\frac12\|\Pi_hf_a\|^2$. In mode language: with the frame mode written as $\tilde b_a=\sum_k(u_{ak}b_k+v_{ak}b_k^\dagger)$ in the independent true normal modes $b_k$, $p'_a=\sum_k|u_{ak}|^2p_k+\sum_k|v_{ak}|^2(1-p_k)$ and $h_a=\|v_a\|^2$, with $1-p_k\ge\frac12$.
\end{proof}

\subsection{The Wick variance of a shot and the per-shot bound}

\begin{lemma}[Wick variance of a shot]\label{lem:FW}
Fix $(\pi,\sigma,s)$. Let $c_P$ be the Nambu vector of the recorded rotated mode of the pair $P=\{a,b\}$: $(f_a+s\omega f_b)/\sqrt2$ for a hopping label and $(f_a+s\omega\bar f_b)/\sqrt2$ for a pairing label, with $\omega=1$ for the real labels ($\mathrm{hr},\mathrm{pr}$) and $\omega=i$ for the imaginary labels ($\mathrm{hi},\mathrm{pi}$); the label constants are $c_{\mathrm{hr}}=+1$, $c_{\mathrm{hi}}=c_{\mathrm{pr}}=c_{\mathrm{pi}}=-1$. Let $d_P:=S\overline{c_P}$ and $H:=\sum_P\gamma_P\big(|c_P\rangle\langle c_P|-|d_P\rangle\langle d_P|\big)$ with $\gamma_P$ real. Then
\[
\mathrm{Var}\Big(\sum_P\gamma_Pn'_P\Bigm|\pi,\sigma,s\Big)=\tfrac12\tr\big(H\mathcal NH(I-\mathcal N)\big)\le\tfrac12\tr(\mathcal KH^2)+\tfrac12\|\Pi_pH\Pi_h\|_F^2,
\]
and consequently
\begin{equation}\tag{FW$'$}\label{eq:FWprime}
\mathrm{Var}\Big(\sum_P\gamma_Pn'_P\Bigm|\pi,\sigma,s\Big)\le5\sum_{P\in\pi}\gamma_P^2(p'_a+p'_b)+2\sum_{P\ \mathrm{pairing}}\gamma_P^2 .
\end{equation}

\end{lemma}

\begin{proof}
Write $\bar{\mathcal N}:=I-\mathcal N$.

\emph{Identity.} Both sides are quadratic forms in $\gamma$. The diagonal coefficient is $p_A(1-p_A)+|c_A^\dagger\mathcal Nd_A|^2$ with $p_A:=\langle n'_A\rangle$, and $c_A^\dagger\mathcal Nd_A=\langle a'_Aa'_A\rangle=0$. For the coefficient of $\gamma_A\gamma_B$ ($A\ne B$; $c_A,d_A\perp c_B,d_B$) the four traces are $-|N'_{AB}|^2$, $-|F'_{AB}|^2$, $-|F'_{AB}|^2$, $-|N'_{AB}|^2$ with the signs of $H$, where $N'_{AB}:=\langle a'^\dagger_Aa'_B\rangle$ and $F'_{AB}:=\langle a'_Aa'_B\rangle$; this gives $2(|F'_{AB}|^2-|N'_{AB}|^2)=2\,\mathrm{Cov}(n'_A,n'_B)$ by (T4). Since the $n'_P$ are $0/1$ variables, $\mathrm{Var}(\sum\gamma n')=\sum\gamma^2p(1-p)+\sum\sum\gamma\gamma\,\mathrm{Cov}$, so the identity is exactly Wick's theorem plus (T4).

\emph{Inequality.} Split $\mathcal N=\mathcal N\Pi_p+\mathcal N\Pi_h$ and $\bar{\mathcal N}=\bar{\mathcal N}\Pi_p+\bar{\mathcal N}\Pi_h$; all four pieces are positive semidefinite, since $\Pi_p$ commutes with $\mathcal N$. Using $\tr(XB)\le\tr(XB')$ for $X\succeq0$ and $B\preceq B'$:
\begin{itemize}
\item $\tr(H\mathcal N\Pi_pH\bar{\mathcal N})\le\tr(H\mathcal N\Pi_pH)=\tr(\mathcal K\Pi_pH^2)$, using $\bar{\mathcal N}\preceq I$;
\item $\tr(H\mathcal N\Pi_hH\bar{\mathcal N}\Pi_h)\le\tr(H\Pi_hH\bar{\mathcal N}\Pi_h)\le\tr(\bar{\mathcal N}\Pi_hH^2)=\tr(\mathcal K\Pi_hH^2)$, using $\mathcal N\Pi_h\preceq\Pi_h$ and then $H\Pi_hH\preceq H^2$;
\item $\tr(H\mathcal N\Pi_hH\bar{\mathcal N}\Pi_p)\le\tr(H\Pi_hH\Pi_p)=\|\Pi_pH\Pi_h\|_F^2$.
\end{itemize}

\emph{Consequence.} The vectors $\{c_P,d_P\}_{P\in\pi}$ are orthonormal: for a pairing label $\langle c_P|d_P\rangle=0$ because $f_a\perp\bar f_a,f_b$ and $\bar f_b\perp\bar f_a,f_b$; disjoint pairs have supports in disjoint sets $\{f_a,f_b,\bar f_a,\bar f_b\}$. So $H^2=\sum\gamma_P^2\big(|c_P\rangle\langle c_P|+|d_P\rangle\langle d_P|\big)$. Then $\tr(\mathcal KH^2)=2\sum\gamma_P^2\langle c_P|\mathcal K|c_P\rangle$ by the particle--hole symmetry of $\mathcal K$, and $\langle c_P|\mathcal K|c_P\rangle\le\langle f_a|\mathcal K|f_a\rangle+\langle f_b|\mathcal K|f_b\rangle$ (Cauchy--Schwarz for a positive semidefinite form; $\langle\bar f_b|\mathcal K|\bar f_b\rangle=\langle f_b|\mathcal K|f_b\rangle$), which is $\le p'_a+p'_b$ since $\mathcal K\preceq\mathcal N$. Next, $\|\Pi_pH\Pi_h\|_F^2\le4\sum\gamma_P^2\|\Pi_hc_P\|^2$, with $\|\Pi_hc_P\|^2\le h_a+h_b$ (hopping) and $\le h_a+1$ (pairing). Lemma~\ref{lem:F0} ($h\le2p'$) then gives \eqref{eq:FWprime}.
\end{proof}

\begin{remark}[The variance of Lemma~\ref{lem:FW} as Pauli blocking]\label{F-rem-blocking}
The covariance rule (T4) used in the proof of Lemma~\ref{lem:FW} is Wick's theorem; it is stated in \S\ref{subsec:furthertools} and proved in Appendix~\ref{app:bg-tools}. Inserting the spectral decomposition $\mathcal N=\sum_i\varrho_i|i\rangle\langle i|$ over the true basis $\{t_k,\bar t_k\}$ ($\varrho=p_k$ on $t_k$ and $1-p_k$ on $\bar t_k$) into the identity of Lemma~\ref{lem:FW} gives
\[
\mathrm{Var}\Big(\sum_P\gamma_Pn'_P\Bigm|\pi,\sigma,s\Big)=\tfrac12\sum_{i,j}|\langle i|H|j\rangle|^2\,\varrho_j(1-\varrho_i).
\]
\end{remark}

Each term is a single-particle transition from an occupied Nambu orbital $|j\rangle$ to an empty one $|i\rangle$ (Pauli blocking); the factor $\frac12$ undoes the double counting of each transition by its particle--hole image. Transitions out of a true particle ($\varrho_j=p_k$) or into a true hole ($1-\varrho_i=p_l$) carry a minority occupation and give $\frac12\tr(\mathcal KH^2)$, noise matched to the metric. Transitions from a true hole into a true particle, the creation of a pair of quasiparticles out of the normal-mode vacuum, carry the weight $(1-p_k)(1-p_l)\ge\frac14$ and give $\frac12\|\Pi_pH\Pi_h\|_F^2$, which is small only if the measured modes have small hole content, tied to their occupation by Lemma~\ref{lem:F0}. The pairing labels, which rotate a frame particle into a frame hole, therefore keep an irreducible noise of order one per pair, the term $2\sum\gamma_P^2$ of \eqref{eq:FWprime}.

\paragraph{Notation for the per-shot bound.}
For $w\in\C^{2m}$ put $\alpha_a:=|\langle f_a|w\rangle|^2$, $\beta_a:=|\langle\bar f_a|w\rangle|^2$, $A_w:=\sum_a\alpha_a$, $B_w:=\sum_a\beta_a=\langle w|\hat\Pi_h|w\rangle$, $Q_w:=\langle w|\mathcal Q_W|w\rangle$ and $N_w:=\langle w|\mathcal N|w\rangle$. Let $\Pi_{(a)}$ be the projector onto $\mathrm{span}\{f_a,\bar f_a\}$ and $w_{(a)}:=\Pi_{(a)}w$; for $b\ne a$ let $\mathcal B_b:=\Pi_{(a)}\mathcal N\Pi_{(b)}$ and $\mathcal A_a:=\Pi_{(a)}\mathcal N\Pi_{(a)}$; for the pair $\{a,b\}$ let $\nu_{ab}:=w^\dagger(\mathcal B_b+\mathcal B_b^\dagger)w=2\,\mathrm{Re}\big(w_{(a)}^\dagger\mathcal B_bw_{(b)}\big)$.

\begin{lemma}[Per-shot bound]\label{lem:R0}
For every frame $W$ and every $w\in\C^{2m}$,
\begin{equation}\tag{R0}\label{eq:R0}
\begin{aligned}
&\mathrm{Var}(w^\dagger Zw)\le640m\|w\|^2Q_w+1024m\|w\|^2B_w+3(N_w+Q_w+B_w)^2+12\|w\|^4,\\
&|w^\dagger Zw-\E|\le4m\|w\|^2;\qquad
\mathrm{Var}(w^\dagger Z_{\mathrm{occ}}w)\le10\big(\|w\|^2Q_w+B_w^2\big),\qquad|w^\dagger Z_{\mathrm{occ}}w-\E|\le\|w\|^2 .
\end{aligned}
\end{equation}
\end{lemma}

\begin{proof}
Write $q:=w^\dagger Zw=\sum_{P\in\pi}\gamma_Pn'_P$ with $\gamma_P=4(m-1)c_\sigma s_Pg_P$ and $g_P:=w^\dagger\mathcal E_{P,\sigma}w\in\R$, and put $x_a:=\alpha_a+\beta_a$, so $\sum_ax_a=\|w\|^2$.

\emph{Label sums.} The two hopping labels give $\sum_\sigma g^2=4|\bar w_aw_b-\overline{\bar w_{\bar a}w_{\bar b}}|^2\le8(\alpha_a\alpha_b+\beta_a\beta_b)$, and the two pairing labels give $\le8(\alpha_a\beta_b+\alpha_b\beta_a)$. By the law of total variance, $\mathrm{Var}(q)\le\E[\mathrm{Var}(q\mid\pi,\sigma,s)]+\E\big[(\E[q\mid\pi,\sigma,s])^2\big]-(\E q)^2$.

\emph{(i) Conditional variance.} Average \eqref{eq:FWprime}, with the constants $9$ and $4$, over $\pi$ and $\sigma$. Each unordered pair lies in $\pi$ with probability $1/(m-1)$ and carries each label with probability $\frac14$, so $\E_{\pi,\sigma}\sum_{P\in\pi}f=(4(m-1))^{-1}\sum_{a<b}\sum_\sigma f$. With $\sum_{a<b}x_ax_b(p'_a+p'_b)=\sum_{a\ne b}x_ax_bp'_a\le\|w\|^2\sum_ax_ap'_a\le\|w\|^2(Q_w+B_w)$ and $\sum_{a<b}(\alpha_a\beta_b+\alpha_b\beta_a)\le A_wB_w$, the result is $\le576m\|w\|^2(Q_w+B_w)+256mA_wB_w$.

\emph{(ii) Sign part.} $\langle n'_P\rangle=m_P+s_P\nu_P$, where $m_P=\frac12(p'_a+p'_b)$ for a hopping label, $m_P\le1$ for a pairing label, and $\nu_P$ is the frame Nambu parameter (Lemma~\ref{lem:A1}). So $\E[q\mid\pi,\sigma,s]=\sum_P4(m-1)c_\sigma g_P(s_Pm_P+\nu_P)$, using $s_P^2=1$. Averaging over the independent signs gives $\sum_P16(m-1)^2g_P^2m_P^2+D_{\mathrm{des}}^2$, where $D_{\mathrm{des}}:=\sum_{P\in\pi}d_P$ and $d_P:=4(m-1)c_\sigma g_P\nu_P$. With $m_P^2\le m_P$ (hopping) and $\le1$ (pairing), the first sum averages to $\le32m\|w\|^2(Q_w+B_w)+64mA_wB_w$.

\emph{(iii) Design part $\E_{\pi,\sigma}D_{\mathrm{des}}^2$.} The diagonal piece $\E_{\pi,\sigma}\sum_Pd_P^2$ gives the same bound as (ii), since $\nu_P^2\le|N'_{ab}|^2\le p'_ap'_b\le\frac12(p'_a+p'_b)$ for hopping labels (Cauchy--Schwarz for $\mathcal N\succeq0$, then AM--GM with $p'\le1$) and $\nu_P^2\le1$ for pairing labels. For the correlated piece, $\E_\sigma d_P=(m-1)\nu_{ab}$, and two disjoint pairs lie in $\pi$ together with probability $1/((m-1)(m-3))$; so the piece equals $\frac{m-1}{m-3}\sum_{P\cap P'=\emptyset}\nu_P\nu_{P'}$. Now $\sum_{\mathrm{disjoint}}=(\sum\nu)^2-\sum_{P,P'\ \text{sharing a mode}}\nu_P\nu_{P'}$ ($P=P'$ included), and the subtracted sum is at most $\sum_a(\sum_b|\nu_{ab}|)^2$ in absolute value. So the piece is $\le\frac{m-1}{m-3}\big[(w^\dagger\mathcal N_{\mathrm{off}}w)^2+\sum_a(\sum_b|\nu_{ab}|)^2\big]$, with $\frac{m-1}{m-3}\le3$ for $m\ge4$, and $|w^\dagger\mathcal N_{\mathrm{off}}w|\le N_w+w^\dagger\mathcal C_W(\mathcal N)w=N_w+\sum\alpha_ap'_a+\sum\beta_a(1-p'_a)\le N_w+Q_w+B_w$. Finally, $\sum_b|\nu_{ab}|\le\|w_{(a)}\|\,\|w\|$: we have $|\nu_{ab}|\le2|w_{(a)}^\dagger\mathcal B_bw_{(b)}|$; since $0\preceq\mathcal N\preceq I$, $\mathcal N-\mathcal N^2\succeq0$, and its $(a)(a)$ block is $\mathcal A_a-\mathcal A_a^2-\sum_{b\ne a}\mathcal B_b\mathcal B_b^\dagger\succeq0$, so $\sum_{b\ne a}\mathcal B_b\mathcal B_b^\dagger\preceq\mathcal A_a-\mathcal A_a^2\preceq\frac14I$, and Cauchy--Schwarz over $b$ gives $\sum_b|\nu_{ab}|\le\|w_{(a)}\|\,\|w\|$, hence $\sum_a(\sum_b|\nu_{ab}|)^2\le\|w\|^4$. Hence the piece is at most $3\big[(N_w+Q_w+B_w)^2+4\|w\|^4\big]$.

\emph{Sum.} $576+32+32=640$ in front of $m\|w\|^2(Q_w+B_w)$; $256+64+64=384$ in front of $mA_wB_w\le m\|w\|^2B_w$; $640+384=1024$ in front of $m\|w\|^2B_w$.

\emph{Ranges.} $\|Z\|\le4(m-1)$ (disjointly supported unit-norm patterns) and $\|\mathcal N_{\mathrm{off}}\|\le2$, so $|w^\dagger Zw-\E|\le(4m-2)\|w\|^2$.

\emph{Occupation shot.} $w^\dagger Z_{\mathrm{occ}}w=\sum\theta_an_a+\mathrm{const}$ with $\theta_a:=\alpha_a-\beta_a$. Lemma~\ref{lem:FW} with $H':=\sum\theta_a(|f_a\rangle\langle f_a|-|\bar f_a\rangle\langle\bar f_a|)$ gives $\mathrm{Var}\le5\sum\theta_a^2p'_a\le10(A_wQ_w+B_w^2)$. Also $\|Z_{\mathrm{occ}}-\E Z_{\mathrm{occ}}\|\le1$.

\emph{Homogeneity.} Every term is of degree $4$ in $w$, so the bound holds for unnormalised $w$.
\end{proof}

\subsection{Proof of Lemma~\ref{lem:FR}}

\begin{proof}[Proof of Lemma~\ref{lem:FR}]
Fix a unit $w$ and put $w':=\Omega_\mu^{-1/2}w$.

\emph{Weights of $w'$.} $\|w'\|^2\le1/\mu$ and $N_{w'}\le1$. Since $(\mu_0/\mu-1)\mathcal N\succeq0$, $\Omega_{\mu_0}\preceq(\mu_0/\mu)\Omega_\mu$; so $Q_{w'}\le C\langle w'|\Omega_{\mu_0}|w'\rangle\le C\mu_0/\mu$, and likewise $B_{w'}=\langle w'|\hat\Pi_h|w'\rangle\le C\mu_0/\mu$ by the hole clause of $G$.

\emph{Term conversion.} Each term of \eqref{eq:R0} is at most its coefficient times $m\mu_0/\mu^2$: the two $m$-terms give $1664C\,m\mu_0/\mu^2$; $3(1+2C\mu_0/\mu)^2\le3(1+2C)^2(\mu_0/\mu)^2\le3(1+2C)^2\mu_0/\mu^2$, using $\mu_0/\mu\ge1$ and $\mu_0\le1$; and $12/\mu^2\le12m\mu_0/\mu^2$, since $m\mu_0\ge m\mu\ge1$. Hence $\mathrm{Var}(w'^\dagger Zw')\le K_1m\mu_0/\mu^2$ with range $\le4m/\mu$, and $\mathrm{Var}(w'^\dagger Z_{\mathrm{occ}}w')\le K_2\mu_0/\mu^2$ with range $\le1/\mu$.

\emph{Bernstein} \eqref{eq:bg-bernstein} at the deviation $t=\eta/4$, with the exponent $nt^2/(4V)$ under the range condition $bt/3\le V$ ($b$ the range, $V$ the variance bound). For the Horvitz--Thompson average the range condition is $\eta\mu\le3K_1\mu_0$, which holds, and the exponent is $n\eta^2\mu^2/(64K_1m\mu_0)\ge10m$. For the occupation average the range condition is $\eta\mu\le12K_2\mu_0$, and the exponent is $n'\eta^2\mu^2/(64K_2\mu_0)\ge10m$.

\emph{Net.} We use the $\frac14$-net of step~(3) in the proof of Lemma~\ref{lem:A2prime}: at most $e^{9m}$ points, and $\|M\|\le2\max_{\mathrm{net}}|w^\dagger Mw|$ for Hermitian $M$. On the event, $|w^\dagger\Omega_\mu^{-1/2}E\,\Omega_\mu^{-1/2}w|\le\eta/2$ on the net ($E:=\hat{\mathcal N}-\mathcal N$), so $\|\Omega_\mu^{-1/2}E\,\Omega_\mu^{-1/2}\|\le\eta$. The failure probability is $\le e^{9m}\cdot2\cdot2e^{-10m}=4e^{-m}$.
\end{proof}

\paragraph{The three factors of the count.} The count $n\asymp m^2\mu_0/(\eta^2\mu^2)$ of Lemma~\ref{lem:FR} is a product of three factors: the Horvitz--Thompson sparsity $m$ of a shot; the entropy $m$ of the unit sphere of $\C^{2m}$, since the deviation must be small in all $e^{9m}$ directions of the net at once, so that the Bernstein exponent must be of order $m$; and the counting noise $\mu_0/\mu^2$ along a direction rescaled by $\Omega_\mu^{-1/2}$, which raises the squared norm by up to $1/\mu$ while, in a frame satisfying $G(\mu_0,C)$, the frame occupation seen along the rescaled direction is at most $C\mu_0/\mu$.

\subsection{Proof of Theorem~\ref{thm:Fprime}}

\begin{proof}[Proof of Theorem~\ref{thm:Fprime}]
Induction on $r$. \emph{Base:} $W_0$ satisfies $G(1,1)\subset G(1,3)$.

\emph{Step.} Suppose $W_{r-1}$ satisfies $G(\mu_{r-1},3)$. $W_{r-1}$ is a deterministic function of the rounds before $r$, through the tie-breaking rule of \S\ref{subsec:F-frames}, and the copies of round $r$ are fresh. So Lemma~\ref{lem:FR} applies conditionally on the past, with $(\mu_0,\mu):=(\mu_{r-1},\mu_r)$ and $C=3$. This needs $1/m\le\mu_r\le\mu_{r-1}\le1$: uncapped, $\mu_r/\mu_{r-1}=\frac14m^{-2^{-r}}<1$ and $\mu_r\ge m^{2^{-r}}/m\ge1/m$; and $\mu_R\le m^{2^{-R}}/m\le2/m$, because $2^R\ge\log_2m$. On $\mathrm{REL}(\mu_r,1/8)$, Lemma~\ref{lem:FS} gives $G(\mu_r,18/7)$.

\emph{Cost.} For $r\le R$, $\mu_{r-1}/\mu_r^2\le4m\cdot4^{-(R-r)}$; uncapped this is an equality, by substituting the schedule. A cap either lowers $\mu_{r-1}$, or forces $\mu_{r-1}=1$ and makes the ratio $1\le4m\cdot4^{-(R-r)}$, since $4^R<4\log_2^2m\le4m$ for $m\ge16$. Hence $\sum_{r\le R}\mu_{r-1}/\mu_r^2\le\frac{16}3m$. The last round has $\mu_R/\mu_{R+1}^2\le2m$. Multiply by $640K_1m^2/\eta^2=640\cdot64\,K_1m^2$ and by $640\cdot64\,K_2m$; the ceilings add $\le2(R+1)$.

\emph{Probability.} $\Prob(\cup_r\mathrm{fail}_r)\le\sum_r\Prob(\mathrm{fail}_r\cap\text{all earlier rounds hold})\le4(R+1)e^{-m}$, which is $\le0.01$ already at $m=16$, $R=2$.
\end{proof}

\section{Proofs for Section~\ref{sec:thmB}}\label{app:proofs-B}

\noindent\emph{Notation.} In this section $J$ denotes the number of occupation bands (the $2\times2$ matrix $J$ of \S\ref{subsec:conventions} appears only inside block formulas such as $\lambda_kJ$); $\mathcal N$ is the Nambu matrix of \S\ref{subsec:F-nambu}, $E:=\hat{\mathcal N}-\mathcal N$ the Nambu form of the Stage-1 error and $\Delta$ its Majorana form; $\mathcal R(A)$ is the higher-order remainder of the far decomposition (\S\ref{subsec:thmB-L1}).

\subsection{Conventions for the proof of Theorem~\ref{thm:B}}\label{subsec:thmB-conventions}

Majoranas, covariance matrices, hopping and pairing blocks and the generator $K_A=\frac i4\sum_{a,b}A_{ab}c_ac_b$ are as in \S\ref{subsec:conventions}; the empty-type normal form is $\Gamma=O\big(\bigoplus_k\lambda_kJ\big)O^T=:ODO^T$ with $\lambda_k=2p_k-1$, $p_k\in[0,\tfrac12]$. \emph{The analysis is done in the true frame}, where $D$ is diagonal. The protocol works in its own frame $W$ (Theorem~\ref{thm:Fprime}). The composite rotation $O':=W^TO$ has rows indexed by frame modes $a$ and columns by true modes $k$, and $\Gamma':=W^T\Gamma W=O'DO'^T$; it transports the measured error $\Delta'$ to the true frame: $\Delta:=O'^T\Delta'O'$. In Nambu form $O'$ is the real form of $\mathcal W=\bigl[\begin{smallmatrix}u&v\\ \bar v&\bar u\end{smallmatrix}\bigr]$ (rows frame modes $a$, columns true modes $k$); its $2\times2$ block $(a,k)$ is the real form of $(u_{ak},v_{ak})$. We write $O'=U+V$ for the particle (hopping-type) and hole (pairing-type) parts, with $\|U\|_\op,\|V\|_\op\le1$ and $\|O'_{ak}\|_F^2=2(|u_{ak}|^2+|v_{ak}|^2)$. The two Gaussian tools used throughout are Lemmas~\ref{lem:wick} and~\ref{lem:MT}.

\subsection{Bands, far and near (type-resolved), the cold group}\label{subsec:thmB-bands}

$J:=\lceil\log_2m\rceil$. Bands by true occupation: $B_j:=\{k:\ p_k\in(2^{-j-1},2^{-j}]\}$ for $1\le j\le J-1$, and the cold band $B_c:=\{k:\ p_k\le2^{-J}\}$ with band index $J$; so there are exactly $J$ bands. Scales $P_j:=2^{-j}$, $P_c:=2^{-J}\in[1/(2m),1/m]$; every scale is $\ge1/(2m)$; $\iota(k)$ denotes the band index of $k$. For $k\in B_j$ ($j<J$), $p_k\in(P_j/2,P_j]$; for $k\in B_c$, $p_k\le P_c$. In every band $b$, $\max(p_k,1/m)\le2P_b$, and every non-cold band has $P_b\ge2^{-(J-1)}\ge1/m$.

\paragraph{Far and near, type-resolved.}
Every $2\times2$ block $X_{kl}$ ($k\ne l$) splits as $X^+_{kl}+X^-_{kl}$ (hopping type in $\mathrm{span}\{I,J\}$, pairing type in $\mathrm{span}\{\sigma_x,\sigma_z\}$). The hopping part of the band block $(b,b')$ is \emph{far} iff $|\iota(b)-\iota(b')|\ge2$ and \emph{near} iff $|\iota(b)-\iota(b')|\le1$; the pairing part of $(b,b')$ is \emph{far} iff $P_b+P_{b'}\le5/8$ (every block except $(1,1)$, $(1,2)$, $(2,1)$) and \emph{near} otherwise. For antisymmetric matrices the diagonal $2\times2$ blocks are multiples of $J$ and are near; in Nambu form the mode-diagonal particle--hole entries vanish (\S\ref{subsec:thmB-class}). $X_{\mathrm{far}}$ and $X_{\mathrm{near}}$ denote the far and near \emph{parts} of $X$ (a block can contribute to both). Gaps in $\lambda$-units: for a far hopping pair $(b,b')$ with $\iota(b)=j<\iota(b')$, $p_k-p_l>P_j/2-P_j/4=P_j/4$, so $|\lambda_k-\lambda_l|>\max(P_b,P_{b'})/2$; for a far pairing pair, $|\lambda_k+\lambda_l|=2-2(p_k+p_l)\ge3/4$. Near hopping pairs have occupation ratio $<4$ (the cold band has no lower bound and is treated separately); near pairing pairs have $p_k,p_l>1/8$; every far hopping pair has $\max P\ge4P_c\ge2/m$.

\paragraph{The cold group.}
$\mathcal P:=B_c$. $X_{\mathcal P}:=$ the hopping parts of the near blocks with an index in $\mathcal P$ --- the blocks $(\mathcal P,\mathcal P)$, $(\mathcal P,B_{J-1})$, $(B_{J-1},\mathcal P)$ --- together with the diagonal blocks of $\mathcal P$; $X_{\mathrm{rest}}:=$ the remaining near parts (hopping parts of equal or adjacent non-$\mathcal P$ blocks, pairing parts of $(1,1)$, $(1,2)$, $(2,1)$, non-$\mathcal P$ diagonal blocks). Every block of $X_{\mathcal P}$ goes to the trace-norm leg (leg~III of \S\ref{subsec:thmB-chain}), never to the $\chi^2$ leg: pure modes have infinite $\chi^2$ weight. Pairing parts of cold blocks are far ($P_c+P\le5/8$) and are removed by the rotation (leg~I).

The far/near rule decides only which bound is valid, not the price: either way an entry is priced with the inverse of the larger occupation of its pair, as a rotation by the angle entry$/$gap where the gap is comparable to that occupation (far, leg~I), and through the $\chi^2$ weight $1/\sqrt{p_kp_l}$ where the occupations are comparable (near, leg~II). A pairing entry has denominator $|\lambda_k+\lambda_l|=2-2(p_k+p_l)$, the cost of creating or removing two particles, which is small only if both modes are nearly half filled; hence pairing is far except among bands $1$--$2$.

\subsection{Nambu objects and the dictionary}\label{subsec:thmB-dict}

The Nambu objects $t_k$, $\bar t_k$, $\Pi_p$, $\Pi_h$, $\mathcal K$, $\Omega_\mu$ are those of \S\ref{subsec:F-nambu}. $V_b$ is the span of the true particle vectors of band $b$ and $h(V_b)$ its hole image.

\paragraph{The dictionary.}
By \eqref{eq:blocks}, applied linearly to errors, a Nambu error $E$ with particle--particle block $E^N$ and anomalous block $E^F$ is sent to the real antisymmetric $X$ with $X_{kk}=2E^N_{kk}J$ and $X_{kl}$ given by the second line of \eqref{eq:blocks} with $(N,F)$ replaced by $(E^N,E^F)$. Identify $aI+bJ\leftrightarrow a+ib$. Then the hopping part is the realification of $z\mapsto2i\,E^Nz$, the pairing part the realification of the antilinear map $z\mapsto2i\,E^F\bar z$ ($\sigma_x$ acts as $z\mapsto i\bar z$, $\sigma_z$ as $z\mapsto\bar z$), and realification preserves operator norms and products, for complex-linear and antilinear block maps alike. We write $\Phi_D$ for the affine map $\mathcal N\mapsto\Gamma$ of \eqref{eq:blocks} and $\Phi_{D,\mathrm{lin}}$ for its linear part.

\emph{Normalisation.} The Stage-1 Nambu estimate is $\hat{\mathcal N}:=N_2^{-1}\sum_tZ_t+N_3^{-1}\sum_tZ_{\mathrm{occ},t}$, from the shots of \S\ref{subsec:F-shots} in the frame $W$ (the counts $N_2,N_3$ are fixed in \S\ref{subsec:thmB-assembly}). The Horvitz--Thompson block $8(m-1)c_\sigma sn'B_{P,\sigma}$ of \S\ref{subsec:thmB-protocol} is the image under \eqref{eq:blocks} of $4(m-1)c_\sigma sn'\mathcal E_{P,\sigma}$, so $Y_t$ is the image of $Z_t$, and $\mathrm{diag}(2\hat p'-1)J$ is the image of $\mathrm{diag}(\hat p',1-\hat p')$. Hence $\tilde\Gamma'=\Phi_D(\hat{\mathcal N})$ and $\Delta':=\tilde\Gamma'-\Gamma'=\Phi_{D,\mathrm{lin}}(E)$ in the frame, and $\tilde\Gamma:=D+\Delta$ in the true frame.

\emph{Covariance.} $\Gamma$ and $\mathcal N$ are related by the fixed Majorana--Nambu map $c=\Omega_{\mathrm M}\alpha$ of \S\ref{subsec:bg-fermions} ($\Omega_{\mathrm M}$ is not the weight $\Omega_\mu$). So $\Phi_D$ is equivariant for every $R\in O(2m)$: under $R$ the Nambu matrix transforms as $\mathcal N\mapsto\bar U_R\mathcal NU_R^T$ (\S\ref{subsec:thmB-transl}), with the unitary $U_R=\Omega_{\mathrm M}^{-1}R\,\Omega_{\mathrm M}$, which preserves the Nambu structure ($S\bar U_RS=U_R$) whether $\det R=+1$ or $-1$ ($\det R=-1$ is a particle--hole swap of one mode). Therefore $\Delta=O'^T\Delta'O'$ is $\Phi_{D,\mathrm{lin}}$ of $E$ written in the true basis $\{t_k,\bar t_k\}$, and for all true bands $b,b'$
\begin{equation}\tag{Dict}\label{eq:Dict}
\begin{gathered}
\|\Delta^+_{bb'}\|_\op=2\|\Pi_{V_b}E\Pi_{V_{b'}}\|_\op,\qquad
\|\Delta^-_{bb'}\|_\op=2\|\Pi_{V_b}E\Pi_{h(V_{b'})}\|_\op,\\
\|\Delta^{\mathrm{diag}}_{bb}\|=2\max_{k\in b}|\langle t_k|E|t_k\rangle|\le2\|\Pi_{V_b}E\Pi_{V_b}\|,
\end{gathered}
\end{equation}
the first including the diagonal $2\times2$ blocks. The pairing block $E^F[b,b']$ has norm $\|\Pi_{V_{b'}}E\Pi_{h(V_b)}\|$, and by the antisymmetry of $E^F$ also $\|\Pi_{V_b}E\Pi_{h(V_{b'})}\|$.

\paragraph{Translation.} In $\mathcal W=\bigl[\begin{smallmatrix}u&v\\ \bar v&\bar u\end{smallmatrix}\bigr]$, $|u_{ak}|=|\langle t_k|f_a\rangle|$ and $|v_{ak}|=|\langle\bar t_k|f_a\rangle|$ in any conjugation convention, $f_a$ the frame particle vectors (\S\ref{subsec:F-frames}). Hence
\begin{equation}\tag{Transl}\label{eq:Transl}
\begin{gathered}
\|v_a\|^2=\|\Pi_hf_a\|^2=h_a,\qquad
\sum_a|v_{ak}|^2=\|\hat\Pi_p\bar t_k\|^2=\|\hat\Pi_ht_k\|^2,\\
\sum_a|u_{ak}|^2p'_a=\langle t_k|\mathcal Q_W|t_k\rangle,\qquad
\|O'_{ak}\|_F^2=2(|u_{ak}|^2+|v_{ak}|^2),
\end{gathered}
\end{equation}
the second by applying $S\circ\mathrm{conj}$, which maps $\hat\Pi_p\mapsto\hat\Pi_h$ and $\bar t_k\mapsto t_k$.

\begin{corollary}[What Theorem~\ref{thm:B} uses]\label{cor:Fprime}
On the event of Theorem~\ref{thm:Fprime}, the output frame $W$ satisfies $G(1/m,3)$, and, by \eqref{eq:Transl} and $p_k+1/m\le2\max(p_k,1/m)$:
\begin{enumerate}
\item[(a)] $\|v_a\|^2=h_a\le2p'_a\le6\max(p'_a,1/m)$ for every frame mode $a$ (Lemma~\ref{lem:F0}), and $\sum_a|v_{ak}|^2=\langle t_k|\hat\Pi_h|t_k\rangle\le3(p_k+1/m)\le6\max(p_k,1/m)$ for every true mode $k$; so $c_F':=6$;
\item[(b)] $\sum_a\|O'_{ak}\|_F^2p'_a=2\sum_a|u_{ak}|^2p'_a+2\sum_a|v_{ak}|^2p'_a\le2\langle t_k|\mathcal Q_W|t_k\rangle+2\sum_a|v_{ak}|^2\le12(p_k+1/m)\le24\max(p_k,1/m)$ (the Frobenius form); so $c_b:=24$; also $\sum_a\|O'_{ak}\|_F^2=2$ ($O'$ is orthogonal);
\item[(G)] the operator frame condition $G(1/m,3)$ itself.
\end{enumerate}
The frame enters the proof of Theorem~\ref{thm:B} only through (a) and (b), in Lemma~\ref{lem:B1}, and through (G), in Lemma~\ref{lem:Htri}.
\end{corollary}

\paragraph{Reading (a) and (b).} In mode language, with the frame modes $\tilde b_a=\sum_k(u_{ak}b_k+v_{ak}b_k^\dagger)$ ($v_{ak}$ the pairing that the frame mixes in; occupation formula in the proof of Lemma~\ref{lem:F0}), (a) says that a frame mode contains true creation operators only in proportion to its occupation, and that a cold true mode is never hidden among the filled modes of the frame; (b) says that the frame modes through which a true mode $k$ is read are, on average with the weights $\|O'_{ak}\|_F^2/2$, as cold as $k$ itself, up to a constant and the floor $1/m$. Together they turn the noise $2(p'_a+p'_b)$ of Lemma~\ref{lem:A1}, which is set by frame occupations, into second moments proportional to $\max(p_k,1/m)+\max(p_l,1/m)$ in the true frame (Lemma~\ref{lem:B1}).

\paragraph{Clipping.} The raw estimate $\tilde\Gamma'$ need not be a valid covariance matrix: $i\tilde\Gamma'$ may have eigenvalues outside $[-1,1]$. $\operatorname{Clip}$ replaces each eigenvalue of the Hermitian matrix $i\tilde\Gamma'$ by the nearest point of $[-1,1]$ and keeps the eigenvectors; the result is real antisymmetric of norm $\le1$, and $\operatorname{Clip}(R\,\cdot\,R^T)=R\operatorname{Clip}(\cdot)R^T$ for every $R\in O(2m)$ by the functional calculus.

\subsection{Second moments in the true frame}\label{subsec:thmB-B1}

\begin{lemma}[Second moments]\label{lem:B1}
On the event of Theorem~\ref{thm:Fprime} (Corollary~\ref{cor:Fprime}~(a), (b), with $c_F'=6$, $c_b=24$), let $C_1':=C_0/2$ with $C_0$ from the proof ($C_0=561\,280$, so $C_1'=280\,640\le10^6$). Then for all true modes $k\ne l$:
\begin{align*}
\E\|\Delta^+_{kl}\|_F^2&\le C_1'\,\frac{8m}{N_2}\big(\max(p_k,1/m)+\max(p_l,1/m)\big)+C_1'\,\frac{s_{kl}}{N_3},\\
\E\|\Delta^-_{kl}\|_F^2&\le2C_1'\,\frac{8m}{N_2}+C_1'\,\frac{s_{kl}}{N_3},\\
\E\|\Delta_{kk}\|_F^2&\le C_1'\max(p_k,1/m)\Big(\frac{8m}{N_2}+\frac1{N_3}\Big),
\end{align*}
where
\[
s_{kl}:=\sum_a\|O'_{ak}\|_F^2\,\|O'_{al}\|_F^2\,\max(p'_a,1/m)
\]
is the weight of the transported occupation-shot error, with
\[
s_{kl}\le2(c_b+2)\min\big(\max(p_k,1/m),\max(p_l,1/m)\big),\qquad\sum_ls_{kl}\le2(c_b+2)\max(p_k,1/m).
\]
In band form, $\E\|\Delta^+_{kl}\|_F^2\le4C_1'(8m/N_2)\max(P_b,P_{b'})+C_1's_{kl}/N_3$ for $k\in b$, $l\in b'$ ($\max(p,1/m)\le2P$ on every band). The $N_3$ terms sum to $O(C_1'm/N_3)$ in every leg of \S\ref{subsec:thmB-chain} and are covered by $N_3\ge Cm/\eps^2$.
\end{lemma}

\begin{proof}
Fix one Horvitz--Thompson shot and condition on its matching $\pi$, labels $\sigma$ and signs $s$; let $q$ be one of the four real entries of the hopping part of the true-frame block $(O'^TY_tO')_{kl}$. Since $Y_t=\sum_{P\in\pi}8(m-1)c_\sigma s_Pn'_P\,\hat B_P$ ($\hat B_P$ the $4\times4$ antisymmetric matrix carrying $B_{P,\sigma}$ at the $(a,b)$ position and $-B_{P,\sigma}^T$ at $(b,a)$, with zero diagonal blocks and $\|\hat B_P\|_\op=1$), we have $q=\sum_{P\in\pi}\gamma_Pn'_P$ with $\gamma_P:=8(m-1)c_\sigma s_P\,\tilde g_P$, where $\tilde g_P$ is the type-projected $(k,l)$ entry of $O'^T_{ak}B_{P,\sigma}O'_{bl}-O'^T_{bk}B_{P,\sigma}^TO'_{al}$: products of a block at $a$ with a block at $b\ne a$ only (the cross structure). Type rule (hop$\times$hop $=$ hop, pair$\times$pair $=$ hop, hop$\times$pair $=$ pair, applied to $(U+V)^T\hat B(U+V)$): for a hopping label the hopping output comes from $U^T\hat BU$ and $V^T\hat BV$, for a pairing label from $U^T\hat BV$ and $V^T\hat BU$. Hence $|\gamma_P|\le8(m-1)g_P$ with
\begin{align*}
g_P&:=\|U_{ak}\|\|U_{bl}\|+\|U_{bk}\|\|U_{al}\|+\|V_{ak}\|\|V_{bl}\|+\|V_{bk}\|\|V_{al}\|&&\text{(hopping label)},\\
g_P&:=\|U_{ak}\|\|V_{bl}\|+\|V_{ak}\|\|U_{bl}\|+\|U_{bk}\|\|V_{al}\|+\|V_{bk}\|\|U_{al}\|&&\text{(pairing label)}
\end{align*}
(Frobenius norms of the $2\times2$ blocks), so
\[
g_P^2\le4\big(\|O'_{ak}\|^2\|O'_{bl}\|^2+\|O'_{bk}\|^2\|O'_{al}\|^2\big)
\]
for both labels, and for pairing labels also
\[
g_P^2\le4\big(\|O'_{ak}\|^2\|V_{bl}\|^2+\|V_{ak}\|^2\|O'_{bl}\|^2+\|O'_{bk}\|^2\|V_{al}\|^2+\|V_{bk}\|^2\|O'_{al}\|^2\big).
\]

\emph{(1) The conditional variance.} Conditionally on $(\pi,\sigma,s)$, $q=\sum_{P\in\pi}\gamma_Pn'_P$ is a real linear combination of the recorded occupations of one shot in the frame $W$, so Lemma~\ref{lem:FW} applies. Its bound \eqref{eq:FWprime}, with $p'\le\max(p',1/m)$, $5\le1+4c_F'$ and $2\le4$, gives
\begin{equation}\label{eq:B11}
\mathrm{Var}(q\mid\pi,\sigma,s)\le(1+4c_F')\sum_{P\in\pi}\gamma_P^2\big(\max(p'_a,1/m)+\max(p'_b,1/m)\big)+4\sum_{P\in\pi,\ \mathrm{pairing}}\gamma_P^2 ,
\end{equation}
every correlation between the pairs of the shot included.

\emph{(2) The conditional mean.} $\E[n'_P\mid\pi,\sigma,s]=m_P+s_P\nu_P$ with $m_P=\frac12(p'_a+p'_b)$ for a hopping label, $m_P=\frac12(p'_a+1-p'_b)\le1$ for a pairing label, and $\nu_P$ a fixed entry of $\Gamma'_{ab}/2$ (Lemma~\ref{lem:A1}); so $\E[q\mid\pi,\sigma,s]=\sum_{P\in\pi}8(m-1)c_\sigma\tilde g_P(s_Pm_P+\nu_P)$, and by the law of total variance, the signs being independent of $(\pi,\sigma)$ and of each other,
\begin{equation}\label{eq:B12}
\begin{gathered}
\mathrm{Var}(q)=\E\,\mathrm{Var}(q\mid\pi,\sigma,s)+\E_{\pi,\sigma}\sum_{P\in\pi}\gamma_P^2m_P^2+\mathrm{Var}_{\pi,\sigma}(D_{\mathrm{des}}),\\
D_{\mathrm{des}}:=\sum_{P\in\pi}d_P,\qquad d_P:=8(m-1)c_\sigma\tilde g_P\nu_P .
\end{gathered}
\end{equation}

\emph{(3) Averaging.} A given unordered pair $(a,b)$ is in $\pi$ with probability $1/(m-1)$ and carries each label with probability $\frac14$.

\emph{First term of \eqref{eq:B11}:}
\begin{align*}
&\frac1{m-1}\cdot\frac14\cdot64(m-1)^2(1+4c_F')\sum_{(a,b),\sigma}g_P^2(\max\nolimits_a+\max\nolimits_b)\\
&\qquad\le16(m-1)(1+4c_F')\cdot16\sum_{a,b}\|O'_{ak}\|^2\|O'_{bl}\|^2\big(\max(p'_a,1/m)+\max(p'_b,1/m)\big)
\end{align*}
(four labels, $g_P^2\le4(\cdots)$, $\sum_{a<b}[f(a,b)+f(b,a)]\le\sum_{a,b}f$), and, factorising the double sum and using (b) (Frobenius form) and $\sum_a\|O'_{ak}\|_F^2=2$,
\[
\sum_{a,b}\|O'_{ak}\|^2\|O'_{bl}\|^2(\max\nolimits_a+\max\nolimits_b)\le2(c_b+2)\big(\max(p_k,1/m)+\max(p_l,1/m)\big).
\] (Every term is a weighted column sum over one frame index times an unweighted column sum over the other; the cross structure of $g_P$ is what excludes a free sum $\sum_bp'_b\asymp m$.) So this term is $\le2^9(1+4c_F')(c_b+2)(m-1)\big(\max(p_k,1/m)+\max(p_l,1/m)\big)$.

\emph{Second term of \eqref{eq:B11}:}
\begin{align*}
\frac1{m-1}\cdot\frac12\cdot64(m-1)^2\cdot4\sum_{(a,b)}g_P^2
&\le2^7(m-1)\cdot4\sum_{a,b}\big(\|O'_{ak}\|^2\|V_{bl}\|^2+\|V_{ak}\|^2\|O'_{bl}\|^2\big)\\
&\le2^{11}c_F'(m-1)\big(\max(p_k,1/m)+\max(p_l,1/m)\big)
\end{align*}
(two pairing labels) by $\sum_a\|O'_{ak}\|_F^2=2$ and the inverse hole content of (a) ($\|V_{ak}\|_F^2=2|v_{ak}|^2$).

\emph{Sign term of \eqref{eq:B12}:} $m_P^2\le m_P\le\max(p'_a,1/m)+\max(p'_b,1/m)$ for hopping labels and $m_P^2\le1$ for pairing labels, so it is bounded by the two sums just computed with $(1+4c_F')\to1$ and $4\to1$: $\le2^9(c_b+2+c_F')(m-1)(\cdots)$.

\emph{Design term of \eqref{eq:B12}:}
\[
\E_{\pi,\sigma}D_{\mathrm{des}}^2=\frac1{4(m-1)}\sum_{(P,\sigma)}d_P^2+\frac{1}{16(m-1)(m-3)}\sum_{\substack{(P,\sigma)\ne(P',\sigma')\\ \text{disjoint}}}d_Pd_{P'},
\]
For the second contribution we first bound the signed sum from above, as below, and then replace its prefactor $1/(16(m-1)(m-3))$ by the larger $1/((m-1)(m-3))$ in the resulting nonnegative upper estimate. With $\sum_\sigma\nu_{P,\sigma}^2=\|\Gamma'_{ab}\|_F^2/8$ and $\sum_{b\ne a}\|\Gamma'_{ab}\|_F^2=8\sum_b(|N'_{ab}|^2+|F'_{ab}|^2)\le8\langle a|\mathcal N^2|a\rangle\le8p'_a$ ($0\preceq\mathcal N\preceq I$):
\begin{align*}
\sum_{(P,\sigma)}d_P^2&\le64(m-1)^2\cdot4\cdot\tfrac18\sum_{a\ne b}\|O'_{ak}\|^2\|O'_{bl}\|^2\|\Gamma'_{ab}\|_F^2\\
&\le32(m-1)^2\sum_a\|O'_{ak}\|^2\cdot2\cdot8p'_a\le2^9c_b(m-1)^2\max(p_k,1/m),
\end{align*}
so the first part is $\le2^7c_b(m-1)(\cdots)$. For the second part, $\sum_{(P,\sigma)}d_P=4(m-1)\,\E q$ with $|\E q|=|(O'^T\Gamma'_{\mathrm{off}}O')_{kl}|\le2$, so
\[
\sum_{\text{disjoint}}d_Pd_{P'}=\Big(\sum d_P\Big)^2-\sum d_P^2-\sum_{\text{overlapping}}d_Pd_{P'}\le64(m-1)^2+(8m-12)\sum_Pd_P^2
\]
(each $(P,\sigma)$ shares a mode with at most $8m-13$ others, the three other labels of the same pair included; $|d_Pd_{P'}|\le(d_P^2+d_{P'}^2)/2$), and divided by $(m-1)(m-3)$ this is $\le2^7+2^{13}c_b(m-1)(\cdots)$ for $m\ge5$ ($(8m-12)/(m-3)\le16$), where $2^7\le2^7(m-1)\big(\max(p_k,1/m)+\max(p_l,1/m)\big)$ since $(m-1)(\max_k+\max_l)\ge2(m-1)/m\ge1$.

Altogether
\begin{gather*}
\mathrm{Var}(q)\le C_0\,(m-1)\big(\max(p_k,1/m)+\max(p_l,1/m)\big),\\
C_0:=2^9(1+4c_F')(c_b+2)+2^{11}c_F'+2^9(c_b+2+c_F')+2^7c_b+2^{13}c_b+2^7,
\end{gather*}
and at $(c_F',c_b)=(6,24)$, $C_0=332\,800+12\,288+16\,384+3\,072+196\,608+128=561\,280$. The Horvitz--Thompson average over $N_2$ independent shots divides by $N_2$, and the four entries of the block (two parameters, each twice) give $\E\|\Delta^+_{kl}\|_F^2\le4C_0(m-1)(\cdots)/N_2\le C_1'(8m/N_2)\big(\max(p_k,1/m)+\max(p_l,1/m)\big)$.

\emph{Pairing entries $(k,l)$.} The type rule exchanges the roles of $U$ and $V$ in $g_P$; the hopping-label contribution now carries one $V$ factor and is bounded as above, the pairing-label contribution is scale-free ($\max_k+\max_l\le1$), and the same $C_0$ serves: $\E\|\Delta^-_{kl}\|_F^2\le2C_1'(8m/N_2)$.

\emph{Diagonal block $(k,k)$.} The Horvitz--Thompson part is the computation above with $l=k$. The $N_3$ occupation shots read all frame occupations jointly; the transported error of the frame diagonal is
\[
\sum_aO'^T_{ak}\big(2(\hat p'_a-p'_a)J\big)O'_{ak}=2\sum_a\det(O'_{ak})\,(\hat p'_a-p'_a)\,J
\]
($M^TJM=\det(M)J$ for a real $2\times2$ matrix $M$), and per shot the entry is $\sum_a\gamma_an_a$ with $\gamma_a:=2\det(O'_{ak})$, $|\gamma_a|\le\|O'_{ak}\|_F^2$ (no random design, no sign term). Lemma~\ref{lem:FW} for the occupation shot, as in the proof of Lemma~\ref{lem:R0} (\emph{Occupation shot}), gives $\mathrm{Var}\le5\sum_a\gamma_a^2p'_a$, so, by $\gamma_a^2\le2\|O'_{ak}\|_F^2$, (b) and $\sum_a\|O'_{ak}\|_F^2=2$, $\mathrm{Var}\le(1+2c_F')\sum_a\gamma_a^2\max(p'_a,1/m)\le2(1+2c_F')(c_b+2)\max(p_k,1/m)$,
including the correlations between the occupations of different frame modes. So
\[
\E\|\Delta^{(N_3)}_{kk}\|_F^2\le4(1+2c_F')(c_b+2)\max(p_k,1/m)/N_3\le C_1'\max(p_k,1/m)/N_3 .
\]

\emph{Off-diagonal blocks $(k\ne l)$} receive the same occupation shots through $2\sum_a\delta_a\,O'^T_{ak}JO'_{al}$, $\delta_a:=\hat p'_a-p'_a$, which is nonzero whenever a frame mode overlaps both $k$ and $l$: each entry of either type is, per shot, $\sum_a\gamma_an_a+\mathrm{const}$ with $\gamma_a:=2\,(O'^T_{ak}JO'_{al})_{\text{entry, type-projected}}$, $|\gamma_a|\le\|O'_{ak}\|_F\|O'_{al}\|_F$, and the same application of Lemma~\ref{lem:FW} gives
\[
\mathrm{Var}\le(1+2c_F')\sum_a\gamma_a^2\max(p'_a,1/m)\le(1+2c_F')\,s_{kl};
\]
the four entries of each of the two types give $\E\|\Delta^{(N_3)}_{kl}\|_F^2\le8(1+2c_F')s_{kl}/N_3\le C_1's_{kl}/N_3$ (the Horvitz--Thompson and occupation parts are independent and zero-mean: no cross term). The bounds on $s_{kl}$: $\|O'_{al}\|_F^2\le2$ gives $s_{kl}\le2\sum_a\|O'_{ak}\|_F^2\max(p'_a,1/m)\le2(c_b+2)\max(p_k,1/m)$ by (b) and $\sum_a\|O'_{ak}\|_F^2=2$, and symmetrically in $l$; and $\sum_ls_{kl}=\sum_a\|O'_{ak}\|_F^2\max(p'_a,1/m)\cdot\sum_l\|O'_{al}\|_F^2=2\sum_a\|O'_{ak}\|_F^2\max(p'_a,1/m)\le2(c_b+2)\max(p_k,1/m)$.
\end{proof}

The co-measurement correlations between the pairs of one shot, and between the occupations of one occupation shot, are included exactly through Lemma~\ref{lem:FW}; a Cauchy--Schwarz bound over the shot would give a scale-free bound and lose the factor $\max(p_k,1/m)+\max(p_l,1/m)$. The frame inputs used are: (a) in the pairing-label term; (b) in (3) and in the $N_3$ part; $\sum_a\|O'_{ak}\|_F^2=2$; and $0\preceq\mathcal N\preceq I$.

\subsection{The far decomposition in the row-lump norm}\label{subsec:thmB-far}

Everything in \S\S\ref{subsec:thmB-far}--\ref{subsec:thmB-L1} is in the Nambu picture, in the true frame; \S\ref{subsec:thmB-transl} translates the output to the Majorana form used by \S\ref{subsec:thmB-chain}.

\paragraph{Nambu objects.} On $\C^{2m}$ with the true particle vectors $t_k$ and holes $\bar t_k$, $\mathcal N=\sum_kp_k|t_k\rangle\langle t_k|+(1-p_k)|\bar t_k\rangle\langle\bar t_k|$, and $n_x$ is the eigenvalue of $\mathcal N$ on the basis vector $x\in\{t_k,\bar t_k\}$. We use the weight $\Omega_{1/m}=\mathcal N+I/m$.
\begin{itemize}
\item \emph{Errors} are Nambu-Hermitian: $Y=Y^*$, $S\bar YS=-Y$.
\item \emph{Generators} are anti-Hermitian with $S\bar AS=A$.
\item $L(A):=[A,\mathcal N]$, so $(L(A))_{xy}=(n_y-n_x)A_{xy}$. This is the Nambu form of the Majorana map $L(A)=[A,D]$ of \S\ref{subsec:thmB-star}; the translation is \S\ref{subsec:thmB-transl}.
\item The dictionary (\S\ref{subsec:thmB-dict}) translates everything to Majorana blocks. It preserves the operator norms of type parts and multiplies Frobenius norms by absolute factors (the constant $C_D$ of \eqref{eq:NF4}).
\end{itemize}

\paragraph{Nambu bands.} These are the particle bands $b$ of \S\ref{subsec:thmB-bands}, with $\iota(b)\in\{1,\dots,J\}$, cold band $\iota=J$ and $P_b=2^{-\iota(b)}\in[1/(2m),\frac12]$, together with \emph{one} hole band $h$, with $\iota(h):=0$ and $P_h:=1$. $\Pi_\beta$ projects onto the band $\beta$, and $b_x$ denotes the particle band of the underlying mode of $x$. For a particle $x$ in band $b$: $p_x\le P_b$, and $p_x>P_b/2$ unless $b$ is cold.

\paragraph{Band-scale weight.} $\Lambda:=\sum_\beta P_\beta\Pi_\beta$. Then $\frac12\Lambda\preceq\Omega_{1/m}\preceq3\Lambda$: on non-cold particles $P_b/2<p+1/m\le P_b+2P_c\le3P_b$; on cold particles $P_c\le1/m\le\Omega\le P_c+1/m\le3P_c$; on holes $\frac12\le\Omega\le1+1/m$.

\paragraph{Warm lump.} $U_\beta:=\bigoplus_{\gamma:\,\iota(\gamma)\le\iota(\beta)}\mathrm{Ran}\,\Pi_\gamma$: the band $\beta$ together with everything at least as warm. In particular $U_h=\mathrm{Ran}\,\Pi_h$.

\paragraph{Far set $\mathcal F$.} A pair $(x,y)$ is in $\mathcal F$ if $x,y$ are both particles or both holes and $|\iota(b_x)-\iota(b_y)|\ge2$, or if one is a particle and the other a hole and $P_{b_x}+P_{b_y}\le5/8$. This is the type-resolved far rule of \S\ref{subsec:thmB-bands} in Nambu form. \emph{Gaps on $\mathcal F$:} $|n_y-n_x|\ge\max(P_{b_x},P_{b_y})/4$ in the same sector and $\ge3/8$ in the mixed sector.

\paragraph{The inverse.} For $B$ Hermitian and supported on $\mathcal F$, $A:=L^{-1}B$ has $A_{xy}=B_{xy}/(n_y-n_x)$. It is anti-Hermitian, and Nambu-symmetric when $B$ is, because $\mathcal F$ is invariant under $x\mapsto\bar x$.

\paragraph{Schur multipliers.}
A \emph{Schur multiplier} with symbol $\varpi$ (a matrix indexed like $X$) is the entrywise product $X\mapsto\varpi\circ X$, $(\varpi\circ X)_{xy}=\varpi_{xy}X_{xy}$; its \emph{Schur norm} is its norm as a map on matrices with the operator norm, $\sup_{\|X\|\le1}\|\varpi\circ X\|$. The inverse $L^{-1}$ is the Schur multiplier with symbol $1/(n_y-n_x)$ on $\mathcal F$. The generator bounds below combine Schur-norm bounds on pieces of this symbol, by the three facts below, with one step that is not a Schur-norm bound: the $n=0$ term of (G1), a column restriction of a row lump (\S\ref{subsec:thmB-nolog}).
\begin{enumerate}
\item[(Sch1)] A rank-one symbol $f(x)g(y)$ has Schur norm $\le\sup|f|\cdot\sup|g|$, since $\varpi\circ X=\mathrm{diag}(f)\,X\,\mathrm{diag}(g)$.
\item[(Sch2)] If $n_y-n_x\ge g>0$ on a block, the symbol $1/(n_y-n_x)$ has Schur norm $\le1/g$ there: $\varpi\circ X=\int_0^\infty\mathrm{diag}(e^{tn_x})\,X\,\mathrm{diag}(e^{-tn_y})\,dt$, a positive combination of rank-one symbols whose integrand has norm $\le e^{t\max n_x-t\min n_y}\|X\|\le e^{-tg}\|X\|$. The case $n_y-n_x\le-g$ is symmetric.
\item[(Sch3)] A Schur multiplier supported on a block partial permutation (one block per block-row and per block-column, e.g.\ the $d$-th upper block diagonal $\{(b,b-d)\}$) has norm equal to the maximum over its blocks: after a permutation of rows and of columns the masked matrix is a direct sum of its blocks, the operator norm of a direct sum is the largest norm of a summand, and each block of $X$ has norm at most $\|X\|$.
\end{enumerate}

\paragraph{Young's inequality.}
For a kernel $k$ on $\{d\ge0\}$ and a sequence $u$ indexed by bands, $(k*u)_\beta:=\sum_dk(d)\,u_{\beta+d}$ satisfies $\|k*u\|_{\ell^2}\le\sum_d|k(d)|\,\|u_{\cdot+d}\|_{\ell^2}\le\|k\|_{\ell^1}\|u\|_{\ell^2}$, by Minkowski's inequality for the sum over $d$. It bounds the sums over colder bands in Lemmas~\ref{lem:G}--\ref{lem:M}; the kernels are listed in \S\ref{subsec:thmB-TM}.

\subsubsection{The norms}\label{subsec:thmB-norms}

\paragraph{Two-sided weighting.} $\tilde X:=\Lambda^{-1/2}X\Lambda^{-1/2}$ and $\|X\|_{\mathrm{rel}}:=\|\Omega_{1/m}^{-1/2}X\Omega_{1/m}^{-1/2}\|$; by $\frac12\Lambda\preceq\Omega_{1/m}\preceq3\Lambda$ the two agree within a factor $3$.

\paragraph{Row lumps (errors; Hermitian $X$).}
\[
r_\beta(X):=\|\Pi_\beta\tilde X\Pi_{U_\beta}\|_\op,\qquad\|X\|_\triangle:=\Big(\sum_\beta r_\beta(X)^2\Big)^{1/2},
\]
an $\ell^2$-sum over the $J+1$ bands with the operator norm inside each lump. For a non-Hermitian $\Psi$ we also use $c_\beta(\Psi):=\|\Pi_{U_\beta}\tilde\Psi\Pi_\beta\|$ ($=r_\beta(\Psi^*)$).

The Sylvester inverse divides an error entry between a cold mode $x$ and a warmer mode $y$ by $n_y-n_x\approx n_y$: the rotation angle of a cold mode into a warmer band is the error measured in units of the warmer occupation. A norm on errors in which $L^{-1}$ is bounded with constants independent of $J$ should therefore weigh each band against everything warmer than it, in units of band scales, which is what the row lump $\Pi_\beta\tilde X\Pi_{U_\beta}$ does. Inside a lump the operator norm is used, because the noise is a random matrix and its operator norm is smaller than its Frobenius norm by the square root of the dimension. Across bands the $\ell^2$-sum is used, because the relative noise levels of the lumps are geometric in the band index (Lemma~\ref{lem:Htri}), so their $\ell^2$-sum is an absolute multiple of the largest.

\paragraph{Frobenius lumps, the bookkeeping norm (Hermitian $Y$).}
\[
y_\beta:=\sqrt{P_\beta}\,\|\Pi_\beta\tilde Y\Pi_{U_\beta}\|_F,\qquad\|Y\|_{\mathrm M}:=\Big(\sum_\beta y_\beta^2\Big)^{1/2}.
\]
(The subscript is a name; it is not the far-free matrix $M$ of \S\ref{subsec:thmB-transl}.)

\paragraph{Generators (far-supported $A$).} $\hat A:=\Lambda^{-1/2}A\Lambda^{1/2}$; the generator norm is $\|A\|_\triangleleft:=\|L(A)\|_\triangle$; and $\|A\|_g:=\big(\sum_\beta\|\Pi_\beta\hat A\|_\op^2\big)^{1/2}$, over full row bands.

\paragraph{Norm properties.} On Hermitian matrices $\|\cdot\|_\triangle$ is a norm: if every $r_\beta$ vanishes, every block $(\beta,\gamma)$ with $\gamma$ at least as warm as $\beta$ vanishes, and the remaining blocks are adjoints of these. On non-Hermitian matrices it is only a seminorm, and it is used only on Hermitian ones ($[A,X]$, $\mathcal R(A)$, $E$ and $L(A)$ are all Hermitian). $\|\cdot\|_{\mathrm M}$ is likewise a norm on Hermitian matrices, and $\|\cdot\|_\triangleleft$ is a norm on far-supported $A$, because $L$ is injective there.

\begin{enumerate}
\item[(NF1)] For Hermitian $X$, $\|\tilde X\|\le2\|X\|_\triangle$.

\emph{Proof.} Write $\tilde X=\tilde X_\le+(\tilde X_>)^*$, where $\tilde X_\le$ consists of the blocks $(\beta,\gamma)$ with $\iota(\gamma)\le\iota(\beta)$ (the row lumps) and $\tilde X_>$ of the strictly warmer part of the row lumps. A matrix partitioned into row bands has $\|X\|^2=\|X^*X\|=\|\sum_\beta X^*\Pi_\beta X\|\le\sum_\beta\|\Pi_\beta X\|^2$, so $\|\tilde X_\le\|$ and $\|\tilde X_>\|$ are both $\le\|X\|_\triangle$.

\item[(NF2)] For Hermitian $Y$, $\|Y\|_{\mathrm M}^2\le\sum_{x,y}|Y_{xy}|^2/\max(\Lambda_x,\Lambda_y)\le2\|Y\|_{\mathrm M}^2$.

\emph{Proof.} $|Y_{xy}|^2/\max(\Lambda_x,\Lambda_y)=|\tilde Y_{xy}|^2\min(\Lambda_x,\Lambda_y)$, and $\min=P_\beta$ for $x\in\beta$, $y\in U_\beta$. The pairs with $y$ strictly colder are the transposes of row-lump pairs of $y$, so the full sum counts each strictly off-lump pair once more.

So $\|\cdot\|_{\mathrm M}$ is, up to $\sqrt2$, the weighted Frobenius norm with weight $1/\max(\Lambda_x,\Lambda_y)$. It has the lump-wise product form $\|Y\|_{\mathrm M}^2=\sum_\beta\|\Pi_\beta Y\Pi_{U_\beta}\Lambda_{U_\beta}^{-1/2}\|_F^2$: a row band against its warm lump, column-weighted, with Hermitian completion understood.

\item[(NF3)] \emph{Masks.} Restricting a Hermitian $Y$ to $\mathcal F$ or to its complement does not increase $\|\cdot\|_{\mathrm M}$, since the norm is entrywise. For $\|\cdot\|_\triangle$ on a particle row lump, $\mathcal F\cap(\beta\times U_\beta)$ and its complement are column subsets (particle columns in bands with $\iota\le\iota(\beta)-2$, and hole columns $\bar l$ with $P_\beta+P_{b_l}\le5/8$, a condition on the column only once $\beta$ is fixed), so $r_\beta$ does not increase. On the hole lump the far mask is ``all minus the block-tridiagonal mask'' in the bands of the underlying modes, of Schur norm $\le1+3=4$. Hence $\|X_{\mathcal F}\|_\triangle$ and $\|X_{\mathcal F^c}\|_\triangle$ are $\le4\|X\|_\triangle$. The cold-group mask of \S\ref{subsec:thmB-bands} is a row deletion on particle lumps (free) and a union of two rectangles on the hole lump (Schur norm $\le3$), so the bound $4$ holds for it too.

\item[(NF4)] \emph{Comparison with the Majorana norms, for Nambu-symmetric $Y$.} With $C_D:=8$,
\begin{equation}\tag{NF4}\label{eq:NF4}
\|Y_{\mathrm{far}}\|_{w'}\le C_D\|Y\|_{\mathrm M},\qquad\|Y_{\mathrm{near}}\|_\chi\le C_D\|Y\|_{\mathrm M},\qquad\|Y\|_{\mathrm M}\le C_D\|Y\|_\star,
\end{equation}
where the left and right sides are computed on the image under the dictionary, with the norms of \S\ref{subsec:thmB-star}.

\emph{Derivation of $C_D\le8$.} Under the dictionary a Majorana $2\times2$ hopping block has $\|Y^+_{kl}\|_F^2=8|Y_{kl}|^2$, with $Y_{kl}$ the Nambu particle--particle entry, and a pairing block has $\|Y^-_{kl}\|_F^2=8|Y_{k\bar l}|^2$. The $w'$ and $\chi$ weights are $\le2/\max(P)$ on adjacent hopping blocks and $\le4$ on near pairing blocks (bands $1$--$2$); elsewhere they equal the weights $1/\max(\Lambda)$ of $\|\cdot\|_{\mathrm M}$, and holes have $\Lambda=1$, which matches the pairing weight $1$. With the factor $2$ of (NF2) this gives $\|Y_{\mathrm{far}}\|_{w'},\|Y_{\mathrm{near}}\|_\chi\le8\|Y\|_{\mathrm M}$. The third inequality holds because $\|Y\|_{\mathrm M}\le\|Y\|_\star$. The hole--hole entries of $\|\cdot\|_{\mathrm M}$ carry the weight $1$ against the particle--particle weight $1/\max(P)\ge2$, so the equivalence uses Nambu symmetry: the hole--hole block is a copy of the particle--particle block. Every matrix to which \eqref{eq:NF4} is applied is Nambu-symmetric: $E$, $\mathcal R(A)$, and their $\mathcal F$- and near-restrictions (\S\ref{subsec:thmB-class}).
\end{enumerate}

\subsubsection{The Nambu class}\label{subsec:thmB-class}

The hole--hole bound of Lemma~\ref{lem:G} and \eqref{eq:NF4} hold on the \emph{Nambu class}: Hermitian $Y$ with $S\bar YS=-Y$, and far-supported anti-Hermitian $A$ with $S\bar AS=A$, for which $L(A)$ is in the class. Off this class the hole--hole symbol $\Lambda_{\bar l}/(p_k-p_l)\asymp1/P_{\max}$ is unbounded, up to $m$. The class contains every source in the chain:
\begin{enumerate}
\item[(i)] $E$: $S\bar ES=-E$, because $\hat{\mathcal N}$ is Nambu-symmetric (\S\ref{subsec:F-shots}).
\item[(ii)] The conjugated matrix and the series: if $S\bar AS=A$, then $e^{A}$ is Nambu-preserving and $M_N:=e^{-A}(\mathcal N+E)e^{A}$ satisfies $S\bar M_NS=I-M_N$; hence $\mathcal R(A)=M_N-\mathcal N-E+L(A)$ and $L(A)=[A,\mathcal N]$ are both in the class $S\bar YS=-Y$.
\item[(iii)] Masks: $\mathcal F$ is invariant under $x\mapsto\bar x$; the near mask and the cold-group mask are defined through the underlying modes and the type, so they are invariant too.
\item[(iv)] Lipschitz step: the differences $\mathcal R(A)-\mathcal R(A')$ and $L(\delta A)$, $\delta A:=A-A'$, are in the class.
\item[(v)] The ball: Lemmas~\ref{lem:G}, \ref{lem:T} and~\ref{lem:M} are applied to $B=L(A)$ with $A$ in the ball $\mathfrak B$ of Lemma~\ref{lem:L1}, which is defined inside the Nambu class; the fixed-point map preserves the class.
\end{enumerate}
In both classes the off-diagonal Nambu block is antisymmetric, so the entries $(t_k,\bar t_k)$ vanish and whether the mode-diagonal particle--hole pair is counted in $\mathcal F$ is immaterial; this matches the fact that the diagonal $2\times2$ blocks of an antisymmetric Majorana matrix are multiples of $J$.

\subsubsection{The Sylvester inverse in the lump norms}\label{subsec:thmB-G}

Throughout \S\S\ref{subsec:thmB-G}--\ref{subsec:thmB-TM}, $B$ is Hermitian, Nambu-symmetric and supported on $\mathcal F$, and $A:=L^{-1}B$, so that $\|A\|_\triangleleft=\|B\|_\triangle$; $s_\beta:=r_\beta(B)$ and $\|s\|:=(\sum_\beta s_\beta^2)^{1/2}=\|B\|_\triangle$; and $\hat A=\Lambda^{-1/2}A\Lambda^{1/2}=\bar\varpi\circ\tilde B$, with the band-scale symbol $\bar\varpi_{xy}:=P_{\beta(y)}/(n_y-n_x)$ on $\mathcal F$. For two bands with $\iota(\gamma)=\iota(x)+d$ we say that $\gamma$ is $d$ colder than $x$.

\begin{lemma}[Generator bounds]\label{lem:G}
\begin{enumerate}
\item[(G1)] For a particle band $\beta$, $\|\Pi_\beta\hat A\Pi_{U_\beta}\|\le9s_\beta$. For the hole band, $\|\Pi_h\hat A\Pi_h\|\le14\max_\gamma s_\gamma$.
\item[(G1$'$)] If $\gamma$ is a particle band $d\ge1$ colder than the band $x$ ($x=h$ allowed), then $\|\Pi_x\hat A\Pi_\gamma\|\le4\cdot2^{-d}s_\gamma$.
\item[(G1$''$)] $\|\hat A\|\le\|A\|_g\le22\|s\|=22\|A\|_\triangleleft$.
\end{enumerate}
\end{lemma}

\begin{proof}
\emph{(G1), particle row band $\beta=b$.}
\begin{itemize}
\item \emph{Particle columns in a band $b'$ that is $d\ge2$ warmer.} $\bar\varpi=(P_{b'}/p_y)\sum_{n\ge0}(p_x/p_y)^n$, with $p_y>P_{b'}/2$ ($b'$ is not cold) and $p_x/p_y\le2^{1-d}\le\frac12$. The $n=0$ term is the right multiplication of the far particle-column restriction of the row lump by $\mathrm{diag}(P_{b'}/p_y)\in[1,2)$; its norm is $\le2s_b$. This is where the triangular truncation $\operatorname{Tri}$ of \S\ref{subsec:thmB-nolog} is absorbed: it costs the row lump, not a Schur norm. The $n\ge1$ terms are rank-one per block, with $\sup\le2\cdot2^{(1-d)n}$ (Sch1), summing to $\le8\cdot2^{-d}$ per block; there is one block per $d$ in a fixed row band, and $\sum_{d\ge2}8\cdot2^{-d}s_b=4s_b$.
\item \emph{Hole columns (mixed far).} $\bar\varpi=1/(1-p_l-p_x)$. For a fixed row band the far hole columns satisfy $p_x+p_l\le5/8$, so (Sch2) gives $\le8/3$.
\item \emph{The adjacent band.} The adjacent band $b-1$ and $b$ itself are inside $U_b$, and $\hat A$ vanishes there because $A$ is far: the $d=1$ pairs do not occur in the Neumann series.
\item \emph{Total:} $\le(2+4+8/3)s_b\le9s_b$.
\end{itemize}

\emph{(G1), hole row, hole columns.} $\Lambda=1$ on holes and $n_{\bar l}-n_{\bar k}=p_k-p_l$. With Nambu symmetry, $B_{\bar k\bar l}=-\overline{B_{kl}}$ and $B_{kl}=\sqrt{P_bP_{b'}}\,\tilde B_{kl}$, so
\[
\hat A_{\bar k\bar l}=-\overline{\tilde B_{kl}}\cdot\sqrt{P_bP_{b'}}/(p_k-p_l).
\]
So the hole--hole block of $\hat A$ is a Schur multiple of the conjugated particle--particle block of $\tilde B$. Per block $(b,b')$, with $b'$ $d$ warmer, the symbol is a constant times $1/(p_k-p_l)$ of one sign, with Schur norm $\le\sqrt{P_bP_{b'}}\cdot4/P_{b'}=4\cdot2^{-d/2}$. Each particle--particle block $\tilde B_{b,b'}$ with $b'$ warmer lies in the row lump of $b$, so $\|\tilde B_{bb'}\|\le s_{\text{colder of }b,b'}\le\max s$; its adjoint covers the lower block. Two block diagonals per $d$ (Sch3) give $\le8\sum_{d\ge2}2^{-d/2}\max s=\frac8{2-\sqrt2}\max s=13.66\max s\le14\max s$.

\emph{(G1$'$), $x$ a particle band} ($d\ge2$; for $d=1$ the pair is near and $\hat A$ vanishes). Here $\bar\varpi=-(P_\gamma/p_x)\sum_n(p_y/p_x)^n$, with $p_x>P_x/2$, $p_y\le P_\gamma$ and ratio $\le2^{1-d}$; by (Sch1) per block this is $\le\sum_{n\ge0}2^{(1-d)(n+1)}\le4\cdot2^{-d}$. And $\|\tilde B_{x,\gamma}\|=\|\tilde B_{\gamma,x}\|\le s_\gamma$, because $x\in U_\gamma$.

\emph{(G1$'$), $x=h$} (hole rows against a colder particle band $\gamma$, $d=\iota(\gamma)$). $\bar\varpi=-P_\gamma/(1-p_k-p_l)$. For fixed $\gamma$ the far hole rows satisfy $P_{b_k}+P_\gamma\le5/8$, which is a rectangle, so the Schur norm is $\le(8/3)P_\gamma\le4\cdot2^{-d}$; and $\|\tilde B_{h,\gamma}\|\le s_\gamma$. The case $d=1$ (band-$1$ particles against holes of bands $\ge3$) is far and is covered.

\emph{(G1$''$).} Particle row band: $\|\Pi_\beta\hat A\|\le9s_\beta+\sum_{d\ge2}4\cdot2^{-d}s_{\beta+d}$; the $\ell^2$-sum over $\beta$, by Minkowski and Young (kernel mass $\sum_{d\ge2}4\cdot2^{-d}=2$), is $\le11\|s\|$. Hole row: $\le14\max s+\sum_\gamma4\cdot2^{-\iota(\gamma)}s_\gamma\le18\|s\|$. So $\|A\|_g\le(11^2+18^2)^{1/2}\|s\|\le21.1\|s\|\le22\|s\|$. Finally $\|\hat A\|\le\|A\|_g$, by the row-band inequality in the proof of (NF1).
\end{proof}

\subsubsection{Commutators with absolute constants}\label{subsec:thmB-TM}

\begin{lemma}[Commutators in the row-lump norm]\label{lem:T}
For $A$ as in \S\ref{subsec:thmB-G} and every Hermitian $X$,
\[
\|[A,X]\|_\triangle\le\kappa_\triangle\,\|L(A)\|_\triangle\,\|X\|_\triangle,\qquad\kappa_\triangle:=2\cdot22+22+8/\sqrt3\le70.62 .
\]
\end{lemma}

\begin{proof}
$\Lambda^{-1/2}[A,X]\Lambda^{-1/2}=\hat A\tilde X-\tilde X(\Lambda^{1/2}A\Lambda^{-1/2})=\Psi+\Psi^*$ with $\Psi:=\hat A\tilde X$, since $\Lambda^{1/2}A\Lambda^{-1/2}=-\hat A^*$ for anti-Hermitian $A$. So $r_\beta([A,X])\le\|\Pi_\beta\Psi\Pi_{U_\beta}\|+\|\Pi_{U_\beta}\Psi\Pi_\beta\|$. Write $x_\beta:=r_\beta(X)$.
\begin{itemize}
\item $\|\Pi_\beta\Psi\Pi_{U_\beta}\|\le\|\Pi_\beta\hat A\|\cdot\|\tilde X\|\le2\|\Pi_\beta\hat A\|\,\|X\|_\triangle$ by (NF1); the $\ell^2$-sum over $\beta$ is $\le2\|A\|_g\|X\|_\triangle\le44\|s\|\,\|X\|_\triangle$.
\item $\Pi_{U_\beta}\Psi\Pi_\beta=\Pi_{U_\beta}\hat A\Pi_{U_\beta}\cdot\Pi_{U_\beta}\tilde X\Pi_\beta+\sum_{\gamma\text{ colder than }\beta}\Pi_{U_\beta}\hat A\Pi_\gamma\cdot\Pi_\gamma\tilde X\Pi_\beta$. The first term is $\le\|\hat A\|\cdot\|\Pi_{U_\beta}\tilde X\Pi_\beta\|=\|\hat A\|x_\beta$ ($X$ Hermitian), which is $\le22\|s\|x_\beta$. In the second term, $\Pi_\gamma\tilde X\Pi_\beta$ is a sub-block of the row lump of $\gamma$ ($\beta\in U_\gamma$), so it is $\le x_\gamma$; by (G1$'$), and since every row band of $U_\beta$ (the hole band included) is strictly warmer than $\gamma$, $\|\Pi_{U_\beta}\hat A\Pi_\gamma\|\le\big(\sum_{\iota(x)\le\iota(\beta)}16\cdot4^{-(\iota(\gamma)-\iota(x))}\big)^{1/2}s_\gamma\le(8/\sqrt3)2^{-d}s_\gamma=4.62\cdot2^{-d}s_\gamma$, $d:=\iota(\gamma)-\iota(\beta)$. The $\ell^2$-sum over $\beta$, with Young, is $\le(22+4.62)\|s\|\,\|X\|_\triangle$.
\end{itemize}
Sum: $\kappa_\triangle\le44+22+4.62=70.62$.
\end{proof}

\begin{lemma}[Commutators in the Frobenius-lump norm]\label{lem:M}
For $A$ as in \S\ref{subsec:thmB-G} and every Hermitian $Y$,
\[
\|[A,Y]\|_{\mathrm M}\le\kappa_{\mathrm M}\,\|L(A)\|_\triangle\,\|Y\|_{\mathrm M},\qquad\kappa_{\mathrm M}:=66.5 .
\]
\end{lemma}

\begin{proof}
Use the same split with $\Psi:=\hat A\tilde Y$, so $y_\beta([A,Y])\le\sqrt{P_\beta}\big(\|\Pi_\beta\Psi\Pi_{U_\beta}\|_F+\|\Pi_{U_\beta}\Psi\Pi_\beta\|_F\big)$; throughout, $\|PQ\|_F\le\|P\|\,\|Q\|_F$.

(a) $\sqrt{P_\beta}\|\Pi_\beta\Psi\Pi_{U_\beta}\|_F\le\|\Pi_\beta\hat A\Pi_{U_\beta}\|\cdot\sqrt{P_\beta}\|\Pi_{U_\beta}\tilde Y\Pi_{U_\beta}\|_F+\sum_{\gamma\text{ colder}}\|\Pi_\beta\hat A\Pi_\gamma\|\cdot\sqrt{P_\beta}\|\Pi_\gamma\tilde Y\Pi_{U_\beta}\|_F$.
\begin{itemize}
\item \emph{The warm square.} On $U_\beta\times U_\beta$ every entry has $\min(\Lambda_x,\Lambda_y)\ge P_\beta$, so $\sqrt{P_\beta}\|\Pi_{U_\beta}\tilde Y\Pi_{U_\beta}\|_F\le\sqrt2\|Y\|_{\mathrm M}$ by (NF2): the whole $\mathrm M$-norm, once per $\beta$. It is the generator side that is $\ell^2$-summed: by (G1), $\sum_\beta\|\Pi_\beta\hat A\Pi_{U_\beta}\|^2\le81\|s\|^2+196\max s^2$. So the $\ell^2$-sum over $\beta$ is $\le\sqrt2\cdot\sqrt{277}\,\|s\|\,\|Y\|_{\mathrm M}\le23.54\|s\|\,\|Y\|_{\mathrm M}\le23.6\|s\|\,\|Y\|_{\mathrm M}$.
\item \emph{A colder intermediate band $\gamma$, $d$ colder.} $\Pi_\gamma\tilde Y\Pi_{U_\beta}$ lies in the row lump of $\gamma$, and $\sqrt{P_\beta}=2^{d/2}\sqrt{P_\gamma}$, so the term is $\le4\cdot2^{-d}s_\gamma\cdot2^{d/2}y_\gamma$. The $\ell^2$-sum with Young (kernel $4\cdot2^{-d/2}$, $d\ge1$, mass $4(1+\sqrt2)\le9.66$) is $\le9.7\|s\|\,\|Y\|_{\mathrm M}$.
\end{itemize}
(b) $\sqrt{P_\beta}\|\Pi_{U_\beta}\Psi\Pi_\beta\|_F\le\|\hat A\|\,y_\beta+\sum_{\gamma\text{ colder}}4.62\cdot2^{-d}s_\gamma\cdot2^{d/2}y_\gamma$; the $\ell^2$-sum is $\le(22+11.2)\|s\|\,\|Y\|_{\mathrm M}$ (kernel $4.62\cdot2^{-d/2}$, mass $4.62(1+\sqrt2)\le11.16$).

Sum: $\le23.6+9.7+33.2=66.5$.
\end{proof}

Below we use the rounded values $\kappa_\triangle\le72$ and $\kappa_{\mathrm M}\le67$.

\paragraph{The only sums over bands.} Besides $\ell^2$-sums of lump norms, every sum over the band index $\iota\in\{0,\dots,J\}$ in Lemmas~\ref{lem:G}--\ref{lem:M} is a convolution with a summable kernel, bounded by Young's inequality (\S\ref{subsec:thmB-far}) with $u_\gamma=s_\gamma x_\gamma$ (or $s_\gamma y_\gamma$) and $\|u\|_2\le\|s\|_\infty\|x\|_2\le\|s\|_2\|x\|_2$. No step sums $J$ equal terms:
\begin{center}\small
\begin{tabular}{@{}lllll@{}}
\toprule
step & kernel $k(d)$ & range & $\ell^1$ mass & constant used\\
\midrule
(G1$''$), particle rows & $4\cdot2^{-d}$ & $d\ge2$ & $2$ & $2$\\
Lemma~\ref{lem:T}, $\Pi_{U_\beta}\hat A\Pi_\gamma$ (rows of $U_\beta$ stacked) & $4.62\cdot2^{-d}$ & $d\ge1$ & $4.62$ & $4.62$\\
Lemma~\ref{lem:M}~(a), $\Pi_\beta\hat A\Pi_\gamma$ against $2^{d/2}$ & $4\cdot2^{-d/2}$ & $d\ge1$ & $9.66$ & $9.7$\\
Lemma~\ref{lem:M}~(b) & $4.62\cdot2^{-d/2}$ & $d\ge1$ & $11.16$ & $11.2$\\
\bottomrule
\end{tabular}
\end{center}
The decay $2^{-d}$ of (G1$'$) is essential: in Lemma~\ref{lem:M}~(a) the colder intermediate band costs the weight ratio $2^{d/2}$ and gains $2^{-d}$, a net $2^{-d/2}$; with only $2^{-d/2}$ from (G1$'$) the kernel would be constant, of $\ell^1$-mass of order $J=\lceil\log_2m\rceil$, and through $e_0$ the count would acquire a power of $\log m$.

\subsubsection{The noise lemma}\label{subsec:thmB-Htri}

\paragraph{Nets for a sesquilinear form.}
Lemmas~\ref{lem:Htri} and~\ref{lem:H0prime} use Bernstein's inequality \eqref{eq:bg-bernstein} and the $\frac14$-nets of \S\ref{subsec:bg-estimation} ($9^d$ points on the unit sphere of $\R^d$; a complex subspace of dimension $n$ has $d=2n$) for a sesquilinear form on two different subspaces, not for a Hermitian quadratic form, so the step from the net to the supremum is slightly different: if $S$ is the supremum of $|x^\dagger My|$ over unit $x,y$ in the two subspaces, replacing $x$ and then $y$ by net points within $\frac14$ gives $S\le\max_{\mathrm{net}}+\frac S4+\frac S4$, so $S\le2\max_{\mathrm{net}}$.

\begin{lemma}[Row-lump noise]\label{lem:Htri}
Let the Stage-1 frame $W$ satisfy $G(1/m,3)$ (Corollary~\ref{cor:Fprime}~(G)), and let $m\ge16$. Let $E:=\hat{\mathcal N}-\mathcal N$ be the Nambu Stage-1 error, from $N_2$ Horvitz--Thompson shots $Z_t$ and $N_3$ occupation shots $Z_{\mathrm{occ},t}$ (\S\ref{subsec:F-shots}) on fresh copies. For $\tau\in(0,1]$ suppose
\[
N_2\ge K_\triangle m^3/\tau^2,\quad N_3\ge K'_\triangle m^2/\tau^2,
\]
with $K_\triangle:=960K_5=960\cdot2.62\cdot10^5\le2.6\cdot10^8$ and $K'_\triangle:=960K_6=960\cdot1.44\cdot10^4\le1.4\cdot10^7$.
Then with probability $\ge1-e^{-m}$, for every Nambu band $\beta$ (the hole band included),
\[
r_\beta(E)\le\tau\,(mP_\beta)^{-1/2}.
\]
Consequently $\|E\|_\triangle\le\tau\big(\sum_\beta1/(mP_\beta)\big)^{1/2}\le\sqrt5\,\tau$ and $\|E\|_{\mathrm{rel}}\le3\|\Lambda^{-1/2}E\Lambda^{-1/2}\|\le6\sqrt5\,\tau$. Here $\sum_\beta1/(mP_\beta)=(1+\sum_{j=1}^J2^j)/m\le(1+2^{J+1})/m\le1/m+4\le5$, using $2^J\le2m$.
\end{lemma}

\begin{proof}
Fix $\beta$, a unit $x\in\mathrm{Ran}\,\Pi_\beta$ and a unit $y\in\mathrm{Ran}\,\Pi_{U_\beta}$; these are deterministic subspaces (the true bands and their warm lump). Put $x':=\Lambda^{-1/2}x$ and $y':=\Lambda^{-1/2}y$. The claim is $|x'^\dagger Ey'|\le\tau(mP_\beta)^{-1/2}$ for all such $x,y$.
\begin{itemize}
\item \emph{Weights.} $\Lambda\ge P_\beta$ on $U_\beta$, so $\|x'\|^2,\|y'\|^2\le1/P_\beta$. $\Omega_{1/m}\preceq3\Lambda$ (both diagonal in the true basis; cold-band and hole cases included), so $\langle x'|\Omega_{1/m}|x'\rangle,\langle y'|\Omega_{1/m}|y'\rangle\le3$.
\item \emph{Polarisation.} $x'^\dagger My'=\frac14\sum_ki^{-k}w_k^\dagger Mw_k$ with $w_k:=x'+i^ky'$, so $|x'^\dagger My'|\le\max_k|w_k^\dagger Mw_k|$; $\|w_k\|^2\le4/P_\beta$ and $\langle w_k|\Omega_{1/m}|w_k\rangle\le12$. By $G(1/m,3)$: $Q_w\le36$, $B_w\le36$ and $N_w\le12$.
\item \emph{Variances, by \eqref{eq:R0}.}
\begin{align*}
\mathrm{Var}(w^\dagger Zw)&\le640m\cdot\frac4{P_\beta}\cdot36+1024m\cdot\frac4{P_\beta}\cdot36+3\cdot84^2+\frac{12\cdot16}{P_\beta^2}\\
&\le(92\,160+147\,456+21\,168+384)\frac m{P_\beta}=261\,168\,\frac m{P_\beta}\le K_5\frac m{P_\beta},
\end{align*}
with $K_5:=2.62\cdot10^5$ (using $1/P_\beta\le2m$ and $m/P_\beta\ge1$); the range is $\le4m\|w\|^2\le16m/P_\beta$. For the occupation shot, $\mathrm{Var}(w^\dagger Z_{\mathrm{occ}}w)\le10(\|w\|^2Q_w+B_w^2)\le10(144/P_\beta+1296)\le K_6/P_\beta$ with $K_6:=1.44\cdot10^4$; the range is $\le4/P_\beta$.
\item \emph{Bernstein} \eqref{eq:bg-bernstein} for each average, at the deviation $t:=\tau(mP_\beta)^{-1/2}/4$. The range conditions $\mathrm{R}\cdot t/3\le\sigma^2$ hold: for the Horvitz--Thompson average $(16m/P_\beta)\,t/3\le(4\sqrt2/3)m/P_\beta\le K_5m/P_\beta$, using $mP_\beta\ge\frac12$, and the occupation average likewise. So the exponents $Nt^2/(4\sigma^2)$ are $N_2\tau^2/(64K_5m^2)\ge15m$ and $N_3\tau^2/(64K_6m)\ge15m$ under the hypotheses.
\item \emph{Nets.} Take $\frac14$-nets of the unit spheres of $\mathrm{Ran}\,\Pi_\beta$ and $\mathrm{Ran}\,\Pi_{U_\beta}$, of real dimensions $\le2m$ and $\le4m$: $\le9^{6m}\le e^{13.2m}$ pairs. For a sesquilinear form, $\sup\le2\max_{\mathrm{net}}$. On the event each net pair has $|x'^\dagger Ey'|\le2t$ (Horvitz--Thompson plus occupation), so the supremum is $\le4t=\tau(mP_\beta)^{-1/2}$.
\item \emph{Both types.} The row lump of a particle band contains, in \emph{one} subspace pair, its hole columns (pairing type) and its warmer particle columns (hopping type), so one net covers both. The hole lump ($h\times h$) is the conjugate copy of the unweighted particle--particle block, at the target $\tau/\sqrt m$.
\item \emph{Union} over $J+1$ bands, $e^{13.2m}$ pairs, $4$ polarisation vectors and $2$ estimators: $\le(J+1)\,e^{13.2m}\cdot4\cdot2\cdot2e^{-15m}=16(J+1)e^{-1.8m}\le e^{-m}$ for $m\ge16$.
\end{itemize}
$W$ is measurable with respect to the past, and the copies are fresh.
\end{proof}

\begin{remark}\label{rem:Htri}
(a) \emph{Per-band versus aggregate.} The per-band conclusion contains the band-pair profile of \S\ref{subsec:thmB-H0}: $\|E_{bb'}\|\le\sqrt{P_bP_{b'}}\,r_b\le\tau\sqrt{P_{b'}/m}$ for $b'\in U_b$, and $\tau/\sqrt m$ for pairing blocks. The aggregate $\|E\|_\triangle\le\sqrt5\tau$ alone does not: a single warm lump may carry all of it. The per-band form is also what makes the $\ell^2$-aggregate $O(\tau)$: the per-band levels $\tau(mP_\beta)^{-1/2}$ are geometric.

(b) \emph{Bernstein versus net.} The matrix Bernstein inequality \cite{Tropp12} applied to $\Lambda^{-1/2}E\Lambda^{-1/2}$ (dimension $2m$) has the same per-shot variance proxy but a dimension factor: $\|E\|_{\mathrm{rel}}\lesssim\tau\sqrt{\log m}$ at $N_2=m^3/\tau^2$, an extra $\log m$ in $N_2$; and it controls only the whole-matrix norm, not the row lumps. The net removes the logarithm, because the exponent $15m$ beats the net entropy $13.2m$, and it gives every row lump separately, at the cost of a union over $J+1$ bands, which is absorbed in $e^{-1.8m}$.

(c) The row-lump level $\tau(mP_\beta)^{-1/2}$ comes from $\mathrm{Var}\lesssim m/P_\beta$: a warm row band against its (even warmer) lump has relative entries of variance $\asymp m/(N_2P_\beta)$, not $\asymp m^2/N_2$. The cold \emph{columns}, which make $\|E\|_{\mathrm{rel}}\asymp\tau$, belong by Hermiticity to the cold row lumps, not to the warm ones.
\end{remark}

\subsubsection{The fixed point at an absolute tolerance}\label{subsec:thmB-L1}

The conjugation series in Nambu form is
\begin{equation}\label{eq:B-series}
e^{-A}(\mathcal N+E)e^{A}=\mathcal N+E-L(A)+\mathcal R(A),\qquad\mathcal R(A)=-[A,E]+\sum_{n\ge2}\frac{(-1)^n}{n!}\ad_A^{n-1}\big(L(A)+[A,E]\big),
\end{equation}
from $e^{-A}Xe^{A}=\sum_n(-1)^n\ad_A^nX/n!$ with $\ad_A\mathcal N=L(A)$. $\mathcal N$ is diagonal in the true basis and has no $\mathcal F$-entries, and $L(A)$ is $\mathcal F$-supported for $\mathcal F$-supported $A$. So the far part vanishes iff $L(A)=(E+\mathcal R(A))_{\mathcal F}$, i.e.\ iff $A=\Phi(A)$ with
\[
\Phi(A):=L^{-1}\big((E+\mathcal R(A))_{\mathcal F}\big).
\]

\begin{lemma}[The far decomposition]\label{lem:L1}
Let $e:=\|E\|_\triangle\le e_0:=\min\{1/5760,\,1/(536\cdot64\cdot C_D^2)\}$; with $C_D=8$ this is $e_0=1/2\,195\,456\approx4.55\cdot10^{-7}$. Then:
\begin{itemize}
\item there is a unique far-supported, Nambu-symmetric, anti-Hermitian $A$ with $\|A\|_\triangleleft\le8e$ such that $e^{-A}(\mathcal N+E)e^{A}$ has no $\mathcal F$-entries;
\item $\|\mathcal R(A)\|_\triangle\le0.53e$;
\item $A_1:=L^{-1}(E_{\mathcal F})$ satisfies $\|A_1\|_\triangleleft=\|E_{\mathcal F}\|_\triangle\le4e$, and $A-A_1=L^{-1}(\mathcal R(A)_{\mathcal F})$.
\end{itemize}
\end{lemma}

\begin{proof}
Work on the ball $\mathfrak B:=\{A\text{ supported on }\mathcal F,\text{ anti-Hermitian},\ S\bar AS=A:\ \|A\|_\triangleleft\le8e\}$, which is closed in a finite-dimensional real space on which $\|\cdot\|_\triangleleft$ is a norm.

\emph{Self-map.} For $A\in\mathfrak B$ put $a:=\|A\|_\triangleleft\le8e$ and $\varphi:=\kappa_\triangle\cdot8e=576e\le1/10$. Lemma~\ref{lem:T} applied to the series gives
\begin{align*}
\|\mathcal R(A)\|_\triangle&\le\kappa_\triangle ae+\sum_{n\ge2}\frac{(\kappa_\triangle a)^{n-1}(a+\kappa_\triangle ae)}{n!}\le\varphi e+\frac{e^\varphi-1-\varphi}\varphi\,(8e+\varphi e)\\
&\le0.1e+0.0518\cdot8.1e\le0.52e\le0.53e,
\end{align*}
using $\kappa_\triangle e=\varphi/8$ and $(e^{0.1}-1.1)/0.1\le0.0518$. So $\|\Phi(A)\|_\triangleleft=\|(E+\mathcal R(A))_{\mathcal F}\|_\triangle\le4\cdot1.53e=6.12e\le8e$, by (NF3); here $L(L^{-1}(B_{\mathcal F}))=B_{\mathcal F}$, since the gaps are nonzero on $\mathcal F$.

\emph{Lipschitz.} Let $A,A'\in\mathfrak B$, $\delta A:=A-A'$ and $\delta:=\|\delta A\|_\triangleleft=\|L(\delta A)\|_\triangle$. Telescoping $\ad_A^k-\ad_{A'}^k=\sum_i\ad_A^i\,\ad_{\delta A}\,\ad_{A'}^{k-1-i}$ and using $\|L(A')\|_\triangle\le8e$, Lemma~\ref{lem:T} gives
\[
\|\mathcal R(A)-\mathcal R(A')\|_\triangle\le\delta\big[\kappa_\triangle e+(e^\varphi-1)(1+\kappa_\triangle e)\big]\le1.2\varphi\delta .
\]
So $\|\Phi(A)-\Phi(A')\|_\triangleleft\le4\cdot1.2\varphi\delta\le0.48\delta\le\frac12\delta$, using the mask factor $4$ of (NF3).

$\Phi$ preserves Nambu symmetry and anti-Hermiticity (\S\ref{subsec:thmB-class}~(ii), (v)), so $\Phi(\mathfrak B)\subset\mathfrak B$, and Banach's fixed-point theorem gives existence and uniqueness in $\mathfrak B$. The bound on $\mathcal R(A)$ is the self-map bound at the fixed point. The second condition in $e_0$ is the condition $\theta=67\cdot8e\le\theta_0=1/(64C_D^2)$ of \S\ref{subsec:thmB-star}.
\end{proof}

\subsubsection{Translation to the Majorana form}\label{subsec:thmB-transl}

The fixed point of Lemma~\ref{lem:L1} is a Nambu generator; the comparison chain of \S\ref{subsec:thmB-chain} uses a Majorana generator. The translation uses only the covariance of the dictionary (\S\ref{subsec:thmB-dict}).
\begin{itemize}
\item \emph{The group map.} An orthogonal $R\in O(2m)$ acts on Nambu matrices by $\mathcal N\mapsto\bar U_R\mathcal NU_R^T$, with $U_R=\Omega_{\mathrm M}^{-1}R\,\Omega_{\mathrm M}$ Nambu-preserving, and $\Phi_D(\bar U_R\mathcal NU_R^T)=R\,\Phi_D(\mathcal N)R^T$. Since $\bar U_R\mathcal NU_R^T=V\mathcal NV^\dagger$ with $V:=\bar U_R$, this is ordinary unitary conjugation. Take $V_t:=e^{-tA}$, with $A$ the generator of Lemma~\ref{lem:L1}. Then $U_{R_t}=e^{-t\bar A}$ and $R_t=\Omega_{\mathrm M}e^{-t\bar A}\Omega_{\mathrm M}^{-1}=e^{-tA_{\mathrm M}}$ with
\[
A_{\mathrm M}:=\Omega_{\mathrm M}\bar A\,\Omega_{\mathrm M}^{-1},
\]
which is real antisymmetric, because $R_t\in O(2m)$ for every $t$.
\item \emph{Conjugation.} $e^{-A_{\mathrm M}}(D+\Delta)e^{A_{\mathrm M}}=\Phi_D\big(e^{-A}(\mathcal N+E)e^{A}\big)=:M$, since $D+\Delta=\Phi_D(\mathcal N+E)$ (\S\ref{subsec:thmB-dict}, \S\ref{subsec:thmB-protocol}).
\item \emph{Commutators.} Differentiate the covariance at $t=0$. For every $Y$, $[A_{\mathrm M},\Phi_{D,\mathrm{lin}}(Y)]=\Phi_{D,\mathrm{lin}}([A,Y])$: the covariance $\Phi_D(VXV^\dagger)=R\,\Phi_D(X)R^T$ holds for every matrix $X$ ($\bar\Omega_{\mathrm M}\bar U_R=R\,\bar\Omega_{\mathrm M}$), and $\Phi_D$ is affine. On the class $S\bar YS=-Y$, $Y$ Hermitian, $\Phi_{D,\mathrm{lin}}(Y)$ is real antisymmetric, which is what the Majorana norms of \S\ref{subsec:thmB-star} require. Hence $L_{\mathrm M}(A_{\mathrm M}):=[A_{\mathrm M},D]=\Phi_{D,\mathrm{lin}}(L(A))$ and $[A_{\mathrm M},\Delta]=\Phi_{D,\mathrm{lin}}([A,E])$; term by term, the Majorana conjugation series is $\Phi_{D,\mathrm{lin}}(\mathcal R(A))$.
\item \emph{Support.} The map $c=\Omega_{\mathrm M}\alpha$ is mode-local, so the block $(k,l)$ of $A_{\mathrm M}$ depends only on the entries of $A$ between the modes $k$ and $l$. Likewise $\Phi_{D,\mathrm{lin}}$ sends the Nambu entries of $(k,l)$ to the Majorana block $(k,l)$: particle--particle and hole--hole entries to the hopping part, particle--hole and hole--particle entries to the pairing part. So $\mathcal F$-support corresponds exactly to far support in the type-resolved sense of \S\ref{subsec:thmB-bands}.
\end{itemize}

\paragraph{Output of the far decomposition.} On the event $\|E\|_\triangle\le e_0$, writing $A$ for $A_{\mathrm M}$ and $L$ for $L_{\mathrm M}$ from here on in Majorana statements:
\begin{enumerate}
\item[(Out1)] a real antisymmetric, far-supported Majorana generator $A$ with $\tilde\Gamma=e^{A}Me^{-A}$, where $M:=e^{-A}(D+\Delta)e^{A}$ has no far parts;
\item[(Out2)] $M-D=\Delta_{\mathrm{near}}+\mathcal R(A)_{\mathrm{near}}$; in Majorana form $M-D=\Phi_{D,\mathrm{lin}}\big(E-L(A)+\mathcal R(A)\big)$, its far part vanishes, and $L(A)$ is far-supported;
\item[(Out3)] $A_1:=L^{-1}(\Delta_{\mathrm{far}})$ and $A-A_1=L^{-1}(\mathcal R(A)_{\mathrm{far}})$, with the Majorana $L=[\cdot,D]$;
\item[(Out4)] the Nambu bounds $\|A\|_\triangleleft\le8e$, $\|\mathcal R(A)\|_\triangle\le0.53e$ and, from \S\ref{subsec:thmB-star}, $\|\mathcal R(A)\|_{\mathrm M}\le1.66\,\theta\|E\|_{\mathrm M}$. The norms $\|\cdot\|_\triangle$, $\|\cdot\|_\triangleleft$, $\|\cdot\|_{\mathrm M}$ are computed on the Nambu form, the norms $\|\cdot\|_w$, $\|\cdot\|_{w'}$, $\|\cdot\|_\chi$, $\|\cdot\|_\star$ on the Majorana form, and \eqref{eq:NF4} connects the two;
\item[(Out5)] \eqref{eq:Whyp} for $X_{\mathrm{rest}}:=(M-D)_{\mathrm{rest}}$ (Lemma~\ref{lem:Whyp-rest}).
\end{enumerate}
Lemma~\ref{lem:B1}, legs I--III to first order, and Lemma~\ref{lem:W} are unaffected: $D$ enters only through $L(A)$ and through $D+\Delta$ inside the commutators.

\subsubsection{The requirement chain}\label{subsec:thmB-req}

Every condition in \S\S\ref{subsec:thmB-far}--\ref{subsec:thmB-W} is a condition on $e=\|E\|_\triangle$ alone; none involves $J$ or $m$:
\begin{center}\small
\begin{tabular}{@{}llll@{}}
\toprule
condition & requirement & value at $e_0$ & where\\
\midrule
self-map and Lipschitz: $\varphi=576e\le1/10$ & $e\le1/5760$ & $2.6\cdot10^{-4}$ & Lemma~\ref{lem:L1}\\
\eqref{eq:Whyp}, first clause: $12\cdot49e\le\frac14$ & $e\le1/2352$ & $2.7\cdot10^{-4}$ & Lemma~\ref{lem:Whyp-rest}\\
\eqref{eq:Whyp}, second clause: $6\cdot49e\le\frac12$ & $e\le1/588$ & $1.3\cdot10^{-4}$ & Lemma~\ref{lem:Whyp-rest}\\
higher order, legs I--III: $\theta=536e\le\theta_0=1/(64C_D^2)$ & $e\le1/(34\,304\,C_D^2)$ & $=\theta_0$ (binding) & \S\ref{subsec:thmB-star}\\
\bottomrule
\end{tabular}
\end{center}
The binding condition, that of \S\ref{subsec:thmB-star}, gives the value of $e_0$ in Lemma~\ref{lem:L1}; the proof reaches $e_0$ through Lemma~\ref{lem:Htri} at $\tau_0:=e_0/\sqrt5$.

\subsection{The comparison chain}\label{subsec:thmB-chain}

Write $M-D=X_{\mathrm{rest}}+X_{\mathcal P}$ (\S\ref{subsec:thmB-bands}); $X_{\mathrm{rest}}$ has no $\mathcal P$ rows or columns. Let $\hat\Gamma:=O'^T\hat\Gamma'O'$ be the output in the true frame (the trace distance is invariant under the frame rotation). $\operatorname{Clip}$ commutes with orthogonal conjugation, so $\hat\Gamma=\operatorname{Clip}(\tilde\Gamma)=e^{A}\operatorname{Clip}(M)e^{-A}$, and with $\mathcal U$ the Gaussian unitary of $e^{A}$ (Lemma~\ref{lem:wick}),
\begin{align*}
T(\rho_D,\rho_{\hat\Gamma})&\le T(\rho_D,\mathcal U\rho_D\mathcal U^\dagger)+T(\mathcal U\rho_D\mathcal U^\dagger,\mathcal U\rho_{\operatorname{Clip}(M)}\mathcal U^\dagger)\\
&=T(\rho_D,\rho_{e^{A}De^{-A}})+T(\rho_D,\rho_{\operatorname{Clip}(M)}),\\
T(\rho_D,\rho_{\operatorname{Clip}(M)})&\le\tfrac12\|\rho_D-\rho_{D+X_{\mathrm{rest}}}\|_1+\tfrac12\|\rho_{D+X_{\mathrm{rest}}}-\rho_{\operatorname{Clip}(M)}\|_1 .
\end{align*}
Validity of every covariance in the chain: $D$ and $e^{A}De^{-A}$ have the spectrum of $D$; $D+X_{\mathrm{rest}}=D_{\mathcal P}\oplus(D_{\mathrm{rest}}+X_{\mathrm{rest}})$, and on the rest factor the display $I+(D_{\mathrm{rest}}+tX_{\mathrm{rest}})^2\succeq\frac12(I+D_{\mathrm{rest}}^2)\succ0$ of Lemma~\ref{lem:W} gives $\|D_{\mathrm{rest}}+tX_{\mathrm{rest}}\|_\op<1$ for all $t\in[0,1]$ (a consequence of \eqref{eq:Whyp}, Lemma~\ref{lem:Whyp-rest}); $\operatorname{Clip}(M)$ is valid by construction.

\paragraph{Leg I (far rotation).}
By Lemma~\ref{lem:MT} and the block form of Lemma~\ref{lem:wick} in the diagonal frame, using $1-\lambda_k\lambda_l=2(p_k+p_l)-4p_kp_l\le4\max(P_b,P_{b'})$ and $1+\lambda_k\lambda_l\le2$,
\begin{align*}
T(\rho_D,\rho_{e^{A}De^{-A}})&\le\sqrt{\mathrm{Var}_{\rho_D}(K_A)}\\
&\le\frac1{\sqrt8}\Big[\sum_{\text{far hop}}4\max(P_b,P_{b'})\|A^+_{bb'}\|_F^2+2\sum_{\text{far pair}}\|A^-_{bb'}\|_F^2\Big]^{1/2}=\frac1{\sqrt8}\|A\|_w
\end{align*}
(the norm $\|\cdot\|_w$ of \S\ref{subsec:thmB-star}). Split $A=A_1+(A-A_1)$. For the first-order part: in the diagonal frame $\|(A_1)^+_{kl}\|_F=\|\Delta^+_{kl}\|_F/|\lambda_l-\lambda_k|\le2\|\Delta^+_{kl}\|_F/\max P$, so $\sum_{\text{far hop}}4\max P\,\|(A_1)_{kl}\|_F^2\le16\sum\|\Delta^+_{kl}\|_F^2/\max P$, whose expectation is $\le16\cdot4C_1'(8m/N_2)\,m^2=O(C_1'm^3/N_2)$ by the band form of Lemma~\ref{lem:B1}; pairing: $2(4/3)^2\,\E\|\Delta^-_{\mathrm{far}}\|_F^2=O(C_1'm^3/N_2)$. The $N_3$ part: $\sum_{\text{far hop}}16s_{kl}/\max P\le\sum_k\big(32/\max(p_k,1/m)\big)\sum_ls_{kl}\le64(c_b+2)m$ ($1/\max P\le2/\max(p_k,1/m)$ for $k\in b$), and the far pairing part likewise; together $O(C_1'm/N_3)$. So $\frac1{\sqrt8}\|A_1\|_w\le\eps/12$ with probability $\ge1-1/40$ for $N_2\ge C_{\mathrm I}m^3/\eps^2$ and $N_3\ge C_{\mathrm I}m/\eps^2$ (Markov on the second moment). The higher-order part satisfies $\frac1{\sqrt8}\|A-A_1\|_w\le\eps/12$ by \S\ref{subsec:thmB-contrib}.

\paragraph{Leg II ($\chi^2$ on the rest factor).}
$\rho_{D+tX_{\mathrm{rest}}}=\rho_{D_{\mathcal P}}\otimes\rho_{D_{\mathrm{rest}}+tX_{\mathrm{rest}}}$; the derivative along the segment factorises, and $\|\cdot\|_1$ of a product with a fixed state is the $\|\cdot\|_1$ of the factor. On the rest factor, Corollary~\ref{cor:Wprime}, which applies because \eqref{eq:Whyp} holds for $X_{\mathrm{rest}}$ (Lemma~\ref{lem:Whyp-rest}) and the base $D_{\mathrm{rest}}$ has finite weights ($1-\lambda_k^2\ge2p_k>P_c\ge1/(2m)$ for non-cold modes), gives
\[
\|\rho_D-\rho_{D+X_{\mathrm{rest}}}\|_1\le\|W_DX_{\mathrm{rest}}W_D\|_F,\qquad W_D:=(I+D^2)^{-1/4}\ \text{on the rest modes},
\]
and
\[
\|W_DXW_D\|_F^2=\sum_{\mathrm{rest}}\frac{\|X_{kl}\|_F^2}{\sqrt{(1-\lambda_k^2)(1-\lambda_l^2)}}\le\sum\frac{\|X_{kl}\|_F^2}{2\sqrt{p_kp_l}}\qquad(1-\lambda^2=4pq\ge2p).
\]
First-order part $X=\Delta_{\mathrm{rest}}$: by Lemma~\ref{lem:B1}, $\E\sum_{\text{near hop, non-}\mathcal P}\|\Delta^+_{kl}\|_F^2/(2\sqrt{p_kp_l})\le C_1'(8m/N_2)\cdot\frac{13}4\cdot3m^2=O(C_1'm^3/N_2)$ (occupation ratio $<4$ and $p\ge1/(2m)$ on these pairs give $(p_k+p_l+2/m)/(2\sqrt{p_kp_l})\le13/4$; at most $3m^2$ near pairs); near pairing parts (bands $1$--$2$: weight $\le8$, noise $2C_1'(8m/N_2)$): $O(C_1'm^3/N_2)$; non-$\mathcal P$ diagonal blocks: $\sum_kC_1'\max(p_k,1/m)(8m/N_2+1/N_3)/(2p_k)=O(C_1'm^2/N_2+C_1'm/N_3)$. The $N_3$ part of the near blocks: $\sum_{\mathrm{near}}s_{kl}/(2\sqrt{p_kp_l})\le\sum_k(1/p_k)\sum_ls_{kl}\le4(c_b+2)m$ (ratio $<4$, $p>1/(2m)$): $O(C_1'm/N_3)$. So $\frac12\|W_D\Delta_{\mathrm{rest}}W_D\|_F\le\eps/12$ with probability $\ge1-1/40$ for $N_2\ge C_{\mathrm{II}}m^3/\eps^2$, $N_3\ge C_{\mathrm{II}}m/\eps^2$. Higher-order part: $\frac12\|W_D\mathcal R(A)_{\mathrm{rest}}W_D\|_F\le\frac12\|\mathcal R(A)_{\mathrm{near}}\|_\chi\le\eps/12$ by \S\ref{subsec:thmB-contrib}.

\paragraph{Leg III (trace norm on the cold group, with the clipping).}
$\operatorname{Clip}(M)=M+C_M$ with $\|C_M\|_1=\sum_i(|\mu_i|-1)_+$ over the eigenvalues $\mu_i$ of $iM$. The eigenvalues $\nu_i$ of $i(D+X_{\mathrm{rest}})$ lie in $[-1,1]$ (validity above) and $iM=i(D+X_{\mathrm{rest}})+iX_{\mathcal P}$ with both summands Hermitian, so by Lidskii's inequality (the vector of eigenvalue differences of two Hermitian matrices is weakly majorised by the eigenvalue vector of their difference \cite{Bhatia97}) $\sum_i(|\mu_i|-1)_+\le\sum_i|\mu_i-\nu_i|\le\|X_{\mathcal P}\|_1$. By \cite[Thm.~1]{BMEL25} ($\|\rho_\Gamma-\rho_{\Gamma'}\|_1\le\frac12\|\Gamma-\Gamma'\|_1$ for any two Gaussian states),
\[
\|\rho_{D+X_{\mathrm{rest}}}-\rho_{\operatorname{Clip}(M)}\|_1\le\tfrac12\|X_{\mathcal P}+C_M\|_1\le\|X_{\mathcal P}\|_1\le2\sqrt{2|\mathcal P|}\,\|X_{\mathcal P}\|_F\le2\sqrt{2m}\,\|X_{\mathcal P}\|_F
\]
($X_{\mathcal P}$ is supported on the $\mathcal P$ rows and columns, of rank $\le4|\mathcal P|$). First-order part: by Lemma~\ref{lem:B1}, summed over the $\le2m|\mathcal P|$ $\mathcal P$-touching near hopping parts and $\mathcal P$ diagonals, on which $p\le2/m$, $\E\|\Delta_{\mathcal P}\|_F^2\le2m\,|\mathcal P|\,C_1'(8m/N_2)\cdot(6/m)=O(C_1'm^2/N_2)$, and the $N_3$ part $\sum_{\mathcal P\text{-touching}}s_{kl}\le2\sum_{k\in\mathcal P}\sum_ls_{kl}\le4(c_b+2)|\mathcal P|\cdot(2/m)=O(c_b)$, times $8m$ from the trace conversion: $O(C_1'm/N_3)$. So $\frac12\cdot2\sqrt{2m}\,\|\Delta_{\mathcal P}\|_F\le\eps/12$ with probability $\ge1-1/40$ for $N_2\ge C_{\mathrm{III}}m^3/\eps^2$ and $N_3\ge C_{\mathrm{III}}m/\eps^2$. Higher-order part: on a $\mathcal P$-touching near block $\sqrt{P_bP_{b'}}\le2P_c\le2/m$, so
\[
\|Y_{\mathcal P}\|_F\le\max_{\mathcal P\text{-touching}}(P_bP_{b'})^{1/4}\|Y_{\mathcal P}\|_\chi\le\sqrt{2/m}\,\|Y_{\mathrm{near}}\|_\chi,
\]
and $\sqrt{2m}\,\|\mathcal R(A)_{\mathcal P}\|_F\le2\|\mathcal R(A)_{\mathrm{near}}\|_\chi\le\eps/12$ by \S\ref{subsec:thmB-contrib}.

\paragraph{The trace-norm bound of leg~III.}
Sketch of the proof of \cite[Thm.~1]{BMEL25}, following \cite[Sec.~IX]{BMEL25}. Along the segment $\Gamma+t(\Gamma'-\Gamma)$ it suffices to show $\|D_X\rho_\Gamma\|_1\le\frac12\|X\|_1$ for every Gaussian $\rho_\Gamma$ and real antisymmetric $X$. The directional derivative is a fixed linear expression in $\rho_\Gamma$, a sum over $a,b$ of $X_{ab}$ times a nested commutator and anticommutator of $\rho_\Gamma$ with $c_a$ and $c_b$, valid at every Gaussian state. At the Fock vacuum it acts on the vacuum, the one-particle sector and the two-particle sector only: a one-particle block $K$, the number $-\tr K$ on the vacuum and a vacuum--pair coupling $L$, all built from the entries of $X$; its trace norm is $\|K\|_1$ plus the trace norm of a $2\times2$ block coupling the vacuum to one pair state, and each of the two terms is at most $\frac14\|X\|_1$. A Gaussian unitary carries the vacuum to any pure Gaussian state and $X$ to $Q^TXQ$, which has the same trace norm; a mixed Gaussian state is a convex combination of pure Gaussian states (its Fock states in the normal frame), and applying the linear derivative formula to each and the triangle inequality gives the claim.

\paragraph{Total.}
$T(\rho_D,\rho_{\hat\Gamma})\le6\cdot\eps/12=\eps/2$, i.e.\ $\|\rho-\hat\rho\|_1\le\eps$.

\subsection{The higher-order terms in the bookkeeping norm}\label{subsec:thmB-star}

The fixed point is controlled in the operator-type norm $\|\cdot\|_\triangle$ at the fixed tolerance $e_0$, but the legs read Frobenius-type quantities, and a second-order term such as $[A,E]$ must be small compared with the first-order error, of order $\eps$, not merely compared with $e_0$, which does not depend on $\eps$. In Lemma~\ref{lem:M} the generator enters through the operator lump norm, where it is $\le8e$, and the error through the Frobenius-lump norm $\|\cdot\|_{\mathrm M}$, where it is of order $\eps$; so the remainder is a small multiple of the first-order error (\S\ref{subsec:thmB-higher}).

The legs of \S\ref{subsec:thmB-chain} read their higher-order terms in the Majorana norms $\|\cdot\|_w$, $\|\cdot\|_{w'}$ and $\|\cdot\|_\chi$. They are controlled through the Nambu bookkeeping norm $\|\cdot\|_{\mathrm M}$: Lemma~\ref{lem:M}, then \eqref{eq:NF4}.

\subsubsection{The Majorana map \texorpdfstring{$L$}{L} and the Majorana norms}\label{subsec:thmB-majnorms}

These norms are applied only to antisymmetric Majorana matrices ($\Delta$, $\mathcal R(A)$ and their parts) and to far-supported generators: every product enters through commutators inside the Nambu calculus of \S\ref{subsec:thmB-far}. All matrices are real $2m\times2m$, in $2\times2$ blocks, in the true frame; $D=\bigoplus_k\lambda_kJ$ with $\lambda_k=2p_k-1\in[-1,0]$; bands, types and far/near are as in \S\ref{subsec:thmB-bands}.

\paragraph{The map $L$.} For a far-supported real antisymmetric $A$ put $L(A):=[A,D]$. Using $JX^+=X^+J$ and $JX^-=-X^-J$, blockwise
\[
[A,D]_{kl}=A_{kl}\lambda_lJ-\lambda_kJA_{kl}=(\lambda_l-\lambda_k)A^+_{kl}J+(\lambda_l+\lambda_k)A^-_{kl}J .
\]
So $L$ preserves support and type and maps far-supported matrices to far-supported matrices; $L(A)$ is antisymmetric, being a commutator of two antisymmetric matrices ($[A,D]^T=DA-AD=-[A,D]$). $L$ is invertible on far-supported matrices:
\[
(L^{-1}B)_{kl}=-\frac{B^+_{kl}J}{\lambda_l-\lambda_k}-\frac{B^-_{kl}J}{\lambda_l+\lambda_k} .
\]
Gaps (\S\ref{subsec:thmB-bands}): far hopping $|\lambda_l-\lambda_k|>\max(P_b,P_{b'})/2$; far pairing $|\lambda_k+\lambda_l|\ge3/4$.

\paragraph{Norms.}
\begin{align*}
\|Y\|_w^2&:=\sum_{\text{far parts}}4\max(P_b,P_{b'})\,\|Y^+_{bb'}\|_F^2+2\,\|Y^-_{bb'}\|_F^2,\\
\|Y\|_{w'}^2&:=\sum_{\text{far parts}}\frac{\|Y^+_{bb'}\|_F^2}{\max(P_b,P_{b'})}+\|Y^-_{bb'}\|_F^2,\qquad
\|Y\|_\chi^2:=\sum_{\text{near parts}}\frac{\|Y_{bb'}\|_F^2}{\sqrt{P_bP_{b'}}},\\
\|Y\|_\star&:=\big(\|Y_{\mathrm{far}}\|_{w'}^2+\|Y_{\mathrm{near}}\|_\chi^2\big)^{1/2}
\end{align*}
(the generator norm with the weight of Lemma~\ref{lem:MT}, for far-supported $Y$; the covariance-change norm; the $\chi^2$ norm, with the $\mathcal P$-touching near parts included at the scale $P_c$; and the $\star$-norm, the $\ell^2$-sum of the far and near norms). The $\chi$-weight $1/\sqrt{P_cP_{J-1}}\asymp m$ on $\mathcal P$-touching near blocks is a bookkeeping weight only: these blocks never enter the $\chi^2$ identity, and leg~III uses them only through the conversion $\|Y_{\mathcal P}\|_F\le\sqrt{2/m}\,\|Y_{\mathrm{near}}\|_\chi$. The type-resolved far/near split makes $\|\Delta\|_\star=O(\eps)$ in second moment: the pairing parts of cold blocks sit in $\|\cdot\|_{w'}$ with weight $1$ and noise $O(8m/N_2)$, never under the $\chi$-weight $\asymp m$.

\begin{lemma}[$L^{-1}$ in these norms]\label{lem:Sprime}
For far-supported $B$, $\|L^{-1}B\|_w\le4\|B\|_{w'}$.
\end{lemma}

\begin{proof}
$L^{-1}$ acts entrywise on $2\times2$ blocks, so Frobenius norms of band blocks scale by the extreme symbol values. For $L^{-1}$, hopping: the $\lambda$-gap is $\ge\max P/2$ and the $w$-weight is $\sqrt{4\max P}$, giving $2\sqrt{\max P}\cdot2\|B\|_F/\max P=4\|B\|_F/\sqrt{\max P}$, i.e.\ $4$ times the $w'$-weighted norm; pairing: $(4/3)\sqrt2\le4$.
\end{proof}

\subsubsection{The higher-order terms}\label{subsec:thmB-higher}

Let $A$ be the fixed point of Lemma~\ref{lem:L1} and $a:=\|A\|_\triangleleft\le8e$. Put
\[
\theta:=\kappa_{\mathrm M}\,a\le67\cdot8e=536e\le\theta_0:=1/(64C_D^2)=1/4096,
\]
which holds since $e\le e_0$; in particular $\theta\le1/10$. Norms are read on the Nambu or on the Majorana form as in (Out4).

The inputs are the series \eqref{eq:B-series} and the fixed-point identity $L(A)=(E+\mathcal R(A))_{\mathcal F}$, so that $\|L(A)\|_{\mathrm M}\le\|E\|_{\mathrm M}+\|\mathcal R(A)\|_{\mathrm M}$ by (NF3). Lemma~\ref{lem:M} costs $\theta$ per commutator in $\|\cdot\|_{\mathrm M}$, hence
\[
\|\mathcal R(A)\|_{\mathrm M}\le\theta\|E\|_{\mathrm M}+c\big(\|E\|_{\mathrm M}+\|\mathcal R(A)\|_{\mathrm M}+\theta\|E\|_{\mathrm M}\big),\qquad c:=\frac{e^\theta-1-\theta}\theta,
\]
with $c\le0.518\,\theta$ ($\le0.0518$ at $\theta\le1/10$),
so
\[
\|\mathcal R(A)\|_{\mathrm M}\le\frac{\theta+c+c\theta}{1-c}\,\|E\|_{\mathrm M}\le1.66\,\theta\,\|E\|_{\mathrm M}\le2\theta\|E\|_{\mathrm M} .
\]
By (Out3), Lemma~\ref{lem:Sprime} and \eqref{eq:NF4},
\begin{align*}
\|A-A_1\|_w&=\|L^{-1}\mathcal R(A)_{\mathrm{far}}\|_w\le4\|\mathcal R(A)_{\mathrm{far}}\|_{w'}\le4C_D\|\mathcal R(A)\|_{\mathrm M}\le8C_D\theta\|E\|_{\mathrm M},\\
\|\mathcal R(A)_{\mathrm{near}}\|_\chi&\le C_D\|\mathcal R(A)\|_{\mathrm M}\le2C_D\theta\|E\|_{\mathrm M} .
\end{align*}

\subsubsection{Second moments, the Markov event, and the three contributions}\label{subsec:thmB-contrib}

$\E\|\Delta\|_\star^2=O\big(C_1'(m^3/N_2+m/N_3)\big)$, by the first-order computations of legs I--III, $N_3$ parts included; the diagonal blocks under the $\chi$-weight $1/P_b$ give $\sum_kC_1'\max(p_k,1/m)/(N_3P_b)\le2C_1'm/N_3$. Markov's inequality gives
\[
\|\Delta\|_\star\le\eps/23\quad\text{with probability }\ge1-1/40\quad\text{for }N_2\ge C_\star'm^3/\eps^2\text{ and }N_3\ge C_\star'm/\eps^2 .
\]
On this event, since $E$ is Nambu-symmetric, \eqref{eq:NF4} gives $\|E\|_{\mathrm M}\le C_D\|\Delta\|_\star\le C_D\eps/23$. With $\theta\le\theta_0=1/(64C_D^2)$, so that $C_D^2\theta\le1/64$, every higher-order contribution in \S\ref{subsec:thmB-chain} is $\le\eps/12$:
\begin{itemize}
\item leg I: $\frac1{\sqrt8}\|A-A_1\|_w\le\frac1{\sqrt8}\cdot8C_D\theta\cdot C_D\eps/23\le\eps/(8\sqrt8\cdot23)\le0.002\,\eps$;
\item leg II: $\frac12\|W_D\mathcal R(A)_{\mathrm{rest}}W_D\|_F\le\frac12\|\mathcal R(A)_{\mathrm{near}}\|_\chi\le C_D^2\theta\eps/23\le\eps/1472$;
\item leg III: $\sqrt{2m}\,\|\mathcal R(A)_{\mathcal P}\|_F\le2\|\mathcal R(A)_{\mathrm{near}}\|_\chi\le4C_D^2\theta\eps/23\le\eps/368$.
\end{itemize}
\begin{remark}[No Schur count over the bands]\label{rem:lumpwise}
By (NF2), $\|\cdot\|_{\mathrm M}$ is the weighted Frobenius norm with weight $1/\max(\Lambda_x,\Lambda_y)$: on a far block $(b,b')$ with $b'$ warmer, $1/\sqrt{\max}$ is the \emph{column} weight of the warmer side, and $\|Y\|_{\mathrm M}^2=\sum_\beta\|\Pi_\beta Y\Pi_{U_\beta}\Lambda_{U_\beta}^{-1/2}\|_F^2$ (a row band against its warm lump, column-weighted, Hermitian completion understood). No diagonal $G_1,G_2$ give it as $\|G_1YG_2\|_F$, so it is not of global product form; it is of lump-wise product form, and that suffices. A row-weight-only norm is equivalent to $w'$ only where the warmer side is the row; $\|\cdot\|_{\mathrm M}$ uses the row-band/warm-lump split, so the weight is always that of the warmer side. In Lemma~\ref{lem:M} the factor $\sqrt{P_\beta}$ of a row band is paid either by the warm square, whose minimum weight is $\ge P_\beta$, or, for a colder intermediate band $\gamma$, by the decay $2^{-d}$ of $\hat A_{\beta,\gamma}$ against the weight loss $2^{d/2}$, a net $2^{-d/2}$, which is summable. So no count of the $J$ bands arises; the generator enters through the \emph{operator} lump norm $\|\cdot\|_\triangleleft$.
\end{remark}

\subsection{Weight stability}\label{subsec:thmB-W}

The $\chi^2$ bound at the base point $D$ controls the integral along the segment from $D$ to $D+X$ only if the weights change little along it. Like any Fisher-type metric they are locally constant only on the scale of the occupations, which is why \eqref{eq:Whyp} asks $X$ to be a relative perturbation, measured two-sidedly in the weights $(1-\lambda^2)^{-1/2}$. The rest part is such a perturbation (Lemma~\ref{lem:Whyp-rest}); the cold group is not, one more reason why it is sent to leg~III.

\begin{lemma}[Weight stability]\label{lem:W}
Let $D=\bigoplus_k\lambda_kJ$ with $|\lambda_k|<1$ and let $X$ be real antisymmetric. For $t\in[0,1]$ put $\Gamma_t:=D+tX$ and $W_t:=(I+\Gamma_t^2)^{-1/4}$. Suppose
\begin{equation}\tag{W-hyp}\label{eq:Whyp}
\big\|(I+D^2)^{-1/2}(DX+XD)(I+D^2)^{-1/2}\big\|_\op\le\tfrac14\qquad\text{and}\qquad\big\|X(I+D^2)^{-1/2}\big\|_\op^2\le\tfrac14 .
\end{equation}
Then $I+\Gamma_t^2\succeq\frac12(I+D^2)\succ0$ for every $t\in[0,1]$ (so $\|\Gamma_t\|_\op<1$ along the segment), and for every $t\in[0,1]$ and every real matrix $Y$,
\[
\|W_tYW_t\|_F\le\sqrt2\,\|W_DYW_D\|_F,\qquad W_D:=W_0=(I+D^2)^{-1/4}.
\]
\end{lemma}

\begin{proof}
$I+\Gamma_t^2=(I+D^2)+P_t$ with $P_t=t(DX+XD)+t^2X^2$, and $(I+D^2)^{-1/2}P_t(I+D^2)^{-1/2}$ has operator norm $\le\frac14+\frac14$: the second term is $(I+D^2)^{-1/2}X^2(I+D^2)^{-1/2}=-\big(X(I+D^2)^{-1/2}\big)^T\big(X(I+D^2)^{-1/2}\big)$, since $X^T=-X$, of norm $\le\|X(I+D^2)^{-1/2}\|_\op^2$. Hence $P_t\succeq-\frac12(I+D^2)$ and $I+\Gamma_t^2\succeq\frac12(I+D^2)\succ0$. The map $x\mapsto x^{-1/2}$ is operator monotone decreasing on positive operators \cite[Ch.~V]{Bhatia97}, so $(I+\Gamma_t^2)^{-1/2}\preceq\sqrt2(I+D^2)^{-1/2}$, i.e.\ $W_t^2\preceq\sqrt2W_D^2$. Then $\|W_tW_D^{-1}\|_\op^2=\|W_D^{-1}W_t^2W_D^{-1}\|_\op\le\sqrt2$, and writing $W_tYW_t=(W_tW_D^{-1})(W_DYW_D)(W_D^{-1}W_t)$ gives $\|W_tYW_t\|_F\le\sqrt2\,\|W_DYW_D\|_F$.
\end{proof}

\begin{corollary}[The $\chi^2$ step with stable weights]\label{cor:Wprime}
If $\Gamma_0=D$ and $\Gamma_1=D+X$ have operator norm $<1$ along the segment and \eqref{eq:Whyp} holds, then by the proof of \cite[Prop.~7.1]{Rub26} (Appendix~A there) --- the $\chi^2$ identity $\|D_X\rho_\Gamma\|_1\le Q(X)^{1/2}$ with $Q(X)=\frac12\|W_\Gamma XW_\Gamma\|_F^2$, $W_\Gamma=(I+\Gamma^2)^{-1/4}$, at each point of the segment --- together with Lemma~\ref{lem:W},
\[
\|\rho_{\Gamma_0}-\rho_{\Gamma_1}\|_1\le\sqrt2\cdot(\tfrac12)^{1/2}\,\|W_DXW_D\|_F=\|W_DXW_D\|_F .
\]
This is the inequality used in leg~II.
\end{corollary}

The $\chi^2$ identity in Corollary~\ref{cor:Wprime} is Lemma~\ref{lem:A-chi2} combined with \eqref{eq:A-chi2id}, both proved in Appendix~\ref{app:bg-tools} for every covariance of operator norm $<1$; here it is applied at each point $\Gamma_t$ of the segment, with the weights controlled through Lemma~\ref{lem:W} instead of by the uniform constant $K_c$ of a class $\mathcal G_c$.

\begin{lemma}[\eqref{eq:Whyp} for the rest part]\label{lem:Whyp-rest}
On the event $\|E\|_\triangle\le e_0$, \eqref{eq:Whyp} holds for $X=X_{\mathrm{rest}}$ on the rest modes, with no further condition and no dependence on $J$ or $m$.
\end{lemma}

\begin{proof}
\emph{Setup.} In Nambu form, $X_{\mathrm{rest}}$ is the near part of $E+\mathcal R(A)$ outside the cold group (\S\ref{subsec:thmB-bands}; band-pair and type masks). Its Majorana image is $(M-D)_{\mathrm{rest}}=\Delta_{\mathrm{rest}}+\mathcal R(A)_{\mathrm{rest}}$, by (Out2). $X_{\mathrm{rest}}$ has no $\mathcal P$ rows or columns, so only non-cold modes carry its weights, and it is Nambu-symmetric (\S\ref{subsec:thmB-class}~(iii)).

\emph{Step 1 (the Nambu weighted norm).} By (NF1), the two masks of (NF3) (near, then cold group) and Lemma~\ref{lem:L1} ($\|E+\mathcal R(A)\|_\triangle\le e+0.53e$):
\[
\|\Lambda^{-1/2}X_{\mathrm{rest}}\Lambda^{-1/2}\|\le2\cdot4\cdot4\cdot(e+0.53e)\le49e .
\]

\emph{Step 2 (the Majorana weight).} In Majorana form $(I+D^2)^{-1/2}=\bigoplus_k(1-\lambda_k^2)^{-1/2}I_2$, and $1-\lambda_k^2=4p_k(1-p_k)\ge2p_k>P_b\ge\Lambda_k$ on non-cold particles, so $(1-\lambda_k^2)^{-1/2}\le\Lambda_k^{-1/2}$. The Majorana weight is mode-wise: it is $\le P_b^{-1/2}$ on \emph{both} $t_k$ and $\bar t_k$, while $\Lambda^{-1/2}$ is $1$ on holes. In Nambu sectors: the particle--particle block and the hole--hole block (its conjugate copy) each have Majorana-weighted norm $\le\|\tilde X_{\mathrm{rest}}\|$, and being block-diagonal they contribute $\le\|\tilde X_{\mathrm{rest}}\|$ together, not twice that; the near particle--hole and hole--particle blocks (bands $1$--$2$) gain $\le2$ on the hole side and contribute $\le2\|\tilde X_{\mathrm{rest}}\|$. So the weighted Nambu matrix is $\le3\|\tilde X_{\mathrm{rest}}\|$, and the Majorana one is $\le6\|\tilde X_{\mathrm{rest}}\|$, by the factor $2$ of the dictionary.

\emph{First clause.} $D$ commutes with the weight and $\|D\|\le1$, so, with $\mathcal T:=(I+D^2)^{-1/2}(DX_{\mathrm{rest}}+X_{\mathrm{rest}}D)(I+D^2)^{-1/2}$,
\[
\|\mathcal T\|_\op\le2\cdot6\cdot\|\Lambda^{-1/2}X_{\mathrm{rest}}\Lambda^{-1/2}\|\le12\cdot49e\le0.11\le\tfrac14,
\]
the factor $2$ coming from $\|D\|\le1$ in $DX+XD$ and the factor $6$ from Step~2.

\emph{Second clause.} $(I+D^2)^{-1/2}\succeq I$, since $1-\lambda_k^2\le1$. Hence
\[
\|X_{\mathrm{rest}}(I+D^2)^{-1/2}\|\le\|(I+D^2)^{-1/2}X_{\mathrm{rest}}(I+D^2)^{-1/2}\|\le6\cdot49e\le0.052\le\tfrac12 .
\]
The numbers $0.11$ and $0.052$ are at $e=1/5760$, the first condition of $e_0$ alone; their values at $e_0$ are in \S\ref{subsec:thmB-req}.
\end{proof}

\subsection{Assembly: probabilities and copies}\label{subsec:thmB-assembly}

\paragraph{Events.} $T(\rho_D,\rho_{\hat\Gamma})\le\eps/2$ holds on the intersection of the following events:
\begin{center}\small
\begin{tabular}{@{}ll@{}}
\toprule
event & failure probability\\
\midrule
Theorem~\ref{thm:Fprime} (Stage F$'$) & $\le4(R+1)e^{-m}\le0.01$\\
Lemma~\ref{lem:Htri} at $\tau=\tau_0$ (Stage 1), giving $\|E\|_\triangle\le\sqrt5\tau_0=e_0$ & $\le e^{-m}\le0.01$\\
the three first-order Markov events of legs I--III & $3/40$\\
the $\star$-Markov event (\S\ref{subsec:thmB-contrib}) & $1/40$\\
\midrule
total & $\le0.12<1/3$\\
\bottomrule
\end{tabular}
\end{center}
On the event of Lemma~\ref{lem:Htri} every deterministic step holds with no further event: Lemma~\ref{lem:L1}, the translation of \S\ref{subsec:thmB-transl}, \S\ref{subsec:thmB-higher}, and Lemma~\ref{lem:Whyp-rest}. The diagonal blocks are inside the event of Lemma~\ref{lem:Htri}, through the occupation part of $E$ in the row lumps.

\emph{Conditioning.} Stage~1 uses fresh copies after Stage~F$'$, and $W$ is measurable with respect to Stage~F$'$. Lemma~\ref{lem:Htri} is applied conditionally on $W$, on the event $G(1/m,3)$. The Markov events use second moments computed conditionally on the frame (Lemma~\ref{lem:B1}), so they hold on the event of Theorem~\ref{thm:Fprime}: $\E[X\mathbf 1_{F'}]\le\sup_{\text{good frames}}\E[X\mid W]$.

\paragraph{Copies.} Stage~F$'$ uses $N_{F'}$ copies and Stage~1 uses $N_2+N_3$, with $N_{F'}$ (Theorem~\ref{thm:Fprime}), $N_2$, $N_3$, $C_N$ and $\tau_0=e_0/\sqrt5\le1$ as in Remark~\ref{rem:B-count}. The hypotheses $N_2\ge K_\triangle m^3/\tau_0^2$ and $N_3\ge K'_\triangle m^2/\tau_0^2$ of Lemma~\ref{lem:Htri} are exactly the second terms of $N_2$ and $N_3$; the first terms $C_Nm^3/\eps^2$ and $C_Nm/\eps^2$ meet the requirements of legs I--III and of \S\ref{subsec:thmB-contrib}; and every condition of \S\S\ref{subsec:thmB-far}--\ref{subsec:thmB-W} is a condition on $e_0$ only (\S\ref{subsec:thmB-req}). The $\eps$-free terms are $K_\triangle/\tau_0^2=5K_\triangle/e_0^2\le5\cdot2.6\cdot10^8\cdot(2\,195\,456)^2=6.27\cdot10^{21}$ in front of $m^3$ and $K'_\triangle/\tau_0^2\le5\cdot1.4\cdot10^7\cdot(2\,195\,456)^2=3.37\cdot10^{20}$ in front of $m^2$.

\emph{Odd $m$.} The Horvitz--Thompson weight is $4m$, $\Prob(P\in\pi)=1/m$ and $\Prob(P,P'\in\pi)=1/(m(m-2))$, so the design factor becomes $m/(m-2)\le5/3$ in place of $(m-1)/(m-3)\le3$, and every constant changes by at most a factor $2$. For Lemma~\ref{lem:Htri} nothing changes: in the proof of \eqref{eq:R0} the even-$m$ display already bounds $m-1$ by $m$, the design factor becomes $m/(m-2)\le3$, and the range $4m+2$ enters only the range condition of Bernstein's inequality. So \eqref{eq:R0} (with the range $(4m+2)\|w\|^2$ in place of $4m\|w\|^2$), $K_5$, $K_6$ and Lemma~\ref{lem:Htri} hold unchanged for odd $m\ge5$, and so do Lemma~\ref{lem:FR} and Theorem~\ref{thm:Fprime}, whose Bernstein range condition becomes $\eta\mu\le12K_1m\mu_0/(4m+2)$ and holds; only $C_N$ is parity-dependent.

\begin{proof}[Proof of Theorem~\ref{thm:B}]
On the intersection of the events above, which has probability $\ge1-0.12>2/3$, the comparison chain of \S\ref{subsec:thmB-chain} gives $T(\rho_D,\rho_{\hat\Gamma})\le\eps/2$, i.e.\ $\|\rho-\hat\rho\|_1\le\eps$. The total number of copies is
\[
N\le2C_N\frac{m^3}{\eps^2}+\big(6.3\cdot10^{21}+3.4\cdot10^{20}+1.6\cdot10^9+3.7\cdot10^7+1\big)m^3\le2C_N\frac{m^3}{\eps^2}+6.7\cdot10^{21}m^3,
\]
using $m/\eps^2\le m^3/\eps^2$, $m^2\le m^3$ and $2(R+1)\le m^3$ (the integer ceilings on $N_2$, $N_3$ are absorbed likewise). So $N\le C_B(m^3/\eps^2+m^3)$ with $C_B:=\max\{2C_N,6.7\cdot10^{21}\}$, and $N\le2C_Bm^3/\eps^2$ for $\eps\le1$. The threshold $m_B=17$ gives $J\ge5$, $m\ge16$ (Theorem~\ref{thm:Fprime}; the union bound of Lemma~\ref{lem:Htri}, $16(J+1)e^{-1.8m}\le e^{-m}$) and $4(R+1)e^{-m}\le0.01$. The lower bound and the range are those of \cite[Thm.~1.1]{Rub26}, as stated.
\end{proof}

\subsection{Constants}\label{subsec:thmB-constants}

\begin{table}[ht]
\centering\small
\begin{tabular}{@{}>{$}l<{$} >{\raggedright\arraybackslash}p{0.45\linewidth} >{\raggedright\arraybackslash}p{0.28\linewidth}@{}}
\toprule
\text{symbol} & value / condition & where \\
\midrule
\eta,\ R & $1/8$, $\lceil\log_2\log_2m\rceil$ & Protocol F$'$ \\
K_1,\ K_2 & $5151$, $120$ (at $C=3$) & Lemma~\ref{lem:FR} \\
\text{frame condition} & $G(1/m,18/7)\subset G(1/m,3)$ & Theorem~\ref{thm:Fprime} \\
c_F',\ c_b & $6$, $24$ & Corollary~\ref{cor:Fprime} \\
C_0,\ C_1' & $561\,280$; $C_0/2\le10^6$ & Lemma~\ref{lem:B1} \\
K_5,\ K_6 & $2.62\cdot10^5$, $1.44\cdot10^4$ & Lemma~\ref{lem:Htri} \\
K_\triangle,\ K'_\triangle & $960K_5\le2.6\cdot10^8$, $960K_6\le1.4\cdot10^7$ & Lemma~\ref{lem:Htri} \\
\|A\|_g/\|A\|_\triangleleft & $\le22$ & Lemma~\ref{lem:G} (G1$''$) \\
\text{(G1), (G1}'\text{)} & $9$ (particle), $14$ (hole); $4\cdot2^{-d}$ & Lemma~\ref{lem:G} \\
\kappa_\triangle,\ \kappa_{\mathrm M} & $70.62$, $66.5$ proven; $72$, $67$ used & Lemmas~\ref{lem:T}, \ref{lem:M} \\
\text{mask constant} & $4$ & (NF3) \\
C_D & $8$ & \eqref{eq:NF4} \\
e_0 & $\min\{1/5760,1/(34\,304\,C_D^2)\}=1/2\,195\,456\approx4.55\cdot10^{-7}$ & \S\ref{subsec:thmB-req} \\
\varphi,\ \text{Lipschitz} & $\varphi=576e\le1/10$; $\|\mathcal R(A)\|_\triangle\le0.53e$; Lip $\le0.48$ & Lemma~\ref{lem:L1} \\
\theta,\ \theta_0 & $\theta=536e\le\theta_0=1/(64C_D^2)=1/4096$; $\|\mathcal R(A)\|_{\mathrm M}\le1.66\theta\|E\|_{\mathrm M}$ & \S\ref{subsec:thmB-higher} \\
\text{(W-hyp)} & $49e$; factors $12$ and $6$; $0.11\le\frac14$, $0.052\le\frac12$ & Lemma~\ref{lem:Whyp-rest} \\
\tau_0 & $e_0/\sqrt5\approx2.04\cdot10^{-7}$ & \S\ref{subsec:thmB-assembly} \\
\eps\text{-free terms} & $\le6.3\cdot10^{21}$ ($m^3$), $\le3.4\cdot10^{20}$ ($m^2$) & \S\ref{subsec:thmB-assembly} \\
C_{\mathrm I},C_{\mathrm{II}},C_{\mathrm{III}},C_\star' & $O(C_1')$, from the second moments of \S\S\ref{subsec:thmB-chain}, \ref{subsec:thmB-contrib}; $C_N$ their maximum & legs I--III, $\star$-norm \\
C_B & $\max\{2C_N,6.7\cdot10^{21}\}$ & \S\ref{subsec:thmB-assembly} \\
K_3,K_4;\ K_H,K_H' & $2.5\cdot10^5$, $7920$; $1.8\cdot10^8$, $5.6\cdot10^6$ & Lemma~\ref{lem:H0prime} \\
\bottomrule
\end{tabular}
\caption{Constants of Theorem~\ref{thm:B}.}\label{tab:B-constants}
\end{table}

\subsection{Why no logarithm remains}\label{subsec:thmB-nolog}

A bookkeeping of the far decomposition in block operator-norm profiles, or in whole-matrix relative norms, loses a factor that grows with the number $J$ of bands. The row-lump norm $\|\cdot\|_\triangle$ does not. This subsection proves the obstruction for whole-matrix relative norms and explains why the row lumps avoid it.

\paragraph{The size of the triangular truncation.}
That $\operatorname{Tri}$ has norm $\Theta(\log n)$ on $n$ blocks is the theorem of Kwapie\'n and Pe\l czy\'nski \cite{KP70}; both directions are elementary. \emph{Upper bound.} Index the blocks by $0,\dots,n-1$; a strictly lower pair $(j,k)$, $j>k$, has a highest binary digit $\ell$ in which $j$ and $k$ differ, where $j$ has $1$ and $k$ has $0$. For fixed $\ell$ the pairs form a block partial permutation (rows and columns sharing the digits above $\ell$), whose masking has Schur norm $1$ by (Sch3); there are $\lceil\log_2n\rceil$ levels, so $\|\operatorname{Tri}\|\le\lceil\log_2n\rceil$. \emph{Lower bound.} The Toeplitz matrix $H=[i/(j-k)]_{j\ne k}$ has $\|H\|\le\pi$: it is a finite section of multiplication on $L^2$ of the circle by the symbol $\sum_{d\ne0}ie^{id\phi}/d=\phi-\pi$, $\phi\in(0,2\pi)$, of modulus $\le\pi$. Its strictly lower triangular part is $i$ times a nonnegative matrix with row sums $\approx\ln n$; since the norm of a nonnegative matrix is at least $\mathbf 1^T\mathcal M\mathbf 1/n$ (test with $\mathbf 1/\sqrt n$) and is monotone in the entries, $\|\operatorname{Tri}(H)\|\ge\ln n-O(1)$, so $\|\operatorname{Tri}\|\ge(\ln n-O(1))/\pi$. This is the mechanism of Proposition~\ref{prop:obstruction}, with $n=J$ bands.

Write $\Omega:=\Omega_{1/m}$, $\tilde B_\Omega:=\Omega^{-1/2}B\Omega^{-1/2}$ and $\hat A_\Omega:=\Omega^{-1/2}A\Omega^{1/2}$, with $A=L^{-1}B$ and $B$ Hermitian on $\mathcal F$ with the Nambu symmetry $S\bar BS=-B$. Entrywise $\hat A_\Omega=\varpi\circ\tilde B_\Omega$, where $\varpi_{xy}:=\Omega_y/(n_y-n_x)$ on $\mathcal F$ and $0$ off $\mathcal F$. Let $\operatorname{Tri}$ be the block-triangular truncation onto the particle--particle pairs $(x,y)$ with $\iota(b_y)\le\iota(b_x)-2$ (colder row, warmer column). Over $J$ blocks, $\operatorname{Tri}$ has norm $\Theta(\log J)$ (above).

\begin{proposition}[The Sylvester inverse in a whole-matrix relative norm]\label{prop:relSylv}
For $m\ge8$ and every Hermitian $B$ supported on $\mathcal F$ with $S\bar BS=-B$, $\alpha(L^{-1}B):=\|\hat A_\Omega\|$ satisfies
\[
\|\operatorname{Tri}(\tilde B_\Omega)\|-12\|\tilde B_\Omega\|\le\|\Pi_p\hat A_\Omega\Pi_p\|\le\alpha(L^{-1}B)\le\|\operatorname{Tri}(\tilde B_\Omega)\|+101\|\tilde B_\Omega\| .
\]
\end{proposition}

\begin{proof}
\emph{Particle--particle, $y$ warmer by $d\ge2$.} $\varpi=1+\Omega_x/(p_y-p_x)$ exactly, since $\Omega_y-(p_y-p_x)=p_x+1/m=\Omega_x$. The ``$1$'' is $\operatorname{Tri}$. For the remainder, $\Omega_x$ is a row diagonal $\le3P_b$ and $1/(p_y-p_x)$ is a positive mixture of rank-one symbols with Schur norm $\le4/P_{b'}$, so the remainder is $\le12\cdot2^{-d}$ per block (Sch1, Sch2); by (Sch3) per $d$-diagonal and $\sum_{d\ge2}12\cdot2^{-d}=6$, it is $\le6\|\tilde B_\Omega\|$.

\emph{Particle--particle, $y$ colder by $d\ge2$.} $|\varpi|\le\Omega_y/\mathrm{gap}\le12\cdot2^{-d}$ per block, giving $\le6\|\tilde B_\Omega\|$.

\emph{Particle--hole ($x=k$, $y=\bar l$).} $\varpi=(1-p_l+1/m)/(1-p_l-p_k)=(1-p_l+1/m)\int_0^\infty e^{-t}e^{tp_k}e^{tp_l}\,dt$. Split into three product pieces, each with $\max p_k+\max p_l\le5/8$: rows in bands $\ge3$ against all holes; row band $2$ against holes of bands $\ge2$; row band $1$ against holes of bands $\ge3$. Each piece has Schur norm $\le(1+1/m)\cdot8/3\le3$ ($m\ge8$), so the total is $\le9\|\tilde B_\Omega\|$.

\emph{Hole--particle ($x=\bar k$, $y=l$).} $|\varpi|\le\Omega_l\cdot8/3\le(\frac12+1/m)\cdot8/3\le3$ ($m\ge8$); with the same split this gives $\le9\|\tilde B_\Omega\|$.

\emph{Hole--hole ($\bar k,\bar l$), with $(k,l)$ far particle--particle.} Here $\varpi=\Omega_{\bar l}/(p_k-p_l)$ is not bounded: the relative weight is $\approx1$ on holes and the gap is $\asymp P_{\max}$. As for the hole rows in (G1) (proof of Lemma~\ref{lem:G}), Nambu symmetry makes the hole--hole block of $\hat A_\Omega$ a Schur multiple of the conjugated particle--particle block, $\hat A_{\bar k\bar l}=-\overline{(\tilde B_\Omega)_{kl}}\cdot\sqrt{\Omega_k\Omega_l}\cdot\sqrt{\Omega_{\bar l}/\Omega_{\bar k}}/(p_k-p_l)$; with $\Lambda$ replaced by $\Omega$, the factors $\Omega\le3\Lambda$ on particles and $\sqrt{\Omega_{\bar l}/\Omega_{\bar k}}\le\sqrt{(1+1/m)/\frac12}\le\sqrt3$ turn the per-block bound $4\cdot2^{-d/2}$ there into $12\sqrt3\cdot2^{-d/2}$, and the total into $24\sqrt3/(2-\sqrt2)=70.97$, i.e.\ $\le71\|\tilde B_\Omega\|$.

\emph{Total remainder} $\le6+6+9+9+71=101$. The lower bound uses only the particle--particle part (remainders $6+6$).
\end{proof}

\begin{proposition}[The obstruction]\label{prop:obstruction}
For every $J\ge4$ there are a far-supported Hermitian Nambu-symmetric $B$ and a near (diagonal) Hermitian Nambu-symmetric $X$ with
\[
\|[L^{-1}B,X]\|_{\mathrm{rel}}\ge c(\ln J-C)\,\|B\|_{\mathrm{rel}}\|X\|_{\mathrm{rel}}\qquad\text{and}\qquad\alpha(L^{-1}B)\ge c(\ln J-C)\,\|B\|_{\mathrm{rel}},
\]
where $c,C$ are absolute ($c\ge1/11$).
\end{proposition}

\begin{proof}
\emph{Construction.} Let $m:=2^J$, so that $J=\lceil\log_2m\rceil$ and $q:=1/m=2^{-J}=P_c$. Modes $j=1,\dots,J$ are active, with $p_j:=\frac34\cdot2^{-j}$ ($j<J$) and $p_J:=0$; the remaining $m-J$ modes are cold spectators ($p=0$) on which $B$ and $X$ vanish, so every norm below may be computed on the active modes. $\Omega_j=p_j+q$. The particle--particle block of $B$ is $\Omega^{1/2}H\Omega^{1/2}$ with $H_{jk}:=i/(j-k)$ for $|j-k|\ge2$ (Hermitian), its hole block is $-\overline{(\cdot)}$, and it has no pairing part. The particle--particle block of $X$ is $\mathrm{diag}(\varsigma_j\Omega_j)$ with $\varsigma_j:=(-1)^j$, and its hole block is $\mathrm{diag}(-\varsigma_j\Omega_j)$. Both are Nambu-legal ($S\bar XS=-X$; $B$ is Hermitian, Nambu-symmetric and supported on the far particle--particle and hole--hole set); $X$ is diagonal, hence near.

\emph{Norms.} $(\tilde B_\Omega)_{pp}=H_{\mathrm{far}}$, and $\|H_{\mathrm{far}}\|\le\pi+2$ (the Hilbert matrix has norm $\le\pi$, and the two removed diagonals have norm $1$ each). The hole block is a Schur multiple of $\bar H$ by $\sqrt{\Omega_k\Omega_l/(\Omega_{\bar k}\Omega_{\bar l})}\le1.02$, so $\|B\|_{\mathrm{rel}}\le1.02(\pi+2)\le5.3$. $\|X\|_{\mathrm{rel}}=\max_j\max(1,\Omega_j/\Omega_{\bar j})=1$.

\emph{Commutator.} $X$ is diagonal, so $[L^{-1}B,X]_{jk}=B_{jk}(\xi_k-\xi_j)/(p_k-p_j)$ with $\xi=\varsigma\Omega$; relatively ($\Omega$-weighted) the $(j,k)$ entry is $H_{jk}(\varsigma_k\Omega_k-\varsigma_j\Omega_j)/(p_k-p_j)$. On rows $j$ even and columns $k$ odd ($\varsigma_j=1$, $\varsigma_k=-1$) this is $-i(\Omega_j+\Omega_k)/(|j-k|\,|p_k-p_j|)$ for both signs of $j-k$: $-i$ times an entrywise nonnegative matrix whose entries are $\ge1/|j-k|$, since $\Omega_j+\Omega_k\ge\max(p_j,p_k)\ge|p_k-p_j|$. For a nonnegative matrix the operator norm is monotone in the entries and $\ge(\mathbf 1^T\mathcal M\mathbf 1)/\sqrt{n_rn_c}$. The even--odd matrix $[1/|j-k|]_{|j-k|\ge3}$ has average row sum $\ge\frac12(\ln J-C)$: the rows $j\in[J/4,3J/4]$ already give $2\sum_{d\text{ odd},\,3\le d\le J/4}1/d\ge\ln J-C$. So $\|[L^{-1}B,X]\|_{\mathrm{rel}}\ge\frac12(\ln J-C)\ge\frac{\ln J-C}{10.6}\|B\|_{\mathrm{rel}}\|X\|_{\mathrm{rel}}$.

\emph{$\alpha$.} On the particle--particle block, $\hat A_\Omega=\varpi\circ H$. For $k$ warmer ($k<j$) the entry is $i(1+\Omega_j/(p_k-p_j))/(j-k)$, $i$ times a positive number; for $k$ colder ($k>j$) it is $(i/(j-k))\,\Omega_k/(p_k-p_j)$, also $i$ times a positive number, since both factors are negative. So $\hat A_{pp}=i\cdot(\text{nonnegative matrix})\ge i\cdot[1/(j-k)]_{j-k\ge2}$ entrywise, and its norm is $\ge$ its average row sum $\ge\ln J-C$.
\end{proof}

\begin{remark}[The obstruction for whole-matrix relative norms]\label{rem:obstruction}
No generator norm $\mathcal G$ admits both $\|L^{-1}B\|_{\mathcal G}\le C\|B\|_{\mathrm{rel}}$ and $\|[A,X]\|_{\mathrm{rel}}\le C'\|A\|_{\mathcal G}\|X\|_{\mathrm{rel}}$ with $C,C'$ independent of $J$: together they would give $\|[L^{-1}B,X]\|_{\mathrm{rel}}\le CC'\|B\|_{\mathrm{rel}}\|X\|_{\mathrm{rel}}$, against Proposition~\ref{prop:obstruction}. The obstruction lies in the bilinear map $(B,X)\mapsto[L^{-1}B,X]$ itself; since the fixed-point source is $(E+\mathcal R(A))_{\mathcal F}$ and $\mathcal R(A)$ contains $[A,E]$, no choice of generator norm repairs it. Weighting the two triangles of the generator differently does not help either: $[A,X]^\sim=\hat A\tilde X+\tilde X\hat A^*$, and with $A=A_{\mathrm{up}}+A_{\mathrm{low}}$ (the cold-row/warm-column part and its adjoint) the only large pieces are $\hat A_{\mathrm{up}}\tilde X+\tilde X\hat A_{\mathrm{up}}^*$, a Lyapunov-type sum with $\hat A_{\mathrm{up}}\approx\operatorname{Tri}(\tilde B)$; for $\tilde X=I$ it collapses to $\tilde B_{\mathrm{far}}$, and for the band-alternating $\tilde X=\mathrm{diag}(\pm1)$ it is $\varpi_{\max}\circ\tilde B$, which is the example above. So the $\log J$ is a property of the error class $\{\|X\|_{\mathrm{rel}}\le1\}$. By Proposition~\ref{prop:relSylv} and $\|[A,X]^\sim\|\le2\alpha\|\tilde X\|$ the loss is also at most $O(\log J)$, so it is $\asymp\log J=\log\log m$. With errors controlled only in the whole-matrix relative norm, the commutator with the far Sylvester generator is unbounded by a factor $\asymp\log J$, whatever norm is put on the generator: the block-triangular truncation in scale is intrinsic to that class. It disappears in the row-lump class $\|\cdot\|_\triangle$, which the Horvitz--Thompson noise satisfies at the same cost $m^3/\tau^2$ (Lemma~\ref{lem:Htri}) and which is closed under the far rotation with absolute constants (Lemmas~\ref{lem:G}--\ref{lem:M}).
\end{remark}

\begin{lemma}[The triangular truncation is a column restriction of the row lumps]\label{lem:Tri}
$\|\operatorname{Tri}(\tilde B)\|\le\big(\sum_b\|\Pi_b\tilde B\Pi_{U_b}\|^2\big)^{1/2}\le\|B\|_\triangle$.
\end{lemma}

\begin{proof}
In each colder row band $b$ the triangular truncation keeps only columns of the warm lump, a sub-block of the row lump (a column restriction); conclude with the row-band inequality in the proof of (NF1).
\end{proof}

$\operatorname{Tri}$ is unbounded on $\{\|\tilde B\|\le1\}$ only because $\|\tilde B\|\le1$ allows every row lump to have size $1$; the lump decomposition, each cold band against its warm lump, is exactly the operator that $\operatorname{Tri}$ is. In the example of Proposition~\ref{prop:obstruction}, $r_j(X)=\Omega_j/P_j\in[\frac34,\frac74]$ on every particle band and $r_h(X)\le\frac7{16}$, so $\|X\|_\triangle\ge\frac34\sqrt J\,\|X\|_{\mathrm{rel}}$, and in fact $\|X\|_\triangle=(\frac34+o(1))\sqrt J$. The warm diagonal entries $\asymp\Omega_b$ of that example are far above the actual noise level $\tau(mP_b)^{-1/2}$ on warm row lumps (Lemma~\ref{lem:Htri}); the $\triangle$-norm prices them at $\asymp\frac34\sqrt J$.

\subsection{A band-pair profile}\label{subsec:thmB-H0}

The event of Lemma~\ref{lem:Htri} contains the following operator-norm profile of the error between band pairs (Remark~\ref{rem:Htri}~(a)).

\begin{lemma}[Band-pair profile]\label{lem:H0prime}
Let $W$ satisfy $G(\mu_0,3)$ with $\mu_0\le1/m$, and let $m\ge16$. Let Stage~1 take $N_2$ Horvitz--Thompson and $N_3$ occupation shots in $W$ with fresh copies, and put $E:=\hat{\mathcal N}-\mathcal N$. For $\tau\in(0,1]$, suppose
\[
N_2\ge K_Hm^3/\tau^2,\quad N_3\ge K_H'm^2/\tau^2,\qquad K_H:=704\cdot2.5\cdot10^5\le1.8\cdot10^8,\quad K_H':=704\cdot7920\le5.6\cdot10^6 .
\]
Then with probability $\ge1-e^{-m}$, for all true bands $b,b'$ (cold band and $b=b'$ included), with $\bar P:=\max(P_b,P_{b'})$,
\[
\|\Pi_{V_b}E\Pi_{V_{b'}}\|\le\tau\sqrt{\bar P/m},\qquad\|\Pi_{V_b}E\Pi_{h(V_{b'})}\|\le\tau/\sqrt m .
\]
\end{lemma}

\begin{proof}
Fix $(b,b')$ and a type. Take unit $x\in V_b$, and unit $y\in V_{b'}$ (hopping) or $y\in h(V_{b'})$ (pairing).

\emph{Polarisation} as in the proof of Lemma~\ref{lem:Htri}, unweighted: $w_k:=x+i^ky$, $|x^\dagger My|\le\max_k|w_k^\dagger Mw_k|$ and $\|w_k\|^2\le4$.

\emph{Hopping weights.} $p\le P_b$ on $V_b$, and $\mu_0\le1/m\le2P_c\le2\bar P$. So $\langle w|\Omega_{\mu_0}|w\rangle\le2(P_b+P_{b'}+2\mu_0)\le12\bar P$; by $G(\mu_0,3)$, $Q_w,B_w\le36\bar P$; and $N_w\le4\bar P$.

\emph{Hopping variances, by \eqref{eq:R0}.} $\mathrm{Var}(w^\dagger Zw)\le(640\cdot4\cdot36+1024\cdot4\cdot36)m\bar P+3(76\bar P)^2+192\le(239\,616+8\,664+384)m\bar P\le2.5\cdot10^5\,m\bar P=:K_3m\bar P$, using $\bar P^2\le\bar P/2\le m\bar P$ and $1\le2m\bar P$; and $\mathrm{Var}(w^\dagger Z_{\mathrm{occ}}w)\le10(4\cdot36\bar P+1296\bar P^2)\le7920\bar P=:K_4\bar P$.

\emph{Pairing weights and variances.} $\langle w|\Omega_{\mu_0}|w\rangle\le2(\frac12+\mu_0+1+\mu_0)\le4$ for $m\ge4$, so $Q_w,B_w\le12$ and $N_w\le4$; $\mathrm{Var}(w^\dagger Zw)\le79\,872m+2\,544\le8.1\cdot10^4m$ and $\mathrm{Var}(w^\dagger Z_{\mathrm{occ}}w)\le10(48+144)=1920$.

\emph{Bernstein and nets} as in the proof of Lemma~\ref{lem:Htri}, for $E_{\mathrm{HT}}:=N_2^{-1}\sum Z_t-\mathcal N_{\mathrm{off}}$ and $E_{\mathrm{occ}}:=N_3^{-1}\sum Z_{\mathrm{occ},t}-\mathcal C_W(\mathcal N)$, with the deviation $t:=\tau\sqrt{\bar P/m}/4$ (hopping) or $\tau/(4\sqrt m)$ (pairing) and the ranges $16m$ and $4$ ($\|w\|^2\le4$). The range conditions hold since $\tau\le1$ and $m\bar P\ge\frac12$, and the exponents $N_2\tau^2/(64K_3m^2)$ and $N_3\tau^2/(64K_4m)$ are $\ge11m$ under the hypotheses (the pairing constants $8.1\cdot10^4$ and $1920$ are smaller). The $\frac14$-nets of the unit spheres of $\mathrm{Ran}\,\Pi_{V_b}$ and of the $y$-space have $\le9^{2(m_b+m_{b'})}\le9^{4m}\le e^{9m}$ pairs, even when $b=b'$, and the supremum is $\le4t$, which is the target.

\emph{Union.} $2J^2$ (pair, type) $\times\,e^{9m}\times4\times2\times2e^{-11m}=32J^2e^{-2m}\le e^{-m}$ for $m\ge16$; measurability is as in Lemma~\ref{lem:Htri}, the subspaces $V_b$ and $h(V_{b'})$ being deterministic.
\end{proof}

By \eqref{eq:Dict}, on the event of Lemma~\ref{lem:H0prime} with $W$ the frame of Theorem~\ref{thm:Fprime}, every band pair has
\[
\|\Delta^+_{bb'}\|_\op\le2\tau\sqrt{\max(P_b,P_{b'})/m},\qquad\|\Delta^-_{bb'}\|_\op\le2\tau/\sqrt m,\qquad\|\Delta^{\mathrm{diag}}_{bb}\|_\op\le2\tau\sqrt{P_b/m}.
\] Operator norms are essential for such profiles: a Frobenius profile would cost $\sqrt m$ on band blocks.

\section{Proofs for Section~\ref{sec:kcopy}}\label{app:proofs-K}

\subsection{Gaussian unitaries on the spinor module}

Put $\pi(X):=\tfrac14\sum_{a,b}X_{ab}c_ac_b=B(X)/(2i)=-iK_X$. By Lemma~\ref{lem:A-rot}, $[\pi(X),c_d]=\sum_aX_{ad}c_a$ and $U_{e^X}:=e^{\pi(X)}$ satisfies $U_Oc_aU_O^\dagger=\sum_bO_{ba}c_b$; every $O\in SO(2m)$ is some $e^X$ (real normal form). Substituting into $B(Y)=\tfrac i2\sum_{a,b}Y_{ab}c_ac_b$ gives $U_OB(Y)U_O^\dagger=B(OYO^T)$, hence $\Tr(U_O\rho U_O^\dagger B(X))=\langle O\Gamma O^T,X\rangle$. If $U,U'$ implement the same $O$, $U^\dagger U'$ commutes with every $c_a$, hence with all operators on $S$, and is a scalar. At $t=1$, $\pi$ is a Lie algebra homomorphism: by the Jacobi identity and $[\pi(X),c_d]=\sum_aX_{ad}c_a$, $[\pi(X),\pi(Y)]-\pi([X,Y])$ commutes with every $c_d$, so it is a scalar; it is traceless, hence zero.

\subsection{Three lemmas and the proof of Proposition~\ref{prop:K4}}

\begin{lemma}[symbol lemma]\label{lem:K1}
For $\varphi\in\mathcal E$ and $X\in\mathfrak{so}(2m)$: $\langle v|\varphi(X)|v\rangle=\langle X_{\xi_\varphi},X\rangle$. Consequently $\langle v|U_O^\dagger\varphi(X)U_O|v\rangle=\langle\mathrm{Ad}_OX_{\xi_\varphi},X\rangle$.
\end{lemma}

\begin{proof}
Write $X=\sum_j\langle X,T_j\rangle T_j+\sum_{\beta\in\Phi}x_\beta E_\beta$. For $\beta\in\Phi$, $\varphi(E_\beta)v$ is zero or has weight $\lambda+\beta\ne\lambda$, so it is orthogonal to $v$ and $\langle v|\varphi(E_\beta)|v\rangle=0$. Hence $\langle v|\varphi(X)|v\rangle=\sum_j\langle X,T_j\rangle(\xi_\varphi)_j=\langle X_{\xi_\varphi},X\rangle$. The second claim follows by equivariance, $U_O^\dagger\varphi(X)U_O=\varphi(\mathrm{Ad}_{O^T}X)$, and orthogonality of $\mathrm{Ad}_O$.
\end{proof}

\begin{lemma}[coherent-state twirls and Parseval]\label{lem:K2}
For $\xi\in\R^m$ define $Y_\xi:\mathfrak{so}(2m)\to\mathrm{Herm}(V)$,
\[
Y_\xi(X):=n\int\langle\mathrm{Ad}_OX_\xi,X\rangle\,U_O|v\rangle\langle v|U_O^\dagger\,dO .
\]
Then (a) $Y_\xi\in\mathcal E$; (b) for every $\varphi\in\mathcal E$, $\langle\varphi,Y_\xi\rangle_{\mathcal E}=n\,\xi\cdot\xi_\varphi$.
\end{lemma}

\begin{proof}
(a) Linearity in $X$ and Hermiticity are clear. For $O'\in SO(2m)$, the substitution $O=O'^TO''$ (left invariance of Haar measure) and orthogonality of $\mathrm{Ad}_{O'}$ give $U_{O'}Y_\xi(X)U_{O'}^\dagger=Y_\xi(\mathrm{Ad}_{O'}X)$.
(b) By Lemma~\ref{lem:K1}, $\Tr(\varphi(E_{ab})U_O|v\rangle\langle v|U_O^\dagger)=\langle v|U_O^\dagger\varphi(E_{ab})U_O|v\rangle=\langle\mathrm{Ad}_OX_{\xi_\varphi},E_{ab}\rangle$, so Parseval in the orthonormal basis $\{E_{ab}\}$ and orthogonality of $\mathrm{Ad}_O$ give $\langle\varphi,Y_\xi\rangle_{\mathcal E}=n\int\langle\mathrm{Ad}_OX_\xi,\mathrm{Ad}_OX_{\xi_\varphi}\rangle\,dO=n\,\xi\cdot\xi_\varphi$.
\end{proof}

\begin{lemma}[generalised Casimir identity]\label{lem:K3}
For every $\varphi\in\mathcal E$,
\[
\sum_{a<b}\varphi(E_{ab})\,d\pi(E_{ab})=\langle\lambda+2\varrho,\xi_\varphi\rangle\,I_V,\qquad\text{hence}\qquad\langle\varphi,d\pi\rangle_{\mathcal E}=n\,(\lambda+2\varrho)\cdot\xi_\varphi .
\]
(For $\varphi=d\pi$ this is the Casimir eigenvalue $C_2(\lambda)=\langle\lambda,\lambda+2\varrho\rangle$.)
\end{lemma}

\begin{proof}
Put $C_\varphi:=\sum_{a<b}\varphi(E_{ab})d\pi(E_{ab})$.

(1) $C_\varphi$ commutes with every $U_O$: $U_OC_\varphi U_O^\dagger=\sum_{a<b}\varphi(\mathrm{Ad}_OE_{ab})d\pi(\mathrm{Ad}_OE_{ab})=C_\varphi$, because $\{\mathrm{Ad}_OE_{ab}\}_{a<b}$ is again an orthonormal basis of $\mathfrak{so}(2m)$ and $\sum_i\varphi(x_i)d\pi(x_i)$, the image of the identity tensor $\sum_ix_i\otimes x_i$, is the same for all orthonormal bases. Since $V$ is irreducible under the $U_O$, Schur's lemma gives $C_\varphi=c_\varphi I_V$ with $c_\varphi=\langle v|C_\varphi|v\rangle$.

(2) The identity tensor in the basis $\{T_j\}\cup\{E_\beta\}$ with dual basis $\{T_j\}\cup\{E_{-\beta}\}$ (by the normalisation of \S\ref{subsec:K-setting}, and $\langle T_j,E_\beta\rangle=0$ by invariance) gives, with the complex-linear extensions,
\[
C_\varphi=\sum_j\varphi(T_j)d\pi(T_j)+\sum_{\beta\in\Phi}\varphi(E_\beta)\,d\pi(E_{-\beta}).
\]

(3) Apply to $v$. $d\pi(T_j)v=\lambda_jv$. For $\beta\in-\Phi^+$, $-\beta\in\Phi^+$ and $d\pi(E_{-\beta})v=0$. For $\beta\in\Phi^+$, $\varphi(E_\beta)v=0$ since $\lambda+\beta$ is not a weight of $V$, so $\varphi(E_\beta)d\pi(E_{-\beta})v=[\varphi(E_\beta),d\pi(E_{-\beta})]v$, and by the infinitesimal equivariance
\[
[\varphi(E_\beta),d\pi(E_{-\beta})]=-[G(E_{-\beta}),\varphi(E_\beta)]=-2i\varphi([E_{-\beta},E_\beta])=2i\varphi([E_\beta,E_{-\beta}]).
\]
Now $[E_\beta,E_{-\beta}]\in\mathfrak t_{\C}$, and invariance of the form gives $\langle[E_\beta,E_{-\beta}],T_j\rangle=\tfrac{\beta_j}{2i}$, so $[E_\beta,E_{-\beta}]=\tfrac1{2i}X_\beta$ and $2i\varphi([E_\beta,E_{-\beta}])=\varphi(X_\beta)$. Therefore
\[
c_\varphi=\langle v|C_\varphi|v\rangle=\sum_j\lambda_j\langle v|\varphi(T_j)|v\rangle+\sum_{\beta\in\Phi^+}\langle v|\varphi(X_\beta)|v\rangle=\lambda\cdot\xi_\varphi+\sum_{\beta\in\Phi^+}\beta\cdot\xi_\varphi=(\lambda+2\varrho)\cdot\xi_\varphi,
\]
using Lemma~\ref{lem:K1} ($\langle v|\varphi(X_\beta)|v\rangle=\langle X_{\xi_\varphi},X_\beta\rangle=\xi_\varphi\cdot\beta$). Taking the trace of $C_\varphi=c_\varphi I_V$ gives $\langle\varphi,d\pi\rangle_{\mathcal E}=\Tr C_\varphi=n\,c_\varphi$.
\end{proof}

\begin{proof}[Proof of Proposition~\ref{prop:K4}]
$D:=Y_\xi-d\pi\in\mathcal E$ by Lemma~\ref{lem:K2}(a). For every $\varphi\in\mathcal E$, Lemmas~\ref{lem:K2}(b) and~\ref{lem:K3} give $\langle\varphi,D\rangle_{\mathcal E}=n\,\xi\cdot\xi_\varphi-n\,(\lambda+2\varrho)\cdot\xi_\varphi=0$ when $\xi=\lambda+2\varrho$ (or under the stated orthogonality). With $\varphi=D$: $\langle D,D\rangle_{\mathcal E}=\sum_{a<b}\Tr(D(E_{ab})^2)=0$, so $D(E_{ab})=0$ for all $a<b$, i.e.\ $D=0$.
\end{proof}

\subsection{The $SU(2)$ case and a worked example}

\paragraph{The $SU(2)$ case of Remark~\ref{rem:K-SU2}.} Let $|\mathbf n\rangle$ be the spin-$j$ coherent state along $\mathbf n$ and $d\mathbf n$ the normalised invariant measure on the sphere, so $(2j+1)\int|\mathbf n\rangle\langle\mathbf n|\,d\mathbf n=I$ and $\langle\mathbf n|\mathbf J|\mathbf n\rangle=j\mathbf n$ (the Q-symbol). The vector operator $\mathbf A:=(2j+1)\int\mathbf n\,|\mathbf n\rangle\langle\mathbf n|\,d\mathbf n$ is $\kappa\mathbf J$ by the Wigner--Eckart theorem, and $\Tr(J_zA_z)=(2j+1)j\int n_z^2\,d\mathbf n=(2j+1)j/3$ against $\kappa\Tr J_z^2=\kappa j(j+1)(2j+1)/3$ gives $\kappa=1/(j+1)$: $(2j+1)\int(j+1)\mathbf n\,|\mathbf n\rangle\langle\mathbf n|\,d\mathbf n=\mathbf J$. The ratio $(j+1)/j$ of the symbols is that of $\langle\mathbf J^2\rangle$ to $|\langle\mathbf J\rangle|^2$. For $\mathrm{Spin}(2m)$, $\dim\mathcal E$ can exceed one (it is two at $m=2$, $\lambda=(2,0)$, one map per factor of $\mathfrak{so}(4)=\mathfrak{su}(2)\oplus\mathfrak{su}(2)$), which is why Lemma~\ref{lem:K3} tests against every $\varphi\in\mathcal E$.

\begin{remark}[a worked example: two modes, two copies]\label{rem:K-m2}
For $m=2$, $\mathfrak{so}(4)=\mathfrak{su}(2)_{\rm L}\oplus\mathfrak{su}(2)_{\rm R}$, spanned by the pairing operators $a_1^\dagger a_2^\dagger$, $a_2a_1$, $L_z:=\tfrac12(n_1+n_2-1)$ and by the hopping operators $a_1^\dagger a_2$, $a_2^\dagger a_1$, $R_z:=\tfrac12(n_1-n_2)$ ($[a_1^\dagger a_2^\dagger,a_2a_1]=2L_z$, $[a_1^\dagger a_2,a_2^\dagger a_1]=2R_z$). In weight coordinates $(L_z,R_z)=\big(\tfrac{\mu_1+\mu_2}4,\tfrac{\mu_1-\mu_2}4\big)$, the positive roots $2e_1\pm2e_2$ raise $L_z$, respectively $R_z$, by one, and $2\varrho=(4,0)$ becomes $(1,1)$: on a component of spins $(j_{\rm L},j_{\rm R})$ the shift is $(j_{\rm L},j_{\rm R})\mapsto(j_{\rm L}+1,j_{\rm R}+1)$, and Proposition~\ref{prop:K4} at $m=2$ is Remark~\ref{rem:K-SU2} in each factor. One copy: $S_+=\mathrm{span}\{|00\rangle,|11\rangle\}=(\tfrac12,0)$, $S_-=\mathrm{span}\{|10\rangle,|01\rangle\}=(0,\tfrac12)$, $\lambda=(1,\pm1)$, $C_2=6=p$; $\lambda+2\varrho=(5,\pm1)$ has squared norm $26$, and $\xi^*=(3,\pm3)$, i.e.\ $(\tfrac32,0)$ on $S_+$, has $18=p(2m-1)$: matchgate shadows, with the qubit factor $3$. Two copies: $S\otimes S=2\,(0,0)\oplus(1,0)\oplus(0,1)\oplus2\,(\tfrac12,\tfrac12)$, and per block, at $I/16$:
\begin{center}\small
\begin{tabular}{@{}lccccc@{}}
\toprule
$\lambda$ & $(j_{\rm L},j_{\rm R})$ & $P(\lambda)$ & $C_2$ & $\lambda+2\varrho$; $\|\cdot\|^2/t^2$ & $\xi^*$; $\|\cdot\|^2/t^2$\\
\midrule
$(0,0)$ & $(0,0)$ & $1/8$ & $0$ & $(4,0)$; $4$ & $0$; $0$\\
$(2,\pm2)$ & $(1,0)$, $(0,1)$ & $3/16$ each & $16$ & $(6,\pm2)$; $10$ & $(4,\pm4)$; $8$\\
$(2,0)$ & $(\tfrac12,\tfrac12)$ & $1/2$ & $12$ & $(6,0)$; $9$ & $(6,0)$; $9$\\
\bottomrule
\end{tabular}
\end{center}
The averages are $35/4$ and $15/2$ per block, $35/2$ and $15$ per copy, against $18$ for matchgate shadows. A singlet outcome carries no information ($O$ is uniformly random and $OX_{(4,0)}O^T/2$ is noise of mean zero, which $\xi^*=0$ removes); on $(1,0)$ the generators of $\mathfrak{su}(2)_{\rm R}$ vanish, and $\xi^*=(4,4)$, i.e.\ $(j_{\rm L}+1,0)$, keeps only the P-symbol of the active factor. The unshifted readout shrinks by $j/(j+1)$, $\tfrac13$ on $(\tfrac12,\tfrac12)$ and $\tfrac12$ on $(1,0)$; no single constant corrects both. With $F_{\max}(2,2)=\tfrac38\cdot8+\tfrac12\cdot4=5$ \cite[Cor.~3.7]{Rub26b} the floor of \S\ref{subsec:K-why} is $2p^2/5=14.4$ per copy, against $p^2/m=18$ for single copies: the ratios $35/36$, $5/6$, $4/5$ of Section~\ref{sec:discussion} are $17.5$, $15$, $14.4$ over $18$.
\end{remark}

\subsection{Proof of Theorem~\ref{thm:K}(i) and (ii)}\label{subsec:K-i-ii}
\begin{proof}[Proof of Theorem~\ref{thm:K}(i)]
Let $\sigma$ be any state on $S^{\otimes t}$, and write $Y^{(\lambda,i)}_\xi$ for the map $Y_\xi$ of Lemma~\ref{lem:K2} with $V=V_{\lambda,i}$. By the outcome density of \S\ref{subsec:K-povm} and Proposition~\ref{prop:K4},
\begin{align*}
\E_\sigma[\widehat\Gamma_{ab}]&=\tfrac1t\sum_{\lambda,i}n_\lambda\int\langle v_{\lambda,i}|U_O^\dagger\sigma U_O|v_{\lambda,i}\rangle\,\langle\mathrm{Ad}_OX_{\lambda+2\varrho},E_{ab}\rangle\,dO=\tfrac1t\sum_{\lambda,i}\Tr\big(\sigma\,Y^{(\lambda,i)}_{\lambda+2\varrho}(E_{ab})\big)\\
&=\tfrac1t\sum_{\lambda,i}\Tr(\sigma\,\Pi_{\lambda,i}G_{ab}\Pi_{\lambda,i})=\tfrac1t\Tr(\sigma G_{ab}),
\end{align*}
since $G_{ab}$ commutes with every $\Pi_{\lambda,i}$ and $\sum_{\lambda,i}\Pi_{\lambda,i}=I$. For $\sigma=\rho^{\otimes t}$, $\Tr(\rho^{\otimes t}G_{ab})=t\,\Tr(\rho B_{ab})=t\Gamma_{ab}$.
\end{proof}

\begin{proof}[Proof of Theorem~\ref{thm:K}(ii)]
$\sum_{a<b}\widehat\Gamma_{ab}^2=t^{-2}\sum_{a<b}\langle\mathrm{Ad}_OX_{\lambda+2\varrho},E_{ab}\rangle^2=t^{-2}\|\mathrm{Ad}_OX_{\lambda+2\varrho}\|^2=t^{-2}\|\lambda+2\varrho\|^2$ (Parseval, orthogonality of $\mathrm{Ad}_O$). Since $0\le\lambda_j\le t$ for $j<m$, $|\lambda_m|\le t$ and $(2\varrho)_j=4(m-j)$, $(\lambda_j+4(m-j))^2\le(t+4(m-j))^2$ for every $j$ (for $j=m$ both sides are $\le t^2$), so $\|\lambda+2\varrho\|^2\le\sum_{i=0}^{m-1}(t+4i)^2=V(m,t)$. Expanding,
\[
V=mt^2+8t\sum_{i<m}i+16\sum_{i<m}i^2=mt^2+4tm(m-1)+\tfrac83(m-1)m(2m-1)\le mt^2+4m^2t+\tfrac{16}3m^3,
\]
and for $t\le m$, $V\le(1+4+\tfrac{16}3)m^3<11m^3$.
\end{proof}

\subsection{The Fisher budget: the shortfall and the step to the rate}

The block budget is close to $mt^2$ for $t\ll m$ \cite[Thm.~3.8]{Rub26b}, by the mechanism of \eqref{eq:K-fluct}. The average Casimir is exactly $2^{-mt}\Tr\sum_{a<b}G_{ab}^2=tp$ (cross terms between copies are traceless), which with $\|\lambda\|^2\le mt^2$ keeps the zero-point part $4\lambda\cdot\delta$ on average within $mt(t-1)$ of its maximum $2tm(m-1)$; a unit shortfall $t-\lambda_j$ ($j<m$) costs $4(m-j)$ in the zero-point part but at most $2t$ in $\|\lambda\|^2$, so on average $\sum_{j<m}(t-\lambda_j)=O(t\sqrt{mt})$ and $\|\lambda\|^2$ falls short of $mt^2$ by $O(t^2\sqrt{mt})$ \cite[proof of Thm.~3.8]{Rub26b}.

\paragraph{From the budget to the rate.} The lower bound \cite[Thm.~3.4]{Rub26b} extends the budget to $2mt^2$ on a ball around $\theta=0$ (viewing $t$ copies as one Gaussian state on $mt$ modes with the parameter repeated $t$ times, \cite[Lemmas~3.1 and~3.2]{Rub26b}), so $n$ copies carry at most $2mkn$ \cite[Lemma~4.1]{Rub26}, and applies the van Trees inequality (Proposition~\ref{prop:bg-vantrees}) with a prior of radius $r=120\sqrt3\,\eps$: $p^2/(2mkn+\pi^2p^2/r^2)\le\E\|\hat\theta-\theta\|_2^2\le0.08\,r^2$ for a learner to trace distance $\eps$ amplified as in \cite[Lemma~6.3]{Rub26}, which uses $n=99N$ copies; so $N\gtrsim p^2/(mk\eps^2)$. A budget $mt^{1+\alpha}$ would give $m^3/(k^\alpha\eps^2)$. The collective bound \cite[Thm.~3.11]{Rub26b} is the same argument with the quantum Fisher information, of trace at most $p$ per copy.

\subsection{Proof of (iii): the Frobenius rate}\label{subsec:K-iii}
\begin{proof}[Proof of Theorem~\ref{thm:K}(iii)]
Write $V:=V(m,t)$.

\emph{Upper bound.} The $B$ block estimates are independent and identically distributed with mean $\Gamma$ (by (i)) and $\E\sum_{a<b}(\widehat\Gamma_{ab}-\Gamma_{ab})^2\le\E\sum_{a<b}\widehat\Gamma_{ab}^2\le V/t^2$ (by (ii)). Hence $\E\|\bar\Gamma-\Gamma\|_F^2=2\sum_{a<b}\mathrm{Var}(\bar\Gamma_{ab})\le2V/(t^2B)=2V/(tN)$, and Markov's inequality gives $\|\bar\Gamma-\Gamma\|_F^2\le6V/(tN)$ with probability $\ge2/3$; so $N\ge6V(m,t)/(t\eps^2)$ suffices. A $k$-copy protocol may use blocks of any size $t\le k$; take $t=\min\{k,m\}$, so $V\le11m^3$ and $N=t\lceil6V/(t^2\eps^2)\rceil\le66\max\{m^3/k,m^2\}/\eps^2+t\le67\max\{m^3/k,m^2\}/\eps^2$. (Confidence $1-\eta$ at cost $O(\log(1/\eta))$: put $r=\eps/3$ and run $B=\lceil8\ln(1/\eta)\rceil$ independent batches of $N_0=t\lceil8V(m,t)/(t^2r^2)\rceil$ copies each, each within $r$ of $\Gamma$ with probability $\ge3/4$ by Markov's inequality, and output any batch estimate within $2r$ of more than half of them: the median trick of \S\ref{subsec:bg-estimation}, as in \cite[Thm.~7.6]{Rub26}.)

\emph{Lower bound.} Let $\mathcal P$ be a $k$-copy protocol with $\|\widehat\Gamma-\Gamma\|_F\le\eps$ with probability $\ge2/3$ on all mixed fermionic Gaussian states, $\eps\le1/3600$. Compose it with the replacement of $\widehat\Gamma$ by the antisymmetric part of its real part (an orthogonal projection onto a subspace containing $\Gamma$, so the Frobenius error does not increase), then with the Frobenius projection $\tilde\Gamma$ onto $\mathcal G_{1/2}$ and the output $\rho_{\tilde\Gamma}$. The hard family of \cite[Thm.~3.4]{Rub26b} consists of states $\rho_\theta$ \cite[Eq.~(1)]{Rub26b} with $\|\theta\|_2\le120\sqrt3\,\eps\le0.058$, whose normal-form values are at most $\tanh(0.058)<1/2$, so it lies in $\mathcal G_{1/2}$; on it, projection does not increase the Frobenius error ((T3) of \S\ref{subsec:furthertools}) and (T1) gives $\|\rho_{\tilde\Gamma}-\rho_\Gamma\|_1\le K_{1/2}\eps$, $K_{1/2}=(2\cdot\tfrac12\cdot\tfrac32)^{-1/2}<1$, i.e.\ trace distance $<\eps/2\le\eps$. The composed protocol is a $k$-copy protocol learning the hard family to trace distance $\eps$ with probability $\ge2/3$, so by \cite[Thm.~3.4]{Rub26b} (which, as stated there, holds on every class containing its hard family) $N>m^3/(8{,}553{,}600\,k\eps^2)$, and by \cite[Thm.~3.11]{Rub26b} (every protocol) $N>m^2/(4{,}276{,}800\,\eps^2)$.
\end{proof}

\subsection{Proof of (iv): trace distance on \texorpdfstring{$\mathcal G_c$}{G\_c}}\label{subsec:K-iv}
\begin{proof}[Proof of Theorem~\ref{thm:K}(iv)]
Write $V:=V(m,t)$.

\emph{Upper bound.} The Frobenius projection onto the closed convex set $\mathcal G_c$ fixes $\Gamma$ and is nonexpansive ((T3)): $\|\tilde\Gamma-\Gamma\|_F\le\|\bar\Gamma-\Gamma\|_F$. By (T1) ($\chi^2$ continuity \cite[Prop.~7.1]{Rub26}, derived after Lemma~\ref{lem:A-chi2}), $\|\rho_{\tilde\Gamma}-\rho_\Gamma\|_1\le K_c\|\tilde\Gamma-\Gamma\|_F$ with $K_c=(2c(2-c))^{-1/2}\le(2c)^{-1/2}$. By \S\ref{subsec:K-iii}, $N\ge6V/(t\cdot2c\eps^2)=3V/(tc\eps^2)$ copies give $\|\bar\Gamma-\Gamma\|_F^2\le2c\eps^2$ with probability $\ge2/3$, hence $\|\rho_{\tilde\Gamma}-\rho_\Gamma\|_1\le\eps$. With $t=\min\{k,m\}$ and $N=t\lceil3V/(t^2c\eps^2)\rceil$: $N\le34\max\{m^3/k,m^2\}/(c\eps^2)$.

\emph{Lower bound.} For $\eps\le\eps_0(c)$ we have $\tanh(120\sqrt3\,\eps)\le1-c$, so the hard family of \cite[Thm.~3.4]{Rub26b} lies in $\mathcal G_c$ (as stated in that theorem) and $N>m^3/(8{,}553{,}600\,k\eps^2)$ for any $k$-copy protocol learning $\mathcal G_c$ to trace distance $\eps$; \cite[Thm.~3.11]{Rub26b} gives $N>m^2/(4{,}276{,}800\,\eps^2)$. (Our upper bound is stated for $\|\cdot\|_1\le\eps$, i.e.\ trace distance $\eps/2$; the factor $2$ is absorbed in the constants.)
\end{proof}

\subsection{Proof of (v), and the crossover in \texorpdfstring{$c$}{c}}\label{subsec:K-v}

Near the vacuum blocks gain only $\sqrt k$. The lower bound in (v) uses $\rho^{\rm F}_w=\Gamma_{\rm a}(w/2)/\det(I+w/2)$, with $w\preceq2I/m$ a density matrix on $\C^m$ and $\Gamma_{\rm a}$ fermionic second quantisation: fewer than $\tfrac12$ particles on average, with the one-particle sector in the state $w$ with probability above $0.3$ \cite[Lemma~5.1]{Rub26b}. Antisymmetrising a random number of copies of the qudit $w$ produces exact copies of $\rho^{\rm F}_w$, and the one-particle sector of a good Gaussian estimate is a good estimate of $w$ \cite[Lemmas~5.2--5.4]{Rub26b}; so a $k$-copy Gaussian learner yields a block learner for $m$-dimensional qudits near $I/m$, which must have $\sup\E\sum_jt_j\sqrt{\min\{t_j,m^2\}}=\Omega(m^3/\eps^2)$ \cite{CMMP26}, \cite[Lemma~2.1]{Rub26b}, whence $N\min\{\sqrt k,m\}=\Omega(m^3/\eps^2)$. Correspondingly, the block Fisher information of $t\le m^2$ qudit copies at $I/m$ is at most of order $m^2t^{3/2}$ \cite[Rem.~3.9]{Rub26b}. Heuristically, by Schur--Weyl duality $t\ll m^2$ copies of $I/m$ occupy mostly irreducible components of $U(m)$ with Young diagrams of row and column lengths $O(\sqrt t)$, so the analogue of $\|\lambda\|^2$ is of order $t^{3/2}$, not $t^2$.
\begin{proof}[Proof of Theorem~\ref{thm:K}(v)]
The near-vacuum family of \cite[Thm.~5.5]{Rub26b} ($\rho^{\rm F}_w$, $w\preceq2I/m$) consists of mixed fermionic Gaussian states, so any protocol for the unrestricted mixed class satisfies $N=\Omega(m^3/(\eps^2\min\{\sqrt k,m\}))$ for $m\ge m_0$, $\eps<\eps_0$ (the constants of that theorem); for $1<k<m^2$ this exceeds the rate of (iv) by the factor $\min\{\sqrt k,\,m/\sqrt k\}$ (for $k\ge m^2$ the two coincide in order). Theorem~\ref{thm:B} gives the single-copy, hence $k$-copy, upper bound $O(m^3/\eps^2)$. On the cold number-conserving classes \cite[Thm.~5.9]{Rub26b} gives $\Theta(m^3/(\eps^2\min\{\sqrt k,m\}))$, and on every $\mathcal G_c$ part (iv) gives $\Theta_c(\max\{m^3/k,m^2\}/\eps^2)$; the two are different functions of $k$, so the $k$-dependence differs between subclasses of the unrestricted class. Part (iii) locates the discrepancy: the covariance is learned at $m^3/k$ everywhere; what changes between the warm and cold regions is the cost of converting Frobenius accuracy into trace distance, which is $K_c\asymp c^{-1/2}$ on $\mathcal G_c$ by (T1) (the order $c^{-1/2}$ is necessary at $c=2/m$ \cite[Rem.~7.5]{Rub26}) and has no dimension-free bound near pure states \cite[Prop.~7.4]{Rub26}.
\end{proof}

\begin{remark}[the crossover in $c$]\label{rem:K-crossover}
\cite[Question~7.3]{Rub26b}, on the crossover in $c$, concerns single copies and is answered by Theorem~\ref{thm:B}: $\Theta(m^3/\eps^2)$, with an upper constant uniform in $c$ and an absolute lower constant \cite{Rub26} on the window $\eps\le\eps_0(c)$ of (iv). What remains open is its $k$-copy analogue, the $k$-dependence on $\mathcal G_c$ when $c=c(m)\to0$. For every $c=o(1)$ the upper constant $34/c$ of (iv) and the absolute lower constant separate. For $c\le1/(2m+1)$ the class $\mathcal G_c$ contains the local near-vacuum family $\{\rho^{\rm F}_w:\|w-I/m\|_\op\le L\eps/m\}$ of \cite[Thm.~5.5]{Rub26b} (its second sentence, with the constant $L$ of that theorem; for $L\eps\le\tfrac12$ every normal-form value is $(2-w_i)/(2+w_i)\le1-1/(2m+1)$, $w_i$ the eigenvalues of $w$). The lower bound there is therefore $\Omega(m^3/(\eps^2\min\{\sqrt k,m\}))$, while the best upper bound, $\min\{34\max\{m^3/k,m^2\}/c,\ C_B(m^3+\eps^2m^3)\}/\eps^2$ from (iv) and Theorem~\ref{thm:B}, does not match it. (The full family $w\preceq2I/m$ lies in no $\mathcal G_c$, $c>0$: a rank-deficient $w$ has pure modes.)
Physically, on $\mathcal G_c$ every minority occupation in the normal-form frame is at least $c/2$, so a state carries at least $mc/2$ excitations on average. For fixed $c$ this is extensive and the copies can be aligned in the $m$ Cartan directions: gain $k$. For $c\le1/(2m+1)$ it can be of order one: $\mathcal G_c$ then contains the near-vacuum family above, fewer than one particle in an unknown one-particle state, the qudit regime, where the gain is at most $\sqrt k$. It is open how the gain interpolates between the two.
\end{remark}

\section{Proofs for Section~\ref{sec:bosonic}}\label{app:proofs-Bos}

Sections~\ref{subsec:Bos-coldframe}--\ref{subsec:Bos-CS} (a rotation estimate and its leakage) serve Theorem~\ref{thm:Bos-sep}(i) and Proposition~\ref{prop:NG-coldhot-k}; Sections~\ref{subsec:Bos-schur}--\ref{subsec:Bos-budget} (deletion of the cold--warm cross block and a split chosen from the data) serve Theorems~\ref{thm:Bos}(i) and~\ref{thm:Bos-energy}.

\subsection{Conventions: derivations and imported results}\label{app:Bos-conv}

\paragraph{Results used.}
Four of the results listed in \S\ref{subsec:Bos-conventions} are stated here, with the reason each holds; the others are stated where they are used: \cite[Thms.~2, 5 and~9]{CMMP26} in \S\ref{subsec:Bos-bands}, \cite[Thm.~4]{KL17} in the proof of Lemma~\ref{lem:Bos-effrank}, \cite[Thm.~4.7.1]{Ver18} in the proof of Lemma~\ref{lem:Bos0}(i), \cite{BJL19} in \eqref{eq:bg-BW} and \S\ref{subsec:Bos-BW}, \cite{PW94} in \S\ref{subsec:Bos-CS}, and \cite{Saf19} with Proposition~\ref{prop:Bos-QFI}.
\begin{itemize}
\item \cite[Lemma~5.8]{Rub26b}: (i) the particle-number law of $\rho_X$ has $P_0=1/Z(X)$ and $P_1=\Tr X/Z(X)$, and its one-particle block is $P_0X$ (derived in the next paragraph); (ii) for bosonic fugacities $X,X'$ with $\|X\|_\op,\|X'\|_\op\le1-\delta'$, $\|\rho_X-\rho_{X'}\|_1\le(2/\delta')\|X-X'\|_1$. Sketch of (ii): along $X_u=X+u(X'-X)$, differentiate $\rho_{X_u}=\Gamma(X_u)/Z(X_u)$ in each photon-number sector. The derivative of $\mathrm{Sym}^\ell(X_u)$ splits into a positive and a negative part, each a compression of a positive operator, so its trace norm is a trace, and these traces sum to $Z\,\Tr((I-X_u)^{-1}|X'-X|)$; with $\frac{d}{du}\log Z=\Tr((I-X_u)^{-1}(X'-X))$ this gives $\|\rho_X-\rho_{X'}\|_1\le2\sup_u\|(I-X_u)^{-1}\|_\op\|X-X'\|_1$, with no factor of the number of modes.
\item \cite[Thm.~5.9 at $k=1$, and the error step of Thm.~5.13]{Rub26b}: the photon-number readout of $s$ cold modes. On each copy the total photon number of the $s$ modes is measured; outcome $0$ has probability $a_0$, and outcome $1$ leaves the copy in its normalised one-photon block, an $s$-level density matrix. Stage~1 estimates the trace $P_1/P_0$ of the vacuum-normalised one-photon block from the frequencies of the outcomes $0$ and $1$; Stage~2 runs single-copy tomography of the $s$-level state on the copies that returned one photon, to accuracy proportional to $\eps$ divided by that trace, at cost $O(s^3/\eps^2)$ (the qudit protocol cited in \cite[Thm.~5.9]{Rub26b}); the total is $O_K((1+s^3)/\eps^2)$, and the output is the passive state whose fugacity is the estimated block. What it uses about the measured state is quoted in \S\ref{subsec:Bos-proof-i}.
\item \cite[Lemma~5.11(i)]{Rub26b}: $P=(SS^T)^{1/2}=e^G$ is symmetric, positive and symplectic, $\Sigma=P\Sigma_pP$ with $\Sigma_p$ passive, $P^2=\mathcal Q^{-1}\Sigma$ with $\mathcal Q=(-(\Sigma\Omega)^2)^{1/2}$, and $\|P\|_\op^2\le2\|\Sigma\|_\op$. Sketch: in the polar form $S=PK$ of the Williamson matrix both factors are symplectic (Appendix~\ref{app:bg-williamson}), so $K=\Phi_{\mathrm{re}}(u)$ is passive and $\Sigma=P(KDK^T)P=P\Sigma_pP$; from $\Sigma\Omega=SD\Omega S^{-1}$, $-(\Sigma\Omega)^2=SD^2S^{-1}$ and $\mathcal Q=SDS^{-1}$, which gives $P^2=\mathcal Q^{-1}\Sigma$ and shows that $P$ is a function of $\Sigma$ alone, although $S$ is not; finally $\Sigma\succeq\tfrac12P^2$.
\item \cite[Lemma~5.12]{Rub26b}: $T(\rho,e^{-iH}\rho e^{iH})\le\sqrt{\operatorname{Var}_\rho(H)}$, which is Lemma~\ref{lem:A-MTgen}, whose proof applies verbatim to unbounded $H$ of finite variance; for a quadratic generator $\tfrac12R^ThR$ on a centred Gaussian state the variance is $\tfrac12\Tr((h\Sigma)^2)+\tfrac18\Tr((h\Omega)^2)$ (\S\ref{subsec:bg-bosons}), and for the displacement generator $v^T\Omega R$ it is $(\Omega v)^T\Sigma(\Omega v)$, so $T(\rho,D(v)\rho D(v)^\dagger)\le\|\Sigma\|_\op^{1/2}\|v\|_2$.
\end{itemize}

\paragraph{Passive states.} We derive the facts about passive states stated in \S\ref{subsec:Bos-conventions}. In an eigenbasis of $Q$, $\rho_Q=\bigotimes_i(1-x_i)x_i^{\hat n_i}$ with $x_i=q_i/(1+q_i)$, which is the displayed number law; equivalently $\rho_Q=\Gamma(X)/Z(X)$ with $X=Q(I+Q)^{-1}=\exp[-\log(I+Q^{-1})]$, $\Gamma(X)=\bigoplus_{\ell\ge0}\mathrm{Sym}^\ell(X)$ the bosonic second quantisation (\S\ref{subsec:bg-reptheory}) and $Z(X)=\Tr\Gamma(X)=\det(I-X)^{-1}=\det(I+Q)$. Since $\Gamma(X)$ is $1$ on the vacuum and $X$ on the one-photon space $\C^m$, the vacuum probability is $P_0=1/Z$, the one-photon block is $P_0X$ and $P_1=\Tr X/Z$; this is \cite[Lemma~5.8(i)]{Rub26b}. For the covariance, $\langle a_ia_j\rangle=0$ and $\langle a_j^\dagger a_i\rangle=Q_{ij}$ give, by a direct computation in the quadratures, $\Sigma=\Phi_{\mathrm{re}}(Q+\tfrac12I)$. Conversely, if the mean is zero and $[\Sigma,\Omega]=0$, then $\Sigma=\Phi_{\mathrm{re}}(M)$ with $M$ Hermitian; diagonalising $M=u\operatorname{diag}(d)u^\dagger$ gives a Williamson form $\Sigma=\Phi_{\mathrm{re}}(u)\operatorname{diag}(d,d)\Phi_{\mathrm{re}}(u)^T$ with an orthogonal symplectic matrix, so the state is $\Gamma(u)$ applied to a product of thermal states, that is $\rho_Q$ with $Q=M-\tfrac12I$.

\paragraph{Heterodyne detection: the outcome law.} For the POVM $\{\pi^{-m}|\alpha\rangle\langle\alpha|\,d^{2m}\alpha\}$ of \S\ref{subsec:bg-bosons} the outcome $z=\alpha$ has the Husimi density $\pi^{-m}\langle\alpha|\rho|\alpha\rangle$, and $Y=\sqrt2(\operatorname{Re}z,\operatorname{Im}z)$. Because $a|\alpha\rangle=\alpha|\alpha\rangle$ and $\pi^{-m}\int|\alpha\rangle\langle\alpha|\,d^{2m}\alpha=I$,
\begin{equation}\label{eq:Bos-het-moments}
\int\alpha_i\bar\alpha_j\,|\alpha\rangle\langle\alpha|\,\frac{d^{2m}\alpha}{\pi^m}=a_ia_j^\dagger,\qquad\int\alpha_i\alpha_j\,|\alpha\rangle\langle\alpha|\,\frac{d^{2m}\alpha}{\pi^m}=a_ia_j,\qquad\int\alpha_i\,|\alpha\rangle\langle\alpha|\,\frac{d^{2m}\alpha}{\pi^m}=a_i :
\end{equation}
heterodyne detection measures anti-normally ordered moments. Hence $\E z=\Tr(\rho a)$ and, for a centred state, $\E z_i\bar z_j=\Tr(\rho\,a_ia_j^\dagger)=\delta_{ij}+Q_{ij}$ and $\E z_iz_j=\Tr(\rho\,a_ia_j)=\mathsf P_{ij}$. Passing to real coordinates converts these into $\E(Y-v)(Y-v)^T=\Phi_{\mathrm{re}}(I+Q)$ plus the pseudocovariance terms, which is $\Sigma+\tfrac12I$ by the covariance identity above (for a passive state; in general by the same computation). That the law is Gaussian follows because the Husimi function is the Wigner function, here the Gaussian $N(v,\Sigma)$, convolved with the vacuum Wigner function $N(0,\tfrac12I)$.

\subsection{The passive metric}\label{app:Bos-QFI}

\begin{proof}[Proof of Proposition~\ref{prop:Bos-QFI}]
For a diagonal tangent, differentiating $\log p_{\boldsymbol s}$ gives
\[
L=\sum_i\frac{H_{ii}}{q_i(1+q_i)}(\hat n_i-q_i),
\]
which is the displayed formula because $d_{ii}=2q_i(1+q_i)$. For $i\ne j$ with $q_i\ne q_j$, realise an off-diagonal tangent by $Q(t)=e^{tA}Qe^{-tA}$, where $A$ is anti-Hermitian and $A_{ij}=H_{ij}/(q_j-q_i)$. Its state derivative is $[a^\dagger Aa,\rho_Q]$. A matrix element connecting $|\boldsymbol s\rangle$ to $|\boldsymbol s+e_i-e_j\rangle$ has thermal probability ratio $x_i/x_j$, where $x_i=q_i/(1+q_i)$. Solving $\dot\rho=(\rho L+L\rho)/2$ on this matrix element gives the coefficient
\[
2A_{ij}\frac{x_j-x_i}{x_j+x_i}=\frac{2H_{ij}}{q_i+q_j+2q_iq_j}.
\]
Within a degenerate eigenspace of $Q$, first diagonalise the corresponding block of $H$ by a fixed unitary; the diagonal calculation then applies, and transforming back gives the same superoperator formula. Tangent directions add linearly. Finally the thermal Wick identity \eqref{eq:Bos-wickvar} (proof of Lemma~\ref{lem:Bos3}) gives $\operatorname{Var}_{\rho_Q}(a^\dagger Ba)=\sum_{ij}|B_{ij}|^2q_i(1+q_j)=\frac12\sum_{ij}d_{ij}|B_{ij}|^2$. Substituting $B_{ij}=2H_{ij}/d_{ij}$ proves \eqref{eq:Bos-QFI}. The finite number of modes and the finite positive occupations make these second moments finite.
\end{proof}

\subsection{A one-sided comparison and further heterodyne estimates}

\begin{proposition}[One-sided comparison]\label{prop:Bos-onesided}
Let $Q\succ0$, $M\succ0$ and $M\succeq\alpha Q$ for some $0<\alpha\le1$. Put $H=M-Q$. Then
\begin{equation}\label{eq:Bos-onesided}
T(\rho_Q,\rho_M)\le\min\Big\{1,\frac{W_Q(H)}{\sqrt2\,\alpha}\Big\}.
\end{equation}
In particular, $M\succeq Q$ gives the constant $1/\sqrt2$, independent of the number of modes and of the extreme occupations.
\end{proposition}

\begin{proof}
Along $Q_t=(1-t)Q+tM$, $Q_t\succeq\alpha Q$. On vectorised matrices the superoperator $\mathcal D_Q$ is $I\otimes Q+Q^T\otimes I+2Q^T\otimes Q$. Tensor products preserve positive order in each positive factor; in particular
\[
Q_t^T\otimes Q_t-\alpha^2Q^T\otimes Q=(Q_t^T-\alpha Q^T)\otimes Q_t+\alpha Q^T\otimes(Q_t-\alpha Q)\succeq0 .
\]
The linear terms dominate $\alpha$ times their counterparts, hence also $\alpha^2$ times them. Therefore $\mathcal D_{Q_t}\succeq\alpha^2\mathcal D_Q$ and $\mathcal D_{Q_t}^{-1}\preceq\alpha^{-2}\mathcal D_Q^{-1}$. In \eqref{eq:Bos-path}, $\dot Q_t=H$. $Q$ and $M$ need not commute.
\end{proof}

The hypothesis $Q\succ0$ is needed: \eqref{eq:Bos-WQ} has zero denominators on vacuum modes, so a cold block requires a different estimate.
Adding noise moves the state away from the pure boundary, where the metric blows up, so the yardstick at $Q$ remains valid along the whole path; a two-sided error can move towards the boundary (only $Q_t\succeq(1-t)Q$ holds), and \eqref{eq:Bos-clip} below pays for it the factor $1+\bar q$. The inflated heterodyne estimator of \cite[Thm.~8]{CMMP26} uses the same one-sided mechanism.

For the cold--warm pairs of the heterodyne risk sum \eqref{eq:Bos-risk}, the bound \eqref{eq:Bos-scalar} gives an upper bound by a constant times $m_cm_w+m_c\sum_bm_b/\nu_b$, even if the cold occupations vanish. When $\nu_{\min}\ge a/m_c$ for a fixed $a>0$,
\begin{equation}\label{eq:Bos-cross-absorb}
m_c\sum_b\frac{m_b}{\nu_b}\le m_c\Big(\sum_b\frac{m_b^2}{\nu_b}\Big)^{1/2}\Big(\sum_b\nu_b^{-1}\Big)^{1/2}\le\sqrt{2/a}\,\sqrt{m_c^3}\,\sqrt{\sum_bm_b^2/\nu_b},
\end{equation}
using $\sum_b\nu_b^{-1}\le2/\nu_{\min}$ for distinct dyadic scales; so the cross-pair cost is absorbed by $m_c^3+\sum_bm_b^2/\nu_b+m^2$.

\paragraph{Noncircular data.}
The same estimator works with noncircular complex Gaussian data. If $z$ is centred with covariance $I+Q$ and pseudocovariance $\mathsf P$, the unbiased estimate $X=L^{-1}\sum_lz_lz_l^\dagger-I$ obeys, in the eigenbasis of $Q$,
\begin{equation}\label{eq:Bos-noncirc}
\E|(X-Q)_{ij}|^2=\frac{(1+q_i)(1+q_j)+|\mathsf P_{ij}|^2}L\le\frac{2(1+q_i)(1+q_j)}L .
\end{equation}
The identity is the complex Isserlis formula for a centred vector, $\E[z_i\bar z_j\bar z_iz_j]=|\E z_i\bar z_j|^2+\E|z_i|^2\E|z_j|^2+|\mathsf P_{ij}|^2$; the identity matrix cancels exactly, $X-Q$ being the sample covariance minus its mean; and the last step is the Cauchy--Schwarz inequality $|\mathsf P_{ij}|^2=|\E z_iz_j|^2\le\E|z_i|^2\E|z_j|^2$. This handles residual squeezing without treating a slightly squeezed state as exactly passive. With an unknown mean, the pair differences of \S\ref{subsec:Bos-conventions} have the same covariance and pseudocovariance.

\subsection{Bures--Wasserstein distance, positive-part repair and coherent-state coupling}\label{subsec:Bos-BW}

For positive semidefinite $A,B$ let $d_{\mathrm{BW}}(A,B)$ be the Bures--Wasserstein distance \eqref{eq:bg-BW} of occupation matrices, over $\C$ with complex unitaries \cite{BJL19}; it is not the Bures distance between the corresponding Gaussian density operators. Its factor representation $d_{\mathrm{BW}}(A,B)=\min_U\|A^{1/2}-B^{1/2}U\|_F$ over unitaries $U$ follows by expanding the square and maximising the cross trace by the polar decomposition; composing optimal alignments gives the triangle inequality, and singular matrices follow by continuity. By the polar decomposition, $d_{\mathrm{BW}}(A,B)\le\|F-G\|_F$ whenever $FF^\dagger=A$ and $GG^\dagger=B$. If $M,Y\succeq0$ and $M^{1/2}YM^{1/2}\succeq M^2$, in particular if $Y\succeq M$, then operator monotonicity of the square root gives $\Tr(M^{1/2}YM^{1/2})^{1/2}\ge\Tr M$, hence
\begin{equation}\label{eq:Bos-BWdom}
d_{\mathrm{BW}}(M,Y)^2\le\Tr Y-\Tr M .
\end{equation}

\begin{lemma}[Positive-part repair]\label{lem:Bos-PSD}
For $Q\succ0$, any Hermitian $X=Q+H$, and $X_+$ its positive part,
\begin{equation}\label{eq:Bos-PSD}
d_{\mathrm{BW}}(Q,X_+)^2\le2\langle H,\mathcal S_Q^{-1}H\rangle=2\sum_{ij}\frac{|H_{ij}|^2}{q_i+q_j}.
\end{equation}
\end{lemma}

\begin{proof}
Let $A=A^\dagger$ solve $AQ+QA=H$, and put
\[
Y=(I+A)Q(I+A)=X+AQA\succeq X,\qquad Y\succeq0,\qquad e=\Tr(AQA)=\tfrac12\langle H,\mathcal S_Q^{-1}H\rangle .
\]
The two factors $Q^{1/2}$ and $(I+A)Q^{1/2}$ give $d_{\mathrm{BW}}(Q,Y)^2\le e$. Write $M=X_+$. Since $M^{1/2}XM^{1/2}=M^2$ and $Y\succeq X$, $M^{1/2}YM^{1/2}\succeq M^2$, and \eqref{eq:Bos-BWdom} gives $d_{\mathrm{BW}}(M,Y)^2\le\Tr Y-\Tr M=e-\Tr X_-\le e$. The triangle inequality gives $d_{\mathrm{BW}}(Q,M)\le2\sqrt e$, which is \eqref{eq:Bos-PSD}.
\end{proof}

\begin{lemma}[Coherent-state coupling]\label{lem:Bos-coupling}
Let $A,B\succeq0$ be occupation matrices, and let $(\alpha,\beta)$ be any pair of random vectors in $\C^m$ with $\alpha$ a proper complex Gaussian of covariance $A$ and $\beta$ a proper complex Gaussian of covariance $B$. Then
\begin{equation}\label{eq:Bos-coupling}
T(\rho_A,\rho_B)\le\big(\E\|\alpha-\beta\|^2\big)^{1/2}.
\end{equation}
In particular $T(\rho_A,\rho_B)\le d_{\mathrm{BW}}(A,B)$.
\end{lemma}

\begin{proof}
By the Glauber--Sudarshan representation of \S\ref{subsec:bg-bosons}, $\rho_A=\E|\alpha\rangle\langle\alpha|$ with $\alpha$ a proper complex Gaussian of covariance $A$; for singular $A$ the zero-variance directions are in the vacuum, the coherent state $|0\rangle$. For any joint law of $(\alpha,\beta)$ with these marginals, $\rho_A-\rho_B=\E(|\alpha\rangle\langle\alpha|-|\beta\rangle\langle\beta|)$, so convexity of the trace distance and the coherent-state overlap $|\langle\alpha|\beta\rangle|^2=e^{-\|\alpha-\beta\|^2}$ give
\[
T(\rho_A,\rho_B)\le\E\sqrt{1-e^{-\|\alpha-\beta\|^2}}\le\sqrt{\E\|\alpha-\beta\|^2}.
\]
For the last assertion couple $\alpha=A^{1/2}z$ and $\beta=B^{1/2}Uz$, with $z$ a standard proper complex Gaussian and $U$ the optimal unitary of the factor representation; then $\E\|\alpha-\beta\|^2=\|A^{1/2}-B^{1/2}U\|_F^2=d_{\mathrm{BW}}(A,B)^2$.
\end{proof}

Thus $X_+$ is an implementable physical reconstruction, and Lemmas~\ref{lem:Bos-PSD} and~\ref{lem:Bos-coupling} give
\begin{equation}\label{eq:Bos-clip}
T(\rho_Q,\rho_{X_+})^2\le2\sum_{ij}\frac{|(X-Q)_{ij}|^2}{q_i+q_j}.
\end{equation}
For $\|Q\|_\op\le\bar q$ the right side is at most $2(1+\bar q)W_Q(X-Q)^2$. The bound weights the \emph{raw} error and controls the \emph{state} error after clipping; the clipped error itself need not be smaller in the weighted norm.
Lemma~\ref{lem:Bos-coupling} needs a positive $P$-function with proper amplitudes; squeezing a near-vacuum mode below the vacuum level destroys it, which is one reason the squeezing is removed first.

\subsection{Removing displacement and squeezing}\label{subsec:Bos-undo}

Fix $K\ge1$ and let $\rho=\rho(v,\Sigma)$ with $\|\Sigma\|_\op\le K$. Throughout this subsection $K',K_V,K_1,K_2,K_4$ denote specific constants depending only on $K$ (polynomial in $K$).

\paragraph{Stage 0.}
Heterodyne $N_0=2n_0$ copies with real outcomes $Y_1,\dots,Y_{N_0}$. Let $\hat v$ be their mean, and
\[
\hat C:=\frac1{n_0}\sum_{i\le n_0}y_iy_i^T,\qquad y_i:=\frac{Y_{2i}-Y_{2i-1}}{\sqrt2}\sim N(0,\Sigma+\tfrac12I)\ \text{i.i.d.},
\]
\[
\hat\Sigma:=\hat C-\tfrac12I+\eta_0I,\qquad\eta_0:=K_V\sqrt{m/n_0}.
\]
If $\hat\Sigma+\tfrac i2\Omega\not\succeq0$ or $\|\hat\Sigma\|_\op>K+1$, output the vacuum and stop. Otherwise compute the squeezing factor $\hat P=e^{\hat G}$ of $\hat\Sigma$ \cite[Lemma~5.11(i)]{Rub26b}, and apply $V_0:=U_{\hat P}^\dagger D(\hat v)^\dagger$ to every later copy.

\begin{lemma}[Stage 0 and the squeezing undo for $\|\Sigma\|_\op\le K$]\label{lem:Bos0}
\begin{enumerate}
\item[(i)] (heterodyne) Suppose $n_0\ge2m$ and $n_0$ is large enough that $\eta_0\le\tfrac12$. With probability at least $0.97$,
\[
\|\hat v-v\|_2\le C_K\sqrt{m/n_0},\quad\|\hat C-(\Sigma+\tfrac12I)\|_F\le C_Km/\sqrt{n_0},\quad\|\hat C-(\Sigma+\tfrac12I)\|_\op\le K_V\sqrt{m/n_0}=\eta_0 .
\]
On this event $\hat\Sigma-\Sigma\succeq0$, so $\hat\Sigma+\tfrac i2\Omega\succeq\Sigma+\tfrac i2\Omega\succeq0$ ($\hat\Sigma$ is a valid covariance), $\|\hat\Sigma\|_\op\le K+2\eta_0\le K+1$, $\|\hat\Sigma-\Sigma\|_\op\le2\eta_0$ and $\|\hat\Sigma-\Sigma\|_F\le C_Km/\sqrt{n_0}$.
\item[(ii)] (the squeezing factor is Lipschitz in both norms) For covariances $\Sigma,\tilde\Sigma$ with $\|\Sigma\|_\op,\|\tilde\Sigma\|_\op\le K+1$ and squeezing factors $P,\tilde P$,
\[
\|P-\tilde P\|\le K_1\|\Sigma-\tilde\Sigma\|
\]
in the Frobenius and in the operator norm.
\item[(iii)] (the undone state) Let $\Sigma=P\Sigma_pP$ with $\Sigma_p=\Phi_{\mathrm{re}}(Q_p+\tfrac12I)$ passive \cite[Lemma~5.11(i)]{Rub26b}, so that the eigenvalues of $Q_p$ are the Williamson occupations of $\Sigma$. Let $\Sigma':=\hat P^{-1}\Sigma\hat P^{-1}$ (the covariance of the copies after $U_{\hat P}^\dagger$), and let $\hat P^{-1}P=P'O$ be the polar decomposition, with $P':=(\hat P^{-1}P^2\hat P^{-1})^{1/2}$ symmetric positive and $O$ orthogonal, both symplectic (the polar factors of a symplectic matrix are symplectic). Then $\Sigma'=P'\Sigma_{p'}P'$ with $\Sigma_{p'}:=O\Sigma_pO^T$ passive ($O=\Phi_{\mathrm{re}}(o)$ for some $o\in U(m)$, and passivity is preserved), whose occupation matrix $Q_{p'}=oQ_po^\dagger$ has the spectrum of $Q_p$. The residual squeezer satisfies, when $K_2\|\hat\Sigma-\Sigma\|_\op\le\tfrac14$,
\[
\|G'\|:=\|\log P'\|\le K_2\|\hat\Sigma-\Sigma\|
\]
in both norms. Consequently, with
\[
\rho':=U_{\hat P}^\dagger D(\hat v)^\dagger\rho\,D(\hat v)U_{\hat P}=D(v'')U_{P'}\rho_{Q_{p'}}U_{P'}^\dagger D(v'')^\dagger,\qquad v'':=\hat P^{-1}(v-\hat v),
\]
\begin{align*}
\theta_{\mathrm{sq}}:=T(\rho',\rho_{Q_{p'}})&\le\|\Sigma'\|_\op^{1/2}\|v''\|_2+\big(\tfrac12\|\Sigma_{p'}\|_\op^2+\tfrac18\big)^{1/2}\|G'\|_F\\
&\le K_4\big(\|v-\hat v\|_2+\|\hat\Sigma-\Sigma\|_F\big).
\end{align*}
A known passive rotation applied afterwards conjugates the residual displacement and squeezer into ones of the same norms.
\end{enumerate}
\end{lemma}

\begin{proof}
(i) Three applications of Markov's inequality, each at level $0.01$. First, $\E\|\hat v-v\|_2^2=\Tr(\Sigma+\tfrac12I)/N_0\le(K+\tfrac12)m/n_0$. Second, by Isserlis' theorem, with $C:=\Sigma+\tfrac12I$,
\[
\E\|\hat C-C\|_F^2=\frac{(\Tr C)^2+\|C\|_F^2}{n_0}\le\frac{2(\Tr C)^2}{n_0},\qquad\Tr C\le2m(K+\tfrac12).
\]
Third, by \cite[Thm.~4.7.1]{Ver18}, $\E\|\hat C-C\|_\op\le K'(\sqrt{2m/n_0}+2m/n_0)\|C\|_\op\le2\sqrt2K'(K+\tfrac12)\sqrt{m/n_0}$ for $n_0\ge2m$; put $K_V:=300K'(K+1)$ and apply Markov's inequality at level $0.01$. On the intersection of the three events, $\hat\Sigma-\Sigma=\hat C-C+\eta_0I\succeq0$ because $\|\hat C-C\|_\op\le\eta_0$; the remaining assertions follow from $\eta_0\le\tfrac12$, $\|\hat\Sigma-\Sigma\|_\op\le\|\hat C-C\|_\op+\eta_0$ and $\|\eta_0I\|_F=\eta_0\sqrt{2m}$.

(ii) This is the argument of \cite[Lemma~5.11(iii)]{Rub26b} with its constants replaced. Put $\kappa:=\|P\|_\op^2\le2(K+1)$, and likewise $\tilde\kappa$. For $\mathcal X:=-(\Sigma\Omega)^2=SD^2S^{-1}$, $D=\operatorname{diag}(d,d)$, one has $\|(\mathcal X+tI)^{-1}\|_\op\le\kappa/(\tfrac14+t)$. The integral representation $\mathcal Q=\sqrt{\mathcal X}=\pi^{-1}\int_0^\infty t^{-1/2}\mathcal X(\mathcal X+tI)^{-1}\,dt$ of the principal square root and the resolvent identity give, in any unitarily invariant norm $\|\cdot\|$,
\[
\|\mathcal Q-\tilde{\mathcal Q}\|\le\kappa\tilde\kappa\,\|\mathcal X-\tilde{\mathcal X}\|\le\kappa\tilde\kappa\,(\|\Sigma\|_\op+\|\tilde\Sigma\|_\op)\,\|\Sigma-\tilde\Sigma\| .
\]
Next, $P^2-\tilde P^2=\mathcal Q^{-1}(\Sigma-\tilde\Sigma)+\mathcal Q^{-1}(\tilde{\mathcal Q}-\mathcal Q)\tilde P^2$ with $\|\mathcal Q^{-1}\|_\op\le2\kappa$, so $\|P^2-\tilde P^2\|\le C_K\|\Sigma-\tilde\Sigma\|$. Finally, for positive definite $A,B$ the solution $Y=\sqrt A-\sqrt B$ of $\sqrt A\,Y+Y\sqrt B=A-B$ satisfies $\|Y\|\le\|A-B\|/(\lambda_{\min}(\sqrt A)+\lambda_{\min}(\sqrt B))$ in every unitarily invariant norm \cite[VII.2]{Bhatia97}, and $\lambda_{\min}(P)=\kappa^{-1/2}\ge(2(K+1))^{-1/2}$.

(iii) $\Sigma'=\hat P^{-1}P\Sigma_pP\hat P^{-1}=P'O\Sigma_pO^TP'$, which is the factorisation stated. Put $\hat\kappa:=\|\hat P\|_\op^2$. Then $\|P'^2-I\|=\|\hat P^{-1}(P^2-\hat P^2)\hat P^{-1}\|\le\hat\kappa\|P^2-\hat P^2\|\le C_K\|\hat\Sigma-\Sigma\|$ by the intermediate bound of (ii), and $\|\log M\|\le2\|M-I\|$ for symmetric $M$ with $\|M-I\|_\op\le\tfrac12$, in both norms, by functional calculus; this gives the bound on $G'$. For the bound on $\theta_{\mathrm{sq}}$, apply \cite[Lemma~5.12]{Rub26b} to the displacement and then to the squeezing; $\|\Sigma'\|_\op\le\hat\kappa\|\Sigma\|_\op$ and $\|\Sigma_{p'}\|_\op\le\|P'^{-1}\|_\op^2\|\Sigma'\|_\op$ are bounded in terms of $K$.
\end{proof}

\paragraph{Bogoliubov blocks.}
For a symmetric positive symplectic $P=e^G$ write $U_P^\dagger aU_P=\mathsf Ua+\mathsf Va^\#$. In complex coordinates $(a,a^\#)$ the generator $G$ is the Hermitian block-off-diagonal matrix $\begin{psmallmatrix}0&g\\\bar g&0\end{psmallmatrix}$ with $g$ complex symmetric, $\|G\|_\op=\|g\|_\op$ and $\|G\|_F=\sqrt2\|g\|_F$. Expanding the exponential, the block $\mathsf U$ contains only even powers of the generator and $\mathsf V$ only odd powers; hence, when $\|G\|_\op$ is small,
\begin{equation}\label{eq:Bos-bogo}
\|\mathsf U-I\|_\op\le C_0\|G\|_\op^2,\qquad\|\mathsf U-I\|_F\le C_0\|G\|_\op\|G\|_F,\qquad\|\mathsf V\|\le C_0\|G\|
\end{equation}
in the operator and in the Frobenius norm.

\begin{lemma}[Removal of displacement and squeezing]\label{lem:Bos-undo}
Fix $K\ge1$. There are $C_K$, $c_K$ and $\eps_1(K)>0$ such that for $0<\eps_0\le\eps_1$ and $n_0=\lceil C_Km^2/\eps_0^2\rceil$, so that Stage~0 uses $L_0=2n_0$ copies, the following holds on an event of probability at least $0.97$. Stage~0 does not stop, and every later copy, after $V_0$, is in the Gaussian state
\begin{equation}\label{eq:Bos-tau}
\tau=D(v'')U_{e^G}\rho_QU_{e^G}^\dagger D(v'')^\dagger,\qquad\|v''\|+\|G\|_F\le c_K\eps_0,\qquad\|G\|_\op^2\le C_Km/L_0,
\end{equation}
with $G:=G'$ and $Q:=Q_{p'}$ of Lemma~\ref{lem:Bos0}(iii); the spectrum of $Q$ is the original occupation spectrum, and all quadrature and occupation operator norms of $\tau$, $\rho_Q$ and $\rho_{Q_\tau}$ below are bounded by constants depending only on $K$. The centred normal occupation matrix of $\tau$ is
\begin{equation}\label{eq:Bos-Qtau}
Q_\tau=\mathsf UQ\mathsf U^\dagger+\mathsf V(I+Q^T)\mathsf V^\dagger,\qquad\|Q_\tau-Q\|_\op\le C_K\|G\|_\op^2\le C_Km/L_0,
\end{equation}
and
\begin{equation}\label{eq:Bos-delta}
T(\tau,\rho_{Q_\tau})\le C_K(\|v''\|+\|G\|_F)=:\delta\le c_K\eps_0 .
\end{equation}
\end{lemma}

\begin{proof}
On the event of Lemma~\ref{lem:Bos0}(i), with $n_0\asymp m^2/\eps_0^2$: $\|G'\|_F\le K_2\|\hat\Sigma-\Sigma\|_F\le C_K\eps_0$; $\|G'\|_\op^2\le K_2^2(2\eta_0)^2\le C_Km/L_0$; and $\|v''\|=\|\hat P^{-1}(v-\hat v)\|\le C_K\sqrt{m/n_0}\le C_K\eps_0/\sqrt m$. The conditions $n_0\ge2m$, $\eta_0\le\tfrac12$ and $K_2\|\hat\Sigma-\Sigma\|_\op\le\tfrac14$ are all implied by $n_0\asymp m^2/\eps_0^2$ for $\eps_0\le\eps_1(K)$. This proves \eqref{eq:Bos-tau} by Lemma~\ref{lem:Bos0}(iii).

For \eqref{eq:Bos-Qtau}, insert $U_{e^G}^\dagger aU_{e^G}=\mathsf Ua+\mathsf Va^\#$ into $(Q_\tau)_{ij}=\langle a_j^\dagger a_i\rangle$: in the passive $\rho_Q$ the moments $\langle aa\rangle$, $\langle a^\dagger a^\dagger\rangle$ vanish and $\langle a_ka_l^\dagger\rangle=(I+Q^T)_{lk}$, and the displacement does not enter the centred normal occupation. Hence $Q_\tau-Q=(\mathsf U-I)Q\mathsf U^\dagger+Q(\mathsf U-I)^\dagger+\mathsf V(I+Q^T)\mathsf V^\dagger$, which by \eqref{eq:Bos-bogo} has operator norm at most $C_K\|G\|_\op^2$: there is no linear term, because $\mathsf U-I$ is even and $\mathsf V$ enters quadratically.

For \eqref{eq:Bos-delta}, Lemma~\ref{lem:Bos0}(iii) bounds $T(\tau,\rho_Q)=\theta_{\mathrm{sq}}$. To compare $\rho_Q$ with $\rho_{Q_\tau}$, couple coherent amplitudes as
\[
\alpha=Q^{1/2}z_1,\qquad\beta=\mathsf UQ^{1/2}z_1+\mathsf V(I+Q^T)^{1/2}z_2
\]
for independent standard proper complex Gaussians $z_1,z_2$. Their covariance matrices are $Q$ and $Q_\tau$, and both are proper. Lemma~\ref{lem:Bos-coupling}, in its coupling form, gives
\[
T(\rho_Q,\rho_{Q_\tau})\le\big(\|(\mathsf U-I)Q^{1/2}\|_F^2+\|\mathsf V(I+Q^T)^{1/2}\|_F^2\big)^{1/2}\le C_K\|G\|_F .\qedhere
\]
\end{proof}

Thus the data from $\tau$ can be analysed at its \emph{exact} normal occupation $Q_\tau$, with its pseudocovariance retained in \eqref{eq:Bos-noncirc}, while $\tau$ itself is within $\delta$ of the passive reference $\rho_{Q_\tau}$. There is no statistical replacement of $L$ copies of $\tau$ by $L$ copies of a nearby state, which would incur an error $L\delta$. Choosing the constant in $L_0$ allows $\eps_0$ to be any fixed multiple of $\eps$, and this stage costs only $m^2/\eps^2$ copies.
\subsection{Relative sampling error}\label{subsec:Bos-effrank}
\begin{lemma}[Effective-rank concentration]\label{lem:Bos-effrank}
Let $z_1,\dots,z_L$ be i.i.d.\ centred complex Gaussian vectors in $\C^m$ with covariance $I+Q$, $Q\succeq0$, and pseudocovariance $\mathsf P$ (possibly nonzero). Put $X:=L^{-1}\sum_lz_lz_l^\dagger-I$ and $E:=X-Q$. Let $D\succ0$ commute with $Q$, with eigenvalues $d_i$ in the eigenbasis of $Q$, and $H_D:=D^{-1/2}(I+Q)D^{-1/2}$. Then
\begin{align}
\E\|D^{-1/2}ED^{-1/2}\|_\op&\le C_0\Big[\sqrt{\frac{\|H_D\|_\op\Tr H_D}L}+\frac{\Tr H_D}L\Big],\label{eq:Bos-effrank}\\
\E\|D^{-1/2}ED^{-1/2}\|_F^2&=\frac1L\sum_{ij}\frac{(1+q_i)(1+q_j)+|\mathsf P_{ij}|^2}{d_id_j}\le\frac{2(\Tr H_D)^2}L,\label{eq:Bos-effrankF}
\end{align}
with equality $(\Tr H_D)^2/L$ in \eqref{eq:Bos-effrankF} when $\mathsf P=0$.
\end{lemma}

\begin{proof}
Put $y=D^{-1/2}z$. $D^{-1/2}$ is complex-linear, so in real coordinates $y$ is the image of the real Gaussian vector $(\operatorname{Re}z,\operatorname{Im}z)$ under a real linear map. Its real covariance $\Sigma_Y$ is one half of the augmented covariance $\begin{psmallmatrix}C_y&\mathsf P_y\\\bar{\mathsf P}_y&\bar C_y\end{psmallmatrix}$, written in a suitable basis, with $C_y=D^{-1/2}(I+Q)D^{-1/2}=H_D$; positivity of the augmented matrix gives $\|\mathsf P_y\|_\op\le\|C_y\|_\op$. Hence $\|\Sigma_Y\|_\op\le\|H_D\|_\op$ and $\Tr\Sigma_Y=\Tr H_D$. \cite[Thm.~4]{KL17}, in the Hilbert-space form with effective rank $\Tr\Sigma_Y/\|\Sigma_Y\|_\op$, gives $\E\|\hat\Sigma_Y-\Sigma_Y\|_\op\le C_0(\sqrt{\|\Sigma_Y\|_\op\Tr\Sigma_Y/L}+\Tr\Sigma_Y/L)$, which is monotone in $\|\Sigma_Y\|_\op$; the pair differences are i.i.d.\ centred Gaussians, as that theorem requires. The complex sample matrix is a linear compression $\Psi(M)=M_{uu}+M_{vv}+i(M_{vu}-M_{uv})$ of the real one: for a unit $x=a+ib$, with $w=(a;b)$ and $Jw=(-b;a)$, $x^\dagger\Psi(M)x=w^TMw+(Jw)^TM(Jw)$, so $\|\Psi(M)\|_\op\le2\|M\|_\op$. The subtracted identity has no sampling error: $D^{-1/2}ED^{-1/2}$ is the complex sample covariance of $y$ minus its mean. This proves \eqref{eq:Bos-effrank}. For \eqref{eq:Bos-effrankF}, $D^{-1/2}ED^{-1/2}$ has entries $E_{ij}/\sqrt{d_id_j}$ in the eigenbasis of $Q$, and \eqref{eq:Bos-noncirc} gives the identity and the bound, since $\sum_i(1+q_i)/d_i=\Tr H_D$.
\end{proof}

Below, the expectation bound is used with Markov's inequality at a constant failure probability. When $\|Q\|_\op\le\bar q$ and $D=Q+\mu I$, one has $\|H_D\|_\op\le(1+\bar q)/\mu$ and $\Tr H_D\le(1+\bar q)\Tr D^{-1}$, so
\begin{equation}\label{eq:Bos-effrankK}
\begin{gathered}
\E\|D^{-1/2}ED^{-1/2}\|_\op\le C_0(1+\bar q)\Big[\sqrt{\frac{\Tr D^{-1}}{\mu L}}+\frac{\Tr D^{-1}}L\Big],\\
\E\|D^{-1/2}ED^{-1/2}\|_F^2\le\frac{2(1+\bar q)^2}L(\Tr D^{-1})^2 .
\end{gathered}
\end{equation}

\subsection{A cold-frame lemma with the profile cost}\label{subsec:Bos-coldframe}

\begin{lemma}[Cold frame]\label{lem:Bos-coldframe}
Let $Q=Q_c\oplus Q_w$ in its true cold--warm frame, with
\[
0\preceq Q_c\preceq\mu I,\qquad Q_w\succeq3\mu I,\qquad\|Q\|_\op\le\bar q,\qquad m_c\mu\le C_c ,
\]
$Q_c$ of size $m_c$. Let $X=Q+E$ be Hermitian, and let $V$ be an isometry whose columns are the $m_c$ eigenvectors of $X$ with the smallest eigenvalues. Set $D=Q+\mu I$ and suppose
\begin{equation}\label{eq:Bos-eta20}
\|D^{-1/2}ED^{-1/2}\|_\op\le\eta:=\tfrac1{20}.
\end{equation}
Split $V=(V_c;V_w)$, and diagonalise $V^\dagger XV=\Lambda$ by a right unitary. Define the warm leakage $\ell=\Tr(V_w^\dagger Q_wV_w)$. Then
\begin{equation}\label{eq:Bos-leak}
\ell\le4\|Q_w^{-1/2}E_{wc}\|_F^2 .
\end{equation}
Also $\|V_w\|_\op^2\le23/57<\tfrac12$, so every squared cosine of a cold--warm principal angle is at least $34/57>\tfrac12$.
\end{lemma}

\begin{proof}
From $-\eta D\preceq E\preceq\eta D$ and min--max on the true cold space, $\lambda_{\max}(\Lambda)\le(1+2\eta)\mu=11\mu/10$. For any unit vector $v$ in the estimated cold space,
\[
(1-\eta)v^\dagger Qv-\eta\mu\le v^\dagger Xv\le(1+2\eta)\mu,
\]
so $v^\dagger Qv\le23\mu/19$. Since $Q_w\succeq3\mu I$, $\|V_w\|_\op^2\le23/57$.

For column $j$ of $V$, the warm rows of the eigenvector equation read
\[
(Q_w+E_{ww}-\lambda_jI)v_{w,j}=-E_{wc}v_{c,j}.
\]
Conjugating its left side by $Q_w^{-1/2}$, its least eigenvalue is at least
\[
1-\frac{11}{30}-\eta\Big(1+\frac13\Big)=\frac{17}{30};
\]
indeed \eqref{eq:Bos-eta20} implies $-\eta(Q_w+\mu I)\preceq E_{ww}\preceq\eta(Q_w+\mu I)$, so $Q_w^{-1/2}E_{ww}Q_w^{-1/2}\succeq-\eta(I+\mu Q_w^{-1})\succeq-\tfrac43\eta I$, and $\lambda_jQ_w^{-1}\preceq\tfrac{11}{30}I$. Hence
\[
\|Q_w^{1/2}v_{w,j}\|\le\frac{30}{17}\|Q_w^{-1/2}E_{wc}v_{c,j}\| .
\]
Square and sum over $j$; $\sum_j\|Q_w^{1/2}v_{w,j}\|^2=\ell$ is invariant under the right unitary, and $\|V_c\|_\op\le1$ gives \eqref{eq:Bos-leak} since $(30/17)^2<4$.
\end{proof}

No spectral gap in $X$ and no bound on $\|Q_w\|_\op$ enter this deterministic lemma.
Lemma~\ref{lem:Bos-coldframe} is a relative Davis--Kahan bound: it controls the warm occupation leaked into the estimated cold space, not the angle between the spaces. The angle need not be small in directions where the warm occupations are small, but then the leak is small too, and the leak is what the state notices (Lemma~\ref{lem:Bos-cross}).

\paragraph{Statistical cost.}
Let $X$ be the raw normal estimate from $L$ centred Gaussian heterodyne samples, possibly noncircular, of a state whose normal occupation is $Q$ with $\|Q\|_\op\le\bar q$. Since $\Tr D^{-1}\le m_c/\mu+\sum_{i\in w}q_i^{-1}$, \eqref{eq:Bos-effrankK} and Markov's inequality show that \eqref{eq:Bos-eta20} holds with any prescribed constant success probability when
\begin{equation}\label{eq:Bos-coldcost}
L\ge C_K\Big(m_c/\mu^2+\mu^{-1}\sum_{i\in w}q_i^{-1}\Big),
\end{equation}
with constants also allowed to depend on an upper bound for $\mu$ (which controls the second term of \eqref{eq:Bos-effrankK}). By \eqref{eq:Bos-noncirc}, in the true eigenframe,
\begin{equation}\label{eq:Bos-leakcost}
\E\|Q_w^{-1/2}E_{wc}\|_F^2=\sum_{i\in w,\,j\in c}\frac{\E|E_{ij}|^2}{q_i}\le\frac{2(1+\bar q)^2m_c}L\sum_{i\in w}q_i^{-1}.
\end{equation}
If the warm eigenvalues lie in dyadic bands with lowest scale at least $a/m_c$, \eqref{eq:Bos-cross-absorb} absorbs $m_c\sum_{i\in w}q_i^{-1}$ into $m_c^3+\sum_bm_b^2/\nu_b$. $D$ and the true frame are used only in the analysis; the implemented estimator is the ordinary spectral decomposition of $X$.

\subsection{The CS bookkeeping and the cross rotation}\label{subsec:Bos-CS}

In the setting of Lemma~\ref{lem:Bos-coldframe}, let $u\in U(m)$ diagonalise $Q$ with the $m_c$ cold eigenvectors first, so $Q=u\operatorname{diag}(q_1,\dots,q_m)u^\dagger$, let $\hat u\in U(m)$ be a unitary whose first $m_c$ columns span the estimated cold space, and let $W:=\hat u^\dagger u$. Take the CS decomposition \cite{PW94} of $W$ with respect to the split $(m_c,m_w)$, written here for $m_c\le m_w$:
\[
W=\operatorname{diag}(U_1,U_2)\,M(\Theta)\,\operatorname{diag}(U_3,U_4)^\dagger,
\]
\[
M(\Theta)=\begin{bmatrix}\cos\Theta&-P_\Theta^\dagger\\ P_\Theta&D_\Theta\end{bmatrix},\qquad
P_\Theta:=\begin{bmatrix}\sin\Theta\\0\end{bmatrix}\in\C^{m_w\times m_c},
\]
with $U_i$ unitary, $\Theta=\operatorname{diag}(\theta_1,\dots,\theta_{m_c})$, $\theta_i\in[0,\pi/2]$, and $D_\Theta:=\operatorname{diag}(\cos\Theta,I)$. (For $m_c>m_w$ the cold--cold corner of $M(\Theta)$ is $\operatorname{diag}(\cos\Theta,I_{m_c-m_w})$ and $P_\Theta=[\sin\Theta,\,0]\in\C^{m_w\times m_c}$, with the same identities.) Moreover $M(\Theta)=e^{A_0}$ with
\[
A_0:=\begin{bmatrix}0&-\tilde\Theta^\dagger\\ \tilde\Theta&0\end{bmatrix},\qquad\tilde\Theta:=\begin{bmatrix}\Theta\\0\end{bmatrix},
\]
an anti-Hermitian generator supported on the cross blocks, with singular values $\theta_i$. Put
\[
\mathsf N_c:=U_3^\dagger\operatorname{diag}(q_1,\dots,q_{m_c})U_3,\qquad\mathsf N_w:=U_4^\dagger\operatorname{diag}(q_{m_c+1},\dots,q_m)U_4 .
\]
Then $\hat u^\dagger Q\hat u=\operatorname{diag}(U_1,U_2)\,M(\Theta)\,(\mathsf N_c\oplus\mathsf N_w)\,M(\Theta)^\dagger\operatorname{diag}(U_1,U_2)^\dagger$ and, by block multiplication,
\begin{equation}\tag{CS}\label{eq:Bos-B}
\begin{aligned}
\big(M(\mathsf N_c\oplus\mathsf N_w)M^\dagger\big)_{cc}&=\cos\Theta\,\mathsf N_c\cos\Theta+P_\Theta^\dagger\mathsf N_wP_\Theta,\\
\big(M(\mathsf N_c\oplus\mathsf N_w)M^\dagger\big)_{ww}&=D_\Theta\,\mathsf N_w D_\Theta+P_\Theta\mathsf N_cP_\Theta^\dagger,\\
\big(M(\mathsf N_c\oplus\mathsf N_w)M^\dagger\big)_{wc}&=P_\Theta\,\mathsf N_c\cos\Theta-D_\Theta\,\mathsf N_wP_\Theta .
\end{aligned}
\end{equation}
The estimated cold columns, written in the true eigenframe, are $u^\dagger\hat u$ restricted to the first $m_c$ columns, $V=\big(U_3\cos\Theta\,U_1^\dagger;\,-U_4P_\Theta U_1^\dagger\big)$. Hence $V_w^\dagger Q_wV_w=U_1P_\Theta^\dagger\mathsf N_wP_\Theta U_1^\dagger$, and the leakage of Lemma~\ref{lem:Bos-coldframe} is
\begin{equation}\label{eq:Bos-leakCS}
\ell=\Tr(P_\Theta^\dagger\mathsf N_wP_\Theta).
\end{equation}

\begin{lemma}[Wick variance of the cross rotation]\label{lem:Bos3}
For the passive product state $\rho_c\otimes\rho_w$ of occupation $\mathsf N_c\oplus\mathsf N_w$ and the cross-supported generator $A_0$ (zero diagonal), with $H$ the quadratic Hamiltonian of $\Gamma(e^{A_0})$,
\[
\operatorname{Var}_{\rho_c\otimes\rho_w}(H)=\sum_{k\in c,\ l\in w}|(A_0)_{kl}|^2\,(q_k+q_l+2q_kq_l),
\]
the entries of $A_0$ being taken in the normal-mode bases of the two groups, and
\[
T\big(\rho_c\otimes\rho_w,\ \Gamma(e^{A_0})(\rho_c\otimes\rho_w)\Gamma(e^{A_0})^\dagger\big)\le\sqrt{\operatorname{Var}_{\rho_c\otimes\rho_w}(H)} .
\]
\end{lemma}

\begin{proof}
The inequality is Lemma~\ref{lem:A-MTgen} (\cite[Lemma~5.12]{Rub26b}). For the variance, $\Gamma(e^{A})=e^{a^\dagger Aa}=e^{-iH}$ with $H=i\,a^\dagger Aa$. For Hermitian $B$ and the thermal product state with occupations $q_i$ in the chosen basis, the Gaussian Wick rule gives $\langle a_i^\dagger a_ja_k^\dagger a_l\rangle=\langle a_i^\dagger a_j\rangle\langle a_k^\dagger a_l\rangle+\langle a_i^\dagger a_l\rangle\langle a_ja_k^\dagger\rangle=\delta_{ij}\delta_{kl}q_iq_k+\delta_{il}\delta_{jk}q_i(1+q_j)$, so
\begin{equation}\label{eq:Bos-wickvar}
\operatorname{Var}(a^\dagger Ba)=\sum_{ij}B_{ij}B_{ji}\,q_i(1+q_j)=\sum_{ij}|B_{ij}|^2q_i(1+q_j).
\end{equation}
For $B=iA_0$, supported on the cross pairs, the two ordered terms of a pair $k\in c$, $l\in w$ give $|(A_0)_{kl}|^2\big(q_k(1+q_l)+q_l(1+q_k)\big)=|(A_0)_{kl}|^2(q_k+q_l+2q_kq_l)$, the rates of the two hops $k\to l$ and $l\to k$. The pair formula is exact for zero-diagonal generators; for a general $A\in\mathfrak u(m)$ the diagonal phases add $\sum_k\varphi_k^2q_k(q_k+1)$.
\end{proof}

\begin{lemma}[Consequences of leakage]\label{lem:Bos-cross}
In the setting of Lemma~\ref{lem:Bos-coldframe}, on the event \eqref{eq:Bos-eta20}, let $\sigma_0$ be the passive product state of occupation $\operatorname{diag}(U_1,U_2)(\mathsf N_c\oplus\mathsf N_w)\operatorname{diag}(U_1,U_2)^\dagger$ and $T_I:=T(\Gamma(\hat u^\dagger)\rho_Q\Gamma(\hat u^\dagger)^\dagger,\sigma_0)$. Then:
\begin{enumerate}
\item[(i)] $\mu\|\sin\Theta\|_F^2\le\ell/3$, and
\begin{equation}\label{eq:Bos-TI}
T_I\le\sqrt{\operatorname{Var}(H)}\le\Big[\frac{\pi^2}4\Big(\frac43+2\mu\Big)\ell\Big]^{1/2};
\end{equation}
\item[(ii)] the cold block $Q'_c:=(\hat u^\dagger Q\hat u)_{cc}$ has $\Tr Q'_c\le\Tr Q_c+\ell$ and $\|Q'_c\|_\op\le\mu+\ell$;
\item[(iii)] the ordered eigenvalues of the warm block $Q'_w:=(\hat u^\dagger Q\hat u)_{ww}$ are at least $34/57>\tfrac12$ times the corresponding ordered eigenvalues of $Q_w$.
\end{enumerate}
\end{lemma}

\begin{proof}
(i) $\mathsf N_w\succeq3\mu I$ and \eqref{eq:Bos-leakCS} give $3\mu\|\sin\Theta\|_F^2\le\ell$. By unitary invariance under the product rotation $\Gamma(\operatorname{diag}(U_1,U_2))$, $T_I$ is the trace distance of Lemma~\ref{lem:Bos3}. In the CS frame the cross block of the generator is $\tilde\Theta$, and $\theta^2\le(\pi^2/4)\sin^2\theta$ on $[0,\pi/2]$ (Jordan's inequality) applied termwise against the nonnegative diagonal entries of $\mathsf N_w$ gives $\Tr(\tilde\Theta^\dagger\mathsf N_w\tilde\Theta)\le(\pi^2/4)\Tr(P_\Theta^\dagger\mathsf N_wP_\Theta)=(\pi^2/4)\ell$ and $\|\tilde\Theta\|_F^2\le(\pi^2/4)\|\sin\Theta\|_F^2$. With the exact variance weight $q_k+q_l+2q_kq_l\le\mu+(1+2\mu)q_l$ for $k$ cold,
\[
\operatorname{Var}(H)\le\frac{\pi^2}4\Big[(1+2\mu)\ell+\mu\|\sin\Theta\|_F^2\Big]\le\frac{\pi^2}4\Big(\frac43+2\mu\Big)\ell .
\]
(ii) By \eqref{eq:Bos-B}, $Q'_c=U_1(\cos\Theta\,\mathsf N_c\cos\Theta+P_\Theta^\dagger\mathsf N_wP_\Theta)U_1^\dagger$, with $\Tr(\cos\Theta\,\mathsf N_c\cos\Theta)\le\Tr\mathsf N_c=\Tr Q_c$, $\|\cos\Theta\,\mathsf N_c\cos\Theta\|_\op\le\mu$, and $\|P_\Theta^\dagger\mathsf N_wP_\Theta\|_\op\le\Tr(P_\Theta^\dagger\mathsf N_wP_\Theta)=\ell$.

(iii) By \eqref{eq:Bos-B}, $Q'_w=U_2(D_\Theta\mathsf N_wD_\Theta+P_\Theta\mathsf N_cP_\Theta^\dagger)U_2^\dagger$, where the smallest singular value of $D_\Theta$ satisfies $\sigma_{\min}(D_\Theta)^2\ge\cos^2\theta_{\max}\ge34/57$ by Lemma~\ref{lem:Bos-coldframe}. Min--max (Ostrowski's theorem) gives $\lambda_k(D_\Theta\mathsf N_wD_\Theta)\ge\sigma_{\min}(D_\Theta)^2\lambda_k(\mathsf N_w)$, and adding a positive semidefinite term does not lower ordered eigenvalues.
\end{proof}

\subsection{Deleting a cross block without a spectral gap}\label{subsec:Bos-schur}

\begin{lemma}[Schur-complement pinching bound]\label{lem:Bos-schur}
Write a positive semidefinite occupation matrix in any orthogonal mode decomposition as
\[
Q=\begin{pmatrix}A&B\\B^\dagger&C\end{pmatrix},\qquad C\succ0,\qquad Q_{\mathrm{pin}}=A\oplus C,\qquad\ell=\Tr(BC^{-1}B^\dagger).
\]
Then
\begin{equation}\label{eq:Bos-schur}
T(\rho_Q,\rho_{Q_{\mathrm{pin}}})\le2\sqrt\ell .
\end{equation}
Neither a separation between the spectra of $A$ and $C$ nor a positive lower bound on $A$ is required.
\end{lemma}

\begin{proof}
The Schur complement $S=A-BC^{-1}B^\dagger$ is positive semidefinite. Put $R=S\oplus C$. The factors
\[
F=\begin{pmatrix}S^{1/2}&BC^{-1/2}\\0&C^{1/2}\end{pmatrix},\qquad H=\begin{pmatrix}S^{1/2}&0\\0&C^{1/2}\end{pmatrix}
\]
satisfy $FF^\dagger=Q$, $HH^\dagger=R$ and $\|F-H\|_F^2=\ell$, so the factor characterisation of the Bures--Wasserstein distance gives $d_{\mathrm{BW}}(Q,R)\le\sqrt\ell$. Also $Q_{\mathrm{pin}}\succeq R$ and $\Tr(Q_{\mathrm{pin}}-R)=\ell$, so \eqref{eq:Bos-BWdom} gives $d_{\mathrm{BW}}(Q_{\mathrm{pin}},R)^2\le\ell$. The triangle inequality for $d_{\mathrm{BW}}$, followed by Lemma~\ref{lem:Bos-coupling}, proves \eqref{eq:Bos-schur}, including singular $S$ and $A$, by the same factors.
\end{proof}

\begin{lemma}[The empirical spectral split]\label{lem:Bos-split-schur}
Let $X=Q+E$ be Hermitian, let $\Pi$ be an orthogonal spectral projector of $X$, and put $\Pi^\perp=I-\Pi$ and $D=Q+\mu I$ with $\mu>0$. Regard $A=\Pi Q\Pi$ and $C=\Pi^\perp Q\Pi^\perp$ as operators on their respective subspaces, and suppose
\[
\|A\|_\op\le a\mu,\qquad C\succeq b\mu I_{\Pi^\perp},\qquad a\ge0,\ b>0 .
\]
Then
\begin{equation}\label{eq:Bos-split-schur}
T(\rho_Q,\rho_{A\oplus C})^2\le4(a+1)(1+b^{-1})\,\mu\,\|D^{-1/2}ED^{-1/2}\|_F^2 .
\end{equation}
Here $A\oplus C$ is written in the $\Pi,\Pi^\perp$ coordinates; the state distance is invariant under the corresponding known passive rotation.
\end{lemma}

\begin{proof}
Since $\Pi$ is spectral for $X$, the cross block satisfies $B=\Pi Q\Pi^\perp=-\Pi E\Pi^\perp$. Two inverse-compression facts are needed.
\begin{enumerate}
\item $\Pi D\Pi\preceq(a+1)\mu\Pi$ on the range of $\Pi$ implies $\Pi\preceq(a+1)\mu D^{-1}$ on the full space: the nonzero eigenvalues of $D^{1/2}\Pi D^{1/2}$ are those of $\Pi D\Pi$ on that range.
\item If $J$ is an isometry onto the range of $\Pi^\perp$, then $J(J^\dagger DJ)^{-1}J^\dagger\preceq D^{-1}$: conjugating by $D^{1/2}$ makes the left side the orthogonal projector onto the range of $D^{1/2}J$. Since $C+\mu I\preceq(1+b^{-1})C$,
\[
JC^{-1}J^\dagger\preceq(1+b^{-1})J(J^\dagger DJ)^{-1}J^\dagger\preceq(1+b^{-1})D^{-1}.
\]
\end{enumerate}
Consequently, with inverses on the range of $\Pi^\perp$ extended by zero,
\begin{align*}
\ell&=\Tr(\Pi E\Pi^\perp C^{-1}\Pi^\perp E\Pi)\le(1+b^{-1})\Tr(ED^{-1}E\Pi)\\
&\le(a+1)(1+b^{-1})\mu\Tr(ED^{-1}ED^{-1})=(a+1)(1+b^{-1})\mu\|D^{-1/2}ED^{-1/2}\|_F^2 .
\end{align*}
Both trace comparisons pair a positive matrix with a Loewner inequality. Applying \eqref{eq:Bos-schur} proves \eqref{eq:Bos-split-schur}.
\end{proof}

This bounds the error of replacing a passive state by the product of its two actual marginals; it replaces the rotation and spectral-gap estimates of \S\ref{subsec:Bos-coldframe}--\S\ref{subsec:Bos-CS} in the proof of Theorem~\ref{thm:Bos}.
A rotation between a nearly degenerate cold and warm direction is poorly determined by the data but barely changes the state, and $\ell$ does not see it: estimating a leaked occupation replaces estimating a rotation, which would need a gap.

\subsection{A split chosen from the data}\label{subsec:Bos-split}

Let $0\le n_1\le\dots\le n_m$ be the original Williamson occupations and $r,r_0,\mu,S_w,\mathcal R(\boldsymbol n)$ as in \eqref{eq:Bos-R}, for the analysis only. After Stage~0 write $Q$ for the exact centred normal occupation of the actual transformed state $\tau$, that is, $Q_\tau$ of Lemma~\ref{lem:Bos-undo}. On the good event of Lemma~\ref{lem:Bos-undo}, its ordered eigenvalues $q_i$ satisfy
\begin{equation}\label{eq:Bos-35}
|q_i-n_i|\le\mu/100,\qquad\|Q\|_\op\le K_\tau,\qquad T(\tau,\rho_Q)\le\delta,
\end{equation}
where $K_\tau$ depends only on $K$. For the first inequality it suffices that the Stage~0 sample count is a sufficiently large $K$-dependent multiple of $m^2/\eps^2$: \eqref{eq:Bos-Qtau} bounds the perturbation by $C_Km/L_0$, Weyl's inequality transfers it to the ordered eigenvalues, and $\mu\ge2/m$. No occupation lower bound enters this stage.

Let $X=Q+E$ be the raw normal-covariance estimate from an independent heterodyne pilot batch, with the mean eliminated by pair differences. Write its eigenvalues as $x_1\le\dots\le x_m$ and set
\begin{equation}\label{eq:Bos-s}
s=\max\big(\{1\}\cup\{j:jx_j\le2\}\big).
\end{equation}
Let $\Pi$ project onto the span of the lowest $s$ eigenvectors of $X$ and $\Pi^\perp=I-\Pi$, and let $A=\Pi Q\Pi$, $C=\Pi^\perp Q\Pi^\perp$ as in Lemma~\ref{lem:Bos-split-schur}. Forcing $s\ge1$ merely simplifies the all-warm case; it costs a bounded one-mode readout. The protocol needs neither $r$, $\mu$, the exact spectrum, nor its eigenbasis to implement \eqref{eq:Bos-s}.

Assume the relative-error event
\begin{equation}\label{eq:Bos-37}
\|D^{-1/2}ED^{-1/2}\|_\op\le\eta:=\tfrac1{100},\qquad D=Q+\mu I,
\end{equation}
equivalently
\begin{equation}\label{eq:Bos-38}
(1-\eta)Q-\eta\mu I\preceq X\preceq(1+\eta)Q+\eta\mu I .
\end{equation}

\begin{lemma}[Properties of the data-chosen split]\label{lem:Bos-split}
Assume \eqref{eq:Bos-35} and \eqref{eq:Bos-37}. Then:
\begin{enumerate}
\item[(i)] (size of the selected cold group)
\begin{equation}\label{eq:Bos-39}
\max\{1,\lfloor r/2\rfloor\}\le s\le4r_0 ;
\end{equation}
\item[(ii)] (cold total occupation and operator norm) with $C_A:=\max\{3,K_\tau\}$ and $a_K:=4+K_\tau$,
\begin{equation}\label{eq:Bos-41}
\Tr A\le C_A,\qquad\|A\|_\op\le a_K\mu ;
\end{equation}
\item[(iii)] (warm minimum) if $s<m$,
\begin{equation}\label{eq:Bos-42}
C\succeq(\mu/8)I_{\Pi^\perp};
\end{equation}
\item[(iv)] (comparison of the remaining eigenvalues) writing $c_1\le\dots\le c_{m-s}$ for the eigenvalues of $C$, for every $j$ with $k=s+j>r$,
\begin{equation}\label{eq:Bos-43}
c_j\ge n_k/2 ,
\end{equation}
while the indices $k\le r$, if any, number at most $r$ and retain the bound $\mu/8$ of \eqref{eq:Bos-42}.
\end{enumerate}
\end{lemma}

\begin{proof}
(i) For $j\ge4r_0$, if such $j$ exists, $j>r$ and $q_j\ge0.49\mu$; hence $x_j\ge0.4751\mu$ and $jx_j>2$. Therefore $s\le4r_0$. For $1\le j\le\lfloor r/2\rfloor$, $q_j\le1.01\mu$ and $x_j\le1.0301\mu$, so $jx_j\le1.0301<2$.

(ii) If the set $\{j:jx_j\le2\}$ is nonempty, $x_s\le2/s$, so $\sum_{j\le s}x_j\le2$. Restricting the left inequality in \eqref{eq:Bos-38} to the range of $\Pi$ gives
\[
\Tr A\le\frac{2+s\eta\mu}{1-\eta}<3,\qquad\|A\|_\op\le\frac{2/s+\eta\mu}{1-\eta}\le4\mu ;
\]
for the last inequality use $s\ge r/3$ when $r\ge2$; the cases $r=0,1$ have $\mu=2$ and $s\ge1$. If the set is empty, $s=1$ and \eqref{eq:Bos-39} implies $r\le1$, since for $r\ge2$ the index $j=1$ would qualify; in this case $\Tr A=\|A\|_\op\le K_\tau$ and $\mu=2$. In every case \eqref{eq:Bos-41} holds.

(iii) If $s<m$, maximality in \eqref{eq:Bos-s} gives $x_{s+1}>2/(s+1)\ge\mu/5$, by \eqref{eq:Bos-39}. The right inequality in \eqref{eq:Bos-38}, restricted to the range of $\Pi^\perp$, implies
\[
C\succeq\frac{1/5-\eta}{1+\eta}\mu I_{\Pi^\perp}\succeq(\mu/8)I_{\Pi^\perp}.
\]
(iv) The same restriction and global min--max give
\[
c_j\ge\frac{x_{s+j}-\eta\mu}{1+\eta}\ge\frac{(1-\eta)q_{s+j}-2\eta\mu}{1+\eta}.
\]
For $k=s+j>r$, $n_k\ge\mu/2$ and \eqref{eq:Bos-35} imply \eqref{eq:Bos-43}.
\end{proof}

If $s=m$, the warm subsystem is omitted and the pinching error is zero. An all-cold input can still give $s<m$; \eqref{eq:Bos-42} and the leftover-cold part of \eqref{eq:Bos-49} below cover that case. Exactly vacuum inputs give $s=m$ on the stated good events. The pilot constant required by \eqref{eq:Bos-39}--\eqref{eq:Bos-43} is large (Remark~\ref{rem:Bos-constants}).
The rule \eqref{eq:Bos-s} may misjudge modes near the threshold, and either misjudgement is cheap: a wrongly cold mode adds a bounded occupation (ii), and the photon-number readout does not care how the cold occupation is distributed; a wrongly warm mode has occupation at least $\mu/8$ (iii), so its heterodyne pairs cost $O(r_0)$ each, for at most $r$ such modes \eqref{eq:Bos-49}.

\subsection{The statistical budget of the gap-free split}\label{subsec:Bos-budget}

Let $I_w=\sum_{i>r}1/n_i$. Since $n_i\ge\mu/2$ on this set,
\begin{equation}\label{eq:Bos-44}
\frac\mu{n_in_j}\le\frac4{n_i+n_j},\qquad\mu I_w^2\le4S_w .
\end{equation}
From \eqref{eq:Bos-35}, $\Tr D^{-1}\le r/\mu+2I_w$. Using \eqref{eq:Bos-44}, $r^2/\mu\le r_0^3/2$ and $\mu^{-3}=r_0^3/8$,
\begin{align}
\mu(\Tr D^{-1})^2&\le C_0(r_0^3+S_w)\le C_0\mathcal R(\boldsymbol n),\label{eq:Bos-45}\\
\mu^{-1}\Tr D^{-1}&\le r/\mu^2+2I_w/\mu\le C_0(r_0^3+S_w)\le C_0\mathcal R(\boldsymbol n).\label{eq:Bos-46}
\end{align}
For the middle inequality in \eqref{eq:Bos-46}, use $I_w\le2\sqrt{S_w/\mu}$ and $2\sqrt{ab}\le a+b$ with $a=S_w$ and $b=\mu^{-3}$. The value $r_0=1$ when $r=0$ is absorbed by $m^2\ge1$.

By \eqref{eq:Bos-effrankK} with $\bar q=K_\tau$, \eqref{eq:Bos-46} and $\mu\le2$, a pilot batch of $L_F\ge C_K\mathcal R(\boldsymbol n)$ samples makes the probability of failure of \eqref{eq:Bos-37} an arbitrarily small fixed constant. Although $\Pi$ is data-dependent, \eqref{eq:Bos-split-schur} holds for that $\Pi$ whenever \eqref{eq:Bos-37} holds, with $a=a_K$ and $b=1/8$ by \eqref{eq:Bos-41}--\eqref{eq:Bos-42}, and its right side does not depend on $\Pi$. Combining \eqref{eq:Bos-split-schur}, \eqref{eq:Bos-45} and the Frobenius line of \eqref{eq:Bos-effrankK} with $\bar q=K_\tau$, and applying Markov's inequality to the nonnegative expression on the right of \eqref{eq:Bos-split-schur}, proves that
\begin{equation}\label{eq:Bos-48}
T(\rho_Q,\rho_{A\oplus C})\le\eps/4
\end{equation}
except with a prescribed small constant probability when $L_F\ge C_K\mathcal R(\boldsymbol n)/\eps^2$. Use a union bound with the failure of \eqref{eq:Bos-37}; the unbiased moments are computed without conditioning on that event.

The warm readout has the same budget. Equations \eqref{eq:Bos-42}--\eqref{eq:Bos-43} imply
\begin{equation}\label{eq:Bos-49}
\sum_{i,j\le m-s}\frac1{c_i+c_j}\le C_0\big(S_w+r^2/\mu+rI_w\big)\le C_0\mathcal R(\boldsymbol n).
\end{equation}
Indeed, warm--warm indices cost at most $2S_w$ by \eqref{eq:Bos-43}; the at most $r$ leftover cold indices cost at most $4r^2/\mu$ among themselves, by \eqref{eq:Bos-42}, and at most $4rI_w$ in cross terms. The latter is absorbed using $rI_w\le\sqrt{(r^2/\mu)(\mu I_w^2)}$ and \eqref{eq:Bos-44}. If $s>r$, the retained warm indices are a subset and the same upper bound holds. If $r=0$, there are no leftover cold indices.

Conditional on the pilot data, use an independent warm batch: heterodyne the modes in the range of $\Pi^\perp$ of $L_w$ further transformed and rotated copies, with pair differences; their centred normal covariance is exactly $I+C$, and residual squeezing enters through the actual pseudocovariance. Let $X_w$ be the raw estimate. By \eqref{eq:Bos-clip}, \eqref{eq:Bos-noncirc} and the bounded occupation $\|C\|_\op\le K_\tau$, the always-physical estimate $\rho_{(X_w)_+}$ satisfies
\begin{equation}\label{eq:Bos-50}
\E\,T(\rho_C,\rho_{(X_w)_+})^2\le\frac{4(1+K_\tau)^2}{L_w}\sum_{i,j\le m-s}\frac1{c_i+c_j}\le\frac{C_K\mathcal R(\boldsymbol n)}{L_w}.
\end{equation}
Finally $s^3\le64r_0^3\le64\,\mathcal R(\boldsymbol n)$ by \eqref{eq:Bos-39}, so the cold readout is within the same budget.
\subsection{Proof of Theorem~\ref{thm:Bos}(i)}\label{subsec:Bos-proof-i}

\begin{proof}[Proof of Theorem~\ref{thm:Bos}(i)]
\emph{Protocol.} Allocate fixed fractions of $N$ to four independent stages: the displacement and squeezing estimation of Stage~0, the pilot $X$ and the split \eqref{eq:Bos-s}, the cold readout, and the warm heterodyne readout. Pair-difference estimates use two copies per centred sample; this fixed factor is absorbed in the allocation. Undo the estimated displacement and squeezer on all later copies, and apply the known passive rotation selected by \eqref{eq:Bos-s}.

\emph{Stage~0.} This is Lemma~\ref{lem:Bos-undo}, with a constant large enough that $\delta\le\eps/A_K$ for a fixed $A_K$ chosen below. Since $N\ge C_Km^2/\eps^2$, \eqref{eq:Bos-35} and the residual trace-distance bound hold except with probability $0.03$.

\emph{Pilot and split.} Sections~\ref{subsec:Bos-split}--\ref{subsec:Bos-budget} give \eqref{eq:Bos-41}, \eqref{eq:Bos-42} and \eqref{eq:Bos-48} with combined further failure probability at most $0.04$.

\emph{Cold readout.} The cold target $\sigma_c=\rho_A$, the marginal of $\rho_Q$ on the range of $\Pi$, has \emph{bounded total occupation}, rather than a prescribed small occupation per mode. Let
\[
F_c=A(I+A)^{-1},\qquad p_0=\langle0|\sigma_c|0\rangle .
\]
Equation \eqref{eq:Bos-41} implies
\begin{equation}\label{eq:Bos-52}
p_0=\det(I+A)^{-1}\ge e^{-C_A},\qquad\|F_c\|_\op\le C_A/(1+C_A)=1-\delta_K,\qquad\delta_K:=(1+C_A)^{-1}>0 .
\end{equation}
Let $\tau_c$ be the actual cold marginal of the transformed and rotated copies. Partial-trace contraction gives $T(\tau_c,\sigma_c)\le T(\tau,\rho_Q)\le\delta$. With $a_0=\langle0|\tau_c|0\rangle$, take $\eps$ small enough that $a_0\ge p_0-\delta\ge e^{-C_A}/2$. Define the vacuum-normalised one-photon matrix $F'=a_0^{-1}\Pi_1\tau_c\Pi_1$, $\Pi_1$ the projector onto the one-photon sector. The one-photon block of $\sigma_c$ is $p_0F_c$ \cite[Lemma~5.8(i)]{Rub26b}. By contraction under photon-number pinching and under the vacuum test, $\|a_0F'-p_0F_c\|_1\le\|\tau_c-\sigma_c\|_1\le2\delta$ and $|a_0-p_0|\le\delta$. With $\Tr F_c\le\Tr A\le C_A$,
\begin{equation}\label{eq:Bos-53}
\|F'-F_c\|_1\le a_0^{-1}\big(2\delta+C_A\delta\big)\le2e^{C_A}(2+C_A)\,\delta .
\end{equation}
The readout is the photon-number and one-particle protocol of \cite[Thm.~5.9]{Rub26b}, Stages~1 and~2, at $k=1$, used as in Stage~2 of the proof of \cite[Thm.~5.13]{Rub26b}. That proof uses about the measured state only this: the particle-number outcomes on fresh copies are independent, with $\Prob(0)=a_0$ and $\Prob(1)=a_0s'$ where $a_0=\langle0|\rho'|0\rangle\ge1/Z_\star$ and $s'=\tr X'$ for the operator $X':=a_0^{-1}\Pi_1\rho'\Pi_1\succeq0$ on $\Fock_1=\C^m$, with $s'$ at most the clipping constant; and the state of a copy on outcome $1$ is exactly $X'/s'$. These hold for every state with $a_0\ge1/Z_\star$, Gaussian or not, with $s'\le1/a_0\le Z_\star$. We apply this with the measured state $\rho':=\tau_c$, $X':=F'$, $Z_\star:=2e^{C_A}$, the passive reference $\sigma_c$ in place of the state $\rho_{p'}$ of that proof, and its $\theta=T(\rho',\rho_{p'})$ replaced by $\theta:=T(\tau_c,\sigma_c)\le\delta$; the two assumptions, $a_0\ge1/Z_\star$ and $\Tr F'\le1/a_0$, hold here. This use is valid even though $\tau_c$ need not be passive, and the qudit is exactly its measured one-photon block. Run with an accuracy parameter that is a fixed $K$-dependent multiple of $\eps$, it returns $\hat F\succeq0$ with $\|\hat F-F'\|_1\le\delta_K\eps/16$ except with probability $0.15$, from $O_K((1+s^3)/\eps^2)$ fresh copies. Choose $A_K$ in Stage~0 so that the right side of \eqref{eq:Bos-53} is also at most $\delta_K\eps/16$. Then $\|\hat F-F_c\|_1\le\delta_K\eps/8$, and for $\eps\le1$,
\[
\|\hat F\|_\op\le1-\delta_K/2,\qquad T(\rho_{\hat F},\sigma_c)\le(2/\delta_K)\|\hat F-F_c\|_1\le\eps/4 ,
\]
where $\rho_{\hat F}$ is the passive state of fugacity $\hat F$; the inequality is \cite[Lemma~5.8(ii)]{Rub26b} with $\delta'=\delta_K/2$, the segment from $F_c$ to $\hat F$ staying in that ball by convexity. This also proves validity of the output on the good event; on any invalid-fugacity branch output a fixed valid Gaussian state. For $s=1$ the one-photon state has dimension one and the frequency estimation alone suffices; the $O_K((1+s^3)/\eps^2)$ bound still holds.

\emph{Warm readout.} Equation \eqref{eq:Bos-50} and Markov's inequality give $T(\sigma_w,\hat\sigma_w)\le\eps/4$, with $\sigma_w=\rho_C$ and $\hat\sigma_w=\rho_{(X_w)_+}$, with failure probability at most $0.04$. If there are no warm modes, omit this stage.

\emph{Assembly.} With $\hat\sigma_c=\rho_{\hat F}$ and $\rho_{A\oplus C}=\sigma_c\otimes\sigma_w$ exactly the product of the two marginals, the triangle inequality through $\rho_Q$ and $\rho_{A\oplus C}$ bounds the error before undoing the known transformations by
\[
T(\tau,\hat\sigma_c\otimes\hat\sigma_w)\le\delta+\eps/4+\eps/4+\eps/4\le\eps .
\] Known inverse transformations preserve this distance and Gaussianity. The total failure probability is at most $0.03+0.04+0.15+0.04=0.26<1/3$. Equations \eqref{eq:Bos-effrankK}, \eqref{eq:Bos-45}--\eqref{eq:Bos-50} and $s^3\le64\mathcal R(\boldsymbol n)$ justify all four allocations. To make the protocol defined also when $N$ is insufficient, abort to a fixed valid state if the selected cold subroutine would exceed its allocated budget; on the good events under \eqref{eq:Bos-thm3} this branch is never taken.
\end{proof}

\begin{remark}[Large constants]\label{rem:Bos-constants}
The constant behind the relative event \eqref{eq:Bos-37} at $\eta=1/100$ is large, and the cold-readout constants are exponential in $K$ through $e^{C_A}$. All of them depend only on $K$; the theorem does not remove the dependence on $K$.
\end{remark}

\subsection{Proof of Theorem~\ref{thm:Bos-sep}(i)}\label{subsec:Bos-sep-upper}

\begin{proof}[Proof of Theorem~\ref{thm:Bos-sep}(i)]
All sample batches below are independent conditional on the earlier classical results. Let $\rho\in\mathcal C_{\rm sep}$, with original occupations $n_i$: $m_c$ of them at most $c/m_c$, the others at least $4c/m_c$.

\emph{1. Undo displacement and squeezing.} Use Lemma~\ref{lem:Bos-undo} with $L_0=C_{c,K}m^2/\eps^2$, the constant chosen so that $\delta$ in \eqref{eq:Bos-delta} is at most $\eps/A$, where $A=A(c,K)$ is fixed after accounting for the cold readout. Let $\tau$ be the actual transformed state, with normal occupation $Q_\tau$, and $Q$ the passive occupation of \eqref{eq:Bos-tau}, with the original spectrum. On the good event, of probability at least $0.97$, \eqref{eq:Bos-Qtau} and $m\ge m_c$ give
\[
\|Q_\tau-Q\|_\op\le C_Km/L_0\le\frac c{20m_c}
\]
for $L_0$ large enough. By Weyl's inequality the $m_c$ cold eigenvalues of $Q_\tau$ are then at most $\mu:=21c/(20m_c)$, its warm eigenvalues are at least $79c/(20m_c)>3\mu$, and each warm eigenvalue differs from the corresponding original one by a relative factor at most $1/80$. By Ky Fan's inequality the sum of the $m_c$ smallest eigenvalues of $Q_\tau$ is at most $c+m_cC_Km/L_0\le c+C_K\eps^2$. For $m_c\ge m_\star(c,K)$ its cold eigenvalues are below $\tfrac14$. Hence the hypotheses of Lemma~\ref{lem:Bos-coldframe} hold for $Q_\tau$, with $\|Q_\tau\|_\op\le K_\tau$ and $m_c\mu=21c/20$. Abort to a fixed valid state if the physicality or norm checks of Stage~0 fail.

\emph{2. Find the cold space.} Heterodyne another $2L_F$ transformed copies, use disjoint pair differences, and form the raw normal estimate $X_F$ (sample covariance minus $I$). Take its lowest $m_c$ eigenvectors as the cold space, completed to a unitary $\hat u$, and apply $\Gamma(\hat u^\dagger)$ to all later copies. On the preceding event, \eqref{eq:Bos-coldcost}--\eqref{eq:Bos-leakcost} with $L_F=C_{c,K}\mathcal R_{\rm sep}/\eps^2$ imply, except with probability at most $0.04$, both \eqref{eq:Bos-eta20} and $\ell\le\eps^2/B^2$, for any fixed required $B=B(c,K)$: the event \eqref{eq:Bos-eta20} costs $m_c/\mu^2+\mu^{-1}\sum_wq_i^{-1}$, and the leakage, by \eqref{eq:Bos-leak}, \eqref{eq:Bos-leakcost} and Markov's inequality, costs $m_c\sum_wq_i^{-1}B^2/\eps^2$; here $m_c/\mu^2\asymp_cm_c^3$, the post-undo warm eigenvalues are within a relative $1/80$ of the original ones, and \eqref{eq:Bos-cross-absorb} with $a=4c$ absorbs $m_c\sum_bm_b/\nu_b$ into $C_c(m_c^3+\sum_bm_b^2/\nu_b)$. The two failures are combined by a union bound, not by conditioning. The passive reference $\rho_{Q_\tau}$, expressed in this frame, has cold and warm marginals $\sigma_c=\rho_{Q'_c}$ and $\sigma_w=\rho_{Q'_w}$. Lemma~\ref{lem:Bos-cross} gives $\Tr Q'_c\le c+O(\eps^2)$, $\|Q'_c\|_\op\le\mu+O(\eps^2)<\tfrac13$, and the warm spectral comparison of Lemma~\ref{lem:Bos-cross}(iii).

\emph{3. Cold readout.} On fresh transformed and rotated copies, measure only the cold modes and run the cold readout of the proof of Theorem~\ref{thm:Bos}(i) (\S\ref{subsec:Bos-proof-i}, from \eqref{eq:Bos-52} to the bound on $T(\rho_{\hat F},\sigma_c)$). It applies verbatim with the cold target $\sigma_c=\rho_{Q'_c}$ in place of $\rho_A$, $Q'_c$ in place of $A$, $m_c$ in place of $s$, and $C_A$ replaced by $C_\star:=c+1$: the only property of $A$ used there is $\Tr A\le C_A$, and here $\Tr Q'_c\le c+O(\eps^2)\le C_\star$; the measured state is the actual cold marginal $\tau_c$, with $T(\tau_c,\sigma_c)\le T(\tau,\rho_{Q_\tau})\le\delta$ by partial-trace contraction; and the constant $A$ of step~1 is chosen, as $A_K$ there, so that the right side of \eqref{eq:Bos-53} is at most $\delta_\star\eps/16$, $\delta_\star:=(1+C_\star)^{-1}$. From $L_c=C_c(1+m_c^3)/\eps^2$ copies it returns a valid passive state $\hat\sigma_c$ with $T(\sigma_c,\hat\sigma_c)\le\eps/4$, except with probability at most $0.15$.

\emph{4. Warm readout.} Heterodyne the warm modes of another $2L_w$ copies and use pair differences. Their normal covariance is exactly $I+Q'_w$; possible residual squeezing is accounted for by \eqref{eq:Bos-noncirc}. Form $X_w$ (sample covariance minus $I$) and output $\hat\sigma_w=\rho_{(X_w)_+}$, which is always physical. As for \eqref{eq:Bos-50}, with $C$ replaced by $Q'_w$, and then by the warm eigenvalue comparison of Lemma~\ref{lem:Bos-cross}(iii) with the relative factor $1/80$ of step~1 (the symmetric sums $\sum_{i,j}1/(q_i+q_j)$ are monotone in each ordered eigenvalue) and \eqref{eq:Bos-dyadic},
\[
\E\,T(\sigma_w,\hat\sigma_w)^2\le\frac{C_K}{L_w}\sum_{i,j\in w}\frac1{n_i+n_j}\le\frac{C_K\mathcal R_{\rm sep}}{L_w}.
\] Taking $L_w=C_{c,K}\mathcal R_{\rm sep}/\eps^2$ gives $T(\sigma_w,\hat\sigma_w)\le\eps/4$ except with probability $0.04$, by Markov's inequality.

\emph{5. Assembly.} Let $\sigma_0$ be the block product state of Lemma~\ref{lem:Bos-cross}, so that $T(\Gamma(\hat u^\dagger)\rho_{Q_\tau}\Gamma(\hat u^\dagger)^\dagger,\sigma_0)=T_I$. Each marginal of the rotated $\rho_{Q_\tau}$ is within $T_I$ of the corresponding marginal of $\sigma_0$, so
\[
T\big(\Gamma(\hat u^\dagger)\rho_{Q_\tau}\Gamma(\hat u^\dagger)^\dagger,\sigma_c\otimes\sigma_w\big)\le T_I+T(\sigma_{0,c},\sigma_c)+T(\sigma_{0,w},\sigma_w)\le3T_I .
\]
Consequently the final error, by unitary invariance, is at most
\[
\delta+3T_I+T(\sigma_c,\hat\sigma_c)+T(\sigma_w,\hat\sigma_w)\le\delta+3C_K\sqrt\ell+\eps/2 ,
\]
using \eqref{eq:Bos-TI} with $\mu$ bounded. Choose $A$ (subject also to step~3) and $B$ to make this at most $\eps$; they are chosen before the constants in $L_0,L_F,L_c,L_w$. The total failure probability is at most $0.03+0.04+0.15+0.04=0.26<1/3$, and the sample counts sum to $C_{c,K}\mathcal R_{\rm sep}/\eps^2$. The protocol uses only $m_c$ and $\mathcal R_{\rm sep}$, never the individual occupations; the cold count $m_c$ is determined by each member of the class because $\nu_{\min}\ge4c/m_c>c/m_c$.
\end{proof}

\subsection{Lower bounds: tensorisation and a direct-sum principle}\label{subsec:Bos-directsum}
Lower bounds add over bands. Every copy contains all blocks, so a learner that gives block $b$ the accuracy $e_b$, with $\sum_be_b^2\lesssim\eps^2$ by Lemma~\ref{lem:Bos-tensor}, needs $\max_bA_b/e_b^2$ copies, where block $b$ alone needs $A_b/e_b^2$; the best allocation equalises the ratios, $e_b^2\propto A_b$, and gives $\sum_bA_b/\eps^2$. The proof of Lemma~\ref{lem:Bos-directsum} uses exactly this allocation.

\begin{lemma}[Trace distance of tensor products]\label{lem:Bos-tensor}
For any finite collection of pairs of quantum states, with $t_b=T(\rho_b,\sigma_b)$,
\begin{equation}\label{eq:Bos-tensor}
T\Big(\bigotimes_b\rho_b,\bigotimes_b\sigma_b\Big)\ge\frac16\min\Big\{1,\sqrt{\textstyle\sum_bt_b^2}\Big\}.
\end{equation}
\end{lemma}

\begin{proof}
A Helstrom measurement on pair $b$ gives binary probabilities $p_b,q_b$ with $|p_b-q_b|=t_b$; label the outcomes so that $p_b\ge q_b$. Put $r_b=(p_b+q_b)/2$ and $a_b=1/[2\max(r_b,1-r_b)]$, so $\tfrac12\le a_b\le1$. A stochastic binary postprocessing with slope $a_b$ and intercept $\tfrac12-a_br_b$ is valid and changes these probabilities to $(1+d_b)/2$ and $(1-d_b)/2$, with $d_b=a_bt_b\in[t_b/2,t_b]$. By contraction it suffices to lower-bound the total variation distance of products of these opposite biased coins.

Let $S=\sum_bd_b^2$ and let $x_b$ be independent uniform signs. That total variation distance equals
\[
\frac12\E\Big|\prod_b(1+d_bx_b)-\prod_b(1-d_bx_b)\Big|=\E|L+Z| ,
\]
where $L=\sum_bd_bx_b$ and $Z$ consists of the odd-degree terms of degree at least three. Orthogonality of sign monomials gives
\[
\E L^2=S,\qquad\E L^4\le3S^2,\qquad\E Z^2\le\sum_{k\ge3}\frac{S^k}{k!}\le S^3e^S/6 .
\]
H\"older interpolation gives $\E|L|\ge\sqrt{S/3}$. If $S\le\tfrac14$, therefore,
\[
\E|L+Z|\ge\sqrt S\Big(1/\sqrt3-S\sqrt{e^S/6}\Big)\ge\sqrt S/3\ge\frac16\sqrt{\textstyle\sum_bt_b^2} .
\]
If $S>\tfrac14$, apply additional independent binary noise reducing all $d_b$ by the common factor $1/(2\sqrt S)$; the resulting $S$ is $\tfrac14$ and its distance is at least $\tfrac16$. Contraction proves \eqref{eq:Bos-tensor} in this case too.
\end{proof}

\begin{lemma}[Direct-sum lower bound]\label{lem:Bos-directsum}
Let a product family of states $\bigotimes_b\rho_b$ over finitely many blocks $b$ be given, each block family compact in trace distance, with independent priors on the blocks. Suppose block $b$ has a constant $A_b>0$ and, for every requested accuracy $e\le e_\star$, a prior such that every adaptive single-copy measurement scheme on block $b$ using $L\le c\,A_b/e^2$ copies has Bayes probability at most $\tfrac15$ of outputting a state $\hat\sigma_b$ with $T(\rho_b,\hat\sigma_b)\le e$, \emph{for arbitrary output states} $\hat\sigma_b$ on block $b$; the constants $c$ and $e_\star$ are uniform in $b$. Then for $\eps<\min\{1/18,e_\star/100\}$ every single-copy adaptive protocol that learns the product family to trace distance $\eps$ with probability $2/3$, even with joint measurements on all blocks within each copy, uses
\begin{equation}\label{eq:Bos-directsum}
L>\frac{c}{10^4}\,\frac{\sum_bA_b}{\eps^2}
\end{equation}
copies.
\end{lemma}

\begin{proof}
Set $S_A=\sum_bA_b$ and $e_b=100\eps\sqrt{A_b/S_A}\le100\eps\le e_\star$, so that $\sum_be_b^2=10^4\eps^2$ and $A_b/e_b^2=S_A/(10^4\eps^2)$ for every $b$. Assume $L\le cS_A/(10^4\eps^2)$; then $L\le cA_b/e_b^2$ for all blocks simultaneously. Take the independent accuracy-dependent hard priors. Replace an arbitrary output state by its nearest point, or an approximate nearest point, $\hat\sigma=\bigotimes_b\hat\sigma_b$ in an $\eps$-net of the compact product family in trace distance. On the original success event $T(\rho,\hat\sigma)\le3\eps$. For $\eps<1/18$, \eqref{eq:Bos-tensor} implies
\begin{equation}\label{eq:Bos-324}
\sum_bT(\rho_b,\hat\sigma_b)^2\le324\eps^2 .
\end{equation}
For each $b$, condition on all other true block parameters and reveal them to the learner. Their copies are then known ancillas: each single-copy POVM element $M_x$ induces $M'_x=\Tr_{-b}[(I\otimes\rho_{-b})M_x]$, a valid POVM on block $b$ with the same outcome statistics, and the adaptive choices and the final output are functions of past outcomes and the revealed parameters. So the original protocol induces a valid single-copy adaptive scheme on block $b$ with $L$ copies, whose output $\hat\sigma_b$ is an arbitrary state. Since the prior is a product, the block Bayes bound applies for each value of the other parameters, and averaging gives
\[
\Pr[T(\rho_b,\hat\sigma_b)>e_b]\ge\tfrac45 .
\]
This does not require independence of the estimated errors. With $Z=\sum_be_b^2\,\mathbf 1\{T(\rho_b,\hat\sigma_b)>e_b\}$ we have $0\le Z\le10^4\eps^2$ and $\E Z\ge8000\eps^2$. On the learner's success event \eqref{eq:Bos-324} implies $Z\le324\eps^2$. If $p$ is the prior-averaged success probability, then
\[
8000\le324p+10000(1-p),\qquad p\le\frac{2000}{9676}<\frac23 ,
\]
contradicting the assumed guarantee, which also holds averaged over the prior.
\end{proof}

\subsection{Lower bounds on dyadic bands}\label{subsec:Bos-bands}
\paragraph{The three hard families.} The lower bounds are assembled from three families of \cite{CMMP26}.

\emph{Cold: \cite[Thm.~2]{CMMP26} at $k=1$.} An $n$-level state $\sigma$ is encoded in the passive state $\rho_\sigma$ of fugacity $\sigma/4$, $\rho_\sigma=\det(I-\sigma/4)\bigoplus_\ell4^{-\ell}\mathrm{Sym}^\ell(\sigma)$. Three properties make learning $\rho_\sigma$ at least as hard as learning $\sigma$. Its mean photon number is at most $\tfrac13$ and, for $\sigma\preceq2I/n$, every occupation is at most $1/(2n-1)$. The normalised one-photon block of any $\eps$-accurate estimate of $\rho_\sigma$ is a $16\eps$-accurate estimate of $\sigma$, for $\eps\le1/100$. And one copy of $\rho_\sigma$ can be prepared exactly from a random number $T$ of copies of $\sigma$ with $\E T\le\tfrac49$: draw $L$ with $\Pr(L=\ell)=\tfrac344^{-\ell}$, project $L$ copies of $\sigma$ onto the symmetric subspace, and retry on failure. Each simulated copy of $\rho_\sigma$ projects a random number $L$ of copies of $\sigma$ jointly onto $\mathrm{Sym}^L$, so a single-copy learner for the cold family yields a learner for $n$-level states $\sigma\preceq2I/n$ that measures entangled blocks of random size, with expected total cost $O(N)$; \cite{CMMP26} (Lemma~1 there) apply the lower bound for variable block sizes of finite-dimensional tomography, whose weighted cost $\sum_jt_j\min\{\sqrt{t_j},n\}$ is $O(N)$, and obtain $\Omega(n^3/\eps^2)$.

\emph{Intermediate: \cite[Thm.~5]{CMMP26}, made explicit in Lemma~\ref{lem:Bos-Bayes}.} The prior has fugacity $r(I+\xi X)$, $r\asymp\nu/(1+\nu)$, with $X$ a traceless GUE matrix conditioned on $\|X\|_\op\le4$; with $\xi\asymp\eps/\sqrt{n\nu}$, learning the state to $\eps$ means learning $X$ to $O(n)$ in trace norm \cite[Lemma~12]{CMMP26}. The prior is spread over a space of dimension about $n^2$, so the posterior can concentrate only after a log-likelihood tilt of order $n^2$, while every adaptive single-copy measurement tilts it by $O(n\nu^2\xi^2)$ per copy \cite[Props.~2 and~3]{CMMP26}, by the photon-number block structure of passive states and Haar moments of the symmetric-power representations. Hence $N\gtrsim n/(\nu^2\xi^2)\asymp n^2/(\nu\eps^2)$. The factor $1/\nu$ is the noise-to-metric ratio of \eqref{eq:Bos-scalar}: heterodyne detection pays it, and this bound shows that every single-copy measurement must.

\emph{Warm: \cite[Thm.~9]{CMMP26}.} A packing of $2^{\Theta(n^2)}$ states $\Gamma(U_a)(\tau_{\nu+\Delta}^{\otimes n/2}\otimes\tau_\nu^{\otimes n/2})\Gamma(U_a)^\dagger$, $\tau_\nu$ thermal of occupation $\nu\ge1$ and $\Delta=C\eps\nu/\sqrt n$, has pairwise trace distance at least $3\eps$ (via their heterodyne distributions and data processing), while the relative entropy between two members is at most $\Delta^2n/\nu^2=O(\eps^2)$ per copy. The Holevo bound then limits the information that $N$ copies can give, under any measurement, to $O(N\eps^2)$, and Fano's inequality needs $\Omega(n^2)$ bits to identify a member. This is the parameter-counting floor $n^2/\eps^2$, valid even for collective measurements.

\begin{lemma}[Finite-constant Bayesian band bound]\label{lem:Bos-Bayes}
There are absolute constants $c_0,e_\star>0$ and $n_\star$ such that for $n\ge n_\star$, $1/n\le\nu\le1$ and $e\le e_\star$ there is a prior on passive $n$-mode states whose occupations all lie in $[\nu,2\nu]$, for which every single-copy adaptive protocol using
\[
L\le c_0\frac{n^2}{\nu e^2}
\]
copies has Bayes probability at most $\tfrac15$ of outputting a state, arbitrary, within trace distance $e$ of the true state.
\end{lemma}

\begin{proof}
This is the proof of \cite[Thm.~5]{CMMP26} with its constants made explicit. Its prior is built from a GUE matrix $X$ conditioned on $\|X\|_\op\le4$, entering with a perturbation size $\xi$; for $\xi\le1/32$ and $\nu\le1$ its occupations lie in $[\nu,2\nu]$, the lower end being attained only on the null set where $X$ has eigenvalue $-4$. The posterior anti-concentration of \cite[Prop.~4]{CMMP26}, which combines their Propositions~1 and~3 (the latter from Proposition~2), has exponent $-c_1n^2+(C_1/\delta_0)Ln\nu^2\xi^2$, with $c_1,C_1$ absolute: every smallness requirement in their Lemma~8 and Propositions~1--3 is of the form $\xi\le$ an absolute constant, and $\E k^2\le12\nu^2n^2$ uses $n\nu\ge1$. Fix $\delta_0=0.01$ and take $L\le c_2n/(\nu^2\xi^2)$, with $c_2$ small enough to retain a negative constant multiple of $n^2$ in the exponent. The ensuing success estimate of \cite[Prop.~5]{CMMP26} is at most $\sqrt{\delta_0+\text{exponentially small terms}}\le\tfrac15$ for $n\ge n_\star$. Choose $\xi=Ce/\sqrt{n\nu}$ with $C$ sufficiently large for the fugacity separation lemma \cite[Lemma~12]{CMMP26}; its smallness requirements hold uniformly for $e\le e_\star$ because $n\nu\ge1$. Then $L\le c_2n/(\nu^2\xi^2)=(c_2/C^2)\,n^2/(\nu e^2)=:c_0n^2/(\nu e^2)$. The proof of \cite[Thm.~5]{CMMP26} starts from an arbitrary output state and projects it to a net of the image of the prior, so the bound holds for arbitrary output states.
\end{proof}

\begin{proof}[Proof of Theorem~\ref{thm:Bos-sep}(ii)]
Apply Lemma~\ref{lem:Bos-directsum} to the bands with $\nu_b\le1$, $m_b\nu_b\ge1$ and $m_b\ge n_\star$, with $A_b=m_b^2/\nu_b$ and the priors of Lemma~\ref{lem:Bos-Bayes}. Their product prior lies in $\mathcal C_{\rm sep}$ when the cold group and the other bands are fixed at an admissible product state (the cold group at the vacuum, the other bands at $\nu_bI$): the band priors have occupations in $[\nu_b,2\nu_b]$, hence covariance at most $\tfrac12+2\nu_{\max}\le K$, and the cold group is the smallest $m_c$ because $\nu_{\min}\ge4c/m_c$. The prior touches a shared dyadic endpoint $\nu_b=2\nu_{b-1}$ only on a null set, so the endpoint convention is immaterial. This gives
\begin{equation}\label{eq:Bos-30}
L=\Omega\Big(\frac{\sum_{b\ \mathrm{eligible}}m_b^2/\nu_b}{\eps^2}\Big).
\end{equation}
Three further contributions are needed.

\emph{Cold group.} The hard family of \cite[Thm.~2]{CMMP26} on $m_c$ modes has occupations at most $1/(2m_c-1)\le c/m_c$ for $c\ge\tfrac23$ \cite[proof of Thm.~5.9]{Rub26b}; it embeds with every other block fixed, and a single-copy learner for the class, run on these products, outputs a state whose cold marginal is $\eps$-accurate by monotonicity of the trace distance under the partial trace. This gives $\Omega(m_c^3/\eps^2)$ for $m_c\ge n_0$ and $\eps<\eps_0$, independently of \eqref{eq:Bos-30}.

\emph{Ineligible low-temperature bands.} If $m_b\nu_b<1$, then $m_b^2/\nu_b<\nu_b^{-3}$. Since the scales are distinct dyadic values and $\nu_{\min}\ge4c/m_c$,
\[
\sum_{b:m_b\nu_b<1}\frac{m_b^2}{\nu_b}\le\frac87\nu_{\min}^{-3}=O_c(m_c^3).
\]
For bands with $m_b<n_\star$, the sum is at most $n_\star^2\sum_b\nu_b^{-1}\le2n_\star^2/\nu_{\min}=O_c(m_c)$, also covered by the cold lower bound.

\emph{Bands above occupation one.} There are at most $\log_2K+1$ such dyadic bands, a constant because $K$ is a fixed class parameter. The hard family of \cite[Thm.~9]{CMMP26} has occupations $\nu$ and $\nu+\Delta$ with $\Delta=C\eps\nu/\sqrt{m_b}\le\nu$, so it embeds inside $[\nu_b,2\nu_b]$, with covariance at most $\tfrac12+2\nu_b\le K$; its proof uses $\nu\ge1$, $\eps<1/1800$ and $m_b$ large, and as a Fano--Holevo bound it holds for arbitrary measurements, a fortiori with known ancillas. For sufficiently large $m_b$ it gives $\Omega(m_b^2/\eps^2)$. Taking the maximum over this bounded number of bands gives their sum up to a $K$-dependent constant. Bands of bounded size have bounded total contribution, already covered by $m_c^3$. For $\nu_b>1$, $m_b^2/\nu_b\le m_b^2$.

Finally the $m^2$ floor and the unweighted band squares in \eqref{eq:Bos-Rsep} are controlled by the weighted sum. With $S=\sum_bm_b^2/\nu_b$, the Cauchy--Schwarz inequality and dyadic summation give
\[
m_w^2\le S\sum_b\nu_b\le2\nu_{\max}S\le KS,\qquad m^2\le2m_c^3+2KS,\qquad\sum_bm_b^2\le\nu_{\max}S .
\]
Therefore $\mathcal R_{\rm sep}$ is equivalent, with $(c,K)$-dependent constants, to $m_c^3+S$. The maximum of the cold lower bound, the \emph{single sum} bound \eqref{eq:Bos-30} and the hot-band bound is at least a constant times their total, which proves $L=\Omega_{c,K}(\mathcal R_{\rm sep}/\eps^2)$.
\end{proof}

\subsection{Proof of Theorem~\ref{thm:Bos}(ii)}\label{subsec:Bos-proof-ii}

\begin{proof}[Proof of Theorem~\ref{thm:Bos}(ii)]
\emph{Upper bound.} Let $r_\ast$ be the canonical count \eqref{eq:Bos-R} of a member of $\mathcal P_K(r;(m_b,\nu_b)_b)$. If $r\ge1$, then $r_\ast\ge r$ because $rn_r\le2$. Also $r_\ast\le2r$: every occupation with index $j>r$ is at least $1/r$, so $jn_j>2$ when $j>2r$. Therefore $r_\ast^3\le8r^3$, and its remaining warm pair sum is no larger than the pair sum over the original prescribed warm groups, which \eqref{eq:Bos-dyadic}, valid for the closed intervals $[\nu_b,2\nu_b]$, bounds by $C_0\sum_bm_b^2/\nu_b$. Apply Theorem~\ref{thm:Bos}(i) with the class budget of \eqref{eq:Bos-thm4}. If $r=0$, $n_i\ge1$ implies $r_\ast\le2$ and $S_w\le m^2/2$, so the same argument gives $O_K(m^2/\eps^2)$.

\emph{Lower bound for $r$ sufficiently large.} The proof of Theorem~\ref{thm:Bos-sep}(ii) applies with $r$ in place of $m_c$, $m-r$ in place of $m_w$, and the lower edge $\nu_{\min}\ge1/r$ of \eqref{eq:Bos-room} in place of $\nu_{\min}\ge4c/m_c$; the room condition \eqref{eq:Bos-room} ensures, as $K\ge2\nu_{\max}+1$ does there, that the full hard families, not only their centre points, satisfy the covariance bound. Two steps change. \emph{Cold group:} the family of \cite[Thm.~2]{CMMP26} on $r$ modes has occupations at most $1/(2r-1)$, which for $r\ge2$ is strictly less than $1/r$ and at most $2/r$, so it lies below all the warm hard blocks and in the present class, and gives $\Omega(r^3/\eps^2)$. \emph{Ineligible bands:}
\[
\sum_{b:m_b\nu_b<1}m_b^2/\nu_b\le(8/7)\nu_{\min}^{-3}\le(8/7)r^3,\qquad\sum_{b:m_b<n_\star}m_b^2/\nu_b\le2n_\star^2/\nu_{\min}\le2n_\star^2r ,
\]
both absorbed by the cold lower bound. The final step gives $(m-r)^2\le2\nu_{\max}S$ and $\sum_bm_b^2\le\nu_{\max}S$ with $S=\sum_bm_b^2/\nu_b$, so $m^2$ and the unweighted band squares are at most $C_K(r^3+S)$, and the maximum of the three lower bounds proves \eqref{eq:Bos-thm4}.

\emph{Bounded $r$, including $r=0$.} Fix the absolute threshold $r_\star$ above which the cold lower bound applies. If $1\le r<r_\star$, then $\nu_{\min}\ge1/r_\star$; if $r=0$, $\nu_{\min}\ge1$. With $\nu_{\max}$ bounded by $K$ there are only $B_K=O_K(1)$ occupied dyadic scales, and the numerator in \eqref{eq:Bos-thm4} is $O_K(m^2)$. For $m$ sufficiently large depending only on $K$, some warm band has size at least $(m-r)/B_K$, exceeding $n_\star$, with $m_b\nu_b\ge1$. Its single-band lower bound, Lemma~\ref{lem:Bos-Bayes} for $\nu_b\le1$ and \cite[Thm.~9]{CMMP26} for $\nu_b\ge1$, is $\Omega_K(m^2/\eps^2)$. The case of an empty warm group has $m=r$ and is covered once $m$ is sufficiently large.
\end{proof}

\subsection{Proof of Corollary~\ref{cor:Bos-twoscale}}\label{subsec:Bos-proof-cor}

\begin{proof}[Proof of Corollary~\ref{cor:Bos-twoscale}]
\emph{(i)} Let $\rho\in\mathcal C$ have occupations $n_1\le\dots\le n_m$. For $j>m_c$, $jn_j>m_c\cdot4c/m_c=4c\ge\tfrac83>2$, so the canonical count of \eqref{eq:Bos-R} satisfies $r\le m_c$ and $r^3\le m_c^3$. In $S_w$, the leftover cold indices $r<i\le m_c$ have $n_i>2/i\ge2/m_c$ by the definition of $r$, and the leftover cold modes cost $O(m_c^3+m_c\sum_wn_i^{-1})$: the pairs among them contribute at most $m_c^2\cdot m_c/4$, and the cross pairs with warm indices at most $2m_c\sum_{i>m_c}1/n_i\le2m_cm_w/\nu\le m_c^3/(4c)+m_w^2/\nu$, using $1/\nu\le m_c/(4c)$. The warm pairs contribute at most $m_w^2/(2\nu)$, and $m^2\le2(m_c^2+m_w^2)$. Hence $\mathcal R(\boldsymbol n)\le C_c(m_c^3+m_w^2(1+1/\nu))$ on $\mathcal C$, and Theorem~\ref{thm:Bos}(i) with the covariance bound $K$ gives (i).

\emph{(ii)} This is the proof of Theorem~\ref{thm:Bos-sep}(ii) for a single warm band, without the direct sum. \emph{Cold group:} its cold-group step, with the warm modes fixed at the thermal product state $\tau_\nu^{\otimes m_w}$ (occupations $\nu$, in $\mathcal C$ since $K\ge\nu+\tfrac12$); the family has occupations $\le1/(2m_c-1)\le c/m_c$ for $c\ge\tfrac23$, $m_c\ge2$, and gives $\Omega(m_c^3/\eps^2)$ for $m_c\ge n_0$ and $\eps<\eps_0$. \emph{Warm group:} with the cold group fixed at the vacuum, the prior of \cite[Thm.~5]{CMMP26} for $1/m_w\le\nu\le1$ (occupations in $[\nu,2\nu]$) and the family of \cite[Thm.~9]{CMMP26} for $\nu\ge1$ (occupations $\nu$ and $\nu+\Delta$, $\Delta=O(\eps\nu/\sqrt{m_w})\le\nu$) lie in $\mathcal C$ because $K\ge2\nu+1$, and the same reduction gives $\Omega(m_w^2/(\nu\eps^2))$, respectively $\Omega(m_w^2/\eps^2)$.

A maximum of lower bounds is a lower bound, and $\max\{a,b\}\ge(a+b)/2$.
\end{proof}

\subsection{Unbounded thermal energy: proof of Theorem~\ref{thm:Bos-energy}}\label{subsec:Bos-energy}

\begin{proof}[Proof of Theorem~\ref{thm:Bos-energy}]
Several sources contribute to the $K$-dependence of Theorem~\ref{thm:Bos}(i), and letting $K\to\infty$ there is not valid. The argument below has absolute constants for arbitrary occupations, unknown passive frame and arbitrary displacement, with no unknown active squeezing. The centred covariance is gauge invariant; its occupation matrix $Q\succeq0$ can have arbitrarily large eigenvalues. The proof also changes the warm reconstruction and permits an empty estimated cold group.
\paragraph{Mean estimation without an energy constant.}
Write the real quadrature covariance as $\Sigma=\Phi_{\mathrm{re}}(Q+I/2)\succeq I/2$ and the displacement as $v$. The heterodyne empirical mean $\hat v$ of $L_m$ copies satisfies
\begin{equation}\label{eq:Bos-57}
\E\,(\hat v-v)^T\Sigma^{-1}(\hat v-v)=\frac{\Tr[\Sigma^{-1}(\Sigma+I/2)]}{L_m}\le\frac{4m}{L_m}.
\end{equation}
For equal-covariance Gaussian states,
\begin{equation}\label{eq:Bos-58}
T\big(D(d)\rho_QD(d)^\dagger,\rho_Q\big)\le\tfrac12\sqrt{d^T\Sigma^{-1}d}.
\end{equation}
This is also \cite[Lemma~18]{CMMP26}. A direct proof avoids an energy-dependent generator estimate. In the passive normal frame, purify each thermal mode of quadrature variance $a=n+\tfrac12$ by a two-mode squeezed state (cross block $c\operatorname{diag}(1,-1)$, $c=\sqrt{n(n+1)}$), and also displace the environment, which does not affect the partial trace, by the $e$ that minimises the variance of the combined displacement generator; the minimum is the Schur complement $a-c^2/a=1/(4a)$ times $|\Omega d|^2$. The Mandelstam--Tamm bound \cite[Lemma~5.12]{Rub26b} on the purification and partial-trace contraction give \eqref{eq:Bos-58}, after summing over the modes and conjugating back by the passive rotation, also at $n=0$, when $c=0$. (The same purification, carried out in the Williamson frame, gives \eqref{eq:Bos-58} for every Gaussian covariance $\Sigma$.) Thus $L_m=Cm/\eps^2$ makes the actual mean-corrected state $\tau$ within $\delta=\eps/A$ of $\rho_Q$ with any prescribed constant success probability, where $A$ is an absolute constant chosen for the cold readout. Apply the estimated displacement on later copies; their exact centred normal occupation is still $Q$. All covariance estimates below use disjoint heterodyne pair differences, so the unknown residual mean causes no bias.

\paragraph{Pilot and cold split, with relative sampling noise.}
Set $D=Q+\mu I$ and $H_D=D^{-1/2}(I+Q)D^{-1/2}$. For passive heterodyne data, $L_F$ independent centred samples and raw error $E$, Lemma~\ref{lem:Bos-effrank} with $\mathsf P=0$ gives \eqref{eq:Bos-effrank} and the equality $L_F\,\E\|D^{-1/2}ED^{-1/2}\|_F^2=(\Tr H_D)^2$ of \eqref{eq:Bos-effrankF}; this retains the actual covariance instead of replacing it by a $K$-dependent norm bound. Since $D$ commutes with $Q$ and $\mu\le2$,
\begin{equation}\label{eq:Bos-60}
\|H_D\|_\op\le\max\{1,\mu^{-1}\}\le3/\mu,\qquad\Tr H_D=m+(1-\mu)\Tr D^{-1}\le m+\Tr D^{-1}.
\end{equation}
Equations \eqref{eq:Bos-44}--\eqref{eq:Bos-46} apply with $q_i=n_i$ exactly, without an upper occupation bound. As $m/\mu=mr_0/2\le m^2/2$, they imply
\begin{equation}\label{eq:Bos-61}
\|H_D\|_\op\Tr H_D\le C_0\mathcal R,\qquad\Tr H_D\le C_0\mathcal R,\qquad\mu(\Tr H_D)^2\le C_0\mathcal R ;
\end{equation}
in the second inequality use $\mu\le2$ and \eqref{eq:Bos-46}, in the third $\mu m^2\le2m^2$ and \eqref{eq:Bos-45}. Thus both the relative operator-norm event \eqref{eq:Bos-37} and a small pinching error \eqref{eq:Bos-split-schur} cost at most $C_0\mathcal R/\eps^2$ copies, with absolute constants.

Modify the empirical rule \eqref{eq:Bos-s} to
\begin{equation}\label{eq:Bos-62}
s=\max\big(\{0\}\cup\{j:jx_j\le2\}\big).
\end{equation}
This is essential: forcing one arbitrarily hot mode into the cold readout would introduce an energy-dependent vacuum probability. If $s>0$, its index satisfies $sx_s\le2$, so the argument of Lemma~\ref{lem:Bos-split}(ii) gives $\Tr A<3$ and $\|A\|_\op\le4\mu$, with no forced-singleton exception. The bounds $s\le4r_0$ and $s\ge\lfloor r/2\rfloor$ still hold. If $s=0$ there is no cold readout and no cross block. If $s<m$, maximality gives $x_{s+1}>2/(s+1)\ge\mu/5$, also when $s=0$; hence \eqref{eq:Bos-42}--\eqref{eq:Bos-43} remain valid. The pinching error is bounded by \eqref{eq:Bos-split-schur} with $a=4$, $b=1/8$, an absolute coefficient. For $s=m$ the warm stage is omitted.

The cold target therefore always has total occupation less than $3$, vacuum probability at least $e^{-3}$ and fugacity eigenvalues at most $\tfrac34$. Equations \eqref{eq:Bos-52}--\eqref{eq:Bos-53} with $C_A$ replaced by $3$ and $\delta_K=\tfrac14$, and the one-photon readout of \S\ref{subsec:Bos-proof-i}, give cold error at most $\eps/4$ from $C(1+s^3)/\eps^2$ copies, with absolute constants. The residual displacement enters only through $\delta$ from \eqref{eq:Bos-57}--\eqref{eq:Bos-58}.

\paragraph{Warm reconstruction: the thermal term retained in the metric.}
Let $C=\Pi^\perp Q\Pi^\perp$ on the selected warm subspace, with dimension $d=m-s$, eigenvalues $c_j$, and $I_C=\sum_j1/c_j$. On the pilot good event,
\begin{equation}\label{eq:Bos-63}
\lambda_{\min}(C)\ge\mu/8,\qquad I_C\le8r/\mu+2I_w,\qquad\sum_{ij}\frac1{c_i+c_j}\le C_0\mathcal R .
\end{equation}
For the middle bound, at most $r$ retained indices come from the canonical cold set, each contributing at most $8/\mu$, and the others satisfy \eqref{eq:Bos-43}; this includes $s=0$ and $r\le1$. The last bound is \eqref{eq:Bos-49}.

Take a fresh raw warm occupation estimate $M=C+E_w$. In its relative error the actual sample covariance is $H_w=C^{-1/2}(I+C)C^{-1/2}=I+C^{-1}$, and \eqref{eq:Bos-46}, \eqref{eq:Bos-63} imply
\begin{equation}\label{eq:Bos-64}
\|H_w\|_\op\Tr H_w\le(1+8/\mu)(m+8r/\mu+2I_w)\le C_0\mathcal R .
\end{equation}
Lemma~\ref{lem:Bos-effrank} with $D=C$ therefore proves, with a prescribed high constant probability,
\begin{equation}\label{eq:Bos-65}
\|C^{-1/2}E_wC^{-1/2}\|_\op\le\tfrac12,\qquad M\succeq C/2\succ0
\end{equation}
from $L_w\ge C_0\mathcal R$ samples. On this event no positive-part repair is necessary. Output $\rho_M$ if $M$ is positive definite and a fixed valid state otherwise; positivity is an implementable check, and the unknown comparison $M\succeq C/2$ is used only in the analysis.

Proposition~\ref{prop:Bos-onesided} with $\alpha=\tfrac12$ gives, on \eqref{eq:Bos-65},
\[
T(\rho_C,\rho_M)^2\le2\sum_{ij}\frac{|(E_w)_{ij}|^2}{c_i+c_j+2c_ic_j}.
\]
By \eqref{eq:Bos-noncirc} with $\mathsf P=0$,
\begin{equation}\label{eq:Bos-66}
\E\sum_{ij}\frac{|(E_w)_{ij}|^2}{c_i+c_j+2c_ic_j}=\frac1{L_w}\sum_{ij}\frac{(1+c_i)(1+c_j)}{c_i+c_j+2c_ic_j}\le\frac1{L_w}\Big(d^2+\sum_{ij}\frac1{c_i+c_j}\Big)\le\frac{C_0\mathcal R}{L_w},
\end{equation}
by the upper bound of \eqref{eq:Bos-scalar} with $s=c_i+c_j$, $p=c_ic_j$.
Markov's inequality on the raw metric error, together with \eqref{eq:Bos-65}, gives warm trace error at most $\eps/4$ at $L_w=C_0\mathcal R/\eps^2$.

\paragraph{Assembly.}
Use independent mean, pilot, cold and warm batches, with fixed fractions of the input budget and the same abort-on-insufficient-budget rule as in \S\ref{subsec:Bos-proof-i}. On the good events,
\[
T(\tau,\hat\sigma_c\otimes\hat\sigma_w)\le\delta+T(\rho_Q,\rho_{A\oplus C})+T(\rho_A,\hat\sigma_c)+T(\rho_C,\hat\sigma_w)\le\eps .
\]
Choose the constants so that the mean event fails with probability at most $0.03$, the pilot relative-error and pinching events together at most $0.04$, the cold readout at most $0.15$, and the warm relative-error and metric events together at most $0.04$; the total is below $1/3$. The mean costs $O(m/\eps^2)$ copies, all remaining stages $O(\mathcal R/\eps^2)$, and $s^3\le64r_0^3\le64\mathcal R$. This proves \eqref{eq:Bos-thm5}.
\end{proof}

\section{Proofs for Section~\ref{sec:unsqueezing}}\label{app:proofs-NG}

\subsection{An explicit single-copy wavelet (affine-group) POVM}\label{subsec:NG-wavelet}

Work in the position representation $L^2(\R)$ of one mode. For $s\in\{+,-\}$ let
\begin{equation}\label{eq:NG-J}
(J_s\psi)(t):=\frac{\psi(s\sqrt t)}{\sqrt2\,t^{1/4}},\qquad t>0 .
\end{equation}
Since $\int_0^\infty|\psi(s\sqrt t)|^2\,dt/(2\sqrt t)=\int_0^\infty|\psi(sx)|^2\,dx$, the map $J_+\oplus J_-$ is a unitary from $L^2(\R)$ onto $L^2(\R_+)\oplus L^2(\R_+)$, with inverse $(J_s^{-1}f)(sx)=\sqrt2\,x^{1/2}f(x^2)$ for $x>0$. On $L^2(\R_+)$ take the seed
\[
\varphi(t):=\frac{t^{5/4}e^{-t/2}}{\sqrt{\Gamma(7/2)}},\qquad\|\varphi\|_2=1,\qquad C_\varphi:=\int_0^\infty|\varphi(t)|^2\,\frac{dt}t=\frac{\Gamma(5/2)}{\Gamma(7/2)}=\frac25,
\]
and for $a>0$, $b\in\R$ put
\begin{equation}\label{eq:NG-M}
\varphi_{a,b}(t):=\sqrt a\,e^{ibt/2}\varphi(at),\qquad M(da\,db):=|\varphi_{a,b}\rangle\langle\varphi_{a,b}|\,\frac{da\,db}{4\pi C_\varphi a^2}.
\end{equation}

\begin{lemma}[normalisation]\label{lem:NG-norm}
For every $f\in L^2(\R_+)$, $\int_{(0,\infty)\times\R}\langle f|M(da\,db)|f\rangle=\|f\|_2^2$. Hence $M$ is a POVM on $L^2(\R_+)$ with outcomes $(a,b)\in(0,\infty)\times\R$.
\end{lemma}

The one-mode POVM $\mathsf M$ is the direct sum of $M$ on the two half-lines: its outcomes are $(s,a,b)$, with effects $J_s^\dagger M(da\,db)J_s$. Recording $s$ is optional; if $s$ is not recorded, the effect of $(a,b)$ is the rank-two operator $\sum_{s=\pm}J_s^\dagger|\varphi_{a,b}\rangle\langle\varphi_{a,b}|J_s\;da\,db/(4\pi C_\varphi a^2)$. In the position representation the normalised effect vector is
\begin{equation}\label{eq:NG-effect}
(J_s^{-1}\varphi_{a,b})(x)=\sqrt{\frac2{\Gamma(7/2)}}\;\mathbf 1_{\{sx>0\}}\,a^{7/4}|x|^3e^{-(a-ib)x^2/2}.
\end{equation}
This is not a Gaussian measurement: the half-line support and the polynomial factor $|x|^3$ are essential to the construction, and the quadratic form of the last exponent does not make it Gaussian. Lemma~\ref{lem:NG-norm} proves that $\mathsf M$ is a valid POVM.

After the substitution $t=x^2$, Lemma~\ref{lem:NG-norm} is the continuous-wavelet resolution of the identity of Grossmann, Morlet and Paul \cite{GMP85} for the affine group of the line in its positive-frequency irreducible representation, with the admissibility constant $C_\varphi$ of Duflo and Moore \cite{DM76}; the proof is included to fix the constants. The dilations $\psi(x)\mapsto r^{1/4}\psi(\sqrt r\,x)$ and the quadratic shears $\psi(x)\mapsto e^{icx^2/2}\psi(x)$, both Gaussian unitaries, map the effect vector of $(a,b)$ to those of $(ra,rb)$ and $(a,b+c)$ respectively. Covariant measurements for a different affine-group action, translation and dilation, were used in \cite{CDS06} to estimate real squeezing and displacement of pure states.

\subsection{The scale law for centred one-mode states}\label{subsec:NG-law}

Write the covariance of a one-mode state as
\[
\Sigma=\begin{pmatrix}\sigma&c\\c&w\end{pmatrix},\qquad\beta:=\frac c\sigma,\qquad h:=w-\frac{c^2}\sigma,\qquad\sigma h=\det\Sigma\ge\tfrac14,
\]
and put $\Delta:=4\det\Sigma=4\sigma h\ge1$ ($\Delta=1$ exactly for pure states) and $\vartheta:=h-1/(4\sigma)\ge0$.

\begin{lemma}[scale law]\label{lem:NG-scale}
The marginal density of $a$ is
\[
f_a(a)=\frac{8}{3\pi\sqrt{2\sigma}}\,\frac{a^{3/2}}{(a+1/(2\sigma))^3},
\]
so $\tau:=2\sigma a$ has density $\frac8{3\pi}\,u^{3/2}(1+u)^{-3}$ on $u>0$, that is, $\tau\sim\mathrm{BetaPrime}(5/2,1/2)$. This law does not depend on $c$ or $h$.
\end{lemma}

The scale estimate thus has a multiplicative error law independent of the energy. Its mean is infinite, so the proofs below use a fixed quantile event and not a first moment of $a$.

Lemma~\ref{lem:NG-scale} has a classical reading, which re-derives it. Write $\mathrm{Gamma}(\alpha,\lambda)$ for the law with density $\lambda^\alpha r^{\alpha-1}e^{-\lambda r}/\Gamma(\alpha)$ on $r>0$, and $\mathrm{BetaPrime}(\alpha,\beta)$ for the law with density $u^{\alpha-1}(1+u)^{-\alpha-\beta}/B(\alpha,\beta)$, which is the law of $s/g$ for independent $s\sim\mathrm{Gamma}(\alpha,1)$, $g\sim\mathrm{Gamma}(\beta,1)$. Integrating $M(da\,db)$ over $b$ leaves, by the Plancherel step of Lemma~\ref{lem:NG-norm}, the multiplication operator $(C_\varphi a)^{-1}\varphi(ax^2)^2\,da$ in position. The scale outcome is thus a randomised readout of position: draw $x$ from the position law of the state, then $a$ with density $(C_\varphi a)^{-1}\varphi(ax^2)^2$. In $s:=ax^2$ this density is $\varphi(s)^2/(C_\varphi s)=s^{3/2}e^{-s}/\Gamma(\frac52)$, so $s\sim\mathrm{Gamma}(\frac52,1)$ independently of $x$, and $a=s/x^2$. For a centred Gaussian state $x^2/(2\sigma)\sim\mathrm{Gamma}(\frac12,1)$, hence
\begin{equation}\label{eq:NG-gammaratio}
\tau=2\sigma a=\frac{s}{x^2/(2\sigma)}\sim\mathrm{BetaPrime}\big(\tfrac52,\tfrac12\big),
\end{equation}
a ratio of a variable supplied by the fiducial and one supplied by the state, neither of which knows the squeezing. The heavy tail $\Pr[\tau>X]\sim\frac{16}{3\pi}X^{-1/2}$ comes from the denominator: a copy found within $\eta\sqrt\sigma$ of the origin, an event of probability of order $\eta$, reports a packet whose position variance is smaller than the state's by a factor of order $\eta^{-2}$.

\subsection{One-shot domination}\label{subsec:NG-NG1}

From the outcome $(a,b)$ define
\begin{equation}\label{eq:NG-PS}
P(a,b):=\frac12\begin{pmatrix}1/a&b/a\\b/a&a+b^2/a\end{pmatrix},\qquad S(a,b):=\begin{pmatrix}a^{-1/2}&0\\ba^{-1/2}&a^{1/2}\end{pmatrix}.
\end{equation}

\begin{lemma}[uniform one-shot domination]\label{lem:NG-dom}
For every centred one-mode Gaussian state $\rho(0,\Sigma)$, the outcome $(a,b)$ of one measurement with $\mathsf M$ satisfies
\[
\Pr\big[P(a,b)\preceq10^6\,\Sigma\big]>0.96 .
\]
\end{lemma}

\begin{remark}[the size of the constant]\label{rem:NG-constant}
The constant in \eqref{eq:NG-N1} is absolute but very large: it is the budget of Lemma~\ref{lem:NG-target} at $C=10^6$, $m=1$, $\delta=0.05$. Its size comes from the domination factor $10^6$, which a single wavelet outcome needs because the scale law of Lemma~\ref{lem:NG-scale} has the heavy tail $\Pr[\tau>x]\sim\frac{16}{3\pi}x^{-1/2}$. It is not optimised.
\end{remark}

\begin{remark}[the order $\eps^{-2}$ is optimal]\label{rem:NG-lower}
Let $\omega_1,\omega_{1+u}$ be the thermal states with mean photon numbers $1$ and $1+u$, $0\le u\le\frac12$. Their vacuum probabilities differ by $\frac12-\frac1{2+u}=\frac u{2(2+u)}$, so $T(\omega_1,\omega_{1+u})\ge u/(2(2+u))$. They commute, and their geometric photon-number laws give
\[
D(\omega_1\|\omega_{1+u})=2\log(1+u/2)-\log(1+u)\le u^2/2,
\]
by integrating the derivative $u/((2+u)(1+u))\le u$. Take $u=16\eps$ with $\eps\le1/100$; then $T(\omega_1,\omega_{1+u})\ge8\eps/(2+16\eps)>2\eps$. A protocol that learns both states to trace distance $\eps$ with probability at least $2/3$ distinguishes them with probability at least $2/3$, so $T(\omega_1^{\otimes N},\omega_{1+u}^{\otimes N})\ge1/3$; by Pinsker's inequality and additivity of the relative entropy, $1/3\le\sqrt{Nu^2/4}=8\eps\sqrt N$, that is, $N\ge1/(576\eps^2)$, for arbitrary collective measurements. Both states are centred, so the one-mode orders of Theorem~\ref{thm:NG1} are matching.
\end{remark}

\subsection{Covariance moments of the multimode covariant measurement}\label{subsec:NG-moments}

The wavelet construction does not extend to many modes by measuring the physical modes separately: an entangled multimode squeezed state may have unsqueezed thermal marginals. We use instead the covariant measurement of \cite[Sec.~8.2]{CGYZ26} with its exact covariance moments (computed below).

In the two-mode squeezed vacuum with squeezing $r$, for instance, the collective quadratures $(q_1-q_2)/\sqrt2$ and $(p_1+p_2)/\sqrt2$ have variance $\frac12e^{-2r}$, while each mode alone is a thermal state of occupation $\sinh^2r$ with no squeezing; the one-mode marginal covariance matrices alone therefore do not identify the squeezed collective quadratures.

\paragraph{The covariant measurement on the Siegel disk.}
Centred pure Gaussian states of $d$ modes are parametrised by the Siegel disk $\mathfrak D_d$ of complex symmetric $d\times d$ matrices $K$ with $KK^\dagger\prec I$, through $|\psi_K\rangle=\det(I-KK^\dagger)^{1/4}\exp(\frac12a^{\dagger T}Ka^\dagger)|0\rangle$. For $k>2d$ copies,
\[
M_k(dK)=\frac1Z\,|\psi_K\rangle\langle\psi_K|^{\otimes k}\,\frac{dK}{\det(I-KK^\dagger)^{d+1}},
\]
with $dK$ the Lebesgue measure and $Z$ a normalising constant, is a POVM on the span of $\{|\psi_K\rangle^{\otimes k}\}$ \cite[Lemma~8.5]{CGYZ26}, completed by the projector onto the orthogonal complement, which plays no role for inputs of the form $\psi^{\otimes k}$. It is covariant, $U_S^{\otimes k}M_k(\mathcal B)U_S^{\dagger\otimes k}=M_k(S\cdot\mathcal B)$ for centred Gaussian unitaries \cite[Lemma~8.13]{CGYZ26}, so the outcome for $|\psi_{S\cdot0}\rangle^{\otimes k}$, transformed by $S^{-1}$, has the law of the outcome for the vacuum \cite[Cor.~8.14]{CGYZ26}. The same POVM is the bosonic Hayashi measurement $\mu_{k,d}$ of \cite[App.~D.4]{CFGLMWW26}: both are a constant times $\int|\phi\rangle\langle\phi|^{\otimes k}$ against a symplectically invariant measure on the orbit of the vacuum, such a measure is unique up to a scalar, and both resolve the projector onto the span of the $k$-fold powers; hence they coincide.

Write $W$ for the covariance of the pure state $\psi_K$ of the outcome and, at the vacuum, $\Theta:=2W$. At the vacuum the outcome density is proportional to $\det(I-KK^\dagger)^\gamma$ with $\gamma:=k/2-d-1$, and the squared singular values $l_i$ of $K$ have the density \eqref{eq:NG-radial} below. By covariance, for $k$ copies of a pure state with covariance $\Sigma_p$, $W$ is an estimator of $\Sigma_p$ with a scalar bias $c_k$ and, for $\gamma>1$, a finite second moment in relative coordinates, independent of the energy, at block size $O(d)$ (Lemmas~\ref{lem:NG-radial} and~\ref{lem:NG-pure}). A single block is not required to reconstruct the state to small trace distance.

The law of $\Theta$ at the vacuum is invariant under conjugation by the orthogonal symplectic matrices, whose commutant is spanned by $I$ and the antisymmetric $\Omega$, so $\E\Theta=c_kI$ and, by covariance, $\E W=c_k\Sigma_p$ for every pure input; $c_k>1$ because a randomly oriented squeezed state has eigenvalues $e^{\pm2r}$ averaging to $\cosh2r$. Near $l=1$ the density \eqref{eq:NG-radial} behaves like $(1-l)^\gamma$, and $x=l/(1-l)=\sinh^2r\approx\frac14e^{2r}$, so the largest outcome squeezing has the power tail $\Pr[x>X]\approx X^{-(\gamma+1)}$, each extra pair of copies raising the exponent by one. The second moment of $\Theta$ is finite for $\gamma>1$, that is $k>2d+4$, and with $d=2m$ the least block is $k=4m+5$, at $\gamma=\frac32$, where one block is a poor estimate ($c_k=k/3$, relative second moment of order $d^2$) but $O(m^2)$ blocks average it to a constant relative error. For $\gamma\ge2(d+1)$, that is $k\ge12m+6$, $c_k\le\frac32$ and the relative second moment of one block is at most $12d$, so $O(m)$ blocks suffice; this is the block size $12m+7$ of Remark~\ref{rem:NG-12m}.

\subsection{Mixed inputs}\label{subsec:NG-mixed}

Let $\rho(0,\Sigma)$ be a centred $m$-mode Gaussian state, $\psi_{\rm std}$ its standard purification on $2m$ modes, which is a pure Gaussian state \cite[Lemma~D.1]{CFGLMWW26}, $\Sigma_p$ the covariance of $\psi_{\rm std}$, and $J_A$ the coordinate projection onto the quadratures of the $m$ system modes, so that $J_A\Sigma_pJ_A^T=\Sigma$. On $k$ copies, apply the bosonic quasi-purification channel $\mathcal Q^{\rm Bose}_{k,m}$ of \cite[App.~D.5, Eq.~(D.29)]{CFGLMWW26} to $\rho(0,\Sigma)^{\otimes k}$, measure the output with the $2m$-mode covariant POVM $M_k$ of \S\ref{subsec:NG-moments} (that is, $\mu_{k,2m}$), and output $W_A:=J_AWJ_A^T$, where $W$ is the covariance of the outcome. The source requires $k\ge2m+1$ for the twirl on the purifying modes \cite[Lemma~D.8]{CFGLMWW26} and $k\ge2(2m)+1=4m+1$ for the $2m$-mode Hayashi measurement; $k=4m+5$, used below, satisfies both.

\paragraph{The reduced-law identity.}
The proof of \cite[Thm.~5.1]{CFGLMWW26}, in App.~D.5 of that paper, shows that for every centred Gaussian state $\rho(0,\Sigma)$, whatever its symplectic spectrum, the law of the reduced estimate obtained in this way is equal, as a measure, to the law obtained by measuring $k$ copies of $\psi_{\rm std}$ with the same POVM and reducing; the identity is restated in the proof of \cite[App.~D.6, Lemma~D.14]{CFGLMWW26}. The protocol uses ordinary copies only; the identity concerns the law of the reduced estimate, and this equality of laws is what the proof of Proposition~\ref{prop:NG-mixed} uses from \cite{CFGLMWW26}. It holds for bounded test functions, and for nonnegative unbounded ones by monotone convergence, with finiteness supplied by \eqref{eq:NG-puremom} and \eqref{eq:NG-puremom32}.

\begin{proposition}[energy-free relative moments for mixed inputs]\label{prop:NG-mixed}
Let $\rho(0,\Sigma)$ be any centred $m$-mode Gaussian state.
(i) For $k=12m+6$,
\begin{equation}\label{eq:NG-mixedmom}
\E W_A=\tfrac32\Sigma,\qquad\E\big\|\Sigma^{-1/2}(W_A-\tfrac32\Sigma)\Sigma^{-1/2}\big\|_F^2\le24m .
\end{equation}
(ii) For $k=4m+5$, with $c_k=(4m+5)/3$,
\begin{equation}\label{eq:NG-mixedmom32}
\E W_A=c_k\Sigma,\qquad\E\big\|\Sigma^{-1/2}(W_A/c_k-\Sigma)\Sigma^{-1/2}\big\|_F^2\le160m^2 .
\end{equation}
In both cases every output $W_A$ is the reduced covariance of a pure Gaussian state, hence a physical covariance: $W_A+\frac i2\Omega\succeq0$. The rescaled output $W_A/c_k$ need not be a physical covariance.
\end{proposition}

\subsection{The gap class \texorpdfstring{$\mathcal K_c$}{K\_c} and a continuity bound in the whitened metric}\label{subsec:NG-continuity}

For $c>0$ let
\[
\mathcal K_c:=\big\{\Sigma:\ \Sigma+\tfrac i2(1+c)\Omega\succeq0\big\},
\]
the covariances whose symplectic eigenvalues are all at least $(1+c)/2$; there is no upper bound. $\mathcal K_c$ is convex and invariant under $\Sigma\mapsto S\Sigma S^T$, $S$ symplectic.

\begin{lemma}[whitened continuity]\label{lem:NG-cont}
Let $c>0$ and $\kappa_c:=(1+c)/(2\sqrt{2c(2+c)})$.
\begin{enumerate}
\item[(i)] For $\Sigma_0,\Sigma_1\in\mathcal K_c$ with $\eta:=\|\Sigma_0^{-1/2}(\Sigma_1-\Sigma_0)\Sigma_0^{-1/2}\|_F<1$ and every mean $v$,
\[
T\big(\rho(v,\Sigma_0),\rho(v,\Sigma_1)\big)\le\kappa_c\,\frac{\eta}{1-\eta}.
\]
\item[(ii)] For $\Sigma_0\in\mathcal K_c$ and means $v,v'$, $T(\rho(v,\Sigma_0),\rho(v',\Sigma_0))\le\frac12\|\Sigma_0^{-1/2}(v-v')\|_2$.
\end{enumerate}
\end{lemma}
Part (i) is a two-sided relative of \cite[Thm.~7]{CMMP26}: if the symplectic eigenvalues of $\bar\Sigma$ are at least $\nu_->\tfrac12$ and $\Sigma'\succeq\bar\Sigma$, then $T(\rho(0,\Sigma'),\rho(0,\bar\Sigma))\le\|\bar\Sigma^{-1/2}(\Sigma'-\bar\Sigma)\bar\Sigma^{-1/2}\|_F/(2\sqrt{2\gamma_-})$ with $\gamma_-=1-1/(2\nu_-)$. There the gap is assumed for one covariance and the other dominates it, and so inherits the gap; here both have a gap, no order is assumed, and the price is the factor $1/(1-\eta)$. For $\nu_-=(1+c)/2$ the constant of \cite{CMMP26} is $\sqrt{1+c}/(2\sqrt{2c})$, and $\kappa_c$ is that constant times $\sqrt{(1+c)/(2+c)}$.

\subsection{Gaussian protocols: the obstruction and the \texorpdfstring{$\log\log E$}{log log E} route}

\paragraph{The Gaussian obstruction.}
In the proof of \cite[Thms.~1.1 and~1.2]{CGYZ26} (Secs.~5--6 there) the hard family is the one-mode squeezed vacua $\Sigma_\theta=\frac12R(\theta)\operatorname{diag}(\lambda,\lambda^{-1})R(\theta)^T$, $R(\theta)$ a phase-space rotation, $\lambda\asymp E$ fixed, $\theta$ uniformly random. Their fidelity $2/\sqrt{4+(\lambda-\lambda^{-1})^2\sin^2(\theta-\theta')}$ shows that trace distance $\eps$ requires the angle to precision $O(\eps/E)$ \cite[Lemma~5.3]{CGYZ26}, hence Fisher information of order $E^2$. A $k$-copy Gaussian measurement has a Gaussian outcome whose covariance is that of the copies plus a seed fixed in advance (vacuum noise $\frac12I$ for heterodyne detection); a seed squeezed along a direction is informative only about angles within $O(1/E)$ of it, and averaged over $\theta$ its Fisher information is at most $4kE$ \cite[Lemma~6.1]{CGYZ26} (there $k(4E'-2)$, with $E'=E+\frac12$ the energy including the vacuum contribution). Adaptivity aligns the next seed with the posterior, whose density after expected Fisher information $\bar J$ is $O(1+\sqrt{\bar J})$ \cite[Lemma~5.5]{CGYZ26}, so $\bar J_t\le\bar J_{t-1}+O(kE)(1+\pi\sqrt{\bar J_{t-1}})$ \cite[Cor.~5.7 and Lemma~5.8, Eq.~(54)]{CGYZ26} and, for single-copy measurements, $\bar J_t$ grows at most as $E,E^{3/2},E^{7/4},\dots$: the residual misalignment factor $E^{2^{-t}}$ is at best square-rooted per round, and $\Omega(\log\log E)$ copies are needed, a $k$-copy measurement being worth at most $k$ adaptive single-copy rounds \cite[Sec.~6.2.2]{CGYZ26} (with $\ell$ rounds of adaptivity the last pays the rest in copies, \cite[Thm.~1.2]{CGYZ26}).

In \cite[Thm.~3; supplement, Lemma~12 and Thm.~14]{BMEM25}, with $V=2\Sigma$, $O(m)$ heterodyne samples estimate $V$ with relative error of order $1+\|V^{-1}\|_\op$ times a small constant, the vacuum noise being resolved against the narrowest quadrature (\S\ref{subsec:NG-target}). Undoing the Williamson frame of the estimate leaves $\|V_{\rm new}^{-1}\|_\op\le C'(\|V^{-1}\|_\op+1)^{1/2}$, $C'$ absolute: the error splits evenly between a quadrature and its conjugate (a symplectic geometric-mean inequality), so each round square-roots the residual squeezing, and as $\|V^{-1}\|_\op\le\|V\|_\op=O(E+m)$, $\lceil\log_2\log_2\|V^{-1}\|_\op\rceil$ rounds precede a final heterodyne round of $O(m^3/\eps^2)$ shots. Feeding $\rho^T$ instead of the vacuum into the second port of the heterodyne beam splitter gives samples of covariance exactly $V$, without vacuum noise, and removes the energy dependence \cite[Thm.~4]{BMEM25}.

\subsection{Heterodyne amplification, normalisation and admissibility}

\paragraph{The factor $2m+\Tr V^{-1}$ of \cite{BMEM25}.} The mechanism of \S\ref{subsec:NG-target}, vacuum noise amplified by the inverse of the narrowest quadrature, also produces the factor $2m+\Tr V^{-1}$ of \cite[supplement, Thm.~10]{BMEM25}, in their convention $V=2\Sigma$: their estimator $\hat V=2\hat\Gamma/(1-\zeta)-I$, with $\zeta$ of order $(\sqrt{2m}+\sqrt{2\log(2/\delta)})/\sqrt N$, satisfies $V\preceq\hat V\preceq V+\frac{2\zeta}{1-\zeta}(V+I)$ with probability $1-\delta$; the covariance part of their perturbation bound \cite[Thm.~2 (supplement, Thm.~8)]{BMEM25}, $\frac{1+\sqrt3}8\Tr((V^{-1}+\hat V^{-1})|\hat V-V|)$, weighs this error by $V^{-1}$ (note $\hat V^{-1}\preceq V^{-1}$), and $\Tr(V^{-1}(V+I))=2m+\Tr V^{-1}$.

\begin{proof}[Proof of Lemma~\ref{lem:NG-target}]
$\Sigma'\succeq C^{-1}S^{-1}PS^{-T}=\frac1{2C}I$. In the convention of \cite{BMEM25} the covariance of $\rho'$ is $V'=2\Sigma'\succeq C^{-1}I$, so $\Tr V'^{-1}\le2mC$. Put $x:=\chi/\sqrt N\le\frac1{10}$ and $\zeta:=2x+2x^2\le0.22$, so that $\zeta/(1-\zeta)\le\frac{110}{39}x$ and the estimator $\hat V=2\hat\Gamma/(1-\zeta)-I$ of \cite[supplement, Thm.~10]{BMEM25} is defined; its proof divides by $1-\zeta$, which is why the condition $N\ge100\chi^2$ is imposed. On the concentration event of that proof, of probability at least $1-\delta$, the empirical heterodyne covariance lies between $(1-\zeta)(V'+I)/2$ and $(1+\zeta)(V'+I)/2$, so $V'\preceq\hat V$ and $\hat V+i\Omega\succeq V'+i\Omega\succeq0$, the uncertainty relation holds and the fallback is not used; and the bound of that proof before its final numerical step gives
\begin{align*}
T(\rho',\hat\rho')&\le\Big(\frac{x}{2\sqrt2}+\frac{1+\sqrt3}2\,\frac{\zeta}{1-\zeta}\Big)\big(2m+\Tr V'^{-1}\big)\\
&\le\Big(\frac1{2\sqrt2}+\frac{55(1+\sqrt3)}{39}\Big)x\,(2m+2mC)<4.3\,(2m+2mC)\,x ,
\end{align*}
which is \eqref{eq:NG-het}; the bound depends on the state only through $\Tr V'^{-1}$. The budgets used in Theorems~\ref{thm:NG1} and~\ref{thm:NG2} satisfy $N\ge100\chi^2$, since there $4.3\,(2m+2mC)\chi/\sqrt N\le\eps<1$ with $C\ge1$ forces $\sqrt N>17\chi$. $U_S$ is unitary, so $T$ is unchanged.
\end{proof}

\begin{proof}[Proof of Lemma~\ref{lem:NG-norm}]
Fix $a>0$. Then $\langle\varphi_{a,b}|f\rangle=\int_0^\infty e^{-ibt/2}g_a(t)\,dt$ with $g_a(t):=\sqrt a\,\varphi(at)f(t)$, which lies in $L^1\cap L^2$ because $\varphi$ is bounded and square integrable. By Plancherel, with the substitution $\omega=b/2$,
\[
\int_\R|\langle\varphi_{a,b}|f\rangle|^2\,db=2\cdot2\pi\int_0^\infty|g_a(t)|^2\,dt=4\pi a\int_0^\infty\varphi(at)^2|f(t)|^2\,dt .
\]
By Tonelli,
\[
\int\langle f|M(da\,db)|f\rangle=\int_0^\infty|f(t)|^2\Big(\frac1{C_\varphi}\int_0^\infty\varphi(at)^2\,\frac{da}a\Big)dt=\|f\|_2^2,
\]
since $\int_0^\infty\varphi(at)^2\,da/a=\int_0^\infty\varphi(s)^2\,ds/s=C_\varphi$ for every $t>0$. For every Borel set $\Delta$ the quadratic form $f\mapsto\int_\Delta\langle f|M(da\,db)|f\rangle$ therefore lies between $0$ and $\|f\|_2^2$, and defines by polarisation an operator $0\preceq M(\Delta)\preceq I$; $M$ of the whole outcome space is $I$, and countable additivity holds weakly by monotone convergence.
\end{proof}

\paragraph{The affine group and admissibility.}
The centred pure one-mode Gaussian states are the wave functions proportional to $e^{-(a-ib)x^2/2}$, $a>0$, $b\in\R$, that is, the points $b+ia$ of the Siegel upper half-plane \cite[Sec.~4.4]{CGYZ26}. Dilations and shears form the affine (``$ax+b$'') group, the factor $AN$ of the Iwasawa decomposition of $\mathrm{Sp}(2,\R)$ complementary to the rotations, which acts simply transitively on this half-plane. In $t=x^2$ the dilation is $f(t)\mapsto\sqrt r\,f(rt)$ and the shear is multiplication by $e^{ict/2}$, a translation after Fourier transformation in $t$: the positive-frequency representation of wavelet analysis. The group is not unimodular (left Haar measure $da\,db/a^2$, right $da\,db/a$), and the orbit of a fiducial vector resolves the identity only if the fiducial is admissible, $C_\varphi<\infty$ \cite{DM76,GMP85}. The covariance relations stated in \S\ref{subsec:NG-wavelet} follow from \eqref{eq:NG-effect}, since $|x|^3$ is homogeneous and $da\,db/a^2$ is invariant under $(a,b)\mapsto(ra,rb)$ and $(a,b)\mapsto(a,b+c)$. In position $C_\varphi=\int_0^\infty|\psi(sx)|^2x^{-2}\,dx$ for $\varphi=J_s\psi$, which is infinite for every wave function that does not vanish at the origin, Gaussian ones included; for the vacuum, $J_s\psi_{\rm vac}\propto t^{-1/4}e^{-t/2}$ and $\int_0^\infty t^{-3/2}e^{-t}\,dt=\infty$ is the one-copy, one-mode case of the divergence that forces $k>2d$ (Siegel-disk paragraph below). A fiducial $t^\alpha e^{-t/2}$, which is $|x|^{2\alpha+1/2}e^{-x^2/2}$ in position, is admissible exactly when $\alpha>0$; $\alpha=\frac54$, the factor $|x|^3$, is chosen so that the shear outcome also has a finite conditional second moment (Lemma~\ref{lem:NG-shear}). Physically, $\mathsf M$ asks which centred squeezed vacuum the copy resembles; the factor $|x|^3$ has no scale of its own, so squeezing the input rescales the answer and shearing it shifts the answer, and its zero at the origin, the non-Gaussian part, is the price of admissibility on a single copy.

\subsection{The chirp, a worked example, and Theorem~\ref{thm:NG1}}

The Wigner function of $\rho(0,\Sigma)$ is $W(q,p)=N(q;0,\sigma)\,N(p;\beta q,h)$, and its position kernel $\rho(x,y)=\int W(\frac{x+y}2,p)\,e^{ip(x-y)}\,dp$ is
\begin{equation}\label{eq:NG-kernel}
\rho(x,y)=(2\pi\sigma)^{-1/2}\exp\Big[-\frac{(x+y)^2}{8\sigma}-\frac{h(x-y)^2}2+\frac{i\beta(x^2-y^2)}2\Big].
\end{equation}
Its diagonal is $N(0,\sigma)$, and $\rho(-x,-y)=\rho(x,y)$. Since $(J_s\rho J_s^\dagger)(t,u)=\rho(s\sqrt t,s\sqrt u)/(2(tu)^{1/4})$, summing the two sign outcomes gives the kernel on $t,u>0$
\begin{equation}\label{eq:NG-K}
K(t,u):=(tu)^{-1/4}\rho(\sqrt t,\sqrt u).
\end{equation}
It is the sum of two half-line kernels, each with prefactor $\tfrac12$. The blocks of $\rho$ between the two half-lines do not enter, because $\mathsf M$ is block diagonal. With $s$ not recorded, the outcome $(a,b)$ of $\mathsf M$ therefore has density $\langle\varphi_{a,b}|K|\varphi_{a,b}\rangle/(4\pi C_\varphi a^2)$.

\begin{proof}[Proof of Lemma~\ref{lem:NG-scale}]
By the Plancherel step of Lemma~\ref{lem:NG-norm}, the integral over $b$ uses only the diagonal $K(t,t)=(2\pi\sigma)^{-1/2}t^{-1/2}e^{-t/(2\sigma)}$, and $f_a(a)=(C_\varphi a)^{-1}\int_0^\infty\varphi(at)^2K(t,t)\,dt$. With $\lambda:=a+1/(2\sigma)$, the gamma integral $\int_0^\infty t^2e^{-\lambda t}\,dt=2/\lambda^3$ gives the formula, and $B(5/2,1/2)=3\pi/8$.
\end{proof}

\begin{lemma}[conditional shear moments]\label{lem:NG-shear}
With $\lambda=a+1/(2\sigma)$,
\begin{equation}\label{eq:NG-shear}
\E[b\,|\,a]=\beta,\qquad\E[(b-\beta)^2\,|\,a]=\lambda^2+\tfrac12\vartheta\lambda .
\end{equation}
\end{lemma}

\begin{proof}
Combining the kernel \eqref{eq:NG-K}, through $(\sqrt t\pm\sqrt u)^2=t+u\pm2\sqrt{tu}$, with the seed factors, $\langle\varphi_{a,b}|K|\varphi_{a,b}\rangle$ is an $a$-dependent positive factor times $\iint F(t,u)\,e^{-i(b-\beta)(t-u)/2}\,dt\,du$, where
\[
F(t,u)=tu\,\exp\big[-A(t+u)+\vartheta\sqrt{tu}\big],\qquad A:=\frac a2+\frac h2+\frac1{8\sigma}.
\]
Put $z=t-u$, $r=(t+u)/2$ and $\Phi(z):=\int F(r+\frac z2,r-\frac z2)\,dr$. The conditional law of $\xi:=(b-\beta)/2$ given $a$ has characteristic function $\Phi(z)/\Phi(0)$, so $\E[\xi\,|\,a]=-i\Phi'(0)/\Phi(0)$ and $\E[\xi^2\,|\,a]=-\Phi''(0)/\Phi(0)$. At $z=0$ the first $z$-derivative of $F(r+\frac z2,r-\frac z2)$ vanishes, and its second derivative divided by $F$ is
\[
-\frac1{2r^2}-\frac\vartheta{4r}.
\]
The normalised diagonal law of $r$, proportional to $F(r,r)=r^2e^{-(2A-\vartheta)r}$ with $2A-\vartheta=\lambda$, is $\mathrm{Gamma}(3,\lambda)$, so $\E[1/r]=\lambda/2$ and $\E[1/r^2]=\lambda^2/2$. Hence $\E[b-\beta\,|\,a]=0$ and $\E[(b-\beta)^2\,|\,a]=4\,\E[\xi^2\,|\,a]=4(\lambda^2/4+\vartheta\lambda/8)$, which is \eqref{eq:NG-shear}. The boundary terms at $t=0$ or $u=0$ vanish: the factor $tu$ and the square of the corresponding first derivative are integrable. One may first integrate on bounded intervals and take limits, or apply Plancherel to the first derivative. The power $t^{5/4}$ in $\varphi$ ensures this; a seed with too small a power need not give a finite conditional second moment of $b$.
\end{proof}

The shear outcome measures the chirp. By the factorisation $W(q,p)=N(q;0,\sigma)\,N(p;\beta q,h)$ above, $\beta=c/\sigma$ is the chirp of the state, its mean momentum at position $q$ being $\beta q$ as for a wave function $e^{i\beta x^2/2}$; $h$ is the momentum variance at fixed position, and $\vartheta=h-1/(4\sigma)=(\Delta-1)/(4\sigma)$ is its excess over a pure state with the same $\sigma$, zero exactly for pure states. A shear $e^{i\kappa q^2/2}$ shifts every momentum by $\kappa q$, hence $\beta$ and, by covariance, the outcome $b$ by $\kappa$. With the chirp removed the kernel \eqref{eq:NG-kernel} is real, so the state is invariant under time reversal, which maps the effect of $(a,b)$ to that of $(a,-b)$; this is why $\E[b\,|\,a]=\beta$ exactly. The conditional spread $\lambda^2+\frac12\vartheta\lambda$, $\lambda=a+1/(2\sigma)$, is on the chirp scale $1/\sigma$ of a packet of width $\sqrt\sigma$, and thermal noise adds to it linearly. The three terms of \eqref{eq:NG-trace} below are the width mismatch in $q$, the width mismatch in the conjugate direction, and the chirp mismatch, each in the units of the state.

$S(a,b)$ is symplectic, $P=\tfrac12SS^T$ and $\det P=\tfrac14$; $P(a,b)$ is the covariance of the pure Gaussian state with wave function proportional to $e^{-(a-ib)x^2/2}$. Using $w-2cb+\sigma b^2=h+\sigma(b-\beta)^2$, elementary inversion gives
\begin{equation}\label{eq:NG-trace}
\Tr(\Sigma^{-1}P)=\frac1{2\sigma a}+\frac a{2h}+\frac{(b-\beta)^2}{2ah}.
\end{equation}
By Lemma~\ref{lem:NG-shear}, with $\tau=2\sigma a$ and $\Delta=4\sigma h\ge1$, and bounding term by term with $\Delta\ge1$,
\begin{equation}\label{eq:NG-condtrace}
\E[\Tr(\Sigma^{-1}P)\,|\,\tau=u]=\frac1u+\frac u\Delta+\frac{(u+1)^2}{\Delta u}+\frac{u+1}{4u}\Big(1-\frac1\Delta\Big)\le2u+\frac94+\frac9{4u}.
\end{equation}

\begin{proof}[Proof of Lemma~\ref{lem:NG-dom}]
Let $G:=\{1/10\le\tau\le10^4\}$. By Lemma~\ref{lem:NG-scale}, bounding $(1+u)^{-3}$ by $1$ below $1/10$ and by $u^{-3}$ above $10^4$,
\[
\Pr[G^c]\le\frac8{3\pi}\Big[\frac25\Big(\frac1{10}\Big)^{5/2}+\frac2{\sqrt{10^4}}\Big]<0.019 .
\]
On $G$ the right side of \eqref{eq:NG-condtrace} is at most $20025$. Markov's inequality applied to $\mathbf 1_G\Tr(\Sigma^{-1}P)$ gives
\[
\Pr[\Tr(\Sigma^{-1}P)>10^6]\le\Pr[G^c]+\frac{\E[\mathbf 1_G\Tr(\Sigma^{-1}P)]}{10^6}<0.019+0.020025<0.04 .
\]
Since $\Sigma^{-1/2}P\Sigma^{-1/2}\succeq0$, its largest eigenvalue is at most its trace $\Tr(\Sigma^{-1}P)$, so $P\preceq10^6\,\Sigma$ with probability greater than $0.96$.
\end{proof}

The constants are absolute and not optimised. The bound is uniform over all $\sigma,c,w$ with $\det\Sigma\ge\frac14$.

\paragraph{A worked example: squeezed thermal light.}
Let $\Sigma=(n+\frac12)\operatorname{diag}(e^{-2r},e^{2r})$, with unknown occupation $n$ and squeezing $r$. Then $\sigma=(n+\frac12)e^{-2r}$, $\beta=0$, $h=(n+\frac12)e^{2r}$, $\Delta=(2n+1)^2$, and by Lemma~\ref{lem:NG-scale}
\[
\log a=2r-\log(2n+1)+\log\tau,\qquad\tau\sim\mathrm{BetaPrime}(\tfrac52,\tfrac12):
\]
in logarithmic coordinates the law of the scale outcome is one fixed profile, rigidly translated by the unknown $2r-\log(2n+1)$; the median of $\tau$ is about $9.5$ and its central $90\%$ interval about $[0.76,\,1.2\cdot10^3]$, at $3$\,dB of squeezing as at $30$\,dB. By Lemma~\ref{lem:NG-shear}, given $a$ the outcome $b$ is centred with $\operatorname{Var}(b/a\,|\,\tau)=(1+\tau^{-1})^2+n(n+1)(1+\tau)\tau^{-2}$, independent of $r$. Undoing $S(a,b)$ gives
\[
\Sigma'=S^{-1}\Sigma S^{-T}=\begin{pmatrix}\tau/2&-(b/a)\,\tau/2\\-(b/a)\,\tau/2&(b/a)^2\tau/2+2(n+\frac12)^2/\tau\end{pmatrix},
\]
whose law depends on $n$ but not on $r$: one copy leaves a random, heavy-tailed residual squeezing whose law is the same at every squeezing, which is Lemma~\ref{lem:NG-dom} in miniature. The domination factor is large because of the heavy tail, not the energy: a large $\tau$ reports a packet much narrower in $q$ than the state, hence, $P$ being pure, much wider in $p$, and this anti-squeezing of the estimate is the term $a/(2h)=\tau/\Delta$ of \eqref{eq:NG-trace} (Remark~\ref{rem:NG-constant}). A hotter state gives a more variable chirp estimate, but its momentum spread $h$ grows in proportion, and the chirp error enters \eqref{eq:NG-trace} divided by $h$; this is why \eqref{eq:NG-condtrace} is uniform in $\Delta$.

\begin{proof}[Proof of Theorem~\ref{thm:NG1}]
(i) Measure one copy with $\mathsf M$ and set $S:=S(a,b)$. The remaining copies are untouched and independent of $(a,b)$; conditionally on the outcome they are i.i.d.\ copies of $\rho(0,\Sigma)$, and $U_S^\dagger$ is applied to each before heterodyne detection. On the event of Lemma~\ref{lem:NG-dom}, \eqref{eq:NG-target} holds with $C=10^6$, and Lemma~\ref{lem:NG-target} with $m=1$ and $\delta=0.05$ shows that $N_1$ copies satisfying \eqref{eq:NG-N1} give $T(\rho,\hat\rho)\le\eps$ with conditional probability at least $0.95$. Undoing $S$ in the classical estimate is part of Lemma~\ref{lem:NG-target}. The total failure probability is below $0.04+0.05=0.09$.

(ii) A balanced beam splitter on two copies maps $\rho(v,\Sigma)\otimes\rho(v,\Sigma)$ to $\rho(0,\Sigma)\otimes\rho(\sqrt2\,v,\Sigma)$ \cite[Lemma~8.1]{CGYZ26}, \cite[App.~D.6, Lemma~D.12]{CFGLMWW26}: the covariance $\Sigma\oplus\Sigma$ is invariant under a passive mixing applied equally to $q$ and $p$, and a Gaussian state with block-diagonal covariance is a product. Measure the centred output with $\mathsf M$, discard the other output, and proceed as in (i) on original single copies. The estimator of \cite[Thm.~10]{BMEM25} estimates the mean together with the covariance, with no assumption on $v$. Only the first step measures two copies jointly, and no transposed state is used.
\end{proof}

\subsection{The radial measurement for pure states: proof of Theorem~\ref{thm:NG-pure}}\label{subsec:NG-radialPOVM}

Throughout this subsection $n:=m+4$, and $\mathrm{Sym}_k$ is the space of real symmetric $k\times k$ matrices with the inner product $\langle X,Y\rangle_F=\Tr(XY)$. For $X\in\mathrm{Sym}_m$ let
\begin{equation}\label{eq:NG-shearmat}
\Lambda_X:=\begin{pmatrix}I&0\\X&I\end{pmatrix}.
\end{equation}
$\Lambda_X$ is symplectic exactly because $X$ is symmetric, $\Lambda_X\Lambda_Y=\Lambda_{X+Y}$, and $U_{\Lambda_X}$ is the multiplication of the position wave function by $e^{iq^TXq/2}$, which maps $p$ to $p+Xq$ and leaves $q$ unchanged. As in the rest of Section~\ref{sec:unsqueezing}, the strength of the Gaussian unitaries applied to the copies and the classical precision are not counted.

\paragraph{Siegel coordinates.}

\begin{lemma}[centred pure states]\label{lem:NG-siegel}
For $A,B\in\mathrm{Sym}_m$ with $A\succ0$ the wave function
\begin{equation}\label{eq:NG-psiAB}
\psi_{A,B}(x):=\frac{(\det A)^{1/4}}{\pi^{m/4}}\exp\Big(-\frac{x^TAx}2+\frac{i\,x^TBx}2\Big),\qquad x\in\R^m,
\end{equation}
is the centred pure Gaussian state with covariance
\begin{equation}\label{eq:NG-SigmaAB}
\Sigma(A,B)=\frac12\,\Lambda_B\begin{pmatrix}A^{-1}&0\\0&A\end{pmatrix}\Lambda_B^T=\frac12\begin{pmatrix}A^{-1}&A^{-1}B\\BA^{-1}&A+BA^{-1}B\end{pmatrix},
\end{equation}
and every centred pure $m$-mode Gaussian state is, up to a phase, $\psi_{A,B}$ for exactly one such pair. Its position covariance is $\Sigma_{qq}=\frac12A^{-1}$.
\end{lemma}

For one mode, \eqref{eq:NG-SigmaAB} is the matrix $P(a,b)$ of \eqref{eq:NG-PS}, and $b+ia$ ranges over the Siegel upper half-plane \cite[Sec.~4.4]{CGYZ26}; for $m$ modes $B+iA$ ranges over the Siegel upper half-space.

\begin{proof}
$\psi_{A,0}$ is the image of the vacuum under the dilation $\psi(x)\mapsto(\det A)^{1/4}\psi(A^{1/2}x)$, a Gaussian unitary. Its position density is that of $N(0,\frac12A^{-1})$; its Fourier transform is $\psi_{A^{-1},0}$, so the momenta have law $N(0,\frac12A)$; and $\langle q_jp_k+p_kq_j\rangle=2\,\mathrm{Re}\,(-i)\!\int\psi_{A,0}\,x_j\partial_k\psi_{A,0}\,dx=0$ because $\psi_{A,0}$ is real. So $\psi_{A,0}$ has covariance $\frac12\operatorname{diag}(A^{-1},A)$, and $\psi_{A,B}=U_{\Lambda_B}\psi_{A,0}$ has covariance \eqref{eq:NG-SigmaAB}; it is pure and centred. Conversely, let $\Sigma=\frac12SS^T$ be a pure centred covariance and write $X:=SS^T=\begin{psmallmatrix}X_1&X_2\\X_2^T&X_3\end{psmallmatrix}$, with $X_1\succ0$. Since $X$ is symmetric and symplectic, $X\Omega X=\Omega$, whose upper left block reads $X_1X_2^T=X_2X_1$; hence $B:=X_2^TX_1^{-1}=X_1^{-1}X_2$ is symmetric, and
\[
\Lambda_{-B}X\Lambda_{-B}^T=\begin{pmatrix}X_1&0\\0&X_3-X_2^TX_1^{-1}X_2\end{pmatrix}
\]
is symmetric and symplectic, which for a block-diagonal matrix forces the lower block to be $X_1^{-1}$. Thus $\Sigma=\Sigma(A,B)$ with $A=X_1^{-1}$. A Gaussian state is determined by its mean and covariance, its Wigner function being the density of $N(v,\Sigma)$, so the state is $\psi_{A,B}$ up to a phase; $A=(2\Sigma_{qq})^{-1}$ and $B=2\Sigma_{pq}A$ are determined by $\Sigma$.
\end{proof}

\paragraph{The radial measurement.}
On $\R^n$ write $x=\sqrt t\,u$ with $t>0$ and $u\in S^{n-1}$, and let $dS$ be the surface measure of the sphere; then $dx=\frac12t^{n/2-1}\,dt\,dS(u)$, so
\begin{equation}\label{eq:NG-Jn}
(J\psi)(u,t):=2^{-1/2}\,t^{(n-2)/4}\,\psi(\sqrt t\,u)
\end{equation}
is a unitary from $L^2(\R^n)$ onto $L^2(S^{n-1}\times\R_+,dS\,dt)$ (for $n=1$, $S^0=\{\pm1\}$, it is the map $J_+\oplus J_-$ of \eqref{eq:NG-J}). On the half-line let
\begin{equation}\label{eq:NG-halfF}
(Ff)(b):=\frac1{\sqrt{4\pi}}\int_0^\infty e^{-ibt/2}f(t)\,dt,\qquad b\in\R,
\end{equation}
and put $\mathcal V:=(I\otimes F)J$.

\begin{lemma}[the radial POVM]\label{lem:NG-Rn}
(i) $\mathcal V$ is an isometry from $L^2(\R^n)$ into $L^2(S^{n-1}\times\R,dS\,db)$, given for Schwartz functions $\psi$, Gaussian ones included, by
\begin{equation}\label{eq:NG-V}
(\mathcal V\psi)(u,b)=\frac1{\sqrt{8\pi}}\int_0^\infty e^{-ibt/2}\,t^{(n-2)/4}\,\psi(\sqrt t\,u)\,dt,
\end{equation}
and $\mathsf R_n(E):=\mathcal V^\dagger\mathbf 1_E\mathcal V$, for Borel sets $E\subseteq S^{n-1}\times\R$, is a POVM. On a pure state $\psi$ its outcome $(u,b)$ has the density $|(\mathcal V\psi)(u,b)|^2$ with respect to $dS\,db$.
(ii) For $B\in\mathrm{Sym}_n$, $\mathcal V(e^{ix^TBx/2}\psi)(u,b)=(\mathcal V\psi)(u,b-u^TBu)$.
(iii) The marginal law of $u$ is the law of $x/|x|$ when $x$ has the density $|\psi(x)|^2$.
\end{lemma}

\begin{proof}
$J$ is unitary by the polar formula for $dx$. For $f\in L^1\cap L^2(\R_+)$, extended by zero to $\R$, $(Ff)(b)=(4\pi)^{-1/2}\hat f(b/2)$ with $\hat f(\omega)=\int e^{-i\omega t}f(t)\,dt$, and Plancherel, $\int|\hat f|^2d\omega=2\pi\|f\|_2^2$, with $db=2\,d\omega$ gives $\|Ff\|_2=\|f\|_2$; $F$ extends by continuity to an isometry of $L^2(\R_+)$ into $L^2(\R)$, and so does $\mathcal V$. Then $\mathsf R_n(E)\succeq0$, $\mathsf R_n(S^{n-1}\times\R)=\mathcal V^\dagger\mathcal V=I$, and countable additivity holds in the strong operator topology because it holds for $E\mapsto\mathbf 1_E$; \eqref{eq:NG-V} is \eqref{eq:NG-Jn} inserted in \eqref{eq:NG-halfF}. (ii) On the ray through $u$ the multiplier is $e^{itu^TBu/2}$, and $F(e^{i\kappa t/2}f)(b)=(Ff)(b-\kappa)$. (iii) By Plancherel on each ray, $\int|(\mathcal V\psi)(u,b)|^2db=\int|(J\psi)(u,t)|^2dt$, and integrating over $u$ in a Borel set $E'\subseteq S^{n-1}$ gives $\int_{\{x/|x|\in E'\}}|\psi(x)|^2dx$.
\end{proof}

The range of $F$ is the subspace of $L^2(\R)$ of functions whose Fourier transform vanishes on a half-line (a Hardy space), so $\mathcal V$ is not onto and $\mathsf R_n$ is a POVM that is not projective. On each ray it is the covariant measurement, in the sense of Holevo \cite[Ch.~4]{Hol11}, of the one-parameter group of radial quadratic phases $e^{i\kappa|x|^2/2}$, whose generator $|x|^2/2$ has spectrum $[0,\infty)$ and therefore no self-adjoint conjugate; for the number operator the same construction gives the canonical phase POVM on which \cite[Thm.~1.5]{CGYZ26} is built. $\mathsf R_n$ acts on one copy together with ancillary modes, and it is not a Gaussian measurement.

\paragraph{The outcome law.}
For real $k>0$ let $\chi^2_k$ be the law $\mathrm{Gamma}(k/2,\frac12)$, in the notation of \S\ref{subsec:NG-law}.

\begin{lemma}[Student law]\label{lem:NG-student}
Let $A,B\in\mathrm{Sym}_n$ with $A\succ0$, let $(u,b)$ be the outcome of $\mathsf R_n$ on $\psi_{A,B}$, and put $a(u):=u^TAu$.
\begin{enumerate}
\item[(i)] $u$ has the law of $A^{-1/2}w/\|A^{-1/2}w\|$, with $w$ uniform on $S^{n-1}$.
\item[(ii)] Given $u$, the outcome $b$ has the density
\begin{equation}\label{eq:NG-student}
p(b\,|\,u)=\frac{\Gamma(\frac{n+2}4)}{\sqrt\pi\,\Gamma(\frac n4)\,a(u)}\Big[1+\frac{(b-u^TBu)^2}{a(u)^2}\Big]^{-(n+2)/4},
\end{equation}
which is the law of $u^TBu+a(u)\,\xi/\sqrt\chi$ for independent $\xi\sim N(0,1)$ and $\chi\sim\chi^2_{n/2}$, independent of $u$: a Student law with $n/2$ degrees of freedom, location $u^TBu$ and scale $a(u)/\sqrt{n/2}$.
\item[(iii)] For $n\ge5$, $\E[b\,|\,u]=u^TBu$ and $\operatorname{Var}(b\,|\,u)=2a(u)^2/(n-4)$.
\end{enumerate}
\end{lemma}

\begin{proof}
(i) By Lemma~\ref{lem:NG-Rn}(iii) $u$ is distributed as $x/|x|$ with $x\sim N(0,\frac12A^{-1})$, the density $|\psi_{A,B}|^2$ (which does not involve $B$); write $x=(2A)^{-1/2}g$ with $g$ standard Gaussian, and $g/|g|$ is uniform on the sphere. (ii) On the ray through $u$, with $b_0:=u^TBu$, $(J\psi_{A,B})(u,t)=c_u\,t^{(n-2)/4}e^{-(a(u)-ib_0)t/2}$ with $c_u$ independent of $t$ and $b$. The gamma integral $\int_0^\infty t^{s-1}e^{-zt}dt=\Gamma(s)z^{-s}$, $\mathrm{Re}\,z>0$, at $s=(n+2)/4$ and $z=(a(u)+i(b-b_0))/2$ gives
\[
|(\mathcal V\psi_{A,B})(u,b)|^2=\frac{c_u^2\,\Gamma(s)^2\,2^{2s}}{4\pi}\,\big(a(u)^2+(b-b_0)^2\big)^{-(n+2)/4}.
\]
As a function of $b$ this is proportional to the bracket in \eqref{eq:NG-student}, and since $\int(1+y^2)^{-(\nu+1)/2}dy=\sqrt\pi\,\Gamma(\nu/2)/\Gamma((\nu+1)/2)$ the normalised conditional density is \eqref{eq:NG-student}. The Student variable $\xi/\sqrt{\chi/\nu}$, $\nu=n/2$, has density proportional to $(1+y^2/\nu)^{-(\nu+1)/2}$, and $y=\sqrt\nu\,(b-b_0)/a(u)$ turns it into \eqref{eq:NG-student}. (iii) For $\nu>2$, $\E[1/\chi]=1/(\nu-2)$, so $\operatorname{Var}(b\,|\,u)=a(u)^2/(\nu-2)=2a(u)^2/(n-4)$, and the mean is $b_0$ by symmetry.
\end{proof}

By Lemma~\ref{lem:NG-Rn}(ii) the unknown quadratic phase only translates the outcome, so the law of $b-u^TBu$ does not involve $B$; this is where the energy drops out. The conditional variance is finite exactly when $n\ge5$. The protocol appends four ancillary modes in the vacuum to the $m$ modes of the copy, which turns $\psi_{A,B}$ into $\psi_{A\oplus I_4,B\oplus0_4}$ on $n=m+4$ modes; without them the variance is infinite for every $m\le4$ (at $m=3$ and $m=4$ the degrees of freedom are $\frac32$ and $2$), and with this estimator at least $\max\{0,5-m\}$ of them are needed. The ancillas are in a known state, as the measurement models of Table~\ref{tab:main} allow.

\paragraph{Angular design.}

\begin{lemma}[design bounds]\label{lem:NG-design}
Let $n\ge5$, $A\in\mathrm{Sym}_n$ with $\frac12I\preceq A\preceq2I$, and let $u$ be distributed as in Lemma~\ref{lem:NG-student}(i). For every $H\in\mathrm{Sym}_n$ with $\|H\|_F=1$,
\begin{equation}\label{eq:NG-design}
\E(u^THu)^2\ge\frac1{8n(n+2)},\qquad\E(u^THu)^4\le6400\,\big(\E(u^THu)^2\big)^2,
\end{equation}
and $\Pr[|u^THu|\ge1/(6n)]\ge\alpha_0:=1/25600$.
\end{lemma}

\begin{proof}
Let $K:=A^{-1/2}HA^{-1/2}$, so that $u^THu=w^TKw/w^TA^{-1}w$ with $w^TA^{-1}w\in[\frac12,2]$; hence $(u^THu)^2\ge\frac14(w^TKw)^2$ and $(u^THu)^4\le16(w^TKw)^4$. For a standard Gaussian $g\in\R^n$, $g=|g|w$ with $|g|$ and $w$ independent, $\E|g|^4=n(n+2)$, $\E|g|^8=n(n+2)(n+4)(n+6)$, and with $X:=g^TKg$, $\mu:=\Tr K$, $s_2:=\Tr K^2$, diagonalisation gives $\E X^2=\mu^2+2s_2$, so
\[
\E(w^TKw)^2=\frac{2\|K\|_F^2+(\Tr K)^2}{n(n+2)}\ge\frac{2\|K\|_F^2}{n(n+2)}\ge\frac{\|H\|_F^2}{2n(n+2)},
\]
since $\|H\|_F\le\|A\|_\op\|K\|_F\le2\|K\|_F$. This gives the first bound. For the second, $X-\mu=\sum_i\lambda_i(g_i^2-1)$ over the eigenvalues $\lambda_i$ of $K$, and with $\E(g_i^2-1)^2=2$, $\E(g_i^2-1)^4=60$,
\[
\E(X-\mu)^4=60\sum_i\lambda_i^4+12\sum_{i\ne j}\lambda_i^2\lambda_j^2=12s_2^2+48\sum_i\lambda_i^4\le60s_2^2 .
\]
By Minkowski's and the Cauchy--Schwarz inequality, $\|X\|_4\le|\mu|+60^{1/4}\sqrt{s_2}\le(1+\sqrt{60}/2)^{1/2}(\mu^2+2s_2)^{1/2}$, so $\E X^4\le(1+\sqrt{60}/2)^2(\E X^2)^2\le25(\E X^2)^2$. As $\E X^4=\E|g|^8\,\E(w^TKw)^4$ and $\E X^2=\E|g|^4\,\E(w^TKw)^2$,
\[
\frac{\E(w^TKw)^4}{(\E(w^TKw)^2)^2}\le25\,\frac{n(n+2)}{(n+4)(n+6)}\le25,
\]
and with the two comparisons of the first line, $\E(u^THu)^4\le16\cdot25\cdot16\,(\E(u^THu)^2)^2$. For the third, apply the Paley--Zygmund inequality \cite{PZ32} to $Z:=(u^THu)^2\ge0$: from $\E Z\le\theta\E Z+\E[Z\mathbf 1_{\{Z>\theta\E Z\}}]\le\theta\E Z+(\E Z^2\Pr[Z>\theta\E Z])^{1/2}$, at $\theta=\frac12$, $\Pr[Z>\frac12\E Z]\ge\frac14(\E Z)^2/\E Z^2\ge1/25600$. On that event $|u^THu|\ge(16n(n+2))^{-1/2}\ge1/(6n)$, the last step because $16n(n+2)\le36n^2$ for $n\ge2$.
\end{proof}

\paragraph{The least-squares estimator.}
Fix an orthonormal basis of $\mathrm{Sym}_n$ and let $\operatorname{vec}_s:\mathrm{Sym}_n\to\R^{D_n}$, $D_n:=n(n+1)/2$, be the coordinate map, so that $\langle X,Y\rangle_F=\operatorname{vec}_s(X)^T\operatorname{vec}_s(Y)$. For $u\in S^{n-1}$ put $z(u):=\operatorname{vec}_s(uu^T)$; then $\|z(u)\|_2=\|uu^T\|_F=1$ and $u^THu=z(u)^T\operatorname{vec}_s(H)$. For outcomes $(u_l,b_l)$, $l=1,\dots,N$, let $z_l:=z(u_l)$ and $G_N:=\sum_lz_lz_l^T$. If $G_N$ is invertible, the estimate is
\begin{equation}\label{eq:NG-LS}
\hat B:=\operatorname{vec}_s^{-1}\Big(G_N^{-1}\sum_{l=1}^Nz_l\,b_l\Big);
\end{equation}
otherwise the protocol outputs the vacuum. Writing $b_l=z_l^T\operatorname{vec}_s(B)+e_l$, the least-squares identity
\begin{equation}\label{eq:NG-LSerror}
\operatorname{vec}_s(\hat B-B)=G_N^{-1}\sum_{l=1}^Nz_l\,e_l
\end{equation}
holds exactly for every invertible $G_N$: the empirical Gram matrix cancels the signal $u_l^TBu_l$, however large $B$ is, and by Lemma~\ref{lem:NG-student} the law of the error, given the directions, does not involve $B$.

\begin{lemma}[Gram matrix]\label{lem:NG-gram}
Let $n\ge5$, $\frac12I\preceq A\preceq2I$, let $u_1,\dots,u_N$ be independent with the law of Lemma~\ref{lem:NG-student}(i), and let $N\ge10^{14}n^2$.
\begin{enumerate}
\item[(i)] With probability at least $0.99$, $G_N\succeq c_0Nn^{-2}I$, $c_0:=\alpha_0/72=1/1843200$.
\item[(ii)] With probability at least $1-D_n\exp(-N/(64n(n+2)))$, $G_N\succeq N(16n(n+2))^{-1}I$.
\end{enumerate}
\end{lemma}

\begin{proof}
(i) For $H\in\mathrm{Sym}_n$ with $\|H\|_F=1$ and $h:=\operatorname{vec}_s(H)$ let $E_H:=\{u:|u^THu|\ge1/(6n)\}=\{u:z(u)^Th\ge\frac1{6n}\}\cup\{u:z(u)^T(-h)\ge\frac1{6n}\}$, and let $\mathcal A$ be the class of these sets. Affine half-spaces of $\R^{D_n}$ form a class of Vapnik--Chervonenkis dimension $D_n+1$ \cite[Ch.~13]{DGL96}, so by the Sauer--Shelah lemma \cite{Sau72,She72} they cut at most $(eN/(D_n+1))^{D_n+1}$ subsets out of $N\ge D_n+1$ points; the trace of $E_H$ on $\{u_1,\dots,u_N\}$ is determined by those of its two half-spaces, so $\mathcal A$ cuts at most $\mathbb S_{\mathcal A}(N):=(eN/(D_n+1))^{2(D_n+1)}$. By the Vapnik--Chervonenkis inequality in the form \cite[Thm.~12.5]{DGL96} (\cite{VC71}),
\[
\Pr\Big[\sup_{E\in\mathcal A}\Big|\frac1N\#\{l:u_l\in E\}-\Pr[u\in E]\Big|>\frac{\alpha_0}2\Big]\le8\,\mathbb S_{\mathcal A}(N)\,e^{-N\alpha_0^2/128}.
\]
Here $D_n+1\le n^2\le N$, and $y\mapsto y\log(eN/y)$ increases on $(0,N]$, so with $x:=N/n^2\ge10^{14}$, $\log(8\,\mathbb S_{\mathcal A}(N))\le\log8+2n^2(1+\log x)$, while $N\alpha_0^2/128=n^2x/(128\cdot25600^2)$; the difference $x/(128\cdot25600^2)-2(1+\log x)$ increases for $x\ge10^{12}$ and exceeds $1125$ at $x=10^{14}$, so the probability is at most $8e^{-1125n^2}<0.01$. On the complementary event, Lemma~\ref{lem:NG-design} gives, for every $H$ of unit norm, at least $N\alpha_0/2$ indices with $|u_l^THu_l|\ge1/(6n)$, so $h^TG_Nh=\sum_l(u_l^THu_l)^2\ge(N\alpha_0/2)(6n)^{-2}=c_0N/n^2$.
(ii) The summands $z_lz_l^T$ are independent, positive semidefinite, of norm $1$, and by Lemma~\ref{lem:NG-design} $\lambda_{\min}(\E G_N)\ge\mu:=N/(8n(n+2))$. The matrix Chernoff bound \cite[Thm.~1.1]{Tropp12} at deviation $\frac12$ gives $\Pr[\lambda_{\min}(G_N)\le\mu/2]\le D_n(\sqrt2\,e^{-1/2})^{\mu}\le D_ne^{-0.153\mu}\le D_ne^{-N/(64n(n+2))}$.
\end{proof}

\paragraph{The coarse step.}
The error of $\hat B$ in operator norm and the window of the covariance after the shear $\Lambda_{-\hat B_{\rm s}}$ are proved in \S\ref{subsec:NG-thermal-proofs} for every centred Gaussian state of bounded Williamson occupations (Lemmas~\ref{lem:NG-opnorm} and~\ref{lem:NG-window}). For $\psi_{A,B}$ the occupations vanish, the outcome law is that of Lemma~\ref{lem:NG-student}, the radial round has $N_r=\lceil10^{14}n^2\rceil$ outcomes, $\|\hat B_{\rm s}-B\|_\op\le\frac1{10}$ holds with probability at least $0.97$, and the window is $0.40\,I\preceq\Sigma'\preceq0.64\,I$.

\paragraph{The pure core.}
For positive definite $X,Z$ let $X\#Z:=X^{1/2}(X^{-1/2}ZX^{-1/2})^{1/2}X^{1/2}$. It is the unique positive definite solution $W$ of $WX^{-1}W=Z$ (put $V:=X^{-1/2}WX^{-1/2}$; then $V^2=X^{-1/2}ZX^{-1/2}$ and $V\succ0$). Hence $(MXM^T)\#(MZM^T)=M(X\#Z)M^T$ for invertible $M$, $(cX)\#(c'Z)=\sqrt{cc'}\,X\#Z$ for $c,c'>0$, $X\#(cI)=\sqrt c\,X^{1/2}$, and $Z\preceq Z'$ implies $X\#Z\preceq X\#Z'$, by operator monotonicity of the square root \cite[Ch.~V]{Bhatia97}. For $Y\succ0$ of size $2m$ let
\begin{equation}\label{eq:NG-purecore}
P(Y):=Y\#\big(\tfrac14\,\Omega Y^{-1}\Omega^T\big).
\end{equation}

\begin{lemma}[pure core]\label{lem:NG-core}
(i) If $Y=S\operatorname{diag}(d,d)S^T$ is a Williamson form, then $P(Y)=\frac12SS^T$, the pure core of \S\ref{subsec:NG-NG2}; in particular $P(Y)$ is a pure covariance and does not depend on the Williamson form chosen. Moreover $P(S'YS'^T)=S'P(Y)S'^T$ for symplectic $S'$, and $P(cY)=P(Y)$ for $c>0$.
(ii) Let $\Sigma$ be a pure covariance and $Y\succ0$ with $\eta:=\|\Sigma^{-1/2}(Y-\Sigma)\Sigma^{-1/2}\|_F\le\frac14$. Then
\begin{equation}\label{eq:NG-corestab}
\big\|\Sigma^{-1/2}(P(Y)-\Sigma)\Sigma^{-1/2}\big\|_F\le\tfrac{44}9\,\eta,\qquad P(Y)\succeq\sqrt{\tfrac{1-\eta}{1+\eta}}\,\Sigma\succeq\sqrt{\tfrac35}\,\Sigma .
\end{equation}
\end{lemma}

\begin{proof}
(i) For symplectic $S'$, $S'\Omega S'^T=\Omega$ gives $\Omega S'^{-T}=S'\Omega$ and $S'^{-1}\Omega^T=\Omega^TS'^T$, so
\[
\Omega(S'YS'^T)^{-1}\Omega^T=S'(\Omega Y^{-1}\Omega^T)S'^T,
\]
and congruence covariance of $\#$ gives $P(S'YS'^T)=S'P(Y)S'^T$; $P(cY)=P(Y)$ by the scaling rule. With $D=\operatorname{diag}(d,d)$, which commutes with $\Omega$, $\frac14\Omega D^{-1}\Omega^T=\frac14D^{-1}$ and $P(D)=D^{1/2}(\frac14D^{-2})^{1/2}D^{1/2}=\frac12I$; so $P(Y)=\frac12SS^T$.
(ii) Write $\Sigma=\frac12S_0S_0^T$ with $S_0$ symplectic. Then $O:=\Sigma^{-1/2}S_0/\sqrt2$ is orthogonal, and $R:=2S_0^{-1}YS_0^{-T}$ satisfies $\Sigma^{-1/2}Y\Sigma^{-1/2}=ORO^T$, so $\|R-I\|_F=\eta$ and $(1-\eta)I\preceq R\preceq(1+\eta)I$. By (i), $P(Y)=S_0P(\frac12R)S_0^T=\frac12S_0\Phi(R)S_0^T$ with $\Phi(R):=R\#(\Omega R^{-1}\Omega^T)$, and $\Sigma^{-1/2}(P(Y)-\Sigma)\Sigma^{-1/2}=O(\Phi(R)-I)O^T$. Let $Z:=\Omega R^{-1}\Omega^T$; $\Omega$ is orthogonal, so $\|Z-I\|_F=\|R^{-1}(I-R)\|_F\le\eta/(1-\eta)$ and $\|Z-R\|_F\le\eta+\eta/(1-\eta)$. With $K:=R^{-1/2}ZR^{-1/2}$, $\Phi(R)-R=R^{1/2}(K^{1/2}-I)R^{1/2}$, and $|\sqrt x-1|\le|x-1|$ for $x>0$ gives $\|K^{1/2}-I\|_F\le\|K-I\|_F\le\|Z-R\|_F/(1-\eta)$. Therefore
\[
\|\Phi(R)-I\|_F\le\eta+\frac{1+\eta}{1-\eta}\Big(\eta+\frac\eta{1-\eta}\Big)\le\Big(1+\frac53\cdot\frac73\Big)\eta=\frac{44}9\,\eta
\]
for $\eta\le\frac14$. For the order bound, $Z\succeq(1+\eta)^{-1}I$, so by monotonicity $\Phi(R)\succeq R\#(1+\eta)^{-1}I=(1+\eta)^{-1/2}R^{1/2}\succeq\sqrt{(1-\eta)/(1+\eta)}\,I$, and $P(Y)=\frac12S_0\Phi(R)S_0^T\succeq\sqrt{(1-\eta)/(1+\eta)}\,\Sigma$.
\end{proof}

\begin{lemma}[trace distance between pure Gaussian states]\label{lem:NG-tdpure}
(i) For pure covariances $\Sigma$ and $P$,
\begin{equation}\label{eq:NG-tdpure}
T\big(\rho(0,\Sigma),\rho(0,P)\big)\le\tfrac14\,\big\|\Sigma^{-1/2}(P-\Sigma)\Sigma^{-1/2}\big\|_F ,
\end{equation}
and the constant $\frac14$ cannot be decreased.
(ii) For a pure covariance $P$ and means $v,v'$, with $x:=\frac14(v-v')^TP^{-1}(v-v')$,
\[
T\big(\rho(v,P),\rho(v',P)\big)=(1-e^{-x})^{1/2}\le\tfrac12\|P^{-1/2}(v-v')\|_2 .
\]
\end{lemma}

\begin{proof}
(i) Write $\Sigma=\frac12SS^T$. $U_S^\dagger$ maps $\rho(0,\Sigma)$ to the vacuum and $\rho(0,P)$ to $\rho(0,P')$, $P':=S^{-1}PS^{-T}$, and as in the proof of Lemma~\ref{lem:NG-core} the whitened difference is $O(2P'-I)O^T$ with $O$ orthogonal. $2P'$ is positive and symplectic; by Proposition~\ref{prop:bg-williamson} (Bloch--Messiah), $2P'=O_1\operatorname{diag}(e^{2r},e^{-2r})O_1^T$ with $O_1$ orthogonal symplectic and $r_j\ge0$, so $\rho(0,P')$ is the image under a passive Gaussian unitary, which fixes the vacuum, of a product of one-mode squeezed vacua of squeezing $r_j$, and $|\langle0|\psi_{P'}\rangle|^2=\prod_j\operatorname{sech}r_j$. For pure states $T=(1-|\langle\cdot|\cdot\rangle|^2)^{1/2}$, so with $1-e^{-y}\le y$ and $\log\cosh r\le r^2/2$ (integrate $\tanh r\le r$)
\[
T^2=1-\exp\Big(-\sum_j\log\cosh r_j\Big)\le\frac12\sum_jr_j^2 .
\]
On the other side $\|2P'-I\|_F^2=\sum_j\big[(e^{2r_j}-1)^2+(e^{-2r_j}-1)^2\big]=\sum_j8\cosh(2r_j)\sinh^2r_j\ge8\sum_jr_j^2$, which proves \eqref{eq:NG-tdpure}. For one mode and $r\to0$ both sides of \eqref{eq:NG-tdpure} equal $(r/\sqrt2)(1+O(r^2))$, so $\frac14$ is attained in the limit. (ii) $U_S^\dagger$, with $P=\frac12SS^T$, maps the two states to coherent states with means $S^{-1}v$ and $S^{-1}v'$, whose overlap is $|\langle\alpha|\alpha'\rangle|^2=e^{-|\alpha-\alpha'|^2}=e^{-\|S^{-1}(v-v')\|^2/2}=e^{-x}$; and $(1-e^{-x})^{1/2}\le\sqrt x$.
\end{proof}

Lemma~\ref{lem:NG-tdpure}(i) has no factor of the dimension: for pure states a relative Frobenius error $\eta$ of the covariance costs trace distance $\eta/4$, while near pure modes a general Frobenius-to-trace conversion fails (\S\ref{subsec:NG-gap}); a bound of the same kind, through the total variation distance of Wigner functions, is \cite[Lemma~1]{CMFLMHCP26}.

\paragraph{The final round.}

\begin{lemma}[heterodyne finisher]\label{lem:NG-finish}
Let $0<\eps<1$ and let $\rho(v',\Sigma')$ be a pure Gaussian state with $\frac14I\preceq\Sigma'\preceq I$, and $\Gamma:=\Sigma'+\frac12I$.
(i) If $D_1,\dots,D_{N_h}$ are independent with law $N(0,\Gamma)$, $N_h\ge358400\,m^2/\eps^2$, $\hat\Gamma:=N_h^{-1}\sum_lD_lD_l^T$ and $\hat\Sigma':=\hat\Gamma-\frac12I$, then with probability at least $0.99$, $\eta:=\|\Sigma'^{-1/2}(\hat\Sigma'-\Sigma')\Sigma'^{-1/2}\|_F\le\eps/4$; on this event $\hat\Sigma'\succ0$, and $\hat P:=P(\hat\Sigma')$ is a pure covariance with $T(\rho(0,\Sigma'),\rho(0,\hat P))\le\frac{11}{36}\eps$ and $\hat P\succeq\frac18I$.
(ii) If moreover $\hat v$ is the mean of $N_v\ge10^4m/\eps^2$ independent samples of $N(v',\Gamma)$, then with probability at least $0.9904$, $\|\hat v-v'\|_2\le\eps/(4\sqrt2)$, and on both events $T(\rho(v',\Sigma'),\rho(\hat v,\hat P))\le\frac{11}{36}\eps+\frac14\eps<\eps$.
\end{lemma}

\begin{proof}
(i) $\E(\hat\Gamma_{ij}-\Gamma_{ij})^2=(\Gamma_{ii}\Gamma_{jj}+\Gamma_{ij}^2)/N_h$, so $\E\|\hat\Gamma-\Gamma\|_F^2=((\Tr\Gamma)^2+\Tr\Gamma^2)/N_h\le(9m^2+\frac92m)/N_h\le14m^2/N_h$, since $\Gamma\preceq\frac32I$. As $\|\Sigma'^{-1/2}M\Sigma'^{-1/2}\|_F\le\|\Sigma'^{-1}\|_\op\|M\|_F\le4\|M\|_F$, $\E\eta^2\le224\,m^2/N_h$, and Markov's inequality at $\eta^2=\eps^2/16$ gives failure probability at most $3584\,m^2/(N_h\eps^2)\le0.01$. On the event, $\hat\Sigma'\succeq(1-\eta)\Sigma'\succ0$; Lemma~\ref{lem:NG-core}(ii) gives a pure $\hat P$ with relative error at most $\frac{44}9\cdot\frac\eps4$ and $\hat P\succeq\sqrt{3/5}\,\Sigma'\succeq\frac18I$, and Lemma~\ref{lem:NG-tdpure}(i) gives $T\le\frac14\cdot\frac{11}9\eps$. (ii) $\E\|\hat v-v'\|_2^2=\Tr\Gamma/N_v\le3m/N_v$, so $\Pr[\|\hat v-v'\|_2>\eps/(4\sqrt2)]\le96m/(N_v\eps^2)\le0.0096$. By Lemma~\ref{lem:NG-tdpure}(ii) and $\hat P^{-1}\preceq8I$, $T(\rho(v',\hat P),\rho(\hat v,\hat P))\le\frac12\sqrt8\,\|\hat v-v'\|_2\le\frac\eps4$; a displacement does not change trace distances, so $T(\rho(v',\Sigma'),\rho(v',\hat P))\le\frac{11}{36}\eps$, and the triangle inequality concludes.
\end{proof}

This round is the mechanism of \cite[Thm.~5]{CMFLMHCP26}: a Gaussian estimate, projected to a pure Gaussian state, and a dimension-free conversion to trace distance for pure states. That theorem learns pure Gaussian states with $\|\Sigma\|_\op\le\bar E$ to trace distance $\eps$ with probability $2/3$ from $O(m^2/\eps^2+(m+\log\log\log\bar E)\log\log\bar E)$ copies with adaptive Gaussian measurements; its hypotheses hold for $\rho(v',\Sigma')$, which is pure with $\|\Sigma'\|_\op\le1$, so $\bar E\le2$ in either normalisation of the vacuum, and it would give the same order $O(m^2/\eps^2)$ here. Lemma~\ref{lem:NG-finish} gives explicit constants and a single round of heterodyne detection; what the first two rounds supply is the window $\frac14I\preceq\Sigma'\preceq I$, with no energy promise on the input.

\begin{proof}[Proof of Theorem~\ref{thm:NG-pure}]
(i) By Lemma~\ref{lem:NG-siegel} the input is $\psi_{A_0,B_0}$ for some $A_0\succ0$ and $B_0\in\mathrm{Sym}_m$, unknown and unbounded, with $\Sigma_{qq}=\frac12A_0^{-1}$.

\emph{Round 1 (homodyne pilot).} Measure all position quadratures of $N_q$ copies; the outcomes $x_l$ are independent with law $N(0,\Sigma_{qq})$. Let $\hat\Sigma_{qq}:=N_q^{-1}\sum_lx_lx_l^T$, positive definite almost surely as $N_q\ge m$. With $g_l:=\Sigma_{qq}^{-1/2}x_l$ standard Gaussian and $\E(g_ig_j-\delta_{ij})^2=1+\delta_{ij}$,
\[
\E\big\|\Sigma_{qq}^{-1/2}(\hat\Sigma_{qq}-\Sigma_{qq})\Sigma_{qq}^{-1/2}\big\|_F^2=\frac{m^2+m}{N_q},
\]
so by Markov's inequality the norm is at most $\frac1{40}$ except with probability $1600(m^2+m)/N_q\le0.01$. Let $L:=(2\hat\Sigma_{qq})^{-1/2}$ and $S_1:=\operatorname{diag}(L,L^{-1})$. The state $U_{S_1}\rho U_{S_1}^\dagger$ has covariance $S_1\Sigma S_1^T=\Sigma(A,B)$ with $A=L^{-1}A_0L^{-1}=\hat\Sigma_{qq}^{1/2}\Sigma_{qq}^{-1}\hat\Sigma_{qq}^{1/2}$ and $B=L^{-1}B_0L^{-1}$, by \eqref{eq:NG-SigmaAB}. $A$ is symmetric and has the eigenvalues of $\Sigma_{qq}^{-1/2}\hat\Sigma_{qq}\Sigma_{qq}^{-1/2}$, so $\|A-I\|_\op\le\|A-I\|_F\le\frac1{40}$; $B$ remains arbitrary.

\emph{Round 2 (radial measurement).} On each of $N_r$ fresh copies apply $U_{S_1}$, append four ancillary modes in the vacuum and measure $\mathsf R_n$. Given round 1 the outcomes are independent, with the law of Lemma~\ref{lem:NG-student} for $\psi_{A\oplus I_4,B\oplus0_4}$, and Lemma~\ref{lem:NG-opnorm} gives $\|\hat B_{\rm s}-B\|_\op\le\frac1{10}$ except with probability $0.03$.

\emph{Round 3 (heterodyne detection).} Let $S_2:=\Lambda_{-\hat B_{\rm s}}S_1$. On each of $N_h$ fresh copies apply $U_{S_2}$ and measure by heterodyne detection. The transformed state is pure and centred with covariance $\Sigma':=\Lambda_{-\hat B_{\rm s}}\Sigma(A,B)\Lambda_{-\hat B_{\rm s}}^T=\frac12\Lambda_X\operatorname{diag}(A^{-1},A)\Lambda_X^T$, $X:=B-\hat B_{\rm s}$, and by Lemma~\ref{lem:NG-window} $\frac14I\preceq\Sigma'\preceq I$. The outcomes are independent $N(0,\Gamma)$ \eqref{eq:bg-het}, and Lemma~\ref{lem:NG-finish}(i), with $N_h\ge3.6\cdot10^5m^2/\eps^2$, gives a pure $\hat P$ with $T(\rho(0,\Sigma'),\rho(0,\hat P))\le\frac{11}{36}\eps$ except with probability $0.01$. Output $\hat\rho:=\rho(0,S_2^{-1}\hat P S_2^{-T})=U_{S_2}^\dagger\rho(0,\hat P)U_{S_2}$, a pure Gaussian state with $T(\hat\rho,\rho)=T(\rho(0,\hat P),\rho(0,\Sigma'))\le\eps$. When $G_{N_r}$ is singular or $\hat\Sigma'\not\succ0$ the output is the vacuum. The failure probability is at most $0.01+0.03+0.01=0.05$.

(ii) For $\rho=\rho(v,\Sigma)$, round 1 measures $2N_q$ single copies and uses the $N_q$ pair differences $(x_{2l}-x_{2l-1})/\sqrt2$, which are independent with law $N(0,\Sigma_{qq})$. In round 2 each of $N_r$ pairs of copies passes through a balanced beam splitter, which maps $\rho(v,\Sigma)^{\otimes2}$ to $\rho(0,\Sigma)\otimes\rho(\sqrt2\,v,\Sigma)$ \cite[Lemma~8.1]{CGYZ26} (as in the proof of Theorem~\ref{thm:NG1}(ii)); the centred port is measured as a copy in (i), by $U_{S_1}$, the four ancillas and $\mathsf R_n$, and the other port is discarded. Round 3 measures $2N_h+N_v$ single copies after $U_{S_2}$; the outcomes $D_l$ are independent $N(S_2v,\Gamma)$, the normalised pair differences $(D_{2l}-D_{2l-1})/\sqrt2$ of the first $2N_h$ are independent $N(0,\Gamma)$ and enter Lemma~\ref{lem:NG-finish}(i), and the last $N_v$ give $\hat v$ in Lemma~\ref{lem:NG-finish}(ii). The output $\rho(S_2^{-1}\hat v,S_2^{-1}\hat PS_2^{-T})$ is within trace distance $\eps$ of $\rho$ except with probability $0.05+0.0096<0.06$. Only round 2 measures two copies jointly.

\emph{Count.} Case (i) uses $N_q+N_r+N_h$ copies and case (ii) $2N_q+2N_r+2N_h+N_v$; since $N_q\le3.2\cdot10^5m^2+1$, $N_r\le2.5\cdot10^{15}m^2+1$ and $\eps<1$, both are at most $6\cdot10^{15}m^2/\eps^2$. No constant depends on the energy, the squeezing or the displacement.

\emph{Lower bound.} Let a protocol learn every centred pure $m$-mode Gaussian state to trace distance $\eps$ with probability $2/3$ from $N$ copies. By the median trick of \S\ref{subsec:bg-estimation}, a fixed number of independent repetitions learns them to trace distance $3\eps$ with probability $0.99$. Replace its output $\hat\rho$ by a centred pure Gaussian state $\phi$ with $T(\phi,\hat\rho)\le\inf_{\phi'}T(\phi',\hat\rho)+\eps$, the infimum over centred pure Gaussian states; then $T(\phi,\psi)\le2T(\hat\rho,\psi)+\eps\le7\eps$, and for pure states $T=(1-|\langle\phi|\psi\rangle|^2)^{1/2}$, so $|\langle\phi|\psi\rangle|^2\ge1-49\eps^2$, with estimates in the centred class. The proof of \cite[Thm.~5.2]{CFGLMWW26} (App.~D.7 there) assumes exactly this, an $N'$-copy measurement whose centred pure estimate has squared overlap at least $1-\eps'^2$ with probability $0.99$ for every centred pure input, and draws the input from a prior on centred pure states with no bound on the energy; it gives $N'\ge k(m^2+m-1)-2m-1$ with $k=\lfloor1/(4\eps'^2)\rfloor$. At $\eps'=7\eps$ this is $\Omega(m^2/\eps^2)$ once $\eps$ is below an absolute $\eps_0$, and the projection uses no copies. The class of pure states contains the centred ones, so the bound applies to (ii) and to every measurement model.
\end{proof}

\begin{remark}[the constants of Theorem~\ref{thm:NG-pure}]\label{rem:NG-pure-constant}
The constants are absolute and very large, $N_r=\lceil10^{14}n^2\rceil$, and are not optimised. They come from the small-ball probability $\alpha_0=1/25600$ of Lemma~\ref{lem:NG-design}, which enters the Vapnik--Chervonenkis bound squared; that lemma uses only $\frac12I\preceq A\preceq2I$, while the pilot gives $\|A-I\|_\op\le\frac1{40}$. The order $n^2$ of the radial round cannot be lowered for this estimator, since $G_N$ is singular for $N<D_n$; it is reached because the operator norm of $\hat B-B$ is a supremum over $e^{O(n)}$ directions, each with variance $O(n/N)$, while its Frobenius norm has mean square of order $n^3/N$ and would cost $m^3$ outcomes.
\end{remark}

\subsection{Bounded thermal occupation: proofs of Theorem~\ref{thm:NG-thermal}, Corollary~\ref{cor:NG-cold-k} and Proposition~\ref{prop:NG-coldhot-k}}\label{subsec:NG-thermal-proofs}

Throughout this subsection $n:=m+4$ and the notation of \S\ref{subsec:NG-radialPOVM} is kept. For a centred Gaussian state $\rho(0,\Sigma)$, on any number of modes, write $\Sigma$ in position and momentum blocks, $\Sigma_{qq}\succ0$, and put
\begin{equation}\label{eq:NG-mixednot}
A:=(2\Sigma_{qq})^{-1},\qquad\beta:=\Sigma_{pq}\Sigma_{qq}^{-1}=B+\beta_a,\qquad\Sigma_{p|q}:=\Sigma_{pp}-\Sigma_{pq}\Sigma_{qq}^{-1}\Sigma_{qp},\qquad\Xi:=\Sigma_{p|q}-\tfrac12A,
\end{equation}
with $B:=\frac12(\beta+\beta^T)$ symmetric and $\beta_a:=\frac12(\beta-\beta^T)$ antisymmetric. The Wigner function, the density of $N(0,\Sigma)$, factorises as $W(q,p)=N(q;0,\Sigma_{qq})\,N(p;\beta q,\Sigma_{p|q})$: $\beta$ is the chirp and $\Sigma_{p|q}$ the momentum covariance at fixed position. For one mode, $\beta$, $\Sigma_{p|q}$ and $\Xi$ are the $\beta$, $h$ and $\vartheta$ of \S\ref{subsec:NG-law}. For $u\in S^{n-1}$ let
\[
\begin{gathered}
a(u):=u^TAu,\qquad\vartheta(u):=u^T\Xi u,\qquad b_0(u):=u^TBu=u^T\beta u,\\
\lambda(u):=a(u)+\vartheta(u)=u^T\big(\Sigma_{p|q}+\tfrac12A\big)u,
\end{gathered}
\]
and for $\nu,\lambda>0$ let
\[
f_{\nu,\lambda}(y):=\frac{\Gamma(\frac{\nu+1}2)}{\sqrt\pi\,\Gamma(\frac\nu2)\,\lambda}\Big(1+\frac{y^2}{\lambda^2}\Big)^{-(\nu+1)/2},
\]
the density of $\lambda\xi/\sqrt\chi$ for independent $\xi\sim N(0,1)$ and $\chi\sim\chi^2_\nu$ (proof of Lemma~\ref{lem:NG-student}(ii)). For an occupation bound $\bar n\ge0$ put $\bar d:=\bar n+\frac12$ and $\lambda_{\bar n}:=2\bar d^2+\frac12=1+2\bar n+2\bar n^2$.

\paragraph{The radial law of a mixed state.}

\begin{lemma}[mixed radial law]\label{lem:NG-mixedlaw}
Let $\rho=\rho(0,\Sigma)$ be a centred Gaussian state on $n$ modes whose Williamson occupations are at most $\bar n$.
\begin{enumerate}
\item[(i)] $\Xi-i\beta_a\succeq0$ and $\Xi+i\beta_a\succeq0$; in particular $\Xi\succeq0$ and $\|\beta_a\|_\op\le\|\Xi\|_\op$. Moreover $\frac12A\preceq\Sigma_{p|q}\preceq2\bar d^2A$, so $0\preceq\Xi\preceq(2\bar d^2-\frac12)A$ and $\Sigma_{p|q}+\frac12A\preceq\lambda_{\bar n}A$. If $\rho$ is pure, then $\Xi=0$, $\beta_a=0$, and $A$ and $B$ are the matrices of Lemma~\ref{lem:NG-siegel}.
\item[(ii)] Let $(u,b)$ be the outcome of $\mathsf R_n$ on $\rho$. Then $u$ has the law of Lemma~\ref{lem:NG-student}(i), and given $u$ the outcome $b$ has the density
\begin{equation}\label{eq:NG-mixture}
p(b\,|\,u)=\sum_{j\ge0}w_j(u)\,f_{\frac n2+j,\lambda(u)}\big(b-b_0(u)\big),\qquad w_j(u):=\frac{(\frac n2)_j}{j!}\Big(\frac{a(u)}{\lambda(u)}\Big)^{n/2}\Big(\frac{\vartheta(u)}{\lambda(u)}\Big)^j,
\end{equation}
where $(x)_j:=\Gamma(x+j)/\Gamma(x)$; the weights are nonnegative with sum one. Equivalently, $b$ has the law of $b_0(u)+\lambda(u)\,\xi/\sqrt\chi$, where $\xi\sim N(0,1)$ is independent of $(u,\chi)$ and $\chi$, given $u$, has the law $\sum_jw_j(u)\,\chi^2_{n/2+j}$. For a pure state $w_0=1$, and this is Lemma~\ref{lem:NG-student}(ii).
\item[(iii)] For $n\ge5$, $\E[b\,|\,u]=b_0(u)$ and $\operatorname{Var}(b\,|\,u)=\lambda(u)^2\,\E[\chi^{-1}\,|\,u]\le2\lambda(u)^2/(n-4)$.
\end{enumerate}
\end{lemma}

\begin{proof}
(i) With $\Omega=\begin{psmallmatrix}0&I\\-I&0\end{psmallmatrix}$ the uncertainty relation $\Sigma+\frac i2\Omega\succeq0$ reads
\[
\begin{pmatrix}\Sigma_{qq}&\Sigma_{qp}+\frac i2I\\\Sigma_{pq}-\frac i2I&\Sigma_{pp}\end{pmatrix}\succeq0 .
\]
Since $\Sigma_{qq}\succ0$, the Schur complement of its upper left block is positive semidefinite. With $\Sigma_{qq}^{-1}\Sigma_{qp}=\beta^T$,
\[
\Sigma_{pp}-\big(\Sigma_{pq}-\tfrac i2I\big)\Sigma_{qq}^{-1}\big(\Sigma_{qp}+\tfrac i2I\big)=\Sigma_{p|q}-\tfrac14\Sigma_{qq}^{-1}-\tfrac i2(\beta-\beta^T)=\Xi-i\beta_a\succeq0 .
\]
Its complex conjugate $\Xi+i\beta_a$ is also positive semidefinite, and their sum is $2\Xi$. As $i\beta_a$ is Hermitian, $-\Xi\preceq i\beta_a\preceq\Xi$ gives $|v^\dagger(i\beta_a)v|\le v^\dagger\Xi v\le\|\Xi\|_\op$ for unit $v\in\C^n$, hence $\|\beta_a\|_\op\le\|\Xi\|_\op$. For the second assertion let $\Sigma=SDS^T$ be a Williamson form, $D=\operatorname{diag}(d,d)$ with $\frac12\le d_j\le\bar d$. As in the proof of Lemma~\ref{lem:NG-core}(i), $\Omega\Sigma^{-1}\Omega^T=SD^{-1}S^T$, and $\frac14D^{-1}\preceq D\preceq\bar d^2D^{-1}$ entrywise; conjugating by $S$,
\[
\tfrac14\,\Omega\Sigma^{-1}\Omega^T\preceq\Sigma\preceq\bar d^2\,\Omega\Sigma^{-1}\Omega^T .
\]
The upper left block of $\Omega M\Omega^T$ is $M_{pp}$, and $(\Sigma^{-1})_{pp}=\Sigma_{p|q}^{-1}$ by the block inverse, so $\frac14\Sigma_{p|q}^{-1}\preceq\Sigma_{qq}\preceq\bar d^2\Sigma_{p|q}^{-1}$; inverting, $\frac14\Sigma_{qq}^{-1}\preceq\Sigma_{p|q}\preceq\bar d^2\Sigma_{qq}^{-1}$, which is $\frac12A\preceq\Sigma_{p|q}\preceq2\bar d^2A$. If $\rho$ is pure, $\Sigma=\Sigma(A',B')$ by Lemma~\ref{lem:NG-siegel}, and \eqref{eq:NG-SigmaAB} gives $A=A'$, $\beta=\frac12B'A'^{-1}\cdot2A'=B'$ and $\Sigma_{p|q}=\frac12(A'+B'A'^{-1}B')-\frac12B'A'^{-1}B'=\frac12A'$.

(ii) In a Williamson form of $\Sigma$, $\rho$ is a Gaussian unitary applied to a product of thermal states, each the marginal of a two-mode squeezed vacuum, so $\rho$ is the marginal on the first $n$ modes of a centred pure Gaussian state $\Psi$ on $2n$ modes. By Lemma~\ref{lem:NG-siegel}, $\Psi$ is a Gaussian function on $\R^n\times\R^n$, with $|\Psi(x,z)|\le C_\Psi e^{-c_\Psi(|x|^2+|z|^2)}$ for some $c_\Psi,C_\Psi>0$. The outcome of $\mathsf R_n$ on $\rho$ has the law of the outcome of $\mathsf R_n\otimes I$ on $\Psi$, so for Borel sets $E$, $\Pr[(u,b)\in E]=\int\|\mathbf 1_E\mathcal V\Psi(\cdot,z)\|^2\,dz$. By \eqref{eq:NG-V} for each Gaussian function $\Psi(\cdot,z)$, and Fubini's theorem, which the Gaussian bound on $\Psi$ allows, $(u,b)$ has the density
\[
p(u,b)=\frac1{8\pi}\int_0^\infty\!\!\int_0^\infty e^{-ib(t-t')/2}\,(tt')^{(n-2)/4}\,\rho(\sqrt t\,u,\sqrt{t'}\,u)\,dt\,dt',\qquad\rho(x,y):=\int\Psi(x,z)\overline{\Psi(y,z)}\,dz .
\]
This kernel of $\rho$ is a continuous Gaussian function, and Fourier inversion of the Wigner function, as in \eqref{eq:NG-kernel}, gives
\begin{equation}\label{eq:NG-mixedkernel}
\begin{aligned}
\rho(x,y)&=\int W\big(\tfrac{x+y}2,p\big)\,e^{ip\cdot(x-y)}\,dp\\
&=c_\rho\exp\Big[-\frac{(x+y)^TA(x+y)}4-\frac{(x-y)^T\Sigma_{p|q}(x-y)}2+\frac{i\,(x-y)^T\beta(x+y)}2\Big]
\end{aligned}
\end{equation}
with $c_\rho>0$. On the ray through $u$, $(\sqrt t\pm\sqrt{t'})^2=t+t'\pm2\sqrt{tt'}$, $u^T\Sigma_{p|q}u=\vartheta+\frac a2$ and $u^T\beta_au=0$ give, with $a,\vartheta,\lambda,b_0$ evaluated at $u$,
\[
\rho(\sqrt t\,u,\sqrt{t'}\,u)=c_\rho\exp\Big[-\frac\lambda2(t+t')+\vartheta\sqrt{tt'}+\frac{i\,b_0}2(t-t')\Big]:
\]
the antisymmetric part $\beta_a$ of the chirp does not enter. Expand $e^{\vartheta\sqrt{tt'}}=\sum_j\vartheta^j(tt')^{j/2}/j!$; since $\vartheta\sqrt{tt'}\le\vartheta(t+t')/2$ and $\lambda-\vartheta=a>0$, dominated convergence allows integration term by term, and each term factorises. With $y:=b-b_0$, $\nu_j:=\frac n2+j$ and the gamma integral of the proof of Lemma~\ref{lem:NG-student},
\[
\Big|\int_0^\infty t^{(\nu_j-1)/2}e^{-(\lambda+iy)t/2}\,dt\Big|^2=\Gamma\big(\tfrac{\nu_j+1}2\big)^2\,2^{\nu_j+1}\,(\lambda^2+y^2)^{-(\nu_j+1)/2}=4\pi\,\Gamma(\nu_j)\,\lambda^{-\nu_j}f_{\nu_j,\lambda}(y),
\]
the last step by the duplication formula $\Gamma(\frac\nu2)\Gamma(\frac{\nu+1}2)=2^{1-\nu}\sqrt\pi\,\Gamma(\nu)$. Hence
\[
p(u,b)=\frac{c_\rho}2\sum_{j\ge0}\frac{\vartheta^j}{j!}\,\Gamma\big(\tfrac n2+j\big)\,\lambda^{-n/2-j}f_{\frac n2+j,\lambda}(y)=\frac{c_\rho}2\,\Gamma\big(\tfrac n2\big)\,a^{-n/2}\sum_{j\ge0}w_j\,f_{\frac n2+j,\lambda}(y),
\]
because $\sum_j\frac{(n/2)_j}{j!}(\vartheta/\lambda)^j=(1-\vartheta/\lambda)^{-n/2}=(\lambda/a)^{n/2}$, which also shows $\sum_jw_j=1$. Integrating over $b$, the density of $u$ is $\frac{c_\rho}2\Gamma(\frac n2)a(u)^{-n/2}$, and dividing gives \eqref{eq:NG-mixture}; a mixture over $j$ of the laws of $\lambda\xi/\sqrt{\chi_j}$, $\chi_j\sim\chi^2_{\nu_j}$, is the law of $\lambda\xi/\sqrt\chi$ with $\chi$ drawn from the mixture. By Lemma~\ref{lem:NG-Rn}(iii), applied to each $\Psi(\cdot,z)$ and integrated over $z$, $u$ is distributed as $x/|x|$ for $x$ with the density $\rho(x,x)$, which is $N(0,\Sigma_{qq})=N(0,\frac12A^{-1})$; this is the law of Lemma~\ref{lem:NG-student}(i).

(iii) For $\nu>2$, $f_{\nu,\lambda}$ is symmetric with variance $\lambda^2\E[1/\chi^2_\nu]=\lambda^2/(\nu-2)$. Since $\nu_j\ge\frac n2>2$, $\E[b\,|\,u]=b_0$ and $\operatorname{Var}(b\,|\,u)=\lambda^2\sum_jw_j/(\frac n2+j-2)\le\lambda^2/(\frac n2-2)$.
\end{proof}

Thermal noise thus enters the radial outcome only through its scale: the degrees of freedom $\frac n2+j$ are at least those of a pure state, the centre $u^TBu$ is unchanged, and the antisymmetric part $\beta_a$, which the radial data do not see, is controlled by $\Xi$, which is bounded through the occupations. Without an occupation bound $\beta_a$ is unbounded, and the window of Lemma~\ref{lem:NG-window} below fails.

\paragraph{Operator-norm error and the covariance window for bounded occupation.}

\begin{lemma}[operator-norm error]\label{lem:NG-opnorm}
Let $\bar n\ge0$, let $\rho(0,\Sigma)$ be a centred $m$-mode Gaussian state with Williamson occupations at most $\bar n$ and $\|A-I\|_\op\le\frac1{40}$, with $A$ and $B$ as in \eqref{eq:NG-mixednot}, let $N:=\lceil10^{14}\lambda_{\bar n}^3n^2\rceil$ outcomes of $\mathsf R_n$ be taken on independent copies of $\rho\otimes|0\rangle\langle0|^{\otimes4}$, and let $\hat B_{\rm s}$ be the upper left $m\times m$ block of \eqref{eq:NG-LS}. Then $\Pr\big[G_N\text{ is invertible and }\|\hat B_{\rm s}-B\|_\op\le\frac1{10}\big]\ge0.97$, for every $B$. For the pure state $\psi_{A,B}$, $\bar n=0$, $\lambda_0=1$, and $N=N_r$.
\end{lemma}

\begin{proof}
Write $\lambda:=\lambda_{\bar n}\ge1$. The state $\rho\otimes|0\rangle\langle0|^{\otimes4}$ on $n$ modes has occupations at most $\bar n$ and the matrices $A\oplus I_4$, $B\oplus0_4$, $\Sigma_{p|q}\oplus\frac12I_4$ and $\Xi\oplus0_4$, with $\|(A\oplus I_4)-I\|_\op\le\frac1{40}$. By Lemma~\ref{lem:NG-mixedlaw}, its directions $u_l$ have the law of Lemma~\ref{lem:NG-student}(i), so Lemmas~\ref{lem:NG-design} and~\ref{lem:NG-gram} apply ($N\ge10^{14}n^2$), and $\lambda_l:=\lambda(u_l)\le\lambda\,u_l^T(A\oplus I_4)u_l\le\frac{41}{40}\lambda\le2\lambda$. Since $\hat B_{\rm s}-B$ is a compression of $\hat B-(B\oplus0_4)$, it suffices to bound the latter, which we call $\hat B-B$. By Lemma~\ref{lem:NG-mixedlaw}(ii) the outcomes have the law of $(u_l,b_l)$ with $b_l=u_l^TBu_l+\lambda_l\xi_l/\sqrt{\chi_l}$, where the pairs $(u_l,\chi_l)$, $l=1,\dots,N$, are independent, $\chi_l$ given $u_l$ has the law $\sum_jw_j(u_l)\chi^2_{n/2+j}$, and the $\xi_l\sim N(0,1)$ are independent of each other and of all $(u_l,\chi_l)$; the estimator is a function of $(u_l,b_l)_l$ only, so probabilities may be computed on this larger space.

\emph{The case $n\ge2048\lambda$.} Let $\mathcal E_1$ be the event of Lemma~\ref{lem:NG-gram}(i) and $\mathcal E_2:=\{\chi_l\ge n/4\ \forall l\}$. For $\chi\sim\chi^2_\nu$, $\E e^{-\chi/2}=2^{-\nu/2}$, so Markov's inequality gives $\Pr[\chi<\nu/2]\le e^{\nu/4}2^{-\nu/2}=e^{-(\frac12\log2-\frac14)\nu}$, at most $e^{-n/32}$ at $\nu=n/2$. A $\chi^2_{n/2+j}$ variable is a $\chi^2_{n/2}$ variable plus an independent nonnegative one, so $\Pr[\chi_l<n/4\,|\,u_l]\le e^{-n/32}$ and $\Pr[\mathcal E_2^c]\le Ne^{-n/32}\le1.01\cdot10^{14}\lambda^3n^2e^{-n/32}$. As $n^2e^{-n/32}$ decreases for $n\ge64$, this is at most $1.01\cdot10^{14}\cdot2048^2\lambda^5e^{-64\lambda}\le7\cdot10^{-8}\lambda^5e^{-64(\lambda-1)}\le7\cdot10^{-8}$, using $\lambda^5\le e^{5(\lambda-1)}$. Conditionally on $(u_l,\chi_l)_l$ in $\mathcal E_1\cap\mathcal E_2$, \eqref{eq:NG-LSerror} is a centred Gaussian vector with covariance
\[
G_N^{-1}\Big(\sum_l\frac{\lambda_l^2}{\chi_l}z_lz_l^T\Big)G_N^{-1}\preceq\frac{16\lambda^2}n\,G_N^{-1}\preceq\frac{16\lambda^2n}{c_0N}\,I ,
\]
so for every unit $v\in\R^n$, $v^T(\hat B-B)v$ is centred Gaussian with variance at most $16\lambda^2n/(c_0N)$. Let $\mathcal N$ be a $\frac14$-net of $S^{n-1}$ with $|\mathcal N|\le9^n$ \cite[Cor.~4.2.13]{Ver18}; with $\|M\|_\op\le2\max_{v\in\mathcal N}|v^TMv|$ for symmetric $M$ \cite[Ex.~4.4.3]{Ver18}, and $\Pr[|N(0,s^2)|>t]\le2e^{-t^2/(2s^2)}$ at $t=\frac1{20}$,
\[
\Pr\Big[\|\hat B-B\|_\op>\tfrac1{10}\,\Big|\,(u_l,\chi_l)_l\Big]\le2\cdot9^n\exp\Big(-\frac{c_0N}{12800\,\lambda^2n}\Big)\le2\cdot9^ne^{-4238\,\lambda n}<0.01 .
\]
The failure probability is at most $0.01+7\cdot10^{-8}+0.01<0.03$.

\emph{The case $5\le n<2048\lambda$.} By Lemma~\ref{lem:NG-gram}(ii), $G_N\succeq N(16n(n+2))^{-1}I$ except with probability $D_ne^{-N/(64n(n+2))}<0.01$, as $D_n\le n^2<4.2\cdot10^6\lambda^2$ and $N/(64n(n+2))>10^{12}\lambda^3$. Given the directions, the $e_l$ are independent and centred with $\operatorname{Var}(e_l)\le2\lambda_l^2/(n-4)\le8\lambda^2/(n-4)\le40\lambda^2/n$ (Lemma~\ref{lem:NG-mixedlaw}(iii)), so by \eqref{eq:NG-LSerror}
\[
\E\big[\|\hat B-B\|_F^2\,\big|\,u_1,\dots,u_N\big]\le\frac{40\lambda^2}n\Tr G_N^{-1}\le\frac{640\,\lambda^2D_n(n+2)}N\le\frac{640\,\lambda^2n^3}N ,
\]
and Markov's inequality gives $\Pr[\|\hat B-B\|_\op>\frac1{10}]\le\Pr[\|\hat B-B\|_F>\frac1{10}]\le64000\,\lambda^2n^3/N\le6.4\cdot10^{-10}\,n/\lambda<1.4\cdot10^{-6}$ on that event. The failure probability is below $0.02$.

For $\psi_{A,B}$, Lemma~\ref{lem:NG-mixedlaw}(i) gives $\Xi=0$ and the matrices of Lemma~\ref{lem:NG-siegel}, and $\lambda_0=1$ gives $N=\lceil10^{14}n^2\rceil=N_r$.
\end{proof}

The two cases share the protocol and differ only in the analysis: for small $n$ the conditional noise is not sub-Gaussian (at $n=5$ it has only two finite moments), and below $n=2048\lambda_{\bar n}$ the Frobenius bound suffices; for large $n$ a Frobenius bound would cost $n^3$ outcomes, and the net, with the conditional Gaussian law given the $\chi_l$, gives $n^2$. The factor $\lambda_{\bar n}^3$ in $N$ pays for the scale $\lambda(u)\le\frac{41}{40}\lambda_{\bar n}$ of the noise and for the threshold $2048\lambda_{\bar n}$ between the two cases; it is a constant of the class.

\begin{lemma}[covariance window]\label{lem:NG-window}
Let $\bar n\ge0$ and $M_{\bar n}:=\frac{11}{10}+\frac{41}{40}(2\bar d^2-\frac12)$. Let $\Sigma$ be the covariance of a centred $m$-mode Gaussian state with Williamson occupations at most $\bar n$ and $\|A-I\|_\op\le\frac1{40}$, with $A$ and $B$ as in \eqref{eq:NG-mixednot}, and let $\hat X\in\mathrm{Sym}_m$ with $\|\hat X-B\|_\op\le\frac1{10}$. Then $\Sigma':=\Lambda_{-\hat X}\Sigma\Lambda_{-\hat X}^T$ satisfies
\begin{equation}\label{eq:NG-window}
\frac{39}{80\,M_{\bar n}^2}\,I\preceq\Sigma'\preceq\frac{82}{39}\,\bar d^2M_{\bar n}^2\,I\preceq K_{\bar n}I ,
\end{equation}
with $K_{\bar n}$ as in Theorem~\ref{thm:NG-thermal}.
For $\bar n=0$ this is $0.40\,I\preceq\Sigma'\preceq0.64\,I$, in particular $\frac14I\preceq\Sigma'\preceq I$.
\end{lemma}

\begin{proof}
Put $Q:=\Sigma_{qq}=\frac12A^{-1}$ and extend \eqref{eq:NG-shearmat} to every real $m\times m$ matrix $Y$, $\Lambda_Y:=\begin{psmallmatrix}I&0\\Y&I\end{psmallmatrix}$, symplectic only for symmetric $Y$; still $\Lambda_Y\Lambda_{Y'}=\Lambda_{Y+Y'}$, $\Lambda_Y^{-1}=\Lambda_{-Y}$ and $\|\Lambda_Y\|_\op\le1+\|Y\|_\op$. Block elimination gives $\Sigma=\Lambda_\beta\operatorname{diag}(Q,\Sigma_{p|q})\Lambda_\beta^T$, since $\beta Q=\Sigma_{pq}$ and $\beta Q\beta^T=\Sigma_{pq}\Sigma_{qq}^{-1}\Sigma_{qp}$; hence
\[
\Sigma'=\Lambda_{\beta-\hat X}\operatorname{diag}(Q,\Sigma_{p|q})\,\Lambda_{\beta-\hat X}^T,\qquad\beta-\hat X=(B-\hat X)+\beta_a .
\]
The unseen antisymmetric part enters here. By Lemma~\ref{lem:NG-mixedlaw}(i), $\|\beta_a\|_\op\le\|\Xi\|_\op\le(2\bar d^2-\frac12)\|A\|_\op\le\frac{41}{40}(2\bar d^2-\frac12)$, so $\|\Lambda_{\beta-\hat X}\|_\op$ and $\|\Lambda_{\beta-\hat X}^{-1}\|_\op$ are at most $M_{\bar n}$. The pilot bound gives $\frac{20}{41}I\preceq Q\preceq\frac{20}{39}I$, and Lemma~\ref{lem:NG-mixedlaw}(i) gives $\frac{39}{80}I\preceq\frac12A\preceq\Sigma_{p|q}\preceq2\bar d^2A\preceq\frac{41}{20}\bar d^2I$; as $\bar d\ge\frac12$, $\frac{39}{80}I\preceq\operatorname{diag}(Q,\Sigma_{p|q})\preceq\frac{82}{39}\bar d^2I$. For $Z\succeq cI$ and invertible $T$, $TZT^T\succeq c\,\|T^{-1}\|_\op^{-2}I$, and $TZT^T\preceq\|Z\|_\op\|T\|_\op^2I$, which gives the first two inequalities of \eqref{eq:NG-window}. For the third, $\frac{41}{40}(2\bar d^2-\frac12)\le4\bar d^2$, so $M_{\bar n}\le\frac{11}{10}+4\bar d^2$, and $\frac{82}{39}\le4$. At $\bar n=0$, $\bar d=\frac12$ and $M_0=\frac{11}{10}$, so the bounds are $0.403$ and $0.636$.
\end{proof}

Only operator norms enter: the window needs $\|\hat B_{\rm s}-B\|_\op\le\frac1{10}$, not a Frobenius bound, which is why the radial round costs $O(m^2)$ and not $O(m^3)$ outcomes. For a pure state the window is two-sided at constants close to $\frac12$; for a mixed state the upper constant grows with the occupation cap, through $\Sigma_{p|q}$ and through $\beta_a$, and the lower one decreases only through $\beta_a$.

\paragraph{Proof of Theorem~\ref{thm:NG-thermal}.}

\begin{proof}[Proof of Theorem~\ref{thm:NG-thermal}]
(i) Let $\rho=\rho(0,\Sigma)$ have occupations $n_j\le\bar n$. Write $K:=K_{\bar n}$.

\emph{Round 1 (homodyne pilot).} This is round 1 of the proof of Theorem~\ref{thm:NG-pure}, on $N_q$ copies; its analysis uses only the law $N(0,\Sigma_{qq})$ of the position outcomes, which holds for every centred Gaussian state. Except with probability $0.01$ it returns $S_1=\operatorname{diag}(L,L^{-1})$ such that $\Sigma_1:=S_1\Sigma S_1^T$ has $\|A-I\|_\op\le\frac1{40}$, $A:=(2(\Sigma_1)_{qq})^{-1}$; $\Sigma_1$ has the occupations of $\Sigma$, and its chirp is arbitrary.

\emph{Round 2 (radial measurement).} On each of $N_r(\bar n)$ fresh copies apply $U_{S_1}$, append four vacuum modes and measure $\mathsf R_n$. By Lemma~\ref{lem:NG-opnorm}, applied to $\rho(0,\Sigma_1)$, $\|\hat B_{\rm s}-B\|_\op\le\frac1{10}$ except with probability $0.03$, $B$ the symmetric part of the chirp of $\Sigma_1$. Put $S_2:=\Lambda_{-\hat B_{\rm s}}S_1$. By Lemma~\ref{lem:NG-window}, $\Sigma'':=S_2\Sigma S_2^T=\Lambda_{-\hat B_{\rm s}}\Sigma_1\Lambda_{-\hat B_{\rm s}}^T$ satisfies $\|\Sigma''\|_\op\le K$.

\emph{Round 3 (Theorem~\ref{thm:Bos}(i)).} Split the remaining budget $N_3:=N-N_q-N_r(\bar n)$ into $54$ batches of $\lfloor N_3/54\rfloor$ copies. On each batch run the protocol of Theorem~\ref{thm:Bos}(i) with covariance bound $K$, accuracy $\eps/3$ and budget $\lfloor N_3/54\rfloor$, on copies transformed by $U_{S_2}$; each run outputs a Gaussian state $\hat\rho''_i$. Return the first $\hat\rho''_i$, in a fixed order, that is within trace distance $\frac23\eps$ of at least $27$ of the other $53$ (a fixed Gaussian state if none is), and output $\hat\rho:=U_{S_2}^\dagger\hat\rho''_iU_{S_2}$. The transformed copies are in the state $\rho'':=\rho(0,\Sigma'')$, which has $\|\Sigma''\|_\op\le K$, any squeezing, and the occupations $n_j$ of $\rho$, symplectic maps preserving Williamson occupations; so its profile rate is $\mathcal R(\boldsymbol n)$. If $\lfloor N_3/54\rfloor\ge9C(K)\mathcal R(\boldsymbol n)/\eps^2$ and $\eps/3\le\eps_\star(K)$, the $54$ runs are independent given rounds 1 and 2, each with $T(\hat\rho''_i,\rho'')\le\eps/3$ with probability at least $\frac23$. By Hoeffding's inequality at least $28$ runs succeed except with probability $e^{-2\cdot9^2/54}=e^{-3}<0.05$. Any two successes are within $\frac23\eps$ of each other, so a success qualifies; the returned estimate and $27$ of its neighbours form $28$ estimates, which meet the at least $28$ successes among $54$, so the returned estimate is within $\frac23\eps$ of a success and within $\eps$ of $\rho''$. This is the selection rule of \cite[proof of Thm.~5.9]{Rub26b}. Then $T(\hat\rho,\rho)=T(\hat\rho''_i,\rho'')\le\eps$, and $\hat\rho$ is a Gaussian state.

\emph{Failure and count.} The failure probability is at most $0.01+0.03+0.05<\frac13$. $N_q\le3.2\cdot10^5m^2+1$ and $N_r(\bar n)\le2.5\cdot10^{15}\lambda_{\bar n}^3m^2+1$, and $\mathcal R(\boldsymbol n)\ge m^2$, so for $\eps\le1$ the conditions hold, in (i) and in (ii) below, whenever $N\ge C(\bar n)\mathcal R(\boldsymbol n)/\eps^2$ with $C(\bar n):=486\,C(K_{\bar n})+6\cdot10^{15}\lambda_{\bar n}^3+58$, and we take $\eps_\star(\bar n):=\min\{1,\eps_\star(K_{\bar n})\}$. No constant depends on $m$, on the energy or on the squeezing.

(ii) For $\rho=\rho(v,\Sigma)$, round 1 measures $2N_q$ single copies and uses the $N_q$ pair differences $(x_{2l}-x_{2l-1})/\sqrt2$, independent with law $N(0,\Sigma_{qq})$. In round 2 each of $N_r(\bar n)$ pairs of copies passes through a balanced beam splitter, which maps $\rho(v,\Sigma)^{\otimes2}$ to $\rho(0,\Sigma)\otimes\rho(\sqrt2\,v,\Sigma)$ \cite[Lemma~8.1]{CGYZ26} (as in the proof of Theorem~\ref{thm:NG1}(ii), which uses only the Gaussian form of the state), and the centred port is measured as a copy in (i). Round 3 runs Theorem~\ref{thm:Bos}(i), which allows arbitrary displacement, on $\rho(S_2v,\Sigma'')$, with single copies. The count is at most doubled in rounds 1 and 2.

\emph{Rates.} For every profile $r\le m$, and every $i>r$ has $n_i>2/i\ge2/m$, so $S_w\le m^3/4$ and $\mathcal R(\boldsymbol n)\le m^2+\frac54m^3$. If every $n_j\ge g$, then $jn_j\le2$ forces $j\le2/g$, so $r^3\le8/g^3$, $S_w\le m^2/(2g)$, and $\mathcal R(\boldsymbol n)\le(1+\frac1{2g})m^2+8/g^3$.

\emph{Lower bound.} The passive family of \cite[Thm.~2]{CMMP26} on $m$ modes consists of centred states whose occupations are at most $1/(2m-1)$ \cite[proof of Thm.~5.9]{Rub26b}, hence at most $\bar n$ for $m\ge(1+\bar n^{-1})/2$; it lies in the class, and \cite[Thm.~2]{CMMP26} at $k=1$ gives $\Omega(m^3/\eps^2)$ for $m\ge n_0$ and $\eps<\eps_0$.
\end{proof}

\paragraph{Proof of Corollary~\ref{cor:NG-cold-k}.}

\begin{proof}[Proof of Corollary~\ref{cor:NG-cold-k}]
Let $\rho=\rho(v,\Sigma)$ have occupations $n_j\le c/m$, and put $\bar n:=c$, $K:=K_c$, $C_\star:=c+1$ and $\delta_\star:=(1+C_\star)^{-1}$. Constants $C_K$, $c_K$ are those of \S\ref{subsec:Bos-undo}, and depend only on $c$ here.

\emph{Coarse rounds.} Rounds 1 and 2 of the proof of Theorem~\ref{thm:NG-thermal} with $\bar n=c$ use $N_q+N_r(c)=O_c(m^2)$ single copies if $v=0$, and twice as many, in pairs, otherwise. Except with probability $0.04$ they return a symplectic $S_2$ with $\|S_2\Sigma S_2^T\|_\op\le K$ (Lemma~\ref{lem:NG-window}). Apply $U_{S_2}$ to every later copy; the transformed state $\rho(S_2v,S_2\Sigma S_2^T)$ has the occupations of $\rho$.

\emph{Squeezing removal.} Apply Lemma~\ref{lem:Bos-undo} with this $K$ and $\eps_0:=\eps/A_\star$, $A_\star=A_\star(c)$ fixed below; its hypotheses are $\|S_2\Sigma S_2^T\|_\op\le K$ and $\eps_0\le\eps_1(K)$, the latter for $\eps\le\eps_\star(c)$. Stage~0 uses $L_0=O_c(m^2/\eps^2)$ single copies and, except with probability $0.03$, every later copy, after the known Gaussian unitary $V_0$, is in a Gaussian state $\tau$ with $T(\tau,\rho_{Q_\tau})\le\delta\le c_K\eps/A_\star$ \eqref{eq:Bos-delta}. By \eqref{eq:Bos-Qtau}, $\|Q_\tau-Q\|_\op\le C_Km/L_0$ with $Q$ of the original occupation spectrum, so $\Tr Q\le c$; with the constant in $L_0$ large enough, $m\|Q_\tau-Q\|_\op\le1$ and $\Tr Q_\tau\le C_\star$.

\emph{Readout.} On fresh copies after $V_0$, run on all $m$ modes the cold readout of the proof of Theorem~\ref{thm:Bos}(i) (\S\ref{subsec:Bos-proof-i}, from \eqref{eq:Bos-52} to the bound on $T(\rho_{\hat F},\sigma_c)$), with the passive reference $\sigma_c:=\rho_{Q_\tau}$ in place of $\rho_A$, $Q_\tau$ in place of $A$, the measured state $\tau$ in place of $\tau_c$, $m$ in place of $s$ and $C_\star$ in place of $C_A$; the only property of $A$ used there is $\Tr A\le C_A$, and $T(\tau,\sigma_c)\le\delta$. Its Stage~2 is that of \cite[Thm.~5.9]{Rub26b} with blocks of at most $k$ copies, the photon number measured on every copy of a block and the one-photon copies grouped within the block, as in Stage~2 of the proof of \cite[Thm.~5.13]{Rub26b}; what it uses about the measured state, quoted in \S\ref{subsec:Bos-proof-i}, does not depend on $k$. With $A_\star$ chosen so that the right side of \eqref{eq:Bos-53} is at most $\delta_\star\eps/16$, it returns, except with probability $0.15$ and from $O_c\big((1+m^3/\min\{\sqrt k,m\})/\eps^2\big)$ copies \cite[Thm.~5.13]{Rub26b}, a passive state $\rho_{\hat F}$ with $T(\rho_{\hat F},\rho_{Q_\tau})\le\eps/4$.

\emph{Assembly.} With the known Gaussian unitary $W:=V_0U_{S_2}$, output $\hat\rho:=W^\dagger\rho_{\hat F}W$. Then $T(\hat\rho,\rho)=T(\rho_{\hat F},\tau)\le\eps/4+\delta\le\eps$ once also $A_\star\ge2c_K$, except with probability $0.04+0.03+0.15<\frac13$. The count is $O_c\big(m^2+m^2/\eps^2+(1+m^3/\min\{\sqrt k,m\})/\eps^2\big)=O_c\big(m^3/(\eps^2\min\{\sqrt k,m\})\big)$, as $m^2\le m^3/\min\{\sqrt k,m\}$. Outside the readout only the radial round of a displaced state measures two copies jointly, so for $k=1$ a centred state is learned from single copies.

\emph{Lower bound.} For $c\ge\frac23$ and $m\ge2$ the passive family of \cite[Thm.~2]{CMMP26} has occupations at most $1/(2m-1)\le c/m$ and lies in the class, and its lower bound $\Omega(m^3/(\eps^2\min\{\sqrt k,m\}))$ holds for every $k$ \cite[Thm.~2]{CMMP26}, \cite[Thm.~5.9]{Rub26b}.
\end{proof}

\paragraph{Proof of Proposition~\ref{prop:NG-coldhot-k}.}

\begin{proof}[Proof of Proposition~\ref{prop:NG-coldhot-k}]
(i) Follow the proof of Theorem~\ref{thm:Bos-sep}(i) (\S\ref{subsec:Bos-sep-upper}), with $m_c$ cold modes and one warm band, and with the following changes.

\emph{Step 1}, with $\mu:=\frac14$ in place of $21c/(20m_c)$. With $L_0=C_{c,K}m^2/\eps^2$ and its constant large enough, $\|Q_\tau-Q\|_\op\le C_Km/L_0\le\frac1{80}$. By Weyl's inequality the $m_c$ cold eigenvalues of $Q_\tau$ are at most $c/m_c+\frac1{80}\le\frac14=\mu$ for $m_c\ge8c$, its warm eigenvalues are at least $\nu-\frac1{80}\ge\frac34=3\mu$ and within a relative factor $\frac1{80}$ of the original ones, as $\nu\ge1$, and by Ky Fan's inequality the sum of its $m_c$ smallest eigenvalues is at most $c+C_K\eps^2$.

\emph{Step 2}, Lemma~\ref{lem:Bos-coldframe} at $\mu=\frac14$. Its proof uses only $Q_c\preceq\mu I$, $Q_w\succeq3\mu I$ and \eqref{eq:Bos-eta20}, and these hold at $\mu=\frac14$. With $D=Q_\tau+\frac14I$, $\Tr D^{-1}\le4m_c+\frac43m_w\le4m$, so \eqref{eq:Bos-effrankK} and Markov's inequality give \eqref{eq:Bos-eta20} with any prescribed constant success probability from $L_F\ge C_Km$ copies; and \eqref{eq:Bos-leakcost}, with $q_i\ge\frac34$ on the warm indices, gives $\E\|Q_w^{-1/2}E_{wc}\|_F^2\le\frac83(1+K_\tau)^2m_cm_w/L_F$, so \eqref{eq:Bos-leak} and Markov's inequality give $\ell\le\eps^2/B_\star^2$ from $L_F=C_{c,K}m^2/\eps^2$, with $B_\star=B_\star(c,K)$ the constant called $B$ in step~2 of \S\ref{subsec:Bos-sep-upper}. No term $m_c/\mu^2\asymp m_c^3$ appears, because $\mu$ is a constant. Lemma~\ref{lem:Bos-cross} holds with $\mu=\frac14$: $T_I\le C_0\sqrt\ell$, $\Tr Q'_c\le c+O(\eps^2)$, and the warm comparison (iii).

\emph{Step 3}, the cold readout with blocks of at most $k$ copies, exactly as in the proof of Corollary~\ref{cor:NG-cold-k}, with $m_c$ in place of $m$, the actual cold marginal $\tau_c$ as the measured state and the cold target $\sigma_c=\rho_{Q'_c}$ of step~3 of \S\ref{subsec:Bos-sep-upper}, with $T(\tau_c,\sigma_c)\le\delta$: $L_c=O_c\big((1+m_c^3/\min\{\sqrt k,m_c\})/\eps^2\big)$ copies, failure probability $0.15$.

\emph{Step 4} is unchanged; since every warm occupation is at least $\nu\ge1$, $\sum_{i,j\in w}1/(n_i+n_j)\le m_w^2/2$ and $L_w=C_{c,K}m_w^2/\eps^2$. \emph{Step 5} is unchanged, $\mu=\frac14$ being bounded, and the failure probability is at most $0.03+0.04+0.15+0.04=0.26$. The total is $O_{c,K}\big(m^2/\eps^2+m_c^3/(\eps^2\min\{\sqrt k,m_c\})+m_w^2/\eps^2\big)$, and $m^2\le2(m_c^2+m_w^2)\le2F_k$ because $m_c^3/\min\{\sqrt k,m_c\}\ge m_c^2$. The protocol uses $m_c$, as in Theorem~\ref{thm:Bos-sep}, which every member of $\mathcal C$ determines because $\nu>c/m_c$.

(ii) \emph{Cold group.} Fix the warm modes at the thermal product $\tau_\nu^{\otimes m_w}$, of covariance $(\nu+\frac12)I\preceq KI$, and vary the cold modes over the passive family of \cite[Thm.~2]{CMMP26} on $m_c$ modes, with occupations at most $1/(2m_c-1)\le c/m_c$. A protocol with blocks of at most $k$ copies on $\rho_c\otimes\tau_\nu^{\otimes m_w}$, with the second factor known, induces one on $\rho_c$: a POVM element $M$ on $t\le k$ copies induces $\Tr_w[(I\otimes(\tau_\nu^{\otimes m_w})^{\otimes t})M]$, as in the proof of Lemma~\ref{lem:Bos-directsum}; and the trace distance contracts under the partial trace. So \cite[Thm.~2]{CMMP26} gives $\Omega(m_c^3/(\eps^2\min\{\sqrt k,m_c\}))$ for $m_c\ge n_0$ and $\eps<\eps_0$. \emph{Warm modes.} Fix the cold modes at the vacuum and use the family of \cite[Thm.~9]{CMMP26} on $m_w$ modes, with occupations $\nu$ and $\nu+\Delta\le2\nu$ and covariance at most $(2\nu+\frac12)I\preceq KI$; as a Holevo--Fano bound it holds for arbitrary measurements (\S\ref{subsec:Bos-bands}) and gives $\Omega(m_w^2/\eps^2)$ for $m_w$ large and $\eps<1/1800$. The larger of the two bounds is at least half their sum.

\emph{Bound on the occupations instead of the covariance.} Run rounds 1 and 2 of the proof of Theorem~\ref{thm:NG-thermal} with this $\bar n$: $O_{\bar n}(m^2)$ single copies for centred states and pairs otherwise, and, except with probability $0.04$, a known $S_2$ with $\|S_2\Sigma S_2^T\|_\op\le K_{\bar n}$. The transformed state has the same occupations, and $K_{\bar n}\ge(2\bar n+1)^2\ge2\nu+1$, so it lies in $\mathcal C(m_c,c;m_w,\nu;K_{\bar n})$; apply (i) with $K=K_{\bar n}$ and transform the output back by $U_{S_2}^\dagger$. The failure probability is at most $0.04+0.26<\frac13$ and the coarse rounds cost $O_{\bar n}(m^2)\le O_{\bar n}(F_k)$. The two families of (ii) are centred and have occupations at most $\max\{c/m_c,2\nu\}\le\bar n$, so they lie in the class, and (ii) holds with $K$ replaced by $2\nu+1$.
\end{proof}

\subsection{Siegel-disk moments and the purification identity}

\paragraph{The Siegel disk.}
By the Takagi factorisation $K=U\operatorname{diag}(t_1,\dots,t_d)U^T$, $U$ unitary, $0\le t_i<1$, the state $\psi_K$ is the passive Gaussian unitary of $U$ applied to a product of one-mode squeezed vacua with $\tanh r_i=t_i$ \cite[Sec.~4.4]{CGYZ26}. So $K=0$ is the vacuum, the boundary of $\mathfrak D_d$ is infinite squeezing, and the covariance of $\psi_K$ has eigenvalues $\frac12e^{\pm2r_i}$, with $l_i=t_i^2$ below. (i) The exponential in $\psi_K$ has vacuum component $1$, so $\langle0|\psi_K\rangle=\det(I-KK^\dagger)^{1/4}$ and the outcome density at $|0\rangle^{\otimes k}$ is $Z^{-1}\det(I-KK^\dagger)^{k/2}\det(I-KK^\dagger)^{-(d+1)}$. (ii) A Gaussian unitary maps $\psi_K$ to $\psi_{S\cdot K}$ up to a phase, which cancels in $|\psi_K\rangle\langle\psi_K|$ \cite[Lemma~8.11]{CGYZ26}, and the measure $dK/\det(I-KK^\dagger)^{d+1}$ is invariant under $K\mapsto S\cdot K$ \cite[Lemma~8.12]{CGYZ26}, \cite{Per86}; the substitution $K'=S\cdot K$ in the integral defining $M_k$ gives the covariance. (iii) For the input $(U_S|0\rangle)^{\otimes k}$, $\Pr[K\in\mathcal B]=\langle0|^{\otimes k}M_k(S^{-1}\cdot\mathcal B)|0\rangle^{\otimes k}$, which is \cite[Cor.~8.14]{CGYZ26}. For $d=1$ and $K=e^{i\phi}\tanh r$ the invariant measure is $\frac12\sinh(2r)\,dr\,d\phi$, the hyperbolic area element: the number of distinguishable squeezed states grows like $e^{2r}$. Since $|\langle0|\psi_K\rangle|^2=\operatorname{sech}r$, $k$ copies of the vacuum overlap $\psi_K^{\otimes k}$ with weight $\operatorname{sech}^kr\approx2^ke^{-kr}$, and the outcome density in $r$ is of order $e^{(2-k)r}$, normalisable exactly when $k>2$. In general, with $\gamma:=k/2-d-1$ the exponent in (i), $\gamma>-1$ is sufficient for $\det(I-KK^\dagger)^{\gamma}$ to be integrable \cite[Lemma~8.4]{CGYZ26}, and necessary because of the factor $(1-l_1)^\gamma$ of \eqref{eq:NG-radial} near $l_1=1$; $\gamma>-1$ is $k>2d$. The resolution of the projector \cite[Lemma~8.5]{CGYZ26} then follows by identifying the span of the $k$-fold powers with a weighted Bergman space on $\mathfrak D_d$, whose reproducing kernel $\det(I-KK'^\dagger)^{-k/2}$ matches the $k$-th power of the overlaps \cite[Sec.~8.2.3]{CGYZ26}.

By (i), at the vacuum the outcome has density proportional to $\det(I-KK^\dagger)^\gamma$, and by the Hua--Takagi Jacobian \cite[Sec.~8.2]{CGYZ26} the squared singular values $l_i\in(0,1)$ of $K$ have density proportional to
\begin{equation}\label{eq:NG-radial}
\prod_i(1-l_i)^\gamma\prod_{i<j}|l_i-l_j| .
\end{equation}

\begin{lemma}[radial moments]\label{lem:NG-radial}
Let $\gamma>1$ and let $(l_1,\dots,l_d)$ have the density \eqref{eq:NG-radial} on $(0,1)^d$. Put $x_i:=l_i/(1-l_i)$, $X:=\sum_ix_i$, $Y:=\sum_ix_i^2$ and $A_d:=d(d+1)/2$. Then
\begin{align}
\E X&=A_d/\gamma,\label{eq:NG-EX}\\
\gamma\,\E X^2&=(A_d+1)\,\E X+\E Y,\qquad(\gamma-\tfrac12)\,\E Y-\tfrac12\,\E X^2=A_d+(d+1-\gamma)\,\E X,\label{eq:NG-EX2}\\
\E Y&=\frac{A_d\,[2\gamma(d+1)+A_d+1]}{\gamma\,(2\gamma^2-\gamma-1)} .\label{eq:NG-EY}
\end{align}
\end{lemma}

\begin{proof}
Let $p$ be the density. The integral of $\sum_i\partial_i(l_ip)$ vanishes. Since $\sum_il_i\partial_i\log p=-\gamma X+\sum_{i<j}(l_i-l_j)/(l_i-l_j)=-\gamma X+d(d-1)/2$, this gives $d+d(d-1)/2-\gamma\,\E X=0$, which is \eqref{eq:NG-EX}. The vector fields $l_iX$ and $x_i$ give \eqref{eq:NG-EX2}: for the first use $\partial_iX=(1+x_i)^2$ and $\sum_il_i(1+x_i)^2=X+Y$; for the second use $\partial_ix_i=(1+x_i)^2$, $x_i/(1-l_i)=x_i(1+x_i)$, $(x_i-x_j)/(l_i-l_j)=1/((1-l_i)(1-l_j))=(1+x_i)(1+x_j)$ and $\sum_{i<j}(1+x_i)(1+x_j)=\frac12[(d+X)^2-(d+2X+Y)]$. Eliminating $\E X^2$ and inserting \eqref{eq:NG-EX} gives \eqref{eq:NG-EY}. To justify the integrations by parts, work on the ordered chamber $0<l_1<\dots<l_d<1$, where $p$ is smooth: $p$ vanishes at collisions, and the symmetric pairwise combinations above have finite limits there; at $l_i=1$ the strongest boundary factor is $(1-l_i)^{\gamma-1}$, which vanishes for $\gamma>1$; and the moments above are finite for $\gamma>1$, since $X^2\le dY$, $|l_i-l_j|\le1$ and $\int_0^1x^2(1-l)^\gamma\,dl<\infty$ exactly when $\gamma>1$. The formal differentiations are therefore justified by truncation. At $d=1$, \eqref{eq:NG-EY} reduces to $2/(\gamma(\gamma-1))$, the second moment of the beta-prime law.
\end{proof}

With $W$ and $\Theta=2W$ as in \S\ref{subsec:NG-moments}, by the covariance formula of \cite[Sec.~4.4]{CGYZ26}, where the vacuum covariance is $I$, $\Theta$ has eigenvalues $e^{\pm2r_i}$ with $\tanh^2r_i=l_i$; since $2\cosh2r_i=2+4x_i$,
\begin{equation}\label{eq:NG-TrTheta}
\Tr\Theta=2d+4X,\qquad\Tr\Theta^2=2d+16X+16Y .
\end{equation}
The vacuum law is invariant under passive Gaussian unitaries, which act on $\Theta$ by conjugation with the orthogonal symplectic matrices, and the real symmetric matrices commuting with all of these are the multiples of $I$. Hence, by \eqref{eq:NG-EX},
\begin{equation}\label{eq:NG-ETheta}
\E\Theta=c_kI,\qquad c_k=1+\frac{d+1}\gamma .
\end{equation}
For $\gamma\ge2(d+1)$, \eqref{eq:NG-EX} gives $\E X\le d/4$; since $A_d+1\le(d+1)^2/2$, $d+1\le\gamma/2$ and $2\gamma^2-\gamma-1\ge\gamma^2$, \eqref{eq:NG-EY} gives $\E Y\le\frac94A_d(d+1)/\gamma^2\le9d/32$. Therefore
\begin{equation}\label{eq:NG-vacmom}
\E\|\Theta-c_kI\|_F^2=\E\Tr\Theta^2-2dc_k^2\le\E\Tr\Theta^2\le12d,\qquad1<c_k\le\tfrac32 .
\end{equation}
For $k=2d+5$, that is $\gamma=\frac32$, \eqref{eq:NG-EX}--\eqref{eq:NG-EY} give $c_k=k/3$, $\E X=\frac23A_d$, $\E Y=\frac13A_d(A_d+3d+4)$ and $\E\Tr\Theta^2=2d+\frac{16}3A_d(A_d+3d+6)$. Since $\E\Tr\Theta=2dc_k$, and since for $d\ge2$ we have $A_d\le\frac34d^2$, $3d+6\le3d^2$ and $c_k\ge\frac23d$,
\begin{equation}\label{eq:NG-vacmom32}
\E\|\Theta/c_k-I\|_F^2=\frac{\E\Tr\Theta^2}{c_k^2}-2d\le\frac9{4d^2}\,(2d+15d^4)\le40d^2 .
\end{equation}

\begin{lemma}[pure inputs]\label{lem:NG-pure}
Let the input be $k$ copies of a centred pure $d$-mode Gaussian state with covariance $\Sigma_p$, and let $W$ be the outcome covariance.
(i) If $k\ge6d+6$, equivalently $\gamma\ge2(d+1)$, then
\begin{equation}\label{eq:NG-puremom}
\E W=c_k\Sigma_p,\qquad\E\big\|\Sigma_p^{-1/2}(W-c_k\Sigma_p)\Sigma_p^{-1/2}\big\|_F^2\le12d .
\end{equation}
(ii) If $d\ge2$ and $k=2d+5$, equivalently $\gamma=\frac32$, then
\begin{equation}\label{eq:NG-puremom32}
\E W=c_k\Sigma_p,\qquad\E\big\|\Sigma_p^{-1/2}(W/c_k-\Sigma_p)\Sigma_p^{-1/2}\big\|_F^2\le40d^2 .
\end{equation}
\end{lemma}

\begin{proof}
Write $\Sigma_p=\frac12SS^T$ with $S$ symplectic. By covariance $W$ has the law of $SW_0S^T$, $W_0$ the outcome covariance at the vacuum, so $\Sigma_p^{-1/2}W\Sigma_p^{-1/2}$ has the law of $O\Theta O^T$ with $O:=(SS^T)^{-1/2}S$, which is orthogonal. Both statements follow from \eqref{eq:NG-ETheta} and \eqref{eq:NG-vacmom}, respectively \eqref{eq:NG-vacmom32}.
\end{proof}

The standard purification is $|\psi_{\rm std}\rangle=(\sqrt\rho\otimes I)\sum_x|x\rangle|x\rangle$, $x$ running over the Fock basis; for a thermal mode of occupation $n$ it is the two-mode squeezed vacuum with $\sinh^2r=n$, and for $\rho=U_S\tau U_S^\dagger$, $\tau$ thermal, the identity $(X\otimes I)\sum_x|x\rangle|x\rangle=(I\otimes X^T)\sum_x|x\rangle|x\rangle$ gives $|\psi_{\rm std}\rangle=(U_S\otimes\bar U_S)(\sqrt\tau\otimes I)\sum_x|x\rangle|x\rangle$, a pure Gaussian state \cite[Lemma~D.1]{CFGLMWW26}. Its copies cannot be prepared without knowing $\rho$. The identity is proved in \cite[App.~D.5]{CFGLMWW26} in three steps. First, $\rho^{\otimes k}$ commutes with the orthogonal mixing of the copy index used in \S\ref{subsec:NG-NG2}, and by the Kashiwara--Vergne (Howe) duality the $k$-copy space is a multiplicity-free sum $\bigoplus_\lambda V_\lambda\otimes W_\lambda$ of irreducible representations of the metaplectic group ($V_\lambda$) and of $O(k)$ ($W_\lambda$); so $\rho^{\otimes k}\cong\bigoplus_\lambda\rho_\lambda\otimes I_{W_\lambda}/\dim W_\lambda$, and $\psi_{\rm std}^{\otimes k}$ is a sum of purifications of $\rho_\lambda$ on $V_\lambda\otimes\tilde V_\lambda$ times maximally entangled vectors on $W_\lambda\otimes\tilde W_\lambda$. Second, $W_A$ is unchanged by Gaussian unitaries on the purifying modes, so its law depends on the input only through the twirl over $\mathrm{Sp}(2m,\R)$ on those modes, defined for $k\ge2m+1$ \cite[Lemma~D.8]{CFGLMWW26}. Third, the twirl keeps of the $V_\lambda\otimes\tilde V_\lambda$ part only its partial trace over $\tilde V_\lambda$, which is $\rho_\lambda$ whatever sits on $\tilde V_\lambda$, times a constant independent of $\rho$. Hence a fixed vector $|v_\lambda\rangle$ on $\tilde V_\lambda$ can stand in for the inaccessible purifying part, and $\mathcal Q^{\rm Bose}_{k,m}$ builds such a state from $\rho^{\otimes k}$ alone: it exploits the blindness of the reduced estimate to the purifying modes.

\begin{proof}[Proof of Proposition~\ref{prop:NG-mixed}]
By the reduced-law identity it suffices to take as input $k$ copies of $\psi_{\rm std}$. For (i), with $d=2m$, $\gamma=k/2-d-1=4m+2=2(d+1)$ and $c_k=\frac32$. Let $L:=\Sigma^{-1/2}J_A\Sigma_p^{1/2}$; then $LL^T=\Sigma^{-1/2}J_A\Sigma_pJ_A^T\Sigma^{-1/2}=I$ and
\[
\Sigma^{-1/2}(W_A-\tfrac32\Sigma)\Sigma^{-1/2}=L\big[\Sigma_p^{-1/2}(W-\tfrac32\Sigma_p)\Sigma_p^{-1/2}\big]L^T .
\]
Compression by $L$, with $\|L\|_\op=1$, does not increase the Frobenius norm, and Lemma~\ref{lem:NG-pure}, with $12d=24m$, gives the second moment in \eqref{eq:NG-mixedmom}; the mean follows from $\E W=\frac32\Sigma_p$ by linearity. For (ii), $\gamma=k/2-d-1=\frac32$; the same compression applied to Lemma~\ref{lem:NG-pure}(ii), with $40d^2=160m^2$, gives the bound, and the mean follows from $\E W=c_k\Sigma_p$.
\end{proof}

\subsection{Blocks linear in \texorpdfstring{$m$}{m}: proof of Theorem~\ref{thm:NG2}}\label{subsec:NG-NG2}

The first step below is the quantum form of the independence of the sample mean and the residuals of a Gaussian sample: the coherent amplitudes of $k+1$ identical copies add in one output port of a balanced multiport and cancel in the others, while the independent, identical fluctuations are unchanged. The \emph{pure core} $P=\frac12SS^T$ of the second step is the covariance of the state obtained by cooling every normal mode of $\hat\Sigma$ to its ground state.

\begin{proof}[Proof of Theorem~\ref{thm:NG2}]
\emph{Unknown displacement.} Each block consists of $k+1=4m+6$ original copies. As in the proof of Theorem~\ref{thm:NG1}(ii), with the beam splitter replaced by $H\otimes I_{2m}$, $H\in O(k+1)$ with first row $(1,\dots,1)/\sqrt{k+1}$ (orthogonal and symplectic, hence a passive Gaussian unitary): the input covariance $I_{k+1}\otimes\Sigma$ is invariant and the output means are $(\sqrt{k+1}\,v,0,\dots,0)$, so the output is exactly $\rho(\sqrt{k+1}\,v,\Sigma)\otimes\rho(0,\Sigma)^{\otimes k}$. Discard the first factor. This costs one extra copy per block. For centred states the step is omitted and blocks of $k=4m+5$ copies suffice.

\emph{Coarse unsqueezing.} Run the measurement of Proposition~\ref{prop:NG-mixed}(ii) independently on $L_0=\lceil64000m^2\rceil$ blocks, let $\bar W$ be the average of the outputs $W_A$, and put $\hat\Sigma:=2\bar W/c_k$. The block outputs are independent, because the blocks use disjoint copies and independent measurements, and by \eqref{eq:NG-mixedmom32}
\[
\E\big\|\Sigma^{-1/2}(\tfrac12\hat\Sigma-\Sigma)\Sigma^{-1/2}\big\|_F^2\le\frac{160m^2}{L_0}\le\frac1{400}.
\]
Markov's inequality at the squared threshold $\frac14$ gives, with probability at least $0.99$, $\|\Sigma^{-1/2}(\frac12\hat\Sigma-\Sigma)\Sigma^{-1/2}\|_\op\le\frac12$, hence
\begin{equation}\label{eq:NG-sandwich32}
\Sigma\preceq\hat\Sigma\preceq3\Sigma .
\end{equation}
Since $2/c_k<1$, $\hat\Sigma$ need not be a physical covariance in general; on \eqref{eq:NG-sandwich32} it is, because $\hat\Sigma+\frac i2\Omega\succeq\Sigma+\frac i2\Omega\succeq0$. As $\hat\Sigma\succ0$ always, it has a Williamson form $\hat\Sigma=S\operatorname{diag}(\nu,\nu)S^T$; on \eqref{eq:NG-sandwich32}, $\nu_j\ge\frac12$, and the pure core $P:=\frac12SS^T$ satisfies $P\preceq\hat\Sigma\preceq3\Sigma$. This is \eqref{eq:NG-target} with $C=3$, and $\Sigma'=S^{-1}\Sigma S^{-T}\succeq\frac16I$.

\emph{Heterodyne round.} Apply $U_S^\dagger$ to each of $N_2$ fresh original copies and measure it by heterodyne detection. Conditionally on $S$ these are i.i.d.\ copies of $\rho(S^{-1}v,\Sigma')$. Lemma~\ref{lem:NG-target} with $C=3$ and $\delta=0.05$, where $4.3\,(2m+6m)=34.4\,m$, gives $T(\rho,\hat\rho)\le\eps$ whenever \eqref{eq:NG-N2} holds; the actual mean is estimated together with the covariance. The total failure probability is at most $0.01+0.05=0.06$.

\emph{Count.} The first round uses $L_0(4m+6)=O(m^3)$ copies in one nonadaptive round, absorbed in the total for $0<\eps<1$; the second uses $N_2=O(m^3/\eps^2)$ copies in one adaptive round. Every constant is absolute: the argument uses exact second moments and Markov's inequality. The value $k=4m+5$ is the least for which the outcome covariance has a finite second moment ($\gamma>1$); the measurement itself exists for $k\ge4m+1$.
\end{proof}

\subsection{Whitened metric and the fixed-gap corollary}

\paragraph{The whitened metric.}
The norm $\|Z\|_F$, $Z:=\Sigma_0^{-1/2}(\Sigma_1-\Sigma_0)\Sigma_0^{-1/2}$, measures a change of covariance in the units of $\Sigma_0$. It is invariant under $\Sigma\mapsto S\Sigma S^T$ for every invertible $S$, since $(S\Sigma_0S^T)^{-1/2}S\Sigma_0^{1/2}$ is orthogonal, and $\frac12\|\Sigma^{-1/2}\dot\Sigma\Sigma^{-1/2}\|_F^2$ is the Fisher information of the classical family $N(v,\Sigma_t)$ (the relative entropy between $N(0,\Sigma_0)$ and $N(0,\Sigma_1)$ is $\frac14\|Z\|_F^2+O(\|Z\|_F^3)$). The quantum Fisher information \eqref{eq:NG-QFI} \cite{Saf19} comes from the symmetric logarithmic derivative of a Gaussian family, a quadratic polynomial in the quadratures. When all symplectic eigenvalues are large the term $\frac14\Omega\otimes\Omega$ is negligible and \eqref{eq:NG-QFI} reduces to the classical value $\frac12\Tr(\Sigma^{-1}\dot\Sigma\Sigma^{-1}\dot\Sigma)+\dot v^T\Sigma^{-1}\dot v$: far from the pure boundary the state is, to first order, as distinguishable as its Wigner function viewed as a classical distribution. Near pure modes $d_jd_l-\frac14\to0$ and the quantum term blows up; the vacuum and the thermal state of occupation $n\ll1$ have fidelity $1/(1+n)$, hence Bures distance about $\sqrt n$, but trace distance $n/(1+n)$: near the pure boundary the Bures length, and with it the method of Lemma~\ref{lem:NG-cont}, overestimates the trace distance.

\begin{remark}[blocks of $12m+7$ with a first round of $O(m^2)$ copies]\label{rem:NG-12m}
With $k=12m+6$ instead, Proposition~\ref{prop:NG-mixed}(i) and $L_0':=\lceil9600m\rceil$ blocks give $\E\|\Sigma^{-1/2}(\bar W-\frac32\Sigma)\Sigma^{-1/2}\|_F^2\le24m/L_0'\le\frac1{400}$, hence with probability at least $0.99$
\begin{equation}\label{eq:NG-sandwich}
\Sigma\preceq\bar W\preceq2\Sigma ;
\end{equation}
$\bar W$ is always physical, its pure core satisfies $P\preceq\bar W\preceq2\Sigma$, and $\Sigma'\succeq\frac14I$. The first round then uses only $\lceil9600m\rceil(12m+7)=O(m^2)$ copies (blocks of $12m+6$ for centred states). This variant is used in Corollary~\ref{cor:NG-gap}, whose total $O_g(m^2/\eps^2)$ requires a first round of $O(m^2)$ copies.
\end{remark}

\begin{proof}[Proof of Lemma~\ref{lem:NG-cont}]
\emph{Fisher information.} For a smooth path $t\mapsto\rho(v_t,\Sigma_t)$ of Gaussian states whose symplectic eigenvalues exceed $\frac12$, the quantum Fisher information of the symmetric logarithmic derivative is, in the convention of this paper \cite{Saf19},
\begin{equation}\label{eq:NG-QFI}
H_t=\tfrac12\operatorname{vec}(\dot\Sigma_t)^T\big(\Sigma_t\otimes\Sigma_t-\tfrac14\Omega\otimes\Omega\big)^{-1}\operatorname{vec}(\dot\Sigma_t)+\dot v_t^T\Sigma_t^{-1}\dot v_t .
\end{equation}

\emph{Whitening.} Let $\Sigma=SDS^T$ be a Williamson form, $D=\operatorname{diag}(d,d)$, $d_j\ge(1+c)/2$. Since $S\Omega S^T=\Omega$, $\Sigma\otimes\Sigma-\frac14\Omega\otimes\Omega=(S\otimes S)(D\otimes D-\frac14\Omega\otimes\Omega)(S\otimes S)^T$, so the first term of \eqref{eq:NG-QFI} equals $\frac12\operatorname{vec}(X')^T(D\otimes D-\frac14\Omega\otimes\Omega)^{-1}\operatorname{vec}(X')$ with $X':=S^{-1}\dot\Sigma S^{-T}$. For each ordered pair of modes $(j,l)$, the operator $D\otimes D-\frac14\Omega\otimes\Omega$ preserves the $2\times2$ block $X'_{jl}$ of $X'$ with rows $q_j,p_j$ and columns $q_l,p_l$, and acts on it as $d_jd_lI_4-\frac14\Omega_1\otimes\Omega_1$, $\Omega_1=\begin{pmatrix}0&1\\-1&0\end{pmatrix}$, with eigenvalues $d_jd_l\pm\frac14$. Put $Y:=D^{-1/2}X'D^{-1/2}$, so that $X'_{jl}=\sqrt{d_jd_l}\,Y_{jl}$. Since $x\mapsto x/(x-\frac14)$ decreases and $d_jd_l\ge(1+c)^2/4$,
\[
\tfrac12\operatorname{vec}(X')^T\big(D\otimes D-\tfrac14\Omega\otimes\Omega\big)^{-1}\operatorname{vec}(X')\le\frac12\sum_{j,l}\frac{d_jd_l}{d_jd_l-\frac14}\,\|Y_{jl}\|_F^2\le\frac{(1+c)^2}{2c(2+c)}\,\|Y\|_F^2,
\]
and $\|Y\|_F^2=\Tr(D^{-1}X'D^{-1}X')=\Tr(\Sigma^{-1}\dot\Sigma\Sigma^{-1}\dot\Sigma)$, because $\Sigma^{-1}=S^{-T}D^{-1}S^{-1}$. The bound is symplectically invariant and does not involve the energy.

\emph{Path.} For (i) let $X:=\Sigma_1-\Sigma_0$ and $\Sigma_t:=\Sigma_0+tX=\Sigma_0^{1/2}(I+tZ)\Sigma_0^{1/2}$, $Z:=\Sigma_0^{-1/2}X\Sigma_0^{-1/2}$, $t\in[0,1]$, at fixed mean. By convexity $\Sigma_t\in\mathcal K_c$, and $I+tZ\succeq(1-\eta)I$ gives $\Sigma_t^{-1}\preceq\Sigma_0^{-1}/(1-\eta)$. As $A\mapsto\Tr(AXAX)$ is monotone on positive matrices for symmetric $X$ (use $A\preceq B$ in each factor), $\Tr(\Sigma_t^{-1}X\Sigma_t^{-1}X)\le\eta^2/(1-\eta)^2$ and
\[
\sqrt{H_t}\le\frac{1+c}{\sqrt{2c(2+c)}}\,\frac\eta{1-\eta}.
\]

\emph{Distance.} The path bound \eqref{eq:bg-path}, $T(\rho,\rho')\le\frac12\int_0^1\sqrt{H_t}\,dt$, gives (i). For (ii) take $v_t=v+t(v'-v)$ at fixed $\Sigma_0$, for which $H_t=(v'-v)^T\Sigma_0^{-1}(v'-v)$.
\end{proof}

\begin{proof}[Proof of Corollary~\ref{cor:NG-gap}]
Run the first round of Remark~\ref{rem:NG-12m}. With probability at least $0.99$ it returns $S$ with $\Sigma':=S^{-1}\Sigma S^{-T}\succeq\frac14I$. The symplectic eigenvalues of $\Sigma'$ are those of $\Sigma$, so $\Sigma'\in\mathcal K_g$. Apply $U_S^\dagger$ to $3N$ fresh copies and measure each by heterodyne detection; the outcomes are i.i.d.\ $N(v',\Gamma)$ with $v':=S^{-1}v$ and $\Gamma:=\Sigma'+\frac12I$. From $2N$ of them form the $N$ pair differences $D_l:=(Y_{2l}-Y_{2l-1})/\sqrt2\sim N(0,\Gamma)$, and put $\hat\Gamma:=\frac1N\sum_lD_lD_l^T$ and $\hat\Sigma':=\hat\Gamma-\frac12I$; let $\hat v$ be the mean of the other $N$ outcomes.

\emph{Relative errors.} With $W_N:=\Gamma^{-1/2}(\hat\Gamma-\Gamma)\Gamma^{-1/2}$, $\E\|W_N\|_F^2=((2m)^2+2m)/N$, since $\E(g_ig_j-\delta_{ij})^2=1+\delta_{ij}$ for a standard Gaussian $g$ in $\R^{2m}$, and the relative covariance error $Z:=\Sigma'^{-1/2}(\hat\Sigma'-\Sigma')\Sigma'^{-1/2}=\Sigma'^{-1/2}\Gamma^{1/2}W_N\Gamma^{1/2}\Sigma'^{-1/2}$ satisfies $\|Z\|_F\le\|\Sigma'^{-1/2}\Gamma\Sigma'^{-1/2}\|_\op\|W_N\|_F=(1+1/(2\lambda_{\min}(\Sigma')))\|W_N\|_F\le3\|W_N\|_F$ (the amplification bound of \S\ref{subsec:NG-target}), independently of the thermal energy. The mean error satisfies $\E\|\Sigma'^{-1/2}(\hat v-v')\|_2^2=\Tr(\Sigma'^{-1}\Gamma)/N=(2m+\frac12\Tr\Sigma'^{-1})/N\le6m/N$.

\emph{Physicality.} Let $\eta:=\|Z\|_F$. If $\eta\le g/(2(1+g))$, then $\hat\Sigma'\succeq(1-\eta)\Sigma'$, and since $\Sigma'+\frac i2(1+g)\Omega\succeq0$, also $\hat\Sigma'+\frac i2(1-\eta)(1+g)\Omega\succeq0$ with $(1-\eta)(1+g)\ge1+\frac g2$. Thus $\hat\Sigma'\in\mathcal K_{g/2}$, and the estimate is physical on that event, without a projection in an unknown metric. The protocol checks the implementable condition $\hat\Sigma'+\frac i2\Omega\succeq0$ and outputs $\hat\rho:=\rho(S\hat v,S\hat\Sigma'S^T)$ if it holds, and a fixed physical state otherwise.

\emph{Error.} On the event $\eta\le g/(2(1+g))$, Lemma~\ref{lem:NG-cont} with $c=g/2$ (note $\Sigma'\in\mathcal K_g\subset\mathcal K_{g/2}$) and the triangle inequality give
\[
T(\rho,\hat\rho)=T\big(\rho(v',\Sigma'),\rho(\hat v,\hat\Sigma')\big)\le\tfrac12\|\Sigma'^{-1/2}(\hat v-v')\|_2+\kappa_{g/2}\,\frac{\eta}{1-\eta}.
\]
Let $\eta_0:=\min\{g/(2(1+g)),\eps/(4\kappa_{g/2})\}<\frac12$. If $\eta\le\eta_0$ and $\|\Sigma'^{-1/2}(\hat v-v')\|_2\le\eps$, then $T(\rho,\hat\rho)\le\eps$. By Markov's inequality these two events fail with probabilities at most $9(4m^2+2m)/(N\eta_0^2)$ and $6m/(N\eps^2)$. Since $\eps<1$, $\eta_0\ge c_g\eps$ with $c_g:=\min\{g/(2(1+g)),1/(4\kappa_{g/2})\}$, so $N=\lceil(1080/c_g^2+120)\,m^2/\eps^2\rceil$ makes each failure probability at most $0.05$. The total failure probability is below $0.01+0.05+0.05<1/3$, and the number of copies is $\lceil9600m\rceil(12m+7)+3N=O_g(m^2/\eps^2)$.
\end{proof}

\end{document}